\documentclass[aos]{imsart}

\usepackage{amsmath}
\usepackage{graphicx}
\usepackage{centernot}
\RequirePackage[authoryear]{natbib}
\usepackage{url} 
\usepackage[shortlabels]{enumitem}
\usepackage{latexsym, epsfig, amssymb, amsthm, mathrsfs, bbm, enumitem}
\usepackage[OT1]{fontenc}
\usepackage{xcolor}
\usepackage[colorlinks,citecolor=blue,linkcolor=blue,urlcolor=blue]{hyperref}
\usepackage{multirow,multicol,booktabs, array}

\makeatletter 

\usepackage[ruled]{algorithm2e}
\usepackage{tikz}
\usepackage{float}
\usepackage{booktabs}
\usepackage{adjustbox}

\DeclareMathOperator*{\argminA}{arg\,min}

\newcommand{\e}{\mathbbm{E}}

\newcommand{\indep}{\perp \!\!\! \perp}
\newcommand{\bigCI}{\mathrel{\text{\scalebox{1.07}{$\perp\mkern-10mu\perp$}}}}
\newcommand{\nbigCI}{\centernot{\bigCI}}

\DeclareMathOperator{\Var}{Var}
\DeclareMathOperator{\Cov}{Cov}
\DeclareMathOperator{\init}{init}

\newtheorem{lemma}{Lemma}
\newtheorem{proposition}{Proposition}
\newtheorem{theorem}{Theorem}
\newtheorem{remark}{Remark}
\newtheorem{setting}{Setting}
\newtheorem{condition}{Condition} 
\renewcommand{\theequation}{
	\arabic{equation}%
}

\newcommand{\ignore}[1]{}{}

\newcommand{\bSigma}{{\bf \Sigma}}
\newcommand{\bOmega}{{\bf \Omega}}
\newcommand{\bTheta}{{\bf \Theta}}

\newcommand{\bb}{\mbox{\bf b}}

\newcommand{\be}{\mbox{\bf e}}

\newcommand{\bx}{\mbox{\bf x}}
\newcommand{\by}{\mbox{\bf y}}
\newcommand{\bz}{\mbox{\bf z}}
\newcommand{\bA}{\mbox{\bf A}}
\newcommand{\bB}{\mbox{\bf B}}
\newcommand{\bC}{\mbox{\bf C}}
\newcommand{\bD}{\mbox{\bf D}}

\newcommand{\bM}{\mbox{\bf M}}

\newcommand{\bQ}{\mbox{\bf Q}}
\newcommand{\bR}{\mbox{\bf R}}

\newcommand{\bU}{\mbox{\bf U}}
\newcommand{\bV}{\mbox{\bf V}}
\newcommand{\bW}{\mbox{\bf W}}
\newcommand{\bX}{\mbox{\bf X}}

\newcommand{\bZ}{\mbox{\bf Z}}

\newcommand{\bveps}{\mbox{\boldmath $\varepsilon$}}

\newcommand{\balpha}{\mbox{\boldmath $\alpha$}}

\newcommand{\bbeta}{\boldsymbol{\beta}}
\newcommand{\brho}{\boldsymbol{\rho}}

\newcommand{\bgamma}{\mbox{\boldmath $\gamma$}}

\newcommand{\bzeta}{\mbox{\boldmath $\zeta$}}

\newcommand{\corr}{\mathrm{corr}}

\newcommand{\veps}{\varepsilon}

\newcommand{\diag}{\mathrm{diag}}
\newcommand{\vecc}{\mathrm{vec}}

\newcommand{\sgn}{\mathrm{sgn}}
\newcommand{\supp}{\mathrm{supp}}

\newcommand\independent{\protect\mathpalette{\protect\independenT}{\perp}}
\def\independenT#1#2{\mathrel{\setbox0\hbox{$#1#2$}%
		\copy0\kern-\wd0\mkern4mu\box0}}

\renewcommand{\ldots}{\cdots}
\renewcommand{\e}{\mathbb{E}}
\renewcommand{\hat}{\widehat}
\renewcommand{\top}{T}

\usepackage{xcolor}
\usepackage[draft,inline,nomargin,index]{fixme}
\fxsetup{theme=color,mode=multiuser}
\FXRegisterAuthor{yf}{ayf}{\color{magenta}}

\DeclareMathOperator{\augg}{aug}
\DeclareMathOperator{\swap}{swap}

\DeclareMathOperator{\FDP}{FDP}
\DeclareMathOperator{\FDR}{FDR} 
\DeclareMathOperator{\FWER}{FWER} 
 
\newcommand\norm[1]{\lVert#1\rVert}

\begin{document}

\begin{frontmatter}
\title{ARK: Robust Knockoffs Inference with Coupling}
\runtitle{ARK}

\begin{aug}
\author[A]{\fnms{Yingying}~\snm{Fan}\ead[label=e1]{fanyingy@marshall.usc.edu}},
\author[B]{\fnms{Lan}~\snm{Gao}\ead[label=e2]{lgao13@utk.edu}}
\and
\author[A]{\fnms{Jinchi}~\snm{Lv}\ead[label=e3]{jinchilv@marshall.usc.edu}}
\address[A]{Data Sciences and Operations Department, Marshall School of Business, 
University of Southern California\printead[presep={,\ }]{e1,e3}}

\address[B]{Department of Business Analytics and Statistics, Haslam College of Business,
The University of Tennessee \printead[presep={,\ }]{e2}}
\end{aug}

\begin{abstract}
We investigate the robustness of the model-X knockoffs framework with respect to the misspecified or estimated feature distribution. We achieve such a goal by theoretically studying the feature selection performance of a practically implemented knockoffs algorithm, which we name as the approximate knockoffs (ARK) procedure, under the measures of the false discovery rate (FDR) and $k$-familywise error rate ($k$-FWER). The approximate knockoffs procedure differs from the model-X knockoffs procedure only in that the former uses the misspecified or estimated feature distribution. A key technique in our theoretical analyses is to couple the approximate knockoffs procedure with the model-X knockoffs procedure so that random variables in these two procedures can be close in realizations. We prove that if such coupled model-X knockoffs procedure exists, the approximate knockoffs procedure can achieve the asymptotic FDR or $k$-FWER control at the target level. We showcase three specific constructions of such coupled model-X knockoff variables, verifying their existence and justifying the robustness of the model-X knockoffs framework. Additionally, we formally connect our concept of knockoff variable coupling to a type of Wasserstein distance. 

\end{abstract}



\end{frontmatter}

\section{Introduction} \label{Sec.1}


The knockoffs inference framework \citep{barber2015controlling, CFJL2018, BarberCandes2019} is a powerful innovative tool for feature selection with controlled error rates. In  particular, the model-X knockoffs \citep{CFJL2018} achieves the false discovery rate (FDR) control at a predetermined level in finite samples without requiring any specific model assumptions on how the response depends on the features, making it an attractive option for feature selection in a wide range of statistical applications. 
The fundamental idea of the knockoffs procedure is to construct knockoff variables that are exchangeable in distribution with the original features but are independent of the response conditional on the original variables. These knockoff variables serve as a control group for the original features, allowing researchers to identify relevant original features for the response. The model-X knockoffs inference has gained increasing popularity since its inception and there have been flourishing developments and extensions of the knockoffs framework and spirits, such as the $k$-familywise error rate ($k$-FWER) control with knockoffs \citep{JS2016}, power analysis for knockoffs procedure \citep{FDLL2020, spector2022powerful, wang2020power, weinstein2020power, FLSU2020}, derandomized knockoffs  \citep{ren2021derandomizing,ren2022derandomized}, knockoffs inference for time series data \citep{chi2021high},
kernel knockoffs procedure \citep{dai2022kernel},
and FDR control by data splitting or creating mirror variables \citep{li2021ggm, dai2022false, cao2021controlling, 
	guo2022threshold}.

A key assumption in the model-X knockoffs inference is that the joint distribution of features is known. However, such information is almost never available in practice. There has been overwhleming empirical evidence that the model-X knockoffs framework is robust to misspecified or estimated feature distributions \citep{CFJL2018, sesia2019,jordon2018knockoffgan,lu2018deeppink,zhu2021deeplink,romano2020deep}.  Yet, the theoretical characterization of its robustness is still largely missing. A notable exception is the recent work of \cite{Barber2020}, where it was formally and elegantly shown that the knockoffs data matrix collecting the knockoff variables can be generated from a distribution, which we name as the \textit{working} distribution for the ease of presentation, that is different from the true underlying feature distribution, and that the resulting FDR inflation can be measured by the empirical Kullback–Leibler (KL) divergence between the true conditional distribution $X_j  | X_{-j}$ and the working conditional distribution. Here, $X_j\in \mathbb R$ stands for the $j$th feature, $X_{-j}\in \mathbb R^{p-1}$ stands for the feature vector with the $j$th feature removed, and $p$ is the feature dimensionality. Two important assumptions in their analyses for ensuring the asymptotic FDR control are 1) the working distribution should be learned independently from the training data used for feature selection and 2) the empirical KL divergence between the two knockoffs data matrices (of diverging dimensionalities) generated from the working and true distributions, respectively, needs to vanish as the sample size increases. Although their results are general and apply to arbitrary dependence structure of the response on features, these two assumptions do not always describe the practical implementation. Our results in the current paper are free of the two assumptions discussed above. 

To put more content into our statements above, especially the one about assumption 2), let us consider the scenario where the true feature matrix has independent and identically distributed (i.i.d.) entries from the $t$-distribution with $\nu$ degrees of freedom, but we misspecify it and use the Gaussian distribution as a working distribution to generate the knockoff variable matrix $\hat\bX\in \mathbb R^{n\times p}$,  where $n$ is the sample size. 
It can be calculated that the empirical KL divergence between $\hat\bX$ and the model-X knockoff variable matrix $\widetilde\bX \in\mathbb R^{n\times p}$ defined in \cite{Barber2020} has mean and variance both at order $  \frac{np}{\nu (\nu + p)}$. 
Thus, only when $\nu^2 \gg n \min (n, p)$ (which is equivalent to $\frac{np}{\nu (\nu + p)} \to 0$), the FDR inflation as derived therein can vanish asymptotically. In contrast, our theory shows that as long as  $\nu^{2} \gg  s^4 (\log p)^{4 + 4/\gamma} $ for some $\gamma \in (0, 1)$ with $s\ll n^{1/2}$ a sparsity parameter, the knockoffs procedure based on the working distribution can achieve the asymptotic FDR control. More details for our results and model assumptions are summarized formally in Section \ref{new-Sec4.1}.
We provide additional comparisons of our results with those of \cite{Barber2020} in various parts of the paper where more specifics can be discussed. We emphasize and acknowledge that \cite{Barber2020} established general robustness results without specific model assumptions, while some of our results rely on certain specific model assumptions. The main point we advocate here is that a different notion of closeness than the KL divergence can be advantageous in studying the robustness of the model-X knockoffs. We also formally connect our concept of closeness to a  type of Wasserstein distance. We provide detailed comparison with some other existing work in the literature in Section \ref{sec:comp}.


The major goal of our paper is to establish a general theory on the robustness of the model-X knockoffs framework for the FDR and $k$-FWER control. We approach the problem by studying the performance of the \textit{approximate knockoffs (ARK)} procedure, an algorithm that is most popularly implemented in practice when applying the knockoffs framework. The ARK procedure differs from the model-X knockoffs in that the former generates the knockoff feature matrix from a working distribution that can be misspecified or learned from the \textit{same} training data for feature selection. By showing that the ARK procedure achieves the asymptotic FDR and $k$-FWER control as sample size increases, we can verify the robustness of the model-X knockoffs. 
An important idea in our technical analyses is \textit{coupling}, where we pair the ARK procedure with the model-X knockoffs procedure in such a way that random variables in these two paired procedures are close in realizations with high probability. Hereafter, we will refer to the model-X knockoffs as the \textit{perfect knockoffs} procedure to emphasize its difference from the approximate knockoffs procedure. It is important to emphasize that we require the realizations of random variables in the paired procedures to be close, instead of the corresponding distributions being close. This is a major distinction from the assumption in \cite{Barber2020}. Our new notion of closeness allows us to justify the robustness of the model-X knockoffs in some broader contexts not covered by studies in the existing literature. We also emphasize that although our conditions are imposed on the perfect knockoff variables, we do \textit{not} need to know or construct them in implementation; the existence of such variables is sufficient for our theoretical robustness analyses. 

We present our theory by first laying out general conditions on the existence of the coupled perfect knockoff statistics and their closeness to the approximate knockoff statistics in Section \ref{Sec.2}, and then provide examples justifying these conditions in Sections \ref{Sec.4} and \ref{Sec.3}.    More specifically, our theory has three layers, related to different stages in applying the knockoffs inference procedure. Our preliminary theory in Section \ref{Sec.2} directly makes assumptions on the quality of the approximate knockoff statistics (cf. \eqref{def.What}) by requiring the existence and closeness of their coupled perfect knockoff statistics. Then under some regularity conditions imposed on the distribution of these perfect knockoff statistics, we prove that the FDR and $k$-FWER are controlled asymptotically using the approximate knockoff statistics. This lays the theoretical foundation for our subsequent analyses in Sections \ref{Sec.4} and \ref{Sec.3}.   

The second layer of our theory, presented in Section \ref{Sec.4}, delves deeper and replaces the coupling condition imposed on the knockoff statistics in Section \ref{Sec.2} with a coupling condition on the approximate knockoff variables generated from some mispecified or estimated feature distribution. Similar in nature to the coupling condition in Section \ref{Sec.2}, this new condition assumes that there exist perfect knockoff variables that can be coupled with approximate knockoff variables so that their realizations are close  to each other with high probability. Since knockoff statistics are known functions of knockoff variables, such alternative condition intuitively and naturally leads to the verification of the coupling condition on knockoff statistics in Section \ref{Sec.2}. Indeed, we showcase using two commonly analyzed knockoff statistics, namely the marginal correlation statistics and the regression coefficient difference (RCD) statistics, that the coupling condition on knockoff variables can guarantee the coupling condition on knockoff statistics. We also verify that for each of these two constructions of knockoff statistics, the other regularity conditions in our preliminary theory in Section \ref{Sec.2} also hold, ensuring the asymptotic FDR and $k$-FWER control. Notably, our theory also reveals that, the marginal correlation is of ``low accuracy," and needs more stringent conditions than RCD to achieve asymptotic FDR control. This message is consistent with \cite{Reconciling2024} when studying the conditional randomization test using the model-X 
framework. 

The last layer of our theory is presented in Section \ref{Sec.3} and showcases three specific constructions of the coupled perfect knockoff variables. 
By imposing conditions on the misspecified or estimated feature distribution, we construct explicitly the coupled perfect knockoff variables and prove that the coupling conditions in the first and second layers of our general theory are satisfied. This gives us a complete theory with conditions imposed on the working distribution for generating knockoff variables and verifies the robustness of the model-X knockoffs inference procedure. Our theory allows high dimensionality of features and allows \textit{in-sample} estimation of the feature distribution. 
The rest of the paper is organized as follows. Section \ref{Sec.2} first introduces the approximate knockoffs procedure and then presents the general conditions and theory for the asymptotic FDR control. We also introduce the coupling idea, a key technique in our theoretical analyses. We illustrate our general theory using two commonly used constructions of knockoff statistics in Section \ref{Sec.4}. Section \ref{Sec.3} further provides three specific constructions of the coupled perfect knockoff variables. 
{We present companion theory for robust $k$-FWER control in Section \ref{sec:k-FWER}.}
We provide detailed discussions on some most related works in Section \ref{sec:comp}, and present some simulation examples in Section \ref{Sec:Simu}.   We conclude our paper by summarizing the key results and discussing some future research directions in Section \ref{Sec.disc}. All the proofs and technical details are provided in the Supplementary Material.

To facilitate the technical presentation, let us introduce some notation that will be used throughout the paper. We use $a_n \ll b_n$ or $a_n = o(b_n)$ to represent $a_n / b_n \to 0$, $a_n \gg b_n $ to represent $a_n / b_n \to \infty$, and $a_n \lesssim b_n $ or $a_n = O(b_n)$ to represent $a_n \leq C b_n$ for an absolute constant $C > 0$. Let $a \land b$ and $a \lor b$ be the minimal and maximal values of $a$ and $b$, respectively. 
For a vector $\bx \in \mathbb{R}^{p}$, denote by $\| \bx \|_1$,  $\| \bx \|_2$, and $\|\bx\|_0$ the $\ell_1$-norm, $\ell_2$-norm, and $\ell_0$-norm, respectively. For $1 \leq j \leq p$, $\bx_j$ is the $j$th component of $\bx$ and $\bx_{-j} $ is a subvector of $\bx$ with the $j$th component removed. For a matrix $\bM \in \mathbb{R}^{n \times p}$, denote by $\bM_{i, j}$ the $(i, j)$th entry of $\bM$, $\bM_{j}$ the $j$th column of $\bM$, and $\bM_{A_1, A_2}$ a submatrix of $\bM$ consisting of $(\bM_{i, j})_{i \in A_1, j \in A_2}$ for sets $A_1 \subset \{1, \ldots, n \}$ and $A_2 \subset \{1, \ldots, p \}$. Let  $\|\bM\|_{\max}$ and $\|\bM\|_2$ be the maximum norm and spectral norm of a matrix $\bM$, respectively.
For $1 \leq j \leq p$, $-j$ represents the set $\{1, \ldots, p \} \setminus \{j\}$, and denote by  $|\mathcal{A}|$ the cardinality of set $\mathcal{A}$. For a positive definite matrix $\bSigma$, let $\lambda_{\min}({\bSigma})$ and $\lambda_{\max}({\bSigma})$ be the smallest and largest eigenvalues of $\bSigma$, respectively.

 \section{Preliminary results on robust knockoffs inference via coupling}  \label{Sec.2}
 
 \subsection{Model setup and model-X knockoffs framework} \label{new.Sec2.1}
 
 Assume that we have $n$ 
 i.i.d. observations $\{ (\bx_i, y_i)\}_{i = 1}^n $ from the population $(X, Y)$, where $X = (X_1, \ldots, X_p)^{\top} $ is the $p$-dimensional feature vector and $ Y \in \mathbb{R}$ is a scalar response. Here, the feature dimensionality $p$ can diverge with the sample size $n$. 
 Adopting the matrix notation, the $n$ i.i.d. observations can be written as the data matrix $\bX =  (\bX_{i, j} )  \in \mathbb{R}^{n \times p}$ collecting the values of all the features and vector $\by = (y_1,\cdots, y_n)^\top \in \mathbb{R}^{n}$ collecting the values of the response. A feature $X_j$ is defined as null (or irrelevant) if and only if it is independent of the response conditional on all the remaining features; that is,  $ Y \indep X_j | X_{-j} $, where $X_{-j}$ is a subvector of $X$ with the $j$th component removed. Denote by $\mathcal{H}_0 = \{1 \leq j \leq p: X_j ~ \mbox{is a null feature} \}$ the set of null features and $ \mathcal{H}_1 = \mathcal{H}_0^c$ that of nonnull (or relevant) features. To ensure the model identifiability and interpretability, we follow  \cite{CFJL2018} and assume that $\mathcal{H}_1$ exists and is unique.  Further assume that the subset of relevant features is sparse such that $p_1 = | \mathcal{H}_1 | = o(n \land p)$, where $|\mathcal{A}|$ stands for the cardinality of a given set. The goal is to select as many relevant features as possible while controlling some error rate measure at the prespecified target level. 
 
 A commonly used measure for evaluating the feature selection performance is FDR \citep{benjamini1995controlling}, where for an outcome $\hat{S}$ of some feature selection procedure, the FDR  is defined as  
 \begin{equation} \label{eq-FDR}
 	\FDR = \e [\FDP]  \  \text{ with } \FDP =  | \hat{S} \cap \mathcal{H}_0 |  /  |  \hat{S} | .
 \end{equation}
 
 The model-X knockoffs framework provides a flexible way for controlling the FDR  at some prespecified target level in finite samples \citep{CFJL2018}, allowing  arbitrary dimensionality of $X$ and arbitrary dependence between response $Y$ and feature vector $X$.   
 A key step of the model-X knockoffs inference \citep{CFJL2018} is to generate the model-X knockoff variables $\widetilde{X} = (\widetilde{X}_1, \ldots, \widetilde{X}_p)^{\top}$ such that $ \widetilde{X} \indep Y | X$ and 
 \begin{equation} \label{eq-1}
 	(X, \widetilde{X})_{\swap(S)} \stackrel{d}{=} (X, \widetilde{X}) \ \mbox{ for each subset } S \subset \{1, \ldots, p\},
 \end{equation}
 where $(X, \widetilde{X})_{\swap(S)}$ is obtained by swapping the components $X_j$ and $\widetilde{X}_j$ in $(X, \widetilde{X})$ for each $j \in S$. 
 
 The construction of the model-X knockoff variables, which we will refer to as the \textit{perfect} knockoff variables in future presentation, requires the exact knowledge of the distribution of feature vector $X$. For example, Algorithm 1 in \cite{CFJL2018} provided a general approach to generating the perfect knockoff variables when such information is available. However, the exact knowledge of feature distribution is usually unavailable in real applications.  Thus, in practical implementation, the problem becomes identifying the relevant subset $\mathcal{H}_1$ with the approximate knockoff variables generated from a feature distribution that can be different from the true underlying one; we name the practical procedure as the \textit{approximate} knockoffs and formally present it in the next section for completeness. As stated in the Introduction, we study the robustness of the model-X knockoffs procedure by investigating the feature selection performance of the approximate knockoffs procedure.

 \subsection{Approximate knockoffs and a roadmap of our analysis} \label{new.Sec2.2}
 
 In practice, the approximate knockoffs inference procedure below is implemented popularly for controlling the FDR.
 
 \begin{enumerate}
 	\item[1)] {\it Generating approximate knockoff variables.} Since the true underlying feature distribution $F(\cdot)$ is generally unavailable, we generate the knockoff variables from some user-specified feature distribution $\hat F(\cdot)$, which can depend on the sample $(\bX, \by)$, using the same algorithm proposed for generating the perfect knockoff variables (e.g., Algorithm 1 in \cite{CFJL2018}). Denote by $\hat \bX = (\hat{\bX}_{i, j}) \in \mathbb R^{n\times p}$ the resulting approximate knockoff variable matrix. 
 	
 	\item[2)] {\it Constructing approximate knockoff statistics.} Pretend that $\hat\bX$ were perfect knockoff variable matrix and follow the same procedure as in \cite{CFJL2018} to calculate the knockoff statistics $\hat W_j$ with $j=1,\cdots, p$. Specifically, we first compute the feature importance statistics   
 	$$ (Z_1, \ldots, Z_p, \hat{Z}_1, \ldots, \hat{Z}_p)^{\top} = t((\bX, \hat{\bX}), \by),$$
 	where $t(\cdot)$ is a measurable function of input $((\bX, \hat{\bX}), \by)$, and $Z_j$ and $\hat{Z}_j$ measure the importance of the $j$th feature and its approximate knockoff counterpart relative to the response, respectively. Then the approximate knockoff statistic $\hat{W}_j$ for the $j$th feature is defined as 
 	\begin{equation}\label{def.What}
 		\hat{W}_j = f_j(Z_j, \hat{Z}_j),
 	\end{equation}
 	where $f_j (\cdot, \cdot)$ is an antisymmetric function satisfying $f_j(x, y) = - f_j (y, x)$. See \cite{barber2015controlling} for examples and characterizations on the valid construction of knockoff statistics.  
 	
 	\item[3)] {\it Selecting relevant features}. Calculate a data-driven threshold $ {T}$ for the knockoff statistics $\{\hat W_j\}_{j=1}^p$ and select the set of important features as $\hat{S} = \{1\leq j \leq p: \hat{W}_j \geq {T}\}$. Denoting $\hat{\mathcal{W}} = \{|\hat{W}_1|, \ldots, |\hat{W}_p|\}$, the threshold for FDR control is defined as
 	\begin{equation} \label{eq-2}
 		{T}  = \min \Big\{t \in \hat{\mathcal{W}}: \frac{\#\{j: \hat{W}_j \leq -t\} } { \#\{\hat{W}_j \geq t\} \lor 1 } \leq q\Big\}
 	\end{equation}
 	where $q\in (0,1)$ is the prespecified level for the FDR.   
 \end{enumerate}
 
 It is seen that the only difference of the algorithm above from the perfect knockoffs procedure \citep{CFJL2018} is how the knockoff variable matrix $\hat \bX$ is generated. The perfect knockoffs procedure based on the true feature distribution $F(\cdot)$ has been shown to control the FDR at the target level \citep{CFJL2018}. For the approximate knockoffs inference, however, it is reasonable to expect some inflation in the FDR control, and the inflation level depends on the qualities of both the approximate knockoff variable matrix $\hat \bX$ and the resulting knockoff statistics $\{\hat{W}_j\}_{j=1}^p$. 
 A desired property is that as the approximate knockoff statistics ``approach" the perfect knockoff statistics, the level of inflation also vanishes. One contribution of our paper is to formally introduce a notion of \textit{closeness} measuring the qualities of the approximate knockoff statistics $\{\hat W_j\}_{j=1}^p$ and knockoff variable matrix $\hat\bX$. As will be discussed in Section \ref{sec:Wasserstein}, our closeness measure is closely related to a type of Wasserstein distance. 
 
 We provide a roadmap of our technical analyses. Our theory has three layers, corresponding reversely to the  steps in the approximate knockoffs procedure described above. To put it into more content, note that the set of selected features $\hat S$ is defined directly as a function of the approximate knockoff statistics $\{\hat W_j\}_{j=1}^p$. Hence, given $\{\hat W_j\}_{j=1}^p$, feature selection can be conducted without the knowledge of $\hat \bX$ or the feature distribution $F(\cdot)$. For this reason, our layer 1 analysis concerns the quality of $\{\hat W_j\}_{j=1}^p$ for achieving  the asymptotic FDR control; see Section \ref{new.Sec2.3} for a characterization on qualified  knockoff statistics. The second layer of our analysis studies the quality of $\hat \bX$ and is built on the first layer. We characterize what kind of  $\hat \bX$ can lead to qualified knockoff statistics $\{\hat W_j\}_{j=1}^p$ satisfying the conditions established in our layer 1 analysis; see Section \ref{Sec.4} for such analysis in our layer 2. Our layer 3 analysis is built on the first two layers and goes all the way to the root of the knockoffs inference; we provide specific examples and conditions on $\hat F(\cdot)$ for ensuring that $\widehat \bX$ satisfies conditions in our layer 2 analysis. The key idea empowering our theoretical investigation is variable coupling behind the approximate knockoffs (ARK) procedure; we formally introduce such idea in the next subsection for laying out preliminary results for our subsequent in-depth analysis.  
 
 

 \subsection{Layer 1 analysis: knockoff statistics coupling} \label{new.Sec2.3}
 
 An important observation is that the perfect knockoff variables in the model-X knockoffs framework are not unique. Consequently, the knockoff statistics are not unique either. Indeed, even with the same algorithm (e.g., Algorithm 1 in \cite{CFJL2018}), the knockoff variables generated from different runs of the algorithm are only identically distributed. Our coupling idea is deeply rooted on such observation. Let us introduce some additional notation to facilitate our formal presentation of the general theory. Following the model-X knockoffs framework, for a realization of the perfect knockoff variable matrix $\widetilde{\bX}$ generated from the true feature distribution $F(\cdot)$, 
 we let  
 $$
 (Z_1^*,\ldots, Z_p^*, \widetilde{Z}_1, \ldots, \widetilde{Z}_p)^{\top} = t((\bX, \widetilde{\bX}), \by)
 $$
 and define the perfect knockoff statistics $\widetilde{W}_j = f_j (Z_j^*, \widetilde{Z}_j)$ for $1 \leq j\leq p$, where functions $t(\cdot)$ and $ f_j(\cdot) $ are identical to the ones in the approximate knockoffs procedure in Section \ref{new.Sec2.2}.


 We now establish preliminary theory on the asymptotic FDR control for the approximate knockoffs inference procedure, with regularity conditions imposed on the $\hat W_j$ values. 
 
 \begin{condition}[Coupling accuracy]  \label{fdr-condition1}
 	There exist perfect knockoff statistics $\{\widetilde{W}_j\}_{j = 1}^p$ such that for some sequence $b_n \to 0 $, 
 	\begin{equation}
 		\mathbb{P} \big( \max_{1 \leq j \leq p } | \hat{W}_j - \widetilde{W}_j  | \geq b_n \big) \to 0.
 	\end{equation}
 \end{condition}
 
 Conditions on the convergence rate $b_n$ for ensuring the asymptotic FDR control will be specified in the subsequent assumptions. Condition \ref{fdr-condition1} above couples each realization of the approximate knockoff statistics $\{\hat W_j\}_{j=1}^p$ with a realization of the perfect knockoff statistics $\{\widetilde W_j\}_{j=1}^p$, and they need to be sufficiently close to each other with high probability. Note that the existence of such $\{\widetilde W_j\}_{j=1}^p$ is required only for the theory, whereas the implementation uses only $\{\hat W_j\}_{j=1}^p$. We will provide examples in later sections verifying the existence of such coupled $\{\widetilde W_j\}_{j=1}^p$.  
 The two conditions below are on the quality of the perfect knockoff statistics $\{\widetilde{W}_j\}_{j=1}^p$ and the signal strength in the data as measured by $\widetilde{W}_j$'s. 
 \begin{condition}[Average concentration of $\widetilde{W}_j$]  \label{fdr-condition2}
 	There exist deterministic quantities $ \{ w_j \}_{j=1}^p
 	$ such that $ p^{-1}\sum_{j = 1}^{p} \mathbb{P} ( | \widetilde{W}_j - w_j | \geq \delta_n )  = o(p^{-1}) $, where $ \delta_n \to 0 $ is  a sequence satisfying $  \delta_n \geq b_n$.  
 \end{condition}

 \begin{condition}[Signal strength] \label{fdr-condition3}
 	Let $ \mathscr{A}_n  = \{j \in \mathcal{H}_1: w_j \geq 5 \delta_n \}$. It holds that  $ a_{n} = | \mathscr{A}_n |  \to \infty$ and
 	$  w_j > - \delta_n $ for $j \in \mathscr{A}_n^c$.
 \end{condition} 
 
 As discussed in  \cite{barber2015controlling} and \cite{CFJL2018}, a desired property of the knockoff statistics is to have a large and positive value for $\widetilde W_j$ if $j\in \mathcal H_1$, and a small and symmetric around zero value for $\widetilde W_j$ if $j\in \mathcal H_0$.    Conditions \ref{fdr-condition2} and \ref{fdr-condition3} together formalize this property. Condition \ref{fdr-condition2}  requires that each perfect knockoff statistic $\widetilde W_j$ is concentrated around some population parameter $w_j$ with rate $\delta_n$ in an average probability sense. By design, $\widetilde W_j$'s and $w_j$'s are  feature importance measures, and Conditions \ref{fdr-condition2} and \ref{fdr-condition3} characterize the desired properties they need to possess. Note that there is no requirement that each individual $w_j$ with $j\in \mathcal H_1$ is positive and large; we only need that there exist enough number (i.e., $a_n$) of $w_j$'s with $j\in\mathcal H_1$ that are positive and large enough. Implicitly, $a_n\rightarrow\infty$ requires that the number of relevant features $|\mathcal H_1|$ diverges with sample size as well.  The condition $\delta_n \geq b_n$ requires that the coupling accuracy $b_n$ should not exceed the order of concentration error so that $\widehat{W}_j$'s are as good as $\widetilde W_j$'s for estimating the population quantities $w_j$'s.

Define $p_0 = |\mathcal{H}_0|$ and  $G(t) = p_0^{-1} \sum_{j\in \mathcal{H}_0} \mathbb{P} (\widetilde{W}_j \geq t) $. By \cite{CFJL2018}, the perfect knockoff statistics $\widetilde{W}_j$ with $j\in \mathcal H_0$ are symmetrically distributed around zero. It follows that $G(t) = p_0^{-1} \sum_{j\in \mathcal{H}_0} \mathbb{P} (\widetilde{W}_j \leq -t)$.  
 We need to impose the technical conditions below on the distribution of the perfect knockoff statistics for our robustness analysis.
 
 \begin{condition}[Weak dependence among nulls] \label{fdr-condition4}
 	For some constants  $ 0 < \gamma < 1$, $ 0< c_1 < 1 $, $C_1 > 0$, and a positive sequence $m_n = o(a_n)$,  it holds that 
 	\begin{equation} \label{condition4-part1}
 		\Var{ \Big( \sum_{j \in \mathcal{H}_0} \mathbbm{1} (\widetilde{W}_j > t)  } \Big) \leq C_1 m_n p_0 G(t) +    o \big( \big( \log p )^{ - 1/\gamma}  [p_0  G (t)]^2 \big)  
 	\end{equation}
 	uniformly over $t \in (0,\, G^{-1} ( \frac { c_1 q a_n  } { p } ) ]$.
 \end{condition}	
 
 \begin{condition}[Distribution of $\widetilde{W}_j$] \label{fdr-condition5}
 	Assume that $G(t)$ is a continuous function.
 	For the same constants $ \gamma$ and $c_1$ as in Condition \ref{fdr-condition4},  it holds that as $n \to \infty$,
 	\begin{equation}  
 		(\log p)^{1/\gamma}  \sup_{t \in (0,\, G^{-1} ( \frac { c_1 q a_n  } { p } ) ] }  \frac { G(t - b_n ) - G(t + b_n) } {  G(t) } \to 0
 		\label{dist-cond1}
 	\end{equation}
 	and
 	\begin{equation} \label{dist-cond2}
 		a_n^{-1} \sum_{j \in \mathcal{H}_1} \mathbb{P} \Big( \widetilde{W}_j < -  G^{-1} ( \frac { c_1 q a_n  } { p } ) + b_n \Big)  \to 0.
 	\end{equation}
 \end{condition}
 
 Condition \ref{fdr-condition4} ensures that the random variable $\sum_{j \in \mathcal{H}_0} \mathbbm{1} (\widetilde{W}_j \geq t)$ has a standard deviation negligible compared to its mean, and thus can concentrate around its mean  $\sum_{j\in \mathcal{H}_0} \mathbb{P} (\widetilde{W}_j \geq t)$. Condition \ref{fdr-condition1} together with \eqref{dist-cond1} in Condition \ref{fdr-condition5} can guarantee that $\sum_{j \in \mathcal{H}_0} \mathbbm{1} (\hat{W}_j \geq t) \approx \sum_{j \in \mathcal{H}_0} \mathbbm{1} (\widetilde{W}_j \geq t)$ in probability, via an application of Markov's inequality.
 Combining these two results we can prove that $\sum_{j \in \mathcal{H}_0} \mathbbm{1} (\hat{W}_j \geq t) \approx \sum_{j \in \mathcal{H}_0} \mathbbm{1} (\widetilde{W}_j \geq t) \approx \sum_{j\in \mathcal{H}_0} \mathbb{P} (\widetilde{W}_j \geq t)$ and similarly $\sum_{j \in \mathcal{H}_0} \mathbbm{1} (\hat{W}_j \leq - t) \approx \sum_{j\in \mathcal{H}_0} \mathbb{P} (\widetilde{W}_j \leq -t)$ uniformly over $0 < t \leq G^{-1} (\frac{c_1 q a_n} {p}) $ with asymptotic probability one. In view of the definition of $T$ in \eqref{eq-2}, assumption \eqref{dist-cond2} ensures that the numerator in the ratio in \eqref{eq-2} is mainly contributed by null features, which together with Conditions \ref{fdr-condition1}--\ref{fdr-condition4} proves that threshold $T$ falls into the range $(0, G^{-1} (\frac{c_1 q a_n} {p})] $ with asymptotic probability one; See Lemma \ref{fdr-lemma1}. Thus, $ \sum_{j \in \mathcal{H}_0} \mathbbm{1} (\hat{W}_j \geq T) \approx \sum_{j \in \mathcal{H}_0} \mathbbm{1} (\hat{W}_j \leq - T)$ with asymptotic probability one by the symmetry of $\{\widetilde{W}_j\}_{j \in \mathcal{H}_0}$. Consequently, the FDR of the approximate knockoffs procedure is asymptotically the same as that of the perfect knockoffs procedure, where the latter has been proved to be controlled at the target level. This ensures that the FDR of the approximate knockoffs procedure can be controlled asymptotically, as formally stated in Theorem \ref{theorem-FDR} below.
 
 Condition \ref{fdr-condition4} above can be easily satisfied if $\widetilde W_j$'s with $j\in \mathcal H_0$ are independent of each other. At the presence of dependence, it imposes an assumption on the strength of correlation among the indicator functions $\mathbbm 1(\widetilde W_j>t)$ with $j\in \mathcal H_0$. The ratio  $\frac { G(t - b_n ) - G(t + b_n) } {  G(t) }$ in  Condition \ref{fdr-condition5} above is closely related to the hazard rate function in survival analysis if $G(t)$ has a probability density function. Loosely speaking, assumption \eqref{dist-cond1} is satisfied for $b_n = o((\log p)^{1/\gamma})$ if the hazard rate function has enough smoothness and is more or less bounded uniformly over the range $t \in (0,\, G^{-1} ( \frac { c_1 q a_n  } { p } ) ]$; it imposes an important condition on coupling accuracy $b_n$. Assumption (\ref{dist-cond2}) is satisfied if 1) only a fast vanishing fraction of $\widetilde W_j$'s for important features take negative values with nonvanishing probabilities, or 2) $\widetilde W_j$'s for important features all take positive values with high probability. 
 
 We are now ready to present our first general theorem on the FDR control for the approximate knockoffs inference procedure. 
 \begin{theorem} \label{theorem-FDR}
 	Under Conditions \ref{fdr-condition1}--\ref{fdr-condition5}, we have  
 	\begin{equation} \label{fdr-result}
 		\limsup_{n \to \infty} \FDR \leq q.
 	\end{equation}
 \end{theorem}
 

 \section{Layer 2 analysis: knockoff variables coupling} \label{Sec.4}

 \subsection{Characterization of approximate knockoff variables} \label{new.Sec3.1}
 
 Section \ref{Sec.2} establishes preliminary theoretical results on the asymptotic FDR control for the approximate knockoffs inference. The key assumption is Condition \ref{fdr-condition1}. Since the knockoff statistics are intermediate results calculated from the knockoff variables, it is important to provide a characterization on the quality of the approximate knockoff variable matrix $\hat\bX$ that can guarantee Condition \ref{fdr-condition1}. The assumption below is imposed for such a purpose. 
 
 \begin{condition} \label{accuracy-knockoffs}
 	For $\hat \bX$ constructed from the approximate knockoffs procedure,  there exists a  perfect knockoff data matrix $\widetilde{\bX}$ and an asymptotically vanishing sequence $\Delta_{n}$ such that 
 	\begin{equation}
 		\mathbb{P} \Big(    \|\widehat{\bX} - \widetilde{\bX} \|_{1, 2}  \geq \Delta_{n} \Big) \to 0,
 	\end{equation}
 	where $\|\widehat{\bX} - \widetilde{\bX} \|_{1, 2} :=  \max_{1 \leq j \leq p} n^{-1/2} \| \hat{\bX}_j - \widetilde{\bX}_j \|_2$, and $\hat{\bX}_j$ and $\widetilde{\bX}_j$ are the $j$th columns of the approximate and perfect knockoff variable matrices $\hat{\bX}$ and $\widetilde{\bX}$, respectively. 
 \end{condition}
 
 Condition \ref{accuracy-knockoffs} above couples each approximate knockoff variable $\hat\bX_j$ with a perfect knockoff variable $\widetilde \bX_j$. Similar to Condition \ref{fdr-condition1}, we need the realizations instead of the distributions of  $\hat\bX_j$ and $\widetilde\bX_j$ to be close, which is a major distinction from the assumption in \cite{Barber2020}. Such distinction allows $\hat{\bX}$ to be constructed using sample $(\bX, \by)$ \textit{without} data splitting under relaxed estimation accuracy assumptions, as will be illustrated in the next two subsections. Later in Section \ref{Sec.3}, we will provide extensive analysis on the coupling order $\Delta_n$ using some specific examples of feature distributions.


 We next show that the closeness between $\hat\bX$ and $\widetilde \bX$ can lead to the closeness between $\hat W_j$'s and $\widetilde W_j$'s as required by Condition \ref{fdr-condition1}.  Since different construction of the knockoff statistics depends on the feature matrix differently, we showcase the theory using two constructions of the knockoff statistics: the marginal correlation knockoff statistics and the regression coefficient difference (RCD) knockoff statistics.  
 
 For clarity, we include Table \ref{table-summary-Sec3} to summarize the sets of assumptions on the model setting, feature distribution, the knockoff statistics, and the corresponding rates for the coupling accuracy in our layer 2 analysis.
 
 \vspace{-0.2cm}
 \begin{table}[htp]
\caption{Summary of key conditions and results for asymptotic FDR control in Layer 2 analysis.}
\label{table-summary-Sec3}
 	\begin{tabular}{l|l|l}
 		\hline
 		Model setting                                                            & \begin{tabular}[c]{@{}c@{}}Nonparametric model \\  \eqref{nonpara-model} in Section \ref{new.Sec3.2}  \end{tabular}                                  & \begin{tabular}[c]{@{}c@{}}Linear model \\ $\by = \bX  \bbeta + \bveps$ in Section \ref{new.Sec3.3} \end{tabular}                                                                           \\ \hline
 		Feature distribution                                                     & $X \stackrel{d}{\sim} N(\bf{0}, \bSigma)$                                                   & sub-Gaussian                        \\ \hline
 		Sparsity assumption                                                      & Condition \ref{marginal_corr_condition2} on $\bSigma^{-1}$ and $\bSigma$ & \begin{tabular}[c]{@{}l@{}}Sparse $\bbeta$; sparse precision \\ matrix for covariates; Condition \ref{debiased-lasso-condition1}\end{tabular} \\ \hline
 		Knockoff statistics                                                                            & Marginal correlation                                                                                             & RCD with debiased Lasso     \\                                          \hline
 		 $\hat W_j$ coupling accuracy & $\Delta_n$& $\Delta_n s\sqrt{(\log p)/n}$ \\
 		\hline
 		  \begin{tabular}[c]{@{}l@{}}$\Delta_n$ requirement\\ for FDR control\end{tabular} & $\Delta_n \sqrt n (\log p)^{1/2 + 1/\gamma}\to 0$                                           & $\Delta_ns(\log p)^{1+1/\gamma}\to 0$       \\
   \hline
 	\end{tabular}
 \end{table}
 \vspace{-0.4cm}

 \subsection{Marginal correlation knockoff statistics} \label{new.Sec3.2}
 
 Marginal correlation is a commonly analyzed measure on variable importance for feature screening due to its simplicity. 
 Given  $\hat\bX$ and $\widetilde \bX$ satisfying Condition \ref{accuracy-knockoffs}, the approximate knockoff statistics based on the marginal correlation difference are defined as 
 \begin{equation} \label{hat-marginal-correlation-new-eq}
 	\hat{W}_j = ( \sqrt{n} \norm{\by}_2 )^{-1} ( | \bX_j^{\top} \by | - | \hat{\bX}_j^{\top} \by | )  \ \text{ for } 1 \leq j \leq p, 
 \end{equation}
 and the coupled perfect knockoff statistics are given by 
 \begin{equation} \label{tilde-marginal-correlation-new-eq}
 	\widetilde{W}_j = ( \sqrt{n} \norm{\by}_2 )^{-1} ( | \bX_j^{\top} \by | - | \widetilde{\bX}_j^{\top} \by | ) \ \text{ for } 1 \leq j \leq p.
 \end{equation}  
 Observe that $  \widetilde{W}_j - \hat{W}_j   =  ( \sqrt{n} \norm{\by}_2 )^{-1}  (  | \hat{\bX}_j^{\top} \by | -  | \widetilde{\bX}_j^{\top} \by | ) $ and thus under Condition \ref{accuracy-knockoffs}, we have that with asymptotic probability one,
 \begin{equation}\label{marginal-w-closeness}
 	\max_{1\leq j\leq p}|\widehat W_j-\widetilde W_j|\leq \Delta_n.
 \end{equation}
 This result is summarized formally in Lemma \ref{pf-thm3-lemma-1} in Section \ref{new.SecA.3} of the Supplementary Material. 

 We consider the flexible nonparametric regression model
 \begin{equation} \label{nonpara-model}
 	Y = f(X_{\mathcal{H}_{1}}) + \varepsilon,
 \end{equation}
 where $f$ is some unknown regression function,  $X_{\mathcal{H}_1} = (X_j)_{j \in \mathcal{H}_1}$ contains all the relevant features for response $Y$, and $\varepsilon$ is the model error satisfying $\varepsilon \independent X$ and $\e(\varepsilon) = 0$. 
 Assume that feature vector $X = (X_1, \ldots, X_p)^{\top} \stackrel{d}{\sim} N(\bf{0}, \bSigma)$ with  $\bSigma$ the positive definite covariance matrix. 
 Moreover, let the distribution of the perfect knockoff variables $\widetilde{X} = (\widetilde{X}_1, \cdots, \widetilde{X}_p)^{\top}$ satisfy that 
 \begin{equation} \label{marginal-normal}
 	(X, \widetilde{X}) =  (X_1, \cdots, X_p, \widetilde{X}_1, \cdots, \widetilde{X}_p)  \stackrel{d}{\sim} N  \Bigg({\bf 0}, \begin{pmatrix}
 		\bSigma &  \bSigma - r I_p  \\
 		\bSigma - r I_p  & \bSigma  \\
 	\end{pmatrix} \Bigg),
 \end{equation}
 where $r > 0$ is a constant such that the above covariance matrix is positive definite.  Here, we consider the equicorrelated construction \citep{CFJL2018} for simpler presentation and the diagonal matrix $r I_p$ can be replaced with a general version $\diag(r_1, \ldots, r_p)$ with possibly distinct diagonal entries $\{ r_j\}_{j = 1}^p$.  Note that the Gaussian distribution assumption is imposed mainly to verify the general Conditions \ref{fdr-condition4} and \ref{fdr-condition5}. If one assumes directly these two conditions, the Gaussian distribution assumption can be removed. 
 
 Furthermore, we make the additional technical assumptions below on the generative model \eqref{nonpara-model} to verify the conditions in our  layer 1 analysis presented  in Section \ref{Sec.2}. 
 
 \begin{condition} \label{marginal_corr_condition1}
 	Y is a sub-Gaussian random variable with sub-Gaussian norm $\| Y \|_{\psi_2}$. 
 \end{condition}
 
 \begin{condition} \label{marginal_corr_condition3}
 	Define $\mathscr{A}_n = \{ j \in \mathcal{H}_1:(\e Y^2)^{ - 1/2} (| \e (X_j Y) | - | \e( \widetilde{X}_j Y) |)| \geq 5 \delta_n \}$ with 
 	\begin{equation}   \label{eq-delta_n}
 		\delta_n = C_{X,Y}\sqrt{  n^{-1 } \log p  },
 	\end{equation}
 	where $C_{X,Y}:=\max\limits_{1 \leq j \leq p} \Big\{   \frac { 16 \sqrt 2 \| X_j \|_{\psi_2} \| Y \|_{\psi_2} } { (\e Y^2)^{1/2} } \lor   \frac { 8\sqrt 2  |w_j| \| Y \|_{\psi_2}^2 } { \e Y^2 } \Big\}$.  
 	It holds that $a_n:= |\mathscr{A}_n| \to \infty $ and $C_{X,Y}$ is a positive constant that is independent of $p$ and $n$. 
 \end{condition}
 
 Denote by $(\bSigma^{-1})_{j}$ the $j$th column of matrix $\bSigma^{-1}$, $\bSigma_{i, j}$ the $(i, j)$th entry of matrix $\bSigma$, and $\bSigma_{\mathcal{H}_1, j}$ a vector given by $ (\bSigma_{i, j})_{i \in \mathcal{H}_1}$.  Recall the definition $G(t) = p_0^{-1} \sum_{j\in \mathcal{H}_0} \mathbb{P} (\widetilde{W}_j \geq t) =   p_0^{-1} \sum_{j\in \mathcal{H}_0} \mathbb{P} (\widetilde{W}_j \leq - t)$.
 
 \begin{condition} \label{marginal_corr_condition2}
 	For some sequence $m_n = o(a_n)$, matrices $\bSigma^{-1}$ and $\bSigma$ are sparse in the sense  that $\max_{1\leq j \leq p} \| (\bSigma^{-1})_j\|_0 \leq m_n$ and $ \sum_{j \in \mathcal{H}_0} \mathbbm{1} (   \bSigma_{\mathcal{H}_1,\, j}  \neq {\bf 0} ) \leq  m_n$.  In addition, $C_1 < r < \min_{1 \leq j \leq p}\bSigma_{j, j} \leq \max_{1 \leq j \leq p} \bSigma_{j, j}< C_2$ for some constants $C_1> 0$ and $C_2 > 0$.
 \end{condition}
 
 \begin{condition} \label{marginal_corr_condition4}
 	It holds that $|\mathcal{H}_1|^{-1} \sum_{j \in \mathcal{H}_1} \mathbb{P} (\widetilde{W}_j < -  t )  \leq 
 	G(t)$ for all $t \in  (0, C_3\sqrt{\frac{\log p}{n}})$ with $C_3>0$ some large constant.
 \end{condition}
 
 Under Conditions \ref{marginal_corr_condition1}--\ref{marginal_corr_condition4}, we can verify that Conditions \ref{fdr-condition2}--\ref{fdr-condition5} are satisfied. This together with Condition \ref{accuracy-knockoffs} and our general theorem on the FDR control (cf. Theorem \ref{theorem-FDR}) leads to the theorem below.   
 \begin{theorem} \label{thm-marginal-corr-fdr}
 	Assume that Conditions \ref{accuracy-knockoffs}--\ref{marginal_corr_condition4} are satisfied. In addition, assume that for some constant $0 < \gamma < 1$, $(\log p)^{1/\gamma} m_n / a_n \to 0 $  and the coupling accuracy $\Delta_n$ in Condition \ref{accuracy-knockoffs} satisfies   $  \sqrt{n} \Delta_n  (\log p)^{1/2 + 1/\gamma} \to 0 $. 
 	Then for the approximate knockoffs inference based on the marginal correlation, we have 
 	\begin{equation*}
 		\limsup_{n \to \infty} \FDR \leq  q. 
 	\end{equation*}	    
 \end{theorem}
 
 Let us make a few remarks on the conditions and result presented in Theorem \ref{thm-marginal-corr-fdr} above. Condition \ref{marginal_corr_condition3} verifies the signal strength assumption in Condition \ref{fdr-condition3} in the specific context of model \eqref{nonpara-model} and marginal correlation knockoff statistics. We show in Lemma \ref{pf-thm3-lemma-2} in Section \ref{new.SecA.3} of the Supplementary Material that Condition \ref{fdr-condition2} holds with  $\delta_n = O(\sqrt{n^{-1} \log p })$. Since we assume Gaussian feature distribution in this section, the dependence among the indicator functions as required by Condition \ref{fdr-condition4} is determined by  covariance matrix $\bSigma$. Hence, Condition \ref{marginal_corr_condition2} is imposed to justify the validity of Condition \ref{fdr-condition4}.   
 It is worth mentioning that the sparse dependence structure assumed in Condition \ref{marginal_corr_condition2} can be replaced with a general assumption that the conditional distribution $X_{\mathcal{H}_0} | X_{\mathcal{H}_1}$ has sparse pairwise dependency and the sequence $ \{ h_j(t; X_{\mathcal{H}_1}): = \e ( \mathbbm{1} (\widetilde{W}_j \geq t)  | X_{\mathcal{H}_1} ) \}_{j \in \mathcal{H}_0} $ has sparse pairwise correlation for each given $t > 0$. 
 Condition \ref{marginal_corr_condition4} is a technical assumption that is intuitive and requires that on average, the probability of a relevant feature having a negative valued $\widetilde W_j$ is smaller than the corresponding probability of an irrelevant feature. Such condition is compatible with our requirement that relevant features should have positive and larger magnitude for $\widetilde{W}_j$.
 
 Note that in this example, $w_j=\e \widetilde W_j$ and the  concentration rate $\delta_n$ as in Condition \ref{fdr-condition2} is $\delta_n\sim \sqrt{(\log p)/n}$. The assumption $  \sqrt{n} \Delta_n  (\log p)^{1/2 + 1/\gamma} \to 0 $ in Theorem \ref{thm-marginal-corr-fdr}  requires that $\Delta_n \ll n^{-1/2} (\log p)^{-1/2 - 1/\gamma}$, and hence, $ \Delta_n\ll \delta_n$. In view of \eqref{marginal-w-closeness}, the requirement of $\Delta_n\ll \delta_n$ indeed restricts that the quality of $\hat W_j$'s, as measured by $\Delta_n$ in the current example, is of an order smaller than $\delta_n$. This also suggests that an independent sample of size $N\gg n$ may be needed to learn the covariate distribution for constructing the approximate knockoff variables in order to achieve the desired accuracy of $\max_{1\leq j\leq p}|\widehat W_j-\widetilde W_j|\leq \Delta_n \ll n^{-1/2} (\log p)^{-1/2 - 1/\gamma} $.  
 
 It is worth mentioning that the bound obtained in \eqref{marginal-w-closeness} may be improved under additional model assumptions. For instance, if additionally the covariates $\{X_j \}_{j=1}^p$ are independent, 
 then under Condition \ref{accuracy-knockoffs}  we can show that $\max_{1 \leq j \leq p } | \hat{W}_j  - \widetilde{W}_j| \leq C \Delta_n \sqrt{n^{-1} \log p} $ (see Lemma \ref{Lemma-corr-indep} in Section \ref{secB.16-Lemma-indep-corr} of the Supplementary Material for details). The improved result is because of the elimination of spurious correlation between null and signal covariates.   In this case, the condition on $\Delta_n$ in Theorem \ref{thm-marginal-corr-fdr} is relaxed to $\Delta_n (\log p)^{1 + 1/\gamma} \to 0$.
 
 The above discussions suggest that knockoff statistics based on marginal correlation are of low quality in the sense that they are less robust to estimation error and model mis-specification. Indeed, we will see in the next section that some other popularly used knockoff statistics such as RCD can achieve asymptotic FDR control under much relaxed assumptions.

 \subsection{Regression coefficient difference with debiased Lasso} \label{new.Sec3.3}
 
 A popularly used construction of the knockoff statistics is RCD. We present our results under the following linear regression model for simplification; the extension to the generalized linear model (GLM) can be found in Section \ref{Supp_Sec_C} of the Supplementary Material. We consider 
 $$
 \by = \bX  \bbeta + \bveps,
 $$
 where $\bbeta = (\beta_j)_{1 \leq j \leq p} \in \mathbb{R}^p$ is the true regression coefficient vector, $\bveps \stackrel{d}{\sim} N({\bf 0}, \sigma^2 I_n)$ is the model error vector, and $\bveps \independent  \bX $. Assume that feature vector $X = (X_1, \ldots, X_p)^{\top}$ has mean ${\textbf 0}_p \in \mathbb{R}^p$ and covariance matrix $\bSigma \in \mathbb{R}^{p \times p }$. Denote by $ \bbeta^{\augg} = (\bbeta^{\top} ,  {\bf 0}_p^{\top})^{\top} \in \mathbb{R}^{2p}$ the augmented true parameter vector. 
 
 Let $\hat{\bbeta} =(\hat{\beta}_j)_{1 \leq j \leq 2p}\in \mathbb{R}^{2p}$ be the debiased Lasso estimator (\cite{ZZ2014}) based on the augmented design
 matrix $\hat{\bX}^{\augg} := [\bX, \hat{\bX}]$, where $\hat{\bX}$ is the approximate knockoff variable matrix. Assume that Condition \ref{accuracy-knockoffs} is satisfied and $\widetilde \bX$ is the coupled perfect knockoffs variable matrix.  
 Similarly, define $\widetilde{\bX}^{\augg} := [\bX, \widetilde{\bX}]$. Then $\hat{\bbeta}$ can be coupled with the debiased Lasso estimator denoted as $\widetilde\bbeta=(\widetilde{\beta}_j)_{1 \leq j \leq 2p}\in \mathbb{R}^{2p}$ based on $\widetilde{\bX}^{\augg}$. Then the RCD knockoff statistics can be defined as 
 \begin{equation}  \label{hat_RCD_knockoff_stat-new}
 	\hat{W}_j = |\hat\beta_j| - |\hat{\beta}_{j+p}|  
 \end{equation}
 \begin{equation} \label{tilde_RCD_knockoff_stat-new}
 	\text{and} \quad    \widetilde{W}_j = |\widetilde\beta_j| - |\widetilde{\beta}_{j+p}|
 \end{equation}
 for the approximate and perfect knockoffs procedures, respectively, for $1 \leq j \leq p$. 
 
 We provide the explicit definition of the debiased Lasso estimator to assist future presentation. For $1\leq j \leq 2p$, the debiased Lasso estimator is a one-step bias correction from some initial estimator $\hat\bbeta^{\init}=(\hat{\beta}_j^{\init})_{1 \leq j \leq 2p}\in \mathbb{R}^{2p}$ and is defined as 
 \begin{equation} \label{eq-debiased-lasso-approx}
 	\hat{\beta}_j = \hat{\beta}^{\init}_j + \frac{ \hat\bz_j^{\top} \big(\by - \hat{\bX}^{\augg} \hat{\bbeta}^{\init} \big) } {\hat\bz_j^{\top} \hat\bX^{\augg}_j},   
 \end{equation}
 where $\hat\bz_j$ is the score vector defined as
 \begin{equation} \label{eq-score-empirical}
 	\hat{\bz}_j = \hat\bX^{\augg}_j - \hat\bX_{-j} \hat{\bgamma}_j
 \end{equation}
 with $\hat{\bgamma}_j := \argminA_{\textbf b} \big\{  (2n)^{-1}  \| \hat\bX^{\augg}_j - \hat\bX^{\augg}_{-j} \textbf b\|_2^2    + \lambda_j \| \textbf b \|_1  \big\}$ and $\{ \lambda_j\}_{j = 1}^{2p}$ the nonnegative regularization parameters. We construct the initial estimator as
 \begin{equation} \label{eq-init-empirical}
 	\hat\bbeta^{\init} := \argminA_{{\textbf b}} \Big\{ (2n)^{-1} \| \by - \hat\bX^{\augg} \textbf b \|_2^2   + \lambda \| \textbf b\|_1 \Big\}
 \end{equation}
 with $\lambda = C \sqrt{n^{-1} \log (2p)}$ the regularization parameter and $C>0$ some constant. 

 Analogously, the coupled debiased Lasso estimator $\widetilde\bbeta$ can be defined componentwisely as
 \begin{equation} \label{eq-debiased-lasso}
 	\widetilde{\beta}_j = \widetilde{\beta}^{\init}_j + \frac{ \widetilde\bz_j^{\top} \big(\by - \widetilde{\bX}^{\augg} \widetilde{\bbeta}^{\init} \big) } {\widetilde\bz_j^{\top} \widetilde\bX^{\augg}_j} \ \mbox{ for }  1 \leq j \leq 2p,   
 \end{equation}
 where 
 \begin{equation} \label{eq-init-perfect}
 	\widetilde\bbeta^{\init} = (\widetilde{\beta}_j^{\init})_{1 \leq j \leq 2p}  := \argminA_{{\textbf b}} \Big\{ (2n)^{-1} \| \by - \widetilde\bX^{\augg} \textbf b \|_2^2   + \lambda \| \textbf b\|_1 \Big\} 
 \end{equation}
 and 
 \begin{equation} \label{eq-score-perfect}
 	\widetilde{\bz}_j = \widetilde\bX^{\augg}_j - \widetilde\bX^{\augg}_{-j} \widetilde{\bgamma}_j  \mbox{ with }  \widetilde{\bgamma}_j := \argminA_{\textbf b} \Big\{ (2n)^{-1}  \| \widetilde\bX^{\augg}_j - \widetilde\bX^{\augg}_{-j} \textbf b\|_2^2  + \lambda_j \| \textbf b \|_1  \Big\}.
 \end{equation}
 It is important to emphasize that the \textit{same} regularization parameters $\lambda$ and $\lambda_j$'s in defining $\hat\bbeta$ should be used as in defining $\widetilde\bbeta$ in \eqref{eq-debiased-lasso} so that their constructions differ only by the used feature matrix; this plays a key role in applying our coupling technique. Indeed, we prove in Lemma \ref{pf-thm5-lemma-1} in Section \ref{new.SecA.5} of the Supplementary Material that the coupling technique together with Condition \ref{accuracy-knockoffs} and some other regularity conditions ensures that with asymptotic probability one,
 \begin{equation}\label{eq: LCD-paring}
 	\max_{1\leq j\leq 2p}|\widetilde\beta_j-\hat\beta_j|\lesssim \Delta_ns\sqrt{ n^{-1} \log p }.  
 \end{equation}
 The above result guarantees that $\hat W_j$'s and $\widetilde W_j$'s are also uniformly close over $1\leq j\leq p$ with  $\max_{1\leq j\leq p}|\hat W_j - \widetilde W_j|\lesssim \Delta_ns\sqrt{n^{-1} \log p }$. As long as $s\Delta_n\rightarrow 0$, this upper bound has a smaller order than the concentration rate $\delta_n$ of $\widetilde W_j$ (cf.  Condition \ref{fdr-condition2}), because here $\delta_n \sim \sqrt{n^{-1}\log p}$ as shown in our Lemma \ref{pf-thm5-lemma-2} in Section \ref{new.SecA.5}. As commented after Theorem \ref{thm-marginal-corr-fdr}, the assumption that the coupling rate of $\max_{1\leq j\leq p}|\widetilde W_j-\hat W_j|$ is of a smaller order than the concentration rate $\delta_n$ plays a key role in establishing our theory on the asymptotic FDR control.

 We next introduce some additional notation and formally present the regularity conditions specific to this section. Observe that by symmetry, the augmented feature vector with the perfect knockoff variables has covariance matrix 
 \begin{equation} \label{augmented-covariance-matrix}
 	\bSigma^{A} = \begin{pmatrix}
 		\bSigma &  \bSigma - \boldsymbol{D}  \\
 		\bSigma - \boldsymbol{D}  & \bSigma  \\
 	\end{pmatrix}, 
 \end{equation}
 where $\boldsymbol{D}$ is a diagonal matrix such that matrix $\bSigma^A$ is positive definite.
 Let $\bOmega^{A}  = (\bSigma^A)^{-1}$ and $ \bgamma_j = (\bgamma_{j, l})_{l \neq j}$ with $\bgamma_{j, l} = -  {\bOmega^{A}_{j, l}} /{\bOmega^{A}_{j, j}}$. It has been shown in  \cite{peng2009partial} that  the residuals
 \begin{equation*}
 	e_j = \widetilde X_j^{\augg} - \widetilde{X}^{\augg}_{-j} \bgamma_j
 \end{equation*}
 satisfy that $\Cov ( e_{j}, \widetilde{X}_{-j}^{\augg}   )  = {\textbf 0}$, $\Var ( e_{j} ) = 1/ \bOmega^A_{j, j} $, and  $  \Cov(e_{j},  e_{l} ) = \frac {\bOmega^A_{j, l}} {\bOmega^A_{j, j} \bOmega^A_{l, l} } $.
 For $1 \leq j \leq 2p$, denote by $ \mathcal{S}_j  = \supp(\bgamma_j) \cup \supp(\widetilde{\bgamma}_j) \cup \supp(\hat{\bgamma}_j)$. 
 Let $J = \supp(\bbeta^{\augg}) \cup \supp(\widetilde{\bbeta}^{
 	\init
 }) \cup \supp(\hat{\bbeta}^{\init})$ and $s := \| \bbeta^{\augg} \|_0 = \| \bbeta  \|_0 = o(n)$.
 We make the technical assumptions below. 
 
 
 \begin{condition} \label{debiased-lasso-condition1}
 	a) For some constant $C_4 > 0$, $\mathbb{P} (|J| \leq C_4 s) \to 1 $. \\
 	b) For some sequence $ m_n \lesssim s$, it holds that $ \max_{1 \leq j \leq 2 p} \| \bOmega^A_j \|_0 \leq m_n $ and $\mathbb{P} ( \max_{1 \leq j \leq 2p } |\mathcal{S}_j| \leq C_5 m_n ) \to 1 $ with some constant $C_5>0$. \\
 	c) $\max_{1 \leq j \leq 2p} \|\bgamma_j \|_2 \leq C_6$ and $C_7 < \lambda_{\min} (\bOmega^{A} ) \leq \lambda_{\max} (\bOmega^A) < C_8$ with some positive constants $C_6$, $C_7$, and $C_8$.
 \end{condition}  
 
 \begin{condition}[Restrictive eigenvalues] \label{debiased-lasso-condition2}
 	Assume that with probability $1 - o(1)$,
 	\begin{equation}
 		\min_{\| \delta\|_0 \leq C_9 s } \frac{ \delta^{\top}[\widetilde\bX^{\augg} ] ^{\top} \widetilde\bX^{\augg} \delta} { n\|\delta\|_2^2 } \geq \kappa_1 
 	\end{equation}
 	for some large enough constant $C_9>0$ and a constant $\kappa_1 > 0$. 
 \end{condition}
 
 \begin{condition} \label{debiased-lasso-condition3}
 	The features $X_j$'s and errors $e_{j}$'s are sub-Gaussian with sub-Gaussian norms $\|X_j \|_{\psi_2} \leq \phi$ and $\|e_j \|_{\psi_2} \leq \phi$ for some constant $\phi > 0$. 
 \end{condition}
 
 \begin{condition} \label{debiased-lasso-condition4}
 	Let $\mathscr{A}_n = \{j \in \mathcal{H}_1: |\beta_j| \gg  \sqrt{n^{-1} \log p}  \}$ and it holds that $a_n :=  |\mathscr{A}_n| \to \infty$. 
 \end{condition}

 \ignore{ 
 	\begin{condition} \label{debiased-lasso-condition2}
 		The features $X_j$'s and the errors $e_j$'s are sub-Gaussian satisfying $\|X_j \|_{\psi_2} \leq \phi$ and $\|e_j \|_{\psi_2} \leq \phi$ for some constant $\phi > 0$ and  with probability $1 - O( p^{-c})$, 
 		\begin{align}
 			&\|\hat{\bgamma}_j - \bgamma_j \|_1 \leq C n^{-1/2 }a_{n,1},~  \| [\bX, \widetilde{\bX}]_{-j} (\hat{\bgamma}_j  - \bgamma_j )  \|_2^2 \leq C  a_{n,2} \label{bias-corrected-cond1} \\
 			&  n^{-1} \| [\bX,  \widetilde{\bX}]^{\top} \bX_j \|_{\max} \leq C, ~ \| \widetilde{\bbeta}^{init}  - \bbeta \|_1 \leq C s \sqrt{\frac{ \log p} {n}}  \label{bias-corrected-cond2},
 		\end{align}
 		where $a_{n,1}$ and $a_{n,2}$ are two possibly diverging sequences. Moreover, $|\sigma^{jk}|/\sqrt{\sigma^{jj} \sigma^{kk}} < c$ for some constant $0< c < 1$. 
 	\end{condition}
 }

 We are now ready to state our results on the FDR control for the approximate knockoffs inference based on the debiased Lasso coefficients. 
 \begin{theorem} \label{thm-debiased-lasso-FDR}
 	Assume that Conditions \ref{accuracy-knockoffs} and  \ref{marginal_corr_condition4}--\ref{debiased-lasso-condition4} hold, $m_n / a_n \to 0 $, and $    \frac{m_n^{1/2} s (\log p)^{3/2 + 1/\gamma}} {\sqrt n} + \Delta_n s (\log p)^{1 + 1 /\gamma} \to 0$ for some constant $0 < \gamma < 1$. Then we have
 	\begin{equation*}
 		\limsup_{n \to \infty} \FDR \leq  q. 
 	\end{equation*}
 \end{theorem}
 
 Similarly as discussed in the last section, Condition \ref{debiased-lasso-condition1} is used to verify the weak dependence assumption in Condition \ref{fdr-condition4}. Condition \ref{accuracy-knockoffs} and the two regularity Conditions \ref{debiased-lasso-condition2}--\ref{debiased-lasso-condition3} are imposed for verifying the coupling accuracy Condition \ref{fdr-condition1}.
 Condition \ref{debiased-lasso-condition4} contributes to verifying the general signal strength requirement in Condition \ref{fdr-condition3}.  
 
 

 \ignore{
 	\begin{theorem} \label{thm-debiased-lasso-FDR}
 		Assume Conditions \ref{accuracy-knockoffs} and \ref{debiased-lasso-condition1}-\ref{debiased-lasso-condition4} are satisfied. When  $(\log p)^{1/\gamma + 1/2} ( n^{1/2} b_n + n^{-1/2} s \log p) \to 0$, we have
 		\begin{equation*}
 			\limsup_{(n, p) \to \infty} \FDR \leq  q.  
 		\end{equation*}
 	\end{theorem}
 }
 
 \ignore{
 	\newpage
 	We start with verification of Condition \ref{fdr-condition1}. Let $J = \supp(\bbeta_0) \cup \supp(\widetilde{\bbeta}) \cup \supp(\hat{\bbeta})$.
 	
 	\begin{proposition}
 		Assume Condition \ref{accuracy-knockoffs} is satisfied. Moreover, with asymptotic probability one, it holds that $|J| \leq m = o(n)$ and the restricted eigenvalue of $n^{-1} [\bX,\widetilde{\bX}]^{\top}_{J'} [\bX,\widetilde{\bX}]_{J'}$ is lower bounded by $\kappa_c$ for any $J'$ with $|J'| \leq m$. Then we have 
 		\begin{equation} 
 			\begin{split}
 				\mathbb{P} \Big(  \max_{1 \leq j \leq p} |\hat{W}_j - \widetilde{W}_j |\lesssim  \kappa_c^{-1}   m^{3/2}\Delta_n (n^{-1} \log p)^{1/2} + m^{1/2}\Delta_n n^{-1/2}\Big)  
 				\geq 1 - o(1).
 			\end{split} 
 		\end{equation}
 	\end{proposition}
 	This result can be improved if we apply another version of condition (take advantage of the specific structure in the distance between approximate and perfect knockoff variables).
 	
 	Cite Lv and Fan's JASA paper on asymptotic equivalence.... 
 	
 	We continue to verify Condition \ref{fdr-condition2}. let $w_j = |\beta_j^0|$ for $1 \leq j \leq p$. Then we can obtain the following result. 
 	\begin{proposition}
 		Assume that $ \| \widetilde{\bbeta}^{init} - \bbeta^0  \|_1 = o_p(1) $. Then we have 
 		\begin{equation}
 			\sum_{j = 1}^p \mathbb{P} \Big( |\widetilde{W}_j - w_j| > C \sqrt{ \frac{\log p} {n}} \Big)   \to 0. 
 		\end{equation}
 	\end{proposition}
 	Next we turn to verification of Condition \ref{fdr-condition5}. 
 	\begin{proposition} \label{prop-bias-corrected-lasso}
 		Suppose that the features $X_j$'s and the errors $e_j$'s are sub-Gaussian satisfying $\|X_j \|_{\psi_2} \leq \phi$ and $\|e_j \|_{\psi_2} \leq \phi$ for some constant $\phi > 0$ and  with probability $1 - O( p^{-c})$, 
 		\begin{align}
 			&\|\hat{\bgamma}_j - \bgamma_j \|_1 \leq C n^{-1/2 }a_{n,1},~  \| [\bX, \widetilde{\bX}]_{-j} (\hat{\bgamma}_j  - \bgamma_j )  \|_2^2 \leq C  a_{n,2} \label{bias-corrected-cond1} \\
 			&  n^{-1} \| [\bX,  \widetilde{\bX}]^{\top} \bX_j \|_{\max} \leq C, ~ \| \widetilde{\bbeta}^{init}  - \bbeta \|_1 \leq C s \sqrt{\frac{ \log p} {n}}  \label{bias-corrected-cond2},
 		\end{align}
 		where $a_{n,1}$ and $a_{n,2}$ are two possibly diverging sequences. Moreover, $|\sigma^{jk}|/\sqrt{\sigma^{jj} \sigma^{kk}} < c$ for some constant $0< c < 1$. 
 		If $(\log p)^{1/\gamma + 1/2}  [ n^{1/2} b_n  + n^{-1/2} s \log p ] \to 0$, then we have \eqref{dist-cond1} in Condition \ref{fdr-condition5} is satisfied.
 	\end{proposition}
 	
 	\begin{remark}
 		Here we assume both the features $X_j$ and the errors $e_j$ are sub-Gaussian for simpler presentation. In fact, if we assume $\|X_j\|_{\psi_2} \leq \phi$ and sparsity of correlations between features, that is, $\| \bgamma_j\|_0 \leq s = o(n)$. In addition, suppose the coefficients $\|\gamma_j\|_{\max} \leq c$ are bounded. Then the errors $e_j$ is also sub-Gaussian with $\| e_j \|_{\psi_2} \leq c s \phi $. (This results follows from the fact: If $X_1$, $X_2$ are random variables such that $X_i$ is $b_i$-sub-Gaussian, then $X_1 + X_2$ is $(b_1 + b_2)$-sub-Gaussian).
 	\end{remark}
 }

 \subsection{Connection of Condition \ref{accuracy-knockoffs} with Wasserstein distance}\label{sec:Wasserstein}
 {\color{black}
 	We detour slightly and discuss the connection of Condition \ref{accuracy-knockoffs} with a type of Wasserstein distance and state a conjecture of ours; it is safe to skip this section and proceed to Section \ref{Sec.3} for knockoff variable coupling. 
 	
 	First recall that the knockoff variable matrix is generated in a rowwise fashion independent of each other. 
 	Given a row $\bx$ of the original data matrix $\bX$, denote by $\hat{\mu}_{\bx} $ the estimated or misspecified  conditional distribution for generating the corresponding row in the approximate knockoff variable matrix $\hat\bX$,  and denote by $\widetilde{\mu}_{\bx} $ its oracle counterpart based on the true feature distribution. Conditional on the original data matrix $\bX$, let $\hat{\mu}^n = \hat{\mu}_{\bx_1} \times \hat{\mu}_{\bx_2}\times \cdots \times \hat{\mu}_{\bx_n} $ and $\widetilde{\mu}^n = \widetilde{\mu}_{\bx_1} \times \widetilde{\mu}_{\bx_2}\times \cdots \times \widetilde{\mu}_{\bx_n}$, where $\bx_i$ is the $i$th row of the original data matrix $\bX$.
 	Define the conditional $(1,2)$-Wasserstein distance between $\hat{\mu}^n$ and $\widetilde{\mu}^n$ as
 	\begin{equation} \label{W_{12}}
 		\mathbb{W}_{1, 2}(\hat{\mu}^{n}, \widetilde{\mu}^n | \bX) = \inf_{\eta \in \Gamma(\hat{\mu}^n, \widetilde{\mu}^n)}   \e_{ ( {\scriptsize \vecc(\hat{\bX}), \vecc(\widetilde{\bX}) } ) \stackrel{d}{\sim} \eta }  [ \|\hat{\bX} - \widetilde{\bX}\|_{1, 2}\vert \bX],
 	\end{equation}
 	where $\Gamma(\hat{\mu}^n, \widetilde{\mu}^n)$ is the set consisting of all couplings of $\hat{\mu}^n $ and $\widetilde{\mu}^n$, $\|\cdot\|_{1,2}$ is the matrix $(1,2)$-norm as defined in Condition \ref{accuracy-knockoffs}, and $\vecc(\hat{\bX})$ stands for vectorization of $\hat{\bX}$ by rows, similarly for $\vecc(\widetilde{\bX})$.  
 	
 	\begin{proposition}\label{prop}
 		Assume that there exists a deterministic sequence $c_n \rightarrow 0$  and a coupling $\eta^*\in \Gamma(\hat{\mu}^n, \widetilde{\mu}^n)$ such that
 		\begin{align}
 			& \mathbb P_{\boldsymbol{X}}(\mathbb{W}_{1, 2}(\hat{\mu}^{n}, \widetilde{\mu}^n | \bX)\geq c_n) \rightarrow 0,\label{eq:wasserstein-1} \\ 
 			&  \e_{ ( {\scriptsize \vecc(\hat{\bX}), \vecc(\widetilde{\bX}) } ) \stackrel{d}{\sim} \eta^* }  [ \|\hat{\bX} - \widetilde{\bX}\|_{1, 2}\vert \bX] \leq C_{\boldsymbol{X}}\mathbb{W}_{1, 2}(\hat{\mu}^{n}, \widetilde{\mu}^n | \bX), \label{eq:wasserstein-2}
 		\end{align}
 		where $C_{\boldsymbol{X}}\geq 1$ depends only on $\bX$ with well-defined expectation $\e_{\boldsymbol{X}}[C_{\boldsymbol{X}}]<\infty$, and $\mathbb P_{\boldsymbol{X}}$ and $\e_{\boldsymbol{X}}$ are probability and expectation taken with respect to $\bX$, respectively. Then as $n\rightarrow \infty$, Condition \ref{accuracy-knockoffs} is satisfied with $\Delta_n$ chosen such that $\e_{\boldsymbol{X}}[C_{\boldsymbol{X}}]c_n \Delta_n^{-1} \rightarrow 0$.
 		
 	\end{proposition}

 	It is seen that assumption \eqref{eq:wasserstein-1} and the existence of $\eta^*$ in Proposition \ref{prop} provide sufficient conditions ensuring Condition \ref{accuracy-knockoffs}. We next verify the existence of $\eta^*$ in a special scenario. 
 	
 	In Section \ref{new.Sec4.2}, we present a concrete construction for coupling of the approximate and perfect knockoff variable matrices under Gaussian distribution, given by \eqref{sec4.1-1} and \eqref{sec4.1-2}, respectively. The lemma below is based on such constructions. The proof of Lemma \ref{le-Gaussian-Wasser} is postponed to Section \ref{Sec_proof_new_le1} of the Supplementary Material. 
 		
 	
 	\begin{lemma}[Gaussian Coupling] \label{le-Gaussian-Wasser} Consider Gaussian knockoffs in Section \ref{new.Sec4.2}.
 		Let $\eta^*$ be the conditional coupling measure used for generating \eqref{sec4.1-1} and \eqref{sec4.1-2}.  Define $\hat{\bD} := (2 r I_p - r^2 \hat{\bOmega})^{1/2} $ and $\bD := (2 r I_p - r^2  {\bOmega})^{1/2}$, where $\bOmega$, $\hat{\bOmega}$, and $r$ are the same as defined in Section \ref{new.Sec4.2}. Let $\bD_j$ and $\hat\bD_j$ be the $j$th columns of $\bD$ and $\hat\bD$, respectively. If $\|\hat{\bD}_j\|_2 \|\bD_j \|_2 - \hat{\bD}_j ^T \bD_j \leq C \|\hat{\bD}_j - \bD_j \|_2^2$ for all $j=1,\cdots, p$ with a constant $C \in (0, 1/2)$, then  \eqref{eq:wasserstein-2} is satisfied with $C_{\boldsymbol{X}} = \frac{2}{1 - 2 C} \big(1 + \sqrt{2 n^{-1}}\big) (r^2 \lor 1)$. 
 	\end{lemma}
 	
 	The condition $\|\hat{\bD}_j\|_2 \|\bD_j \|_2 - \hat{\bD}_j ^T \bD_j \leq C \|\hat{\bD}_j - \bD_j \|_2^2$ can be satisfied if the covariates are close to independent, i.e., $\bOmega$ close to diagonal. In particular, when $\bOmega$ and $\hat{\bOmega}$ are both diagonal, it holds that $\hat{\bD}_j^T \bD_j - \| \hat{\bD}_j\|_2 \| \bD_j \|_2 = 0 \leq \|\hat{\bD}_j - \bD_j \|_2^2 $. We conjecture that for more general $\bOmega$ and $\hat{\bOmega}$, the coupling measure used for generating \eqref{sec4.1-1} and \eqref{sec4.1-2} could still satisfy  \eqref{eq:wasserstein-2}. Proving the existence of $\eta^*$ in the general scenario is highly challenging and left for future research.


\section{Layer 3 analysis: construction of coupled knockoff variables} \label{Sec.3} 


In this section, we present three specific constructions for the coupled perfect knockoff variables and verify that they satisfy Condition \ref{accuracy-knockoffs} with the desired convergence rate. 

\subsection{Knockoffs for multivariate $t$-distribution} \label{new-Sec4.1}

In this example, we will construct knockoffs for multivariate $t$-distributed features by leveraging only information of the first two moments; the knowledge of the $t$-distribution will \textit{not} be utilized in the approximate knockoffs construction. Assume that the underlying true feature distribution for $X = (X_1, \ldots, X_p)^{\top} $ is the multivariate centered $t$-distribution $t_{\nu} (\boldsymbol{0}, \bOmega^{-1})$ with unknown parameters $\nu$ and $\bOmega^{-1}$. We construct the approximate knockoff variables from the Gaussian distribution with the attempt to match the first two moments of feature vector $X$. It is seen that the working distribution $\hat F$ is misspecified. It has been a common practice to use the multivariate Gaussian distribution to construct knockoff variables in practice; see, e.g., \cite{CFJL2018, bai2021kimi}.

Assume that there is an effective estimator $ \hat\bTheta $ for the precision matrix $\bTheta := [\Cov(X)]^{-1} = \frac{\nu - 2}{\nu  } \bOmega $ constructed using data matrix $\bX$.
We construct the approximate knockoffs variable matrix $\hat{\bX}$ from the misspecified Gaussian distribution as
\begin{equation} \label{t-appro-knock}
	\hat{\bX} = \bX (I_p - r \hat\bTheta ) + \bZ (2 r I_p - r^2 \hat\bTheta)^{1/2}, 
\end{equation}
where $r$ is a constant such that $2 r I_p - r^2 \hat\bTheta $ is positive definite,  and $\bZ \in \mathbb{R}^{n \times p}$ is independent of $(\bX, \by)$ and consists of i.i.d. standard Gaussian entries. 

Before suggesting our coupled perfect knockoff variables, it is necessary to review some properties of the multivariate $t$-distribution. Note that an alternative representation of the $i$th row of $\bX$ is $\bx_i = \frac{ \eta_i } { \sqrt{Q_i / \nu} }$, where $\nu > 0$ is the degrees of freedom, $\eta_i \stackrel{d}{\sim} N({\bf 0}, \bOmega^{-1})$, $Q_i \stackrel{d}{\sim} \chi_{\nu}^2$, and $\eta_i \indep Q_i$.  Here, $\chi_{\nu}^2$ is the chi-square distribution with $\nu$ degrees of freedom. 
When $\nu$ is large, the distribution of $\bx_i$ is close to the Gaussian distribution $ N({\bf 0},  (\frac{\nu - 2}{\nu}\bOmega)^{-1}) $.  Using this alternative representation, the design matrix $\bX$ can be written as 
\begin{equation}\label{eq:t-alternative-rep}
	\bX  =\diag(\frac{1}{\sqrt{\bQ /\nu} })\boldsymbol{\eta}, 
\end{equation}
where $\boldsymbol{\eta}$ is the matrix with rows $\{\eta_i\}_{i=1}^n$, and  $\diag(\frac{1}{\sqrt{\bQ /\nu} }) = \diag( \frac{1} {\sqrt{Q_1 / \nu}}, \ldots, \frac{1} {\sqrt{Q_n / \nu}})$.

\ignore{\color{blue}We are ready to introduce our construction of the coupled perfect knockoff variable matrix:
	\begin{equation} \label{t-perf-knock}
		\widetilde{\bX} = \bX(I_p - r \bOmega) + \diag(\frac{1}{\sqrt{\bQ /\nu} } ) \bZ (2 r I_p - r^2 \bOmega)^{1/2},
	\end{equation}
	where $\diag(\frac{1}{\sqrt{\bQ /\nu} }) = \diag( \frac{1} {\sqrt{Q_1 / \nu}},\frac{1} {\sqrt{Q_2 / \nu}}, \ldots, \frac{1} {\sqrt{Q_n / \nu}})$ with $\{Q_i\}_{i = 1}^n$ i.i.d. random variables sampled from the conditional distribution $Q|X$. Let $\boldsymbol{\eta} = (\eta_1, \ldots, \eta_n)$ be sampled from the conditional distribution $\eta | X$, and $r $ and $\bZ$ the identical realizations to those used in \eqref{t-appro-knock}.  By construction, we can see that 
	\begin{equation*}
		\begin{split}
			(\bX, \widetilde{\bX}) 
			& = \diag(\frac{1}{\sqrt{\bQ /\nu} }) \big(\boldsymbol{\eta}, \boldsymbol{\eta}(I_p - r \bOmega) +   \bZ (2 r I_p - r^2 \bOmega)^{1/2} \big) \\
			& \stackrel{d}{=}  \diag(\frac{1}{\sqrt{\bQ /\nu} }) ( \boldsymbol{\eta}, \widetilde{\boldsymbol{\eta}} ),
		\end{split}
	\end{equation*}
	where $(\boldsymbol{\eta}, \widetilde{\boldsymbol{\eta}})$ have i.i.d. rows that follow a common Gaussian distribution $N({\bf 0}, \bSigma^{\augg})$ with
	\begin{equation} \label{eq-cov-aug}
		\bSigma^{\augg} = \begin{pmatrix}
			\bOmega^{-1} &  \bOmega^{-1} - r I_p \\
			\bOmega^{-1} - r I_p & \bOmega^{-1} 
		\end{pmatrix}.
	\end{equation} 
	Thus, this verifies that $\widetilde{\bX}$ forms a perfect knockoff data matrix for $\bX$. }

	
	We are ready to introduce our construction of the coupled perfect knockoff variable matrix  
	\begin{equation} \label{t-perf-knock}
		\widetilde{\bX} = \bX(I_p - r \bOmega) + \diag(\frac{1}{\sqrt{\bQ /\nu} } ) \bZ (2 r I_p - r^2 \bOmega)^{1/2},
	\end{equation}
	where $\bQ$, $\nu$, and $\bOmega$ are identical to the ones in \eqref{eq:t-alternative-rep}, and $r$ and $\bZ$  are identical to the ones in \eqref{t-appro-knock}. Thus, $\bZ$ is independent of $\bQ$ and $\boldsymbol{\eta}$. In view of \eqref{eq:t-alternative-rep}, we can see that 
	\begin{equation*}
		\begin{split}
			(\bX, \widetilde{\bX}) 
			& = \diag(\frac{1}{\sqrt{\bQ /\nu} }) \big(\boldsymbol{\eta}, \boldsymbol{\eta}(I_p - r \bOmega) +   \bZ (2 r I_p - r^2 \bOmega)^{1/2} \big) \\
			& {:=}  \diag(\frac{1}{\sqrt{\bQ /\nu} }) ( \boldsymbol{\eta}, \widetilde{\boldsymbol{\eta}} ),
		\end{split}
	\end{equation*}
	where $(\boldsymbol{\eta}, \widetilde{\boldsymbol{\eta}})$ have i.i.d. rows that follow the Gaussian distribution $N({\bf 0}, \bSigma^{\augg})$ with
	\begin{equation} \label{eq-cov-aug}
		\bSigma^{\augg} = \begin{pmatrix}
			\bOmega^{-1} &  \bOmega^{-1} - r I_p \\
			\bOmega^{-1} - r I_p & \bOmega^{-1} 
		\end{pmatrix}.
	\end{equation} 
	Thus, this verifies that $\widetilde{\bX}$ forms a perfect knockoff variable matrix for $\bX$.

	The proposition below verifies that the coupling assumption in Condition \ref{accuracy-knockoffs} holds. 
	
	\begin{proposition} \label{prop-t}
		Assume that $C_{l} \leq \| \bOmega^{-1}\|_2 \leq C_u$
		and $ \|(2 r I_p - r^2 \bOmega)^{-1} \|_2\leq C_u$ for some constants $C_u> 0$ and $C_{l} > 0$.  Assume further that $ \bOmega $ and $\hat\bTheta$ are both sparse in the sense that $ \max_{1 \leq j \leq p} (\| \bOmega_j \|_0 + \|\hat\bTheta_j \|_0 ) \leq \rho_n$ almost surely with $ \rho_n (n^{-1} \log p)^{1/2} \to 0 $ and $\rho_n \nu^{-1/2} \to 0$, and that there exists a constant $C > 0$ such that 
		\begin{equation} \label{prop-t-cond1}
			\mathbb{P}\big ( \| \hat\bTheta -  \bTheta \|_2  \geq C \rho_n  (n^{-1} \log p)^{1/2}   \big) \to 0.
		\end{equation}
		Then as $ \nu \geq 9 $ and $\log p = o(n^{1 - 4/\nu})$,  we have that for some constant $C > 0$,
		\begin{equation} \label{re-prop3}
			\mathbb{P} \bigg( \| \hat{\bX} - \widetilde{\bX} \|_{1,2} \leq C \big( \rho_n  (n^{-1} \log p)^{1/2} +  \nu^{-1/2} \big) \bigg) \to 1.
		\end{equation}
	\end{proposition}
	
	The assumed convergence rate of $\rho_n (n^{-1} \log p)^{1/2}$ for precision matrix estimation in \eqref{prop-t-cond1} has been verified in many existing works (e.g., \cite{Cai2011constrained}, \cite{fan2016overview}, and \cite{FanLv2016}) under the sparsity assumption. Proposition \ref{prop-t} above indicates that the knockoffs procedure can potentially achieve the asymptotic FDR control even when the working distribution is misspecified but with the first two moments matched.

	We next compare our results to those in \cite{Barber2020}. For simplicity, let us further assume that $\bOmega = I_p$ and is known. Then $X \stackrel{d}{\sim} t_{\nu} ({\bf 0}, I_p)$ and the constructed approximate knockoff variables $\hat X \stackrel{d}{\sim} N({\bf 0}, \frac{\nu} {\nu - 2} I_p)$. We set $r = 1$ in \eqref{t-appro-knock} and \eqref{t-perf-knock} when constructing the approximate and perfect knockoff matrices, and hence the augmented covariance matrix in \eqref{eq-cov-aug} is given by $\bSigma^{\augg} = I_{2p}$. In such case, Proposition \ref{prop-t} guarantees that 
	\begin{equation*} 
		\mathbb{P} \bigg( \max_{1 \leq j \leq p } n^{-1/2} \| \hat{\bX}_j - \widetilde{\bX}_j \|_2 \leq C \nu^{-1/2} \bigg) \to 1. 
	\end{equation*}
	This implies that Condition \ref{accuracy-knockoffs} is satisfied with $\Delta_n = C \nu^{-1/2} $. 
	Observe that $X_j = \frac{Z_j} {\sqrt{\mathcal{X}_{\nu}^2 / \nu}}$ with $Z_j \stackrel{d}{\sim} N(0, 1)$ and the denominator satisfies that for an absolute constant $C > 0$ and $ \nu \gg \log (np) $,
	$$ 
	\mathbb{P} \bigg( |\mathcal{X}_{\nu}^2 / \nu - 1 | \geq C \sqrt{ \nu^{-1} \log (np )  } \bigg) = O( (np)^{- C^2 / 8} ). 
	$$
	These indicate that the multivariate $t$-distribution is asymptotically close to the standard Gaussian distribution when $ \nu \gg \log (np) $. 
	Thus, under Conditions \ref{marginal_corr_condition4}--\ref{debiased-lasso-condition2} and \ref{debiased-lasso-condition4} for the setting of the linear model, if we construct the knockoff statistics as RCD based on the debiased Lasso, we can prove  similarly as Theorem \ref{thm-debiased-lasso-FDR} that 
	$$\limsup_{n \to \infty} \FDR \leq q,$$ 
	when $ \nu^{1/2} \gg s (\log p)^{1 + 1/\gamma}  $   and $\frac{s (\log p)^{3/2 + 1/\gamma} } {\sqrt n} \to 0 $ for some $0 < \gamma < 1$. 
	
	\cite{Barber2020} also derived an upper bound on the FDR inflation. Directly applying their result and calculating the KL divergence in their upper bound under the specific model setting stated above, we can obtain the lemma below. 
	\begin{lemma} \label{KL-multi-t}
		By applying Theorem 1 in \cite{Barber2020}, it requires at least $ \nu^2 \gg n  \min(n, p)$ for $\limsup_{n \to \infty} \FDR \leq q$. 
	\end{lemma}

	The intuition behind Lemma \ref{KL-multi-t} above is that Theorem 1 in \cite{Barber2020} requires the empirical KL divergence $\max_{j\in \mathcal{H}_0} \hat{KL}_j$ converging to zero in probability, where
	\begin{equation*}
		\begin{split}
			\hat{KL}_{j} & = \sum_{i = 1}^n \bigg[ \frac{\bX_{i, j} ^2 (\nu - 2)} {2 \nu} - \frac{\nu + p}{2} \log \bigg(1 + \frac{\bX_{i, j} ^2} {\nu +  \| \bX_{i, -j} \|_2^2} \bigg) \\
			& \quad - \bigg(\frac{\hat{\bX}_{i, j}^2 (\nu - 2)} {2 \nu } - \frac{\nu + p}{2} \log \bigg(1 + \frac{\hat{\bX}_{i, j}^2} {\nu +  \| \bX_{i, -j} \|_2^2} \bigg) \bigg)\bigg].
		\end{split}
	\end{equation*}
	Here, $\bX = (\bX_{i, j} ) \in \mathbb{R}^{n \times p}$ consists of i.i.d. rows sampled from $t_{\nu} ({\bf 0}, I_p)$, while $\hat{\bX} = (\hat{\bX}_{i, j})\in \mathbb{R}^{n \times p}$ consists of i.i.d. rows sampled from $N({\bf 0}, I_p)$. 
	As shown in the proof of Lemma \ref{KL-multi-t} in Section \ref{new.SecB.1} of the Supplementary Material, $\hat{KL}_j$ is a sum of i.i.d. random variables with positive mean of order $\frac{C p}{\nu (\nu + p)}$. Hence, $\hat{KL}_j$ is concentrated at $\frac{C n p}{\nu (\nu + p)}$ and to ensure that $\hat{KL}_j \stackrel{d}{\to} 0$, we need at least $  \frac{np}{\nu (\nu + p)} \to 0 $, or equivalently, $\nu^2 \gg n \min(n, p)$. Such condition is stronger than our requirement $ \nu^{1/2} \gg s (\log p)^{1 + 1/\gamma}   $ derived from the coupling technique when $s = o(\sqrt n)$ and $p \geq n$.

	
	\vspace{-0.2cm}
	\begin{table}[]
 	\caption{Summary of key conditions and results in Layer 3 analysis}
        \label{table-2-sumamry-layer3}
		\begin{tabular}{l|l|l|l}
			\hline
			Covariate distribution                                                              & $t_{\nu}(\bf{0},\bOmega^{-1})$                                                                  & $N(\bf 0,\bOmega^{-1})$                                                        & Nonparanormal                                                                \\
			\hline
			\begin{tabular}[c]{@{}l@{}}Source of error in\\ constructing $\hat\bX$\end{tabular} & \begin{tabular}[c]{@{}l@{}}Misspecified distribution\\ and estimated $\bOmega$\end{tabular} & Estimated $\bOmega$                                                       & Estimated $\bOmega$                                                          \\
			\hline
			\begin{tabular}[c]{@{}l@{}}Verified coupling\\ rate $\Delta_n$\end{tabular}         & $ \rho_n  (n^{-1} \log p)^{1/2} +  \nu^{-1/2}$                                        & $\rho_n\sqrt{\frac{\log p} {n}}$ & $\rho_n \sqrt{\frac{\log p}{n}} + \sqrt{  \frac{ p \rho_n (\log n)^3} {n} }$ \\
			\hline 
			$p\gg n$? & Yes & Yes & No\\
			\hline
			\begin{tabular}[c]{@{}l@{}}Marginal correlation\\ knockoff statistics\end{tabular}  & \multicolumn{3}{l}{\begin{tabular}[c]{@{}l@{}} Out-sample estimation needed for general $\bOmega$;\\ In-sample estimation allowed for diagonal $\bOmega$\end{tabular}}                                                    \\
			\hline
			\begin{tabular}[c]{@{}l@{}}RCD with \\ debiased Lasso\end{tabular}                  & \multicolumn{3}{l}{In-sample estimation allowed for general $\bOmega$ with sparsity} \\                                                                          \hline               
		\end{tabular}
	
  \vspace{-0.1cm}
	\end{table}

	\subsection{Gaussian knockoffs} \label{new.Sec4.2}
	
	We now study the commonly used example of Gaussian knockoffs with the correctly specified distribution family. Assume that feature vector $X = (X_1, \ldots, X_p)^{\top}\stackrel{d}{\sim} N(\textbf{0}, \bOmega^{-1})$ with unknown precision matrix $\bOmega$, and we have an effective estimate $\hat{\bOmega}$ that may be constructed using in-sample observations. A popularly used approximate knockoff variable matrix is
	\begin{equation} \label{sec4.1-1}
		\hat{\bX} = \bX (I_p - r \hat{\bOmega}) + \bZ (2 r I_p - r^2 \hat{\bOmega})^{1/2}, 
	\end{equation}
	where $r > 0$ is some constant such that $2 r I_p - r^2 \hat{\bOmega}$ is positive definite, and $\bZ = (Z_{i, j}) \in \mathbb{R}^{n \times p}$ is independent of $(\bX, \by)$ with i.i.d. entries $Z_{i,j} \stackrel{d}{\sim} N(0, 1)$. Note that the approximate knockoff variable matrix in \eqref{sec4.1-1} uses the correctly specified distribution family for $\bX$ (i.e., the Gaussian distribution).
	
	We couple  $\hat\bX$ with the perfect knockoff variable matrix
	\begin{equation} \label{sec4.1-2}
		\widetilde{\bX}  = \bX (I_p - r  \bOmega) + \bZ (2 r I_p - r^2 \bOmega)^{1/2},
	\end{equation}
	where importantly, $\bZ$ and $r$ are identical to those  used in constructing $\hat{\bX}$. We present the result below regarding the accuracy of the approximate knockoff variables.
	\begin{proposition} \label{prop-gaussian}
		Assume that $C_{l} \leq \| \bOmega^{-1}\|_2 \leq C_u$
		and $ \|(2 r I_p - r^2 \bOmega)^{-1} \|_2 \leq C_u$ for some constants $C_u> 0$ and $C_{l} > 0$.  Assume further that precision matrix $\bOmega$ and its estimator $\hat{\bOmega}$ are both sparse in the sense that $ \max_{1 \leq j \leq p} \| (\bOmega_j \|_0 + \| \hat{\bOmega}_j \|_0) \leq   \rho_n$ almost surely with  $\rho_n (n^{-1} \log p )^{1/2} \to 0$, and that there exists a constant $C > 0$ such that  
		\begin{equation} \label{eq-5}
			\mathbb{P} \big(\| \hat{\bOmega} - \bOmega\|_2 \geq C \rho_n (n^{-1} \log p )^{1/2} \big) \to 0.
		\end{equation}
		Then we have that for some constant $C > 0$, 
		\begin{equation}\label{eq:Gaussian-knockoff-variable-matching}
			\mathbb{P} \big(\| \hat{\bX} - \widetilde{\bX} \|_{1,2} \leq C  \rho_n (n^{-1} \log p )^{1/2} \big)\to 1.
		\end{equation}
	\end{proposition}

	Proposition \ref{prop-gaussian} above implies that Condition \ref{accuracy-knockoffs} is satisfied with coupling accuracy $\Delta_n = C \rho_n (n^{-1} \log p )^{1/2}$, where $\rho_n$ represents the sparsity level of $\bOmega$ and its estimator. We discuss the implication on FDR control utilizing the previously studied two knockoff statistics, namely the marginal correlation and RCD statistics, by applying Theorems \ref{thm-marginal-corr-fdr}--\ref{thm-debiased-lasso-FDR}, and then compare with the relevant results in \cite{Barber2020}.
	
	First consider the linear model and the RCD knockoff statistics based on the debiased Lasso.  It follows from Theorem \ref{thm-debiased-lasso-FDR} that under Conditions \ref{marginal_corr_condition4}--\ref{debiased-lasso-condition2} and \ref{debiased-lasso-condition4}, we have $\limsup_{n \to \infty} \FDR \leq q$ provided that $ s \rho_n (\log p )^{3/2 + 1/\gamma} = o(\sqrt {n})$ for some $0 < \gamma < 1$. Our technical analyses do \textit{not} require data splitting or an independent pretraining sample.  In comparison, the results in \cite{Barber2020} require an independent unlabeled pretraining data set with sample size $N$ to estimate the unknown precision matrix. Specific to the model setting considered in this section, 
	their results indicate that $\limsup_{n \to \infty} \FDR \leq q$ when $N\gg n  \rho_n (\log p )^2$. This again shows the advantage of our coupling technique in the robustness analyses. 
	
	
	Next we move to the marginal correlation statistics. In view of \eqref{eq:Gaussian-knockoff-variable-matching}, \eqref{marginal-w-closeness}, and Theorem \ref{thm-marginal-corr-fdr}, it is seen that in-sample estimation generally cannot meet the required condition of $\sqrt n \Delta_n(\log p)^{1/2+1/\gamma} \rightarrow 0$ in Theorem \ref{thm-marginal-corr-fdr}, and hence there is no guarantee of asymptotic FDR control even using our coupling idea. This message is consistent with \cite{Reconciling2024}, where the model-X 
 framework for conditional independence test is investigated; see Section \ref{sec:comp} for more detailed discussion. In the special case of independent features as discussed at the end of Section \ref{new.Sec3.2}, in-sample estimation can achieve asymptotic FDR control if $ (\log p)^{1+1/\gamma}\max_j|\hat\sigma_j^{-2} -\sigma_j^{-2}| = o_p(1)$, where $\hat\sigma_j^2$ and $\sigma_j^2$ are estimated and true variance for the $j$th feature, respectively.  
	
	We next compare with the relevant results in \cite{Barber2020} for marginal correlation statistics. It is discussed in their Section 3.2.1 that the KL divergence in their FDR inflation upper bound can be replaced with some $E_j$ defined on summary statistics, such as
	\begin{align}
		E_j= E_j(\bX_j^T\by, \hat\bX_j^T\by), \text{ with } E_j(a,b)=\log \Big(\frac{\mathbb P((\bX_j^T\by, \hat\bX_j^T\by)=(a,b)| \bX_{-j},\hat \bX_{-j}, \by)}{\mathbb P((\bX_j^T\by, \hat\bX_j^T\by)=(b,a)| \bX_{-j},\hat \bX_{-j}, \by)}\Big),
	\end{align}
	and that their FDR inflation upper bound remains to hold. An independent pretraining sample is required for generating their $\hat\bX_j$'s. Note that $E_j$ above depends on the ``closeness" of $\hat{\bX}_j^T\by$ to $\bX_j^T\by$. For the FDR inflation in their upper bound to asymptotically vanish,  it is required that $\max_j |E_j| = o_p(1)$. 
	It is unclear how $\max_j |E_j| = o_p(1)$  can be translated into the explicit bound on the estimation accuracy of $\bOmega$ when the covariate dependence is most general. In the simpler case of independent covariates, the condition reduces to $\max_j |E_j| = O_p( (\log p) \max_{1 \leq j \leq p } |\hat\sigma_j^{-2} - \sigma_j^{-2} | ) $. Comparing to our condition of $ (\log p)^{1+1/\gamma}\max_j|\hat\sigma_j^{-2} -\sigma_j^{-2}| = o_p(1)$ discussed above, the additional term of $(\log p) ^{1/\gamma}$ is the price we pay for in-sample estimation.
	
	\subsection{Nonparanormal knockoffs} \label{new.Sec4.3}
	
	We further investigate a much more general distribution family, that is, the Gaussian copula distributions. Assume that $X = (X_1, \ldots, X_p)^{\top} $ has marginal distributions $X_j \stackrel{d}{\sim} F_j(\cdot)$ and satisfies that $(\Phi^{-1} (F_1(X_1)), \ldots, \Phi^{-1}(F_p(X_p)))^{\top} \stackrel{d}{\sim} N({\bf 0}, \bOmega^{-1})$, where the diagonal entries of $\bOmega^{-1}$ are all one. Further assume that we have effective estimators $\hat{F}_j$ for $ F_j$ and $\hat{\bOmega}$ for $\bOmega$.
	Define $ \hat{\bV} = (\hat{\bV}_{i, j}) \in \mathbb{R}^{n\times p}$ with $ \hat{\bV}_{i,j} = \Phi^{-1} (\hat{F}_j (\bX_{i, j}) ) $ and $ \widetilde{\bV} = (\widetilde{\bV}_{i, j}) \in \mathbb{R}^{n\times p}$ with $ \widetilde{\bV}_{i,j} = \Phi^{-1} ({F}_j (\bX_{i, j}) ) $. Let $\hat{\bU} = (\hat{\bU}_{i, j}) \in \mathbb{R}^{n \times p}$ be given by
	\begin{equation} \label{eq-6}
		\hat{\bU} =  \hat{\bV} (I_p - r \hat{\bOmega}) + \bZ (2 r I_p - r^2 \hat{\bOmega})^{1/2},
	\end{equation}
	where $r > 0$ is some constant such that $2 r I_p - r^2 \hat{\bOmega}$ is positive definite, and $\bZ = (\bZ_{i, j}) \in \mathbb{R}^{n \times p}$ is independent of $(\bX, \by)$ with i.i.d. entries  $Z_{i,j} \stackrel{d}{\sim} N(0, 1)$. We construct the approximate knockoff variable matrix as $ \hat{\bX} = (\hat{\bX}_{i, j}) \in \mathbb{R}^{n \times p}$ with 
	\begin{equation}
		\hat{\bX}_{i, j} =  \hat{F}_j^{-1} (\Phi(\hat{\bU}_{i, j})). 
	\end{equation}
	It is seen that this example also uses the correctly specified distribution family for $X$, i.e., the Gaussian copula.

	We suggest to construct the coupled perfect knockoff variable matrix as $\widetilde \bX = (\widetilde{\bX}_{ij})$ with
	\begin{equation} \label{gaussian-copula-knockoffs}
		\widetilde{\bX}_{i,j} = F_{j}^{-1} (\Phi(\widetilde{\bU}_{i,j})),
	\end{equation}
	where $\widetilde{\bU}_{i,j}$ represents the $(i,j)$th entry of matrix 
	\begin{equation}
		\widetilde{\bU} = \widetilde{\bV} (I_p - r \bOmega) + \bZ (2 r I_p - r^2 \bOmega)^{1/2}
	\end{equation}
	with $\bZ$ and $r$ identical in values to the ones used in \eqref{eq-6}.
	The proposition below characterizes the coupling rate between  $\hat{\bX}$ and $\widetilde{\bX}$. 
	
	\begin{proposition}  \label{prop-gaussian-copula}
		Assume that \eqref{eq-5} is satisfied and both $\bOmega$ and $\hat{\bOmega}$ are sparse in the sense that $\max_{1\leq j \leq p} (\|\bOmega_j \|_0 +\|\hat{\bOmega}_j \|_0)   \leq \rho_n$ with $ p \rho_n = o(n /(\log n)^3)$ almost surely. 
		Assume further that for $1 \leq j \leq p$, the distribution estimators satisfy $\frac{1}{2n} \leq \hat{F}_j (x) \leq 1 - \frac{1}{2n} $ for each $x \in \supp(X_j)$, $ \supp(X_{j}) \subset [-b, b]$ for some constant $b >0$, and there exists a constant $M > 0$ such that
		\begin{equation} \label{p2-cond3}
			\mathbb{P} \Big( \max_{1 \leq j \leq p} \sup_{ x  \in [ 2M n^{-1} \log n, 1 - 2M n^{-1} \log n] } \big| \hat{F}_j^{-1} (x) -  {F}_j^{-1} (x) \big|   \geq   (M n^{-1 }  \log n )^{1/2}  \Big) \to 0,
		\end{equation}
		\begin{equation} \label{p2-cond4}
			\begin{split}
				&\mathbb{P} \Big( \max_{1 \leq j \leq p} \sup_{x \in (F_j^{-1}(2M n^{-1} \log n), F_j^{-1} (1 - 2M n^{-1} \log n) )} \frac{ | \hat{F}_j(x) - F_j(x) | } { F_j(x) [1 - F_j(x)] } \geq   (M n^{-1 } \log n )^{1/2} \Big) \\
				&\quad \to 0,
			\end{split}
		\end{equation}
		\begin{equation} \label{p2-cond2}
			\mathbb{P} \Big( \max_{1 \leq j \leq p} \sup_{ x, y \in (0, 1) } \frac{\big| \hat{F}_j^{-1} (x) - \hat{F}_j^{-1} (y) \big| } {  |x - y| +  ( n^{-1 } (\log n) |x - y| )^{1/2} +    n^{-1} \log n } \geq M   \Big) \to 0.
		\end{equation}
		Then we have 
		\begin{equation}
			\mathbb{P} \bigg(  \| \hat{\bX} - \widetilde{\bX} \|_{1,2} \leq C \Big( \rho_n \sqrt{\frac{\log p}{n}} + \sqrt{  \frac{ p \rho_n (\log n)^3} {n} } \Big) \bigg) \to 1.
		\end{equation}    
	\end{proposition}

	\begin{remark}
		When estimators $\{\hat{F}_j\}_{j = 1}^p$ are the empirical distribution functions and $p = o(n)$, it can be shown that \eqref{p2-cond3}, \eqref{p2-cond4}, and \eqref{p2-cond2} can be satisfied when the density function $f_{X_j}$ is uniformly bounded on the support. 
	\end{remark}
	
	See, e.g., \cite{liu2009nonparanormal, LiuHanYuanLaffertyWasserman2012} for the estimation of nonparanormal distributions, and we opt not to discuss it here due to the space constraint. We also remark that the bounded support assumption of $ \supp(X_{j}) \subset [-b, b]$ is to simplify the technical proofs and may be removed by applying the truncation technique and letting $b$ slowly diverge with $n$. Since such technical relaxation is not the main focus of the current paper, we choose not to explore it here.
	
	\section{Robust knockoffs for $k$-FWER control}\label{sec:k-FWER}
	
	Model-X knockoffs framework has also been explored for the purpose of $k$-FWER control \citep{lehmannRomano2005}, where the goal is to control
	\begin{equation} \label{eq-FWER}
		k\text{-}\FWER = \mathbb{P} ( |\hat{S} \cap \mathcal{H}_0| \geq k ) 
	\end{equation}
below a prespecified target level $q\in (0,1)$.     
	Given the approximate knockoff statistics $\{\hat{W}_j\}_{j=1}^p$, the set of selected features is
	$\hat{S} = \{1 \leq j \leq p: \hat{W}_j \geq T_v \}$, where the threshold is defined as 
	\begin{equation} \label{eq-3}
		{T}_v = \sup \Big\{t \in \hat{\mathcal{W}}: \# \{j: - \hat{W}_j \geq t\} = v \Big\}
	\end{equation} 
	with $v$ the largest integer such that
	\begin{equation} \label{eq-4}
		\sum_{i = k}^{\infty} 2^{-(i+v)} {i+v-1 \choose i} \leq q.
	\end{equation}

	When the true feature distribution is known,  \cite{JS2016} showed that the perfect knockoffs inference procedure provides precise finite-sample control on the $k$-FWER.  We now establish the companion theory for the approximate knockoffs inference procedure. 
	
	
	Denote by $\hat{V} = | \hat{S} \cap \mathcal{H}_0|$ the number of false discoveries. Similar to the FDR analysis, we assume that the number of relevant features $|\mathcal H_1|\rightarrow \infty$ as $n\rightarrow \infty$. Further, we consider the scenario where $k$ diverges very slowly with $n$. Our layer 1 theory will again build on the key Condition \ref{fdr-condition1} that there exist coupled perfect knockoff statistics that are sufficiently close to the approximate knockoff statistics. However, different from the FDR study where Conditions \ref{fdr-condition2}--\ref{fdr-condition5} are needed, we assume instead the two technical conditions below and their interpretations are similar to Conditions \ref{fdr-condition4}--\ref{fdr-condition5}. Recall the definition that $G(t) = p_0^{-1} \sum_{j \in \mathcal{H}_0} \mathbb{P}(\widetilde{W}_j \geq t)$ and $p_0 = |\mathcal{H}_0|$.
	
	\begin{condition}[Weak dependence] \label{fwer-condition1}
		For constants $ 0 < \gamma < 1$ and $C > 0$, and a positive sequence $m_n = o(k)$,  it holds that 
		\begin{equation} \label{condition4-part1}
			\Var{ \Big( \sum_{j \in \mathcal{H}_0} \mathbbm{1} (\widetilde{W}_j > t)  } \Big) \leq C m_n p_0 G(t) +  o\big( (\log k )^{ - 1/\gamma}  [p_0  G (t)]^2  \big)  
		\end{equation}
		uniformly over $t \in (G^{-1} ( \frac {3 k  } { 2 p } ),\, G^{-1} ( \frac { k  } { 2 p } ) )$.
	\end{condition}	
	
	\begin{condition} \label{kFWER-cond1} 
		Assume that $G(t)$ is a continuous function. It holds that as $n \to \infty$,
		\begin{align}
			\sup_{t \in \big( G^{-1} (\frac {3k} {2p}),  G^{-1} (\frac {k} {2p})\big) } \frac {G (t - b_n) - G(t + b_n)} {G(t)} \to 0  \label{kWER-cond1-1}
			\intertext{and}
			k^{-1} \sum_{j \in \mathcal{H}_1} \mathbb{P} \Big(\widetilde{W}_j < - G^{-1} (\frac {3 k } {2 p}) \Big) \to 0  \label{KWER-cond1-2}
		\end{align}
	\end{condition}
	
	Now we are ready to present our general theorem on the $k$-FWER control for the approximate knockoffs procedure. 
	\begin{theorem}  \label{thm-FWER}
		Assume that Conditions \ref{fdr-condition1}, \ref{fwer-condition1}, and \ref{kFWER-cond1} are satisfied, $k \to \infty$, and $m_n / k \to 0 $ as $n\rightarrow\infty$. Then for each $\varepsilon > 0$, we have
		\begin{equation} \label{eq11.new}
			\limsup_{n \to \infty}\, \mathbb{P} ( \hat{V} \geq k(1 + \varepsilon)  ) \leq q.
		\end{equation}
	\end{theorem}
	
	The main idea for proving Theorem \ref{thm-FWER} is to compare the approximate knockoff statistics $\{\hat{W}_j\}_{j=1}^p$ with their coupled perfect counterparts $\{ \widetilde{W}_j\}_{j=1}^p$ and show that the approximate threshold $T_v$ satisfies $|T_v - \widetilde{T}_v| \leq b_n$ as long as $\max_{1 \leq j \leq p }|\hat{W}_j - \widetilde{W}_j| \leq b_n$, where $\widetilde{T}_v$ is the corresponding threshold from the perfect knockoff statistics. Moreover, we can show that  for each $\veps > 0$,  with high probability it holds that $ \widetilde{T}_{v + M_v + 1} <   \widetilde{T}_v - 2b_n \leq \widetilde{T}_{v + M_v} $ for some integer $M_v \leq  k \veps$. Therefore, the probability of the approximate knockoffs inference procedure making at least $k$ false discoveries can be related to that of the $k$-FWER control with the perfect knockoff statistics, which establishes the desired result in Theorem \ref{thm-FWER}. 
	
	Similar to the layer 2 FDR analysis in Section \ref{Sec.4}, we showcase the general theory using two constructions of the knockoff statistics: the marginal correlation  and the RCD knockoff statistics, under the coupling accuracy assumption in Condition \ref{accuracy-knockoffs}. 
	
	With the marginal correlation knockoff statistics, under the same model setting of Section \ref{new.Sec3.2},  the following result on the $k$-FWER control can be established.
	\begin{theorem} \label{thm-marginal-corr-kFWER}
		Assume the same model setting \eqref{nonpara-model} as in Section \ref{new.Sec3.2} and the marginal correlation knockoff statistics \eqref{hat-marginal-correlation-new-eq}. Further, assume that Conditions \ref{accuracy-knockoffs}, \ref{marginal_corr_condition2}, and \ref{marginal_corr_condition4}  are satisfied, $k \to \infty$, $m_n / k \to 0$, and $  \Delta_n \sqrt{n \log p}   \to 0 $.
		Then for each $\varepsilon > 0$, we have
		\begin{equation*} 
			\limsup_{n  \to \infty}\, \mathbb{P} ( \hat{V} \geq k(1 + \varepsilon)  ) \leq q.
		\end{equation*}
	\end{theorem}
	
	Analogously, with the RCD knockoff statistics, under the same setting of Section \ref{new.Sec3.3}, we have the parallel theorem for the $k$-FWER control below. 
	\begin{theorem} \label{thm-debiased-lasso-FWER} 
		Assume the same linear model setting as in Section \ref{new.Sec3.3} and the RCD knockoff statistics \eqref{hat_RCD_knockoff_stat-new}.  Further, assume that Conditions \ref{accuracy-knockoffs},  \ref{marginal_corr_condition4}, and \ref{debiased-lasso-condition1}--\ref{debiased-lasso-condition3} are satisfied, $k \to \infty$, $m_n / k \to 0$, and $  \frac{ m_n^{1/2} s (\log p)^{3/2} (\log k)^{1/\gamma}} {\sqrt n} + \Delta_n s \log p  \to 0$ for some constant $0 < \gamma < 1$. Then for each $\varepsilon > 0$, we have
		\begin{equation} \label{eq11.new}
			\limsup_{n \to \infty}\, \mathbb{P} ( \hat{V} \geq k(1 + \varepsilon)  ) \leq q.
		\end{equation}
	\end{theorem}


	\section{Connection with literature}\label{sec:comp}
	
	We now provide more detailed comparison with three additional existing works  \cite{FDLL2020}; \cite{FLSU2020}; \cite{Reconciling2024}. 
	
	\cite{FLSU2020} investigated the power and robustness of knockoffs inference in the  linear model setting where the features follow a latent factor model with parametric idiosyncratic noise. In-sample estimation is allowed for constructing their approximate knockoff variables. Condition 4 therein for robustness analysis is essentially a preliminary form of our knockoff variable coupling condition under their parametric model assumption, and Condition 6 therein is loosely comparable to the proved results in our Theorem \ref{theorem-FDR}; these two conditions are model specific and directly assumed therein without theoretical justification. \cite{FDLL2020} provided theoretical guarantee for the asymptotic FDR control for the approximate knockoffs procedure under an assumption that the FDR function is Lipschitz with respect to feature covariance matrix when the feature distribution is jointly Gaussian. In their paper, the feature distribution and model sparsity are learned by balanced sample splitting, and the dependence of response $Y$ on covariates in $X$ can be  nonlinear and arbitrary. Their Lipschitz assumption on FDR function is comparable to the proved results in our Theorem \ref{theorem-FDR}.  
	
	\cite{Reconciling2024} studied the robustness of the conditional randomization test
	(CRT) and demonstrated that, when the feature distribution is learned in sample, type-I error control cannot be attained for arbitrary test statistics. In their Section 3, a test statistic that is closely related to the marginal correlation test was investigated and it was shown that its type-I error can be arbitrarily inflated when in-sample feature distribution is learned. This message is similar to ours in the sense that marginal correlation statistics have low accuracy (see Section \ref{new.Sec3.2}).   \cite{Reconciling2024} also established an interesting double-robustness phenomenon: errors in fitting the distribution of the features can be compensated for by using a test statistic that more accurately captures the distribution of the response given the features. Since for FDR or $k$-FWER control, there is only one source of error caused by estimated/misspecified covariate distribution, the double-robustness may not be a relevant property in our study. 
	
	Comparing to these existing works, a major innovation of our paper is the introduction of a new closeness measure for evaluating the qualities of the approximate knockoff variables and knockoff statistics. This new measure is closely related to the $(1,2)$-Wasserstein distance. The coupling idea for knockoffs robustness analysis and the $(1,2)$-Wasserstein distance are both new to the literature; they equip us with a much more powerful tool for better understanding the practical robustness of the model-X framework. Indeed, as revealed in our analysis,  the robustness of model-X procedure goes beyond the scenarios already revealed in the literature.  
	The connection to the $(1,2)$-Wasserstein distance also suggests that the robustness of model-X can be a general phenomenon beyond the covariate distribution examples provided in our current paper. 
	
	There exist some other less related works in the literature that contribute to relaxing the assumption of fully known feature distribution in the model-X knockoffs framework.  
	For instance, \cite{huang2020relaxing} relaxed such assumption via assuming the existence of sufficient statistic for the model and proposing an alternative conditional exchangeability for knockoffs given the sufficient statistic.

	\section{Simulation studies}\label{Sec:Simu}
	In this section, we examine the finite-sample performance of the approximate knockoffs inference using the approximate or misspecified feature distribution through some simulation examples.

	\subsection{Approximate feature distribution} \label{Sec:Simu_sub1}
	Our first simulation example considers Gaussian feature vector $ X \stackrel{d}{\sim} N({\bf 0}, \boldsymbol{\Omega}^{-1})$, where the precision matrix $\bOmega = (\omega_{ij}) \in \mathbb{R}^{p \times p}$ is unknown and sparse with entries  $\omega_{ij} = 0.2^{|i - j|}$ for $|i-j| < 10$ and $\omega_{ij} = 0$ for $|i-j| \geq 10$. We apply the James--Stein-type shrinkage estimator for the covariance matrix (as in the R Package `knockoff') and examine the FDR control of the approximate knockoffs inference procedure with estimated covariance matrix. In-sample estimation is used for learning the feature covariance matrix.  We consider two settings: the linear regression model and logistic  regression model. 
	\begin{setting} \label{simu-linear}
		Assume that $Y = X \bbeta + \varepsilon$, where $\varepsilon  $ is a random error with $\varepsilon \stackrel{d}{\sim} N(0, 1)$. 
		Let the coefficient $\bbeta \in \mathbb{R}^p$ be sparse with 50 nonzero components, where the nonzero locations are randomly selected and each nonzero coefficient is randomly generated from $\{\pm 3 \}$. 
	\end{setting}
	
	\begin{setting} \label{simu-logistic}
		Assume that the response $Y$ depends on $X$ through a logistic regression model. Let the regression coefficient $\bbeta \in \mathbb{R}^p$ be sparse with 30 nonzero components, where the nonzero locations are randomly selected and each nonzero coefficient is randomly generated from $\{\pm 3\}$. 
	\end{setting}
	
	We consider the construction of knockoff statistics using the debiased Lasso regression coefficient difference. We set $p = 400$ and $n \in  \{150, 250, 350, 500\} $. From the numerical results in Table \ref{table-simu-approx_2}, it is seen that for a few settings of sample size, the FDR is marginally inflated above the target level $q = 0.2$ due to the estimated feature distribution and Monte Carlo error. 
	Overall, the approximate knockoffs inference procedure demonstrates robust FDR control across various values of sample size $n$, which verifies our theoretical analysis that the knockoff statistics based on the debiased Lasso regression coefficient difference can guarantee the asymptotic FDR control with in-sample learned feature distribution.

	\ignore{
		\begin{table}[htp]
			\label{table-simu-approx}
			\small
			\centering
			\begin{tabular}{ c | c c c c || c | c c c c} 
				\hline
				\multicolumn{5}{c||}{Setting \ref{simu-linear}} & \multicolumn{5}{c}{Setting \ref{simu-logistic}} \\
				\hline
				$n$ & 150 & 250 & 500 & 800 & $n$ & 150 & 250 & 500 & 800  \\ 
				\hline
				Lasso RCD & 0.171  &   0.200 &  0.201  & 0.194   & Lasso RCD & 0.096  
				&   0.184  &   0.191  &  0.190 \\ 
				D-Lasso RCD & 0.167 & 0.233 & 0.215  & 0.221 & D-Lasso RCD  & 0.131  &  0.177  & 0.197 & 0.206 \\ 
				\hline 
			\end{tabular}
			\caption{FDR control for the approximate knockoffs procedure using estimated feature distribution under Settings \ref{simu-linear} and \ref{simu-logistic}, with a targeted FDR level $q = 0.2$. Results are based on 100 replications. }
		\end{table}
	}
	
	\vspace{-0.2cm}
	\begin{table}[htp]
 	\caption{FDR control for the approximate knockoffs procedure using estimated feature distribution under Settings \ref{simu-linear} and \ref{simu-logistic}, with a targeted FDR level $q = 0.2$. Results are based on 100 replications.}
		\label{table-simu-approx_2}
		\begin{tabular}{ c | c c c c || c | c c c c} 
			\hline
			\multicolumn{5}{c||}{Setting \ref{simu-linear}} & \multicolumn{5}{c}{Setting \ref{simu-logistic}} \\
			\hline
			$n$ & 150 & 250 & 350 & 500 & $n$ & 150 & 250 & 350 & 500  \\ 
			\hline
			FDR &  0.186  & 0.211  & 0.203  & 0.189  & FDR  &  0.142  & 0.205  & 0.207  & 0.205 \\ 
			\hline 
		\end{tabular}
	
	\end{table}
	\vspace{-0.4cm}

	\subsection{Misspecified feature distribution}
	In the second simulation example, we consider a feature vector $X \in \mathbb{R}^p$ generated from a multivariate $t$-distribution $t_{\nu} ({\bf {0}, \bSigma})$ with covariance matrix  $\boldsymbol\Sigma = (\sigma_{ij}) \in \mathbb{R}^{p \times p}$ and $\sigma_{ij} = 0.5^{|i - j|}$. To examine the effect of misspecified feature distribution, we generate knockoff variables using the misspecified Gaussian distribution $N({\bf 0}, \frac{\nu}{\nu - 2}\bSigma)$ with matched first two moments and explore the FDR control of approximate knockoffs inference procedure as the number of degrees of freedom $\nu$ changes. We fix the sample size as $n = 300$ and the dimensionality as $p = 400$. Again the linear model in Setting \ref{simu-linear} and logistic model in Setting \ref{simu-logistic} are considered. We investigate the FDR control using knockoff statistics constructed from the debiased Lasso coefficient difference. 
	
	The number of degrees of freedom $\nu$ determines the closeness between the approximate and coupled perfect knockoff procedures, as demonstrated in layer 3 analysis in Section \ref{new-Sec4.1}. We examine the behavior of the approximate knockoffs procedure for $\nu \in \{5, 10, 20, 50 \}$.
	It is observed from Table \ref{table-simu-misspe} that the approximate knockoffs procedure can have slightly inflated FDR for a small value of $\nu = 5$, while achieving desired FDR control almost always for larger values of $\nu = 10$, $20$, and $50$. This again verifies our theoretical analysis. 
	\vspace{-0.2cm}
	\begin{table}[htp]
 	\caption{FDR control for the approximate knockoffs procedure using misspecified feature distribution under Settings \ref{simu-linear} and \ref{simu-logistic}, with a targeted FDR level $q = 0.2$. Results are based on 100 replications. }
		\label{table-simu-misspe}	
		
		\begin{tabular}{ c | c c c c || c | c c c c} 
			\hline
			\multicolumn{5}{c||}{Setting \ref{simu-linear}} & \multicolumn{5}{c}{Setting \ref{simu-logistic}} \\
			\hline
			$\nu$ & 5 & 10 & 20 & 50 &  $\nu$ & 5 & 10 & 20 & 50 \\ 
			\hline
			FDR & 0.238 & 0.190 & 0.206 & 0.195 & FDR & 0.175 & 0.162 & 0.186 &0.169  \\
			\hline
			\hline
		\end{tabular}	
	\end{table}
\vspace{-0.3cm}

	\ignore{ 
		\begin{table}[htp]
			\small
			\centering
			\begin{tabular}{ c | c c c c || c | c c c c} 
				\hline
				\multicolumn{5}{c||}{Setting \ref{simu-linear}} & \multicolumn{5}{c}{Setting \ref{simu-logistic}} \\
				\hline
				$\nu$ & 5 & 10 & 20 & 50 &  $\nu$ & 5 & 10 & 20 & 50 \\ 
				\hline
				FDR & 0.247 & 0.211 & 0.206 & 0.196 & FDR & 0.178 & 0.185 & 0.177 & 0.197 \\
				\hline
				\hline
			\end{tabular}
			\caption{FDR control for the approximate knockoffs procedure using misspecified feature distribution under Settings \ref{simu-linear} and \ref{simu-logistic}, with a targeted FDR level $q = 0.2$. Results are based on 100 replications. $\rho = 0.2$}
			
		\end{table}
	}	

	\section{Discussions} \label{Sec.disc}
	
	We have investigated in this paper the robustness of the model-X knockoffs framework introduced in \cite{CFJL2018} by characterizing the feature selection performance of the approximate knockoffs (ARK) procedure, a popularly implemented version of the model-X knockoffs framework in practice.  The approximate knockoffs procedure differs from the model-X knockoffs procedure in that it uses the misspecified or estimated feature distribution to generate the knockoff variables \textit{without} the use of sample splitting. We have proved formally that the approximate knockoffs procedure can achieve the asymptotic FDR and $k$-FWER control as the sample size diverges in the high-dimensional setting. A key idea empowering our technical analysis is coupling, where we pair statistics in the approximate knockoffs procedure with those in the model-X knockoffs procedure so that they are close in realizations with high probability. The knockoff variable coupling has been investigated under some specific distribution assumptions in the current work. An interesting future study is to investigate the coupling idea under a broader class of or even general feature distributions.

\begin{acks}[Acknowledgments]
The authors would like to thank the anonymous referees, an Associate Editor, and the Editor for their constructive comments that improved the quality of this paper.
\end{acks}

\begin{funding}
YF was supported in part by NIH Grant 1R01GM131407 and NSF grant 2310981. JL was supported in part by NSF Grants EF-2125142 and DMS-2324490.
\end{funding}

\begin{supplement}
\stitle{Supplement to ``ARK: Robust Knockoffs Inference with Coupling"}
\sdescription{The Supplementary Material \cite{FGL2024} contains all the proofs and technical details, and an extension of the analysis in Section \ref{new.Sec3.3} to the setting of the generalized linear model.}
\end{supplement}



\bibliographystyle{imsart-nameyear} 
\bibliography{references}       

\newpage

	\appendix
	\setcounter{page}{1}
	\setcounter{section}{0}
	\renewcommand{\theequation}{A.\arabic{equation}}
	\renewcommand{\thesubsection}{A.\arabic{subsection}}
	\setcounter{equation}{0}
	
	\begin{center}{\bf \large  Supplement to ``ARK: Robust Knockoffs Inference with Coupling''}
		
		\bigskip
		
		Yingying Fan, Lan Gao and Jinchi Lv 
		\medskip
	\end{center}

	\noindent  This Supplementary Material contains the proofs of Theorems \ref{theorem-FDR}--\ref{thm-debiased-lasso-FWER}, Propositions \ref{prop}--\ref{prop-gaussian-copula}, and some key technical lemmas. All the notation is the same as defined in the main body of the paper.
	Section \ref{Sec.A} presents the Proofs of Theorems \ref{theorem-FDR}--\ref{thm-debiased-lasso-FWER} and Propositions \ref{prop}--\ref{prop-gaussian-copula}. 
	We provide the proofs of the key lemmas and additional technical details in Section \ref{proof-lemmas}. In Section \ref{Supp_Sec_C}, we extend the analysis in Section \ref{new.Sec3.3} for knockoff statistics constructed with the regression coefficient difference  to the setting of the generalized linear model (GLM). 
	Throughout the Supplement, $ C  $ stands for some positive constant whose value may change from line to line.

	\renewcommand{\thesubsection}{A.\arabic{subsection}}
	
	\section{Proofs of Theorems \ref{theorem-FDR}--\ref{thm-debiased-lasso-FWER} and Propositions \ref{prop}--\ref{prop-gaussian-copula}} \label{Sec.A}
	
	\subsection{Proof of Theorem \ref{theorem-FDR}} \label{proof-thm1}
	
	It has been shown in \cite{CFJL2018} that the model-X knockoffs inference procedure achieves the exact FDR control when the perfect knockoff statistics are employed. Note that the approximate knockoff statistics $ \{ \hat{W}_j \}$ are expected to provide a reliable approximation to the perfect knockoff statistics $ \{ \widetilde{W}_j \} $, as assumed in  Condition \ref{fdr-condition1}. The main idea of the proof is to establish the FDR control for the approximate knockoffs inference procedure through a comparison of the approximate knockoff statistics and a certain realization of the perfect knockoff statistics. 
	The two lemmas below provide a sketch of the proof and can be established under Conditions \ref{fdr-condition1}--\ref{fdr-condition5}. 
	
	\begin{lemma} \label{fdr-lemma2}
		Assume that Conditions \ref{fdr-condition1}, \ref{fdr-condition4}, and \ref{fdr-condition5} are satisfied. When $a_n \to \infty$ and $m_n / a_n \to 0 $, we have that for some constant $0 < c_1 < 1$,
		\begin{equation}  \label{prop_b1} 
			\begin{split}
				\sup_{t \in \big(0, \, G^{-1} (  \frac { c_1 q a_n } { p} ) \big]     } \Bigg|  \frac { \sum_{j \in \mathcal{H}_0} \mathbbm{1} (\hat{W}_j \geq t) } { \sum_{j \in \mathcal{H}_0}  \mathbb{P} ( \widetilde{W}_j \geq t ) }  - 1 \Bigg|  = o_p(1), 
			\end{split}
		\end{equation}
		\begin{equation}  \label{prop_b1_2}
			\begin{split}
				\sup_{t \in \big(0, \, G^{-1} (  \frac { c_1 q a_n } { p} ) \big]     } \Bigg|  \frac { \sum_{j \in \mathcal{H}_0} \mathbbm{1} (\hat{W}_j \leq -t) } { \sum_{j \in \mathcal{H}_0}  \mathbb{P} ( \widetilde{W}_j \leq - t ) }  - 1 \Bigg| = o_p(1).    
			\end{split}
		\end{equation}
		
		\ignore{
			\begin{align}
				& \sup_{t \in \big(0, \, G^{-1} (  \frac { c_1 q a_n } { p} ) \big)     } \Bigg|  \frac { \sum_{j \in \mathcal{H}_0} \mathbbm{1} (\hat{W}_j \geq t) } { \sum_{j \in \mathcal{H}_0}  \mathbb{P} ( \widetilde{W}_j \geq t ) }  - 1 \Bigg| = o_p(1),\\
				& \sup_{t \in \big(0, \, G^{-1} (  \frac { c_1 q a_n } { p} ) \big)     } \Bigg|  \frac { \sum_{j \in \mathcal{H}_0} \mathbbm{1} (\hat{W}_j \leq - t) } { \sum_{j \in \mathcal{H}_0}  \mathbb{P} ( \widetilde{W}_j \leq  - t ) }  - 1 \Bigg| = o_p(1) .
			\end{align}
		}
	\end{lemma}

	\begin{lemma} \label{fdr-lemma1}
		Under Conditions  \ref{fdr-condition1}--\ref{fdr-condition5}, we have that for some constant $0 < c_1 < 1$, $ \mathbb{P} (  T \leq G^{-1} (  \frac { c_1 q a_n } { p} ) ) \to 1 $. 
	\end{lemma}
	
	\ignore{
		\begin{proposition} \label{fdr-prop3} 
			\rm{Under Condition \ref{fdr-condition5}, we have
				\begin{align}
					\sup_{t \in  (0, M_{n, p})   }   \frac { \sum_{j \in H_0} \mathbbm{1} (t - b_n \leq  \widetilde{W}_j  \leq t + b_n) } { \sum_{j \in H_0}  \mathbb{P} ( \widetilde{W}_j \geq t ) }  = o_p(1)  \label{prop_b2}
				\end{align}
			}
		\end{proposition} 
	}
	
	We present the proofs of Lemmas \ref{fdr-lemma2} and \ref{fdr-lemma1} in Sections \ref{proof.lem2} and \ref{proof-prop1}, respectively. 
	Now we are ready to prove Theorem \ref{theorem-FDR}.	Let us define two events $\mathcal{B}_1 = \{T \leq G^{-1} ( \frac {c_1 q a_n} {p} ) \}$ 
	and $$ \mathcal{B}_{2, \epsilon}  = \Big\{ \sup_{t \in (0, G^{-1} ( \frac {c_1 q a_n} {p} )]}  \Big(\Big| \frac {\sum_{j \in \mathcal{H}_0} \mathbbm{1} (\hat{W}_j \geq t )} {\sum_{j \in \mathcal{H}_0} \mathbb{P} (\widetilde{W}_j \geq t ) } - 1 \Big| \lor \Big| \frac {\sum_{j \in \mathcal{H}_0} \mathbbm{1} (\hat{W}_j \leq - t )} {\sum_{j \in \mathcal{H}_0} \mathbb{P} (\widetilde{W}_j \leq - t ) } - 1 \Big| \Big)\leq \epsilon \Big\}$$ 
	for $\epsilon > 0$. Lemmas \ref{fdr-lemma2} and \ref{fdr-lemma1} above have shown that $ \mathbb{P}(\mathcal{B}_1^c  ) \to 0   $  and $ \mathbb{P}(\mathcal{B}_{2, \epsilon}^c  ) \to 0 $ for each $\epsilon > 0$.  In addition, it holds naturally that $0 \leq  \FDP \leq 1  $. Then it follows that
	\begin{equation} \label{thm1-bound0}
		\begin{split}
			\FDR 
			& \leq \e \bigg( \frac { \sum_{j \in \mathcal{H}_0} \mathbbm{1} ( \hat{W}_j \geq T ) } { 1 \lor \sum_{ j  = 1}^p \mathbbm{1} (\hat{W}_j \geq T)  } \cdot \mathbbm{1} (\mathcal{B}_1)\mathbbm{1} (\mathcal{B}_{2, \epsilon}) \bigg)  + \mathbb{P}(\mathcal{B}_1^c  ) +  \mathbb{P}(\mathcal{B}_{2, \epsilon}^c  ) \\
			& =   \e \bigg( \frac { \sum_{j \in \mathcal{H}_0} \mathbbm{1} ( \hat{W}_j \geq T ) } { 1 \lor \sum_{ j  = 1}^p \mathbbm{1} (\hat{W}_j \geq T)  } \cdot  \mathbbm{1} (\mathcal{B}_1)\mathbbm{1} (\mathcal{B}_{2, \epsilon}) \bigg) +  o(1) .
		\end{split} 	
	\end{equation}
	In view of the definition of  threshold $T$ in \eqref{eq-2}, we can deduce that 
	\begin{equation}
		\begin{split}
			& \frac { \sum_{j \in \mathcal{H}_0} \mathbbm{1} ( \hat{W}_j \geq T ) } { 1 \lor \sum_{ j  = 1}^p \mathbbm{1} (\hat{W}_j \geq T)  } \cdot  \mathbbm{1} (\mathcal{B}_1)\mathbbm{1} (\mathcal{B}_{2, \epsilon}) \\
			& = \frac { \sum_{j \in \mathcal{H}_0} \mathbbm{1} ( \hat{W}_j \geq T ) } {   \sum_{j \in \mathcal{H}_0 } \mathbbm{1} (\hat{W}_j \leq -T) } \cdot \frac { \sum_{j \in \mathcal{H}_0 } \mathbbm{1} (\hat{W}_j \leq - T) } { 1 \lor \sum_{ j  = 1}^p \mathbbm{1} (\hat{W}_j \geq T)  }\cdot  \mathbbm{1} (\mathcal{B}_1)\mathbbm{1}(\mathcal{B}_{2, \epsilon})  \\
			& \leq q \cdot \frac { \sum_{j \in \mathcal{H}_0} \mathbbm{1} ( \hat{W}_j \geq T ) } {   \sum_{j \in \mathcal{H}_0 } \mathbbm{1} (\hat{W}_j \leq  - T ) }\cdot  \mathbbm{1} (\mathcal{B}_1)\mathbbm{1}(\mathcal{B}_{2, \epsilon}).
		\end{split}
	\end{equation}
	
	Furthermore, it is easy to see that on event $\mathcal{B}_1 \cap \mathcal{B}_{2, \epsilon} $, we have 
	\begin{align*}    	
		\frac { \sum_{j \in \mathcal{H}_0} \mathbbm{1} ( \hat{W}_j \geq T ) } {  \sum_{j \in \mathcal{H}_0 } \mathbbm{1} (\hat{W}_j \leq  - T ) } 
		& \leq \sup_{t \in (0, G^{-1} ( \frac {c_1 q a_n} {p} )]} \frac { \sum_{j \in \mathcal{H}_0} \mathbbm{1} ( \hat{W}_j \geq t ) } {  \sum_{j \in \mathcal{H}_0 } \mathbbm{1} (\hat{W}_j \leq  - t ) } \\
		& \leq \frac {1 + \epsilon} {1 - \epsilon} \sup_{t \in (0, G^{-1} ( \frac {c_1 q a_n} {p} )]} \frac { \sum_{j \in \mathcal{H}_0} \mathbb{P} ( \widetilde{W}_j \geq t ) } {\sum_{j \in \mathcal{H}_0} \mathbb{P} ( \widetilde{W}_j \leq - t )} \\
		& = \frac {1 + \epsilon} {1 - \epsilon},
	\end{align*}
	where the last equation above is obtained by the symmetry of the perfect knockoff statistics $\{\widetilde{W}_j \}_{j \in \mathcal{H}_0 }$ that $\mathbb{P} ( \widetilde{W}_j \geq t ) = \mathbb{P} ( \widetilde{W}_j \leq - t )$. Therefore, we can obtain that for any $\epsilon > 0$, 
	\begin{equation}
		\FDR \leq q \cdot \frac{1 + \epsilon} {1 - \epsilon} + o(1),
	\end{equation}
	which yields the desired result \eqref{fdr-result}. This completes the proof of Theorem \ref{theorem-FDR}.
	
	\subsection{Proof of Theorem \ref{thm-marginal-corr-fdr}} \label{new.SecA.3}
	The main idea of the proof is to directly apply Theorem \ref{theorem-FDR} by verifying Conditions \ref{fdr-condition1}--\ref{fdr-condition5} involved. We will show in the lemmas below that Conditions \ref{fdr-condition1}--\ref{fdr-condition5} are satisfied for the marginal correlation knockoff statistics under Conditions \ref{accuracy-knockoffs}--\ref{marginal_corr_condition4} and the setting of nonparametric regression model \eqref{nonpara-model} with normal features. Proofs of Lemmas \ref{pf-thm3-lemma-1}--\ref{pf-thm3-lemma-4} are presented in Sections \ref{new.SecB.7}--\ref{new.SecB.10}.
	\begin{lemma} \label{pf-thm3-lemma-1}
		Assume that Condition \ref{accuracy-knockoffs} is satisfied. Then we have that  
		\begin{equation}
			\mathbb{P} \Big( \max_{1 \leq  j\leq p } | \hat{W}_j - \widetilde{W}_j  | \geq  \Delta_n \Big) \to 0.
		\end{equation}
	\end{lemma}
	
	Lemma \ref{pf-thm3-lemma-1} above shows that Condition \ref{fdr-condition1} is satisfied with sequences $b_n := \Delta_n$. Define $w_j = (\e Y^2)^{-1/2} ( |\e (X_j Y)| - |\e(\widetilde{X}_j Y)| ) $ for $1 \leq j \leq p$. Note that $w_j = 0$ for $j \in \mathcal{H}_0$ since $(X_j, X_{\mathcal{H}_1}) \stackrel{d}{=} (\widetilde{X}_j, X_{\mathcal{H}_1})$ for $j \in \mathcal{H}_0$ by the exchangeability between $X_j$ and $\widetilde{X}_j$. Recall from the definition in \eqref{eq-delta_n} that 
	$$
	\delta_n = \sqrt{\frac {\log p} {n} } \max\limits_{1 \leq j \leq p} \Big\{   \frac { 16 \sqrt 2 \| X_j \|_{\psi_2} \| Y \|_{\psi_2} } { (\e Y^2)^{1/2} } \lor   \frac { 8\sqrt 2  |w_j| \| Y \|_{\psi_2}^2 } { \e Y^2 } \Big\}.
	$$
	We have the concentration inequality below for $\widetilde{W}_j$ under the sub-Gaussian assumption in Condition \ref{marginal_corr_condition1}. 
	\begin{lemma} \label{pf-thm3-lemma-2}
		Assume that Condition \ref{marginal_corr_condition1} is satisfied. When $\log p = o(n)$, we have that 
		\begin{equation} \label{verify-prop3-result}
			\sum_{j = 1}^p \mathbb{P}  ( |\widetilde{W}_j - w_j | \geq \delta_n ) \leq 6 p^{-1} + p \exp\Big\{ - \frac { n (\e Y^2)^2 } {8 \e Y^4 } \Big\}.
		\end{equation}
	\end{lemma}
	Lemma \ref{pf-thm3-lemma-2} above indicates that Condition \ref{fdr-condition2} related to the concentration rate of $\widetilde{W}_j$ is satisfied with $\delta_n $ defined in \eqref{eq-delta_n} and that $\Delta_n \leq \delta_n$, where $\Delta_n $ is the approximation accuracy of the approximate knockoff statistics obtained in Lemma \ref{pf-thm3-lemma-1}. In addition, from the definition of $w_j$, under Condition \ref{marginal_corr_condition3} we have that the general Condition \ref{fdr-condition3} on the signal strength is also satisfied. Next we will turn to the verification of Conditions \ref{fdr-condition4}--\ref{fdr-condition5}. 
	\begin{lemma} \label{pf-thm3-lemma-3}
		Assume that Condition \ref{marginal_corr_condition2} is satisfied. Then we have that for each $t \geq 0$,
		\begin{equation} \label{pf-thm3-lemma-3-result}
			\frac{ \Var \big( \sum_{j \in \mathcal{H}_0} \mathbbm{1} (\widetilde{W}_j \geq t)   \big)  } { p_0  G (t) } \leq 2 m_n.
		\end{equation} 
	\end{lemma}
	
	\begin{lemma} \label{pf-thm3-lemma-4}
		Assume that Conditions \ref{marginal_corr_condition2} and \ref{marginal_corr_condition4} are satisfied. Then when $(\log p)^{1/\gamma} m_n / a_n \to 0 $ and $\sqrt n \Delta_n (\log p)^{1/2 + 1/\gamma} \to 0$ for some constant $0< \gamma < 1$, 
		we have that 
		\begin{equation}  \label{lemma10-1} 
			(\log p)^{1/\gamma}  \sup_{t \in (0,\, G^{-1} ( \frac { c_1 q a_n  } { p } ) ] }  \frac { G(t - \Delta_n ) - G(t + \Delta_n) } {  G(t) } \to 0	 
		\end{equation}
		and 
		\begin{equation} \label{lemma10-2}
			a_n^{-1} \sum_{j \in \mathcal{H}_1} \mathbb{P} \Big( \widetilde{W}_j < -  G^{-1} ( \frac { c_1 q a_n  } { p } ) + \Delta_n \Big)  \to 0
		\end{equation}
		as $n \to \infty$.
	\end{lemma}
	
	Lemma \ref{pf-thm3-lemma-3} above shows that Condition \ref{fdr-condition4} is satisfied, while Lemma \ref{pf-thm3-lemma-4} above implies that Condition \ref{fdr-condition5} is satisfied. Finally, the conclusion of Theorem \ref{thm-marginal-corr-fdr} can be obtained by directly applying the general Theorem \ref{theorem-FDR}. This completes the proof of Theorem \ref{thm-marginal-corr-fdr}.
	
	\subsection{Proof of Theorem \ref{thm-debiased-lasso-FDR}} \label{new.SecA.5}
	The main idea of the proof is to directly apply Theorem \ref{theorem-FDR} by verifying Conditions \ref{fdr-condition1}--\ref{fdr-condition5} for the knockoff statistics constructed from the debiased Lasso coefficients. A key observation is that the debiased Lasso coefficients are asymptotically normal.
	Denote by $$\tau_j = \|\widetilde\bz_j \|_2 / |\widetilde\bz_j^{\top} \widetilde\bX^{\augg}_j|.$$ The debiased Lasso coefficient can be written as 
	\begin{equation} \label{normal-decomp}
		\sqrt n ( \widetilde{\beta}_j - \beta_j^{\augg} ) =  \frac{\widetilde\bz_j^{\top} \bveps} { \|\widetilde\bz_j \|_2 } \cdot \sqrt n \tau_j + \sum_{k \neq j} \frac{ \sqrt n  \widetilde\bz_j^{\top} \widetilde\bX^{\augg}_k (\beta_k^{\augg} - \widetilde{\beta}_k^{\init})} { \widetilde\bz_j^{\top} \widetilde\bX^{\augg}_j }  . 
	\end{equation}
	Observe that $\frac{\widetilde{\bz}_j^{\top} \bveps} { \|\widetilde{\bz}_j \|_2 } \sim N(0, \sigma^2) $, $\sqrt n\tau_j =O_p(1)$, and the remainder term in \eqref{normal-decomp} above is of order $o_p(1)$. Thus, the debiased Lasso 
	estimator is asymptotically normal in the sense that $$\tau_j^{-1} (\widetilde{\beta}_j - \beta_j^{\augg}) \stackrel{d}{\to} N(0, \sigma^2).$$ Our proof will build mainly on such intuition. Throughout the proof below, constant $C$ may take different values from line to line.
	
	We first present two lemmas below about the consistency of Lasso estimators $\widetilde\bbeta^{\init}$ and $\widetilde\bgamma_j $. We omit the proofs of Lemmas \ref{pf-thm5-lemma_00} and \ref{pf-thm5-lemma_01} here to avoid redundancy since they are well-known results for the consistency of Lasso estimators in the literature.
	\begin{lemma} \label{pf-thm5-lemma_00}
		Under Conditions \ref{debiased-lasso-condition1}--\ref{debiased-lasso-condition3}, we have that with probability $1 - o(p^{-3})$, 
		\begin{align}
			\| \widetilde{\bbeta}^{\init} - \bbeta^{\augg} \|_1 \leq C s \sqrt{ \frac{ \log p} {n} }, \label{lasso-1} \\
			\| \widetilde{\bbeta}^{\init} - \bbeta^{\augg} \|_2 \leq C \sqrt{ \frac{ s\log p} {n} },  \\
			\| \widetilde{\bX}^{\augg} (\widetilde{\bbeta}^{\init} - \bbeta^{\augg} ) \|_2 \leq C \sqrt { s\log p} .
		\end{align}
	\end{lemma}
	
	\begin{lemma} \label{pf-thm5-lemma_01}
		Under Conditions \ref{debiased-lasso-condition1}--\ref{debiased-lasso-condition3}, we have that with probability $1 - o(p^{-3})$, 
		\begin{align}
			\max_{1 \leq  j \leq 2p } \| \widetilde{\bgamma}_j - \bgamma_j \|_1 \leq C m_n \sqrt{ \frac{ \log p} {n} } ,   \\
			\max_{1 \leq  j \leq 2p } \|  \widetilde{\bgamma}_j - \bgamma_j  \|_2 \leq  C \sqrt{ \frac{ m_n\log p} {n} },  \\
			\max_{1 \leq j \leq 2p }  \| \widetilde{\bX}_{-j}^{\augg} (\widetilde{\bgamma}_j - \bgamma_j) \|_2 \leq C \sqrt { m_n \log p}. 
		\end{align}
		In addition, when $\frac{m_n \log p}{n} \to 0$ we have that with probability $1 - o(p^{-3})$, 
		\begin{align}
			& |\sqrt n \tau_j - (\e e_j^2)^{-1/2} | \leq C \sqrt{\frac{m_n \log p}{n}},  \label{bdd-tau}\\
			& \big|\widetilde{\bz}_j^{\top} \widetilde{\bz}_l - \Cov(e_j, e_l) \big| \leq C \sqrt{\frac{m_n \log p} {n}}. \label{bdd-zz}
		\end{align}
	\end{lemma}
	
	The four lemmas below outline the proof for verifying the general Conditions \ref{fdr-condition1}--\ref{fdr-condition5}. Proofs of Lemma \ref{pf-thm5-lemma-1}--\ref{pf-thm5-lemma-4} are provided in Sections \ref{new.SecB.11}--\ref{new.SecB.14}, respectively.
	
	\begin{lemma} \label{pf-thm5-lemma-1} 
		Assume that Conditions \ref{accuracy-knockoffs} and \ref{debiased-lasso-condition1}--\ref{debiased-lasso-condition3} are satisfied. Then as $\Delta_n s^{1/2}   \to 0$ and $  \sqrt{\frac{s \log p} {n}}  \to 0$, we have that 
		\begin{equation}
			\mathbb{P} \bigg( \max_{1 \leq j \leq 2p} | \widetilde{\beta}_j - \hat{\beta}_j | \geq C \Delta_n s  \sqrt{\frac{\log p}{n}} \bigg) \to 0.
		\end{equation}
	\end{lemma}
	Lemma \ref{pf-thm5-lemma-1} above indicates that Condition \ref{fdr-condition1} is satisfied with  sequences $b_n :=  C \Delta_n s \sqrt{\frac{\log p} {n}}$. 
	Let us define $w_j = |\beta_j| $.
	\begin{lemma} \label{pf-thm5-lemma-2}
		Assume that Conditions \ref{debiased-lasso-condition1}--\ref{debiased-lasso-condition3} are satisfied. Then as $   s \sqrt{\frac{m_n \log p} {n}} \to 0 $, we have that for some $C > 0$, $ \sum_{j= 1}^p \mathbb{P} (|\widetilde{W}_j - w_j | \geq C \sqrt{n^{-1}\log p} ) \to 0$.
	\end{lemma}

	Lemma \ref{pf-thm5-lemma-2} above shows that Condition \ref{fdr-condition2} related to the concentration rate of $\widetilde{W}_j$ is satisfied with $\delta_n = C \sqrt{n^{-1} \log p} $.  In addition, it holds that $ b_n \ll C \sqrt{n^{-1} \log p}  $ due to the assumption $\Delta_n s \to 0 $ in Theorem \ref{thm-debiased-lasso-FDR}. In addition, in light of the definition of $w_j$, under Condition \ref{debiased-lasso-condition4} we have that the general Condition \ref{fdr-condition3} on the signal strength is also satisfied. We next turn to the verification of Conditions \ref{fdr-condition4}--\ref{fdr-condition5}.

	\begin{lemma} \label{pf-thm5-lemma-3}
		Assume that Conditions \ref{debiased-lasso-condition1}--\ref{debiased-lasso-condition3} are satisfied. Then as $  \frac{ m_n^{1/2}  s (\log p)^{3/2 + 1/\gamma } } { \sqrt n }  \to 0 $, we have that $ \Var{ \big( \sum_{j \in \mathcal{H}_0} \mathbbm{1} (\widetilde{W}_j > t)  } \big) \leq  V_1 (t) + V_2 (t) $, where for some $ 0 < \gamma < 1$ and $ 0< c_1 < 1 $,  
		\begin{equation} 
			(\log p )^{1/\gamma}  \sup_{t \in (0,\, G^{-1} ( \frac { c_1 q a_n  } { p } ) ] }  \frac{V_1 (t) } { [p_0  G (t)]^2  } \to 0 
		\end{equation}
		and 
		\begin{equation} 
			\sup_{t \in (0,\, G^{-1} ( \frac { c_1 q a_n  } { p } ) ] }   \frac{ V_2(t)   } { p_0  G (t) }  \lesssim  m_n.
		\end{equation} 
		
	\end{lemma}

	\begin{lemma} \label{pf-thm5-lemma-4}
		Assume that Conditions \ref{accuracy-knockoffs}, \ref{marginal_corr_condition4}, and \ref{debiased-lasso-condition1} --\ref{debiased-lasso-condition3} are satisfied. Then when \\$    \frac{m_n^{1/2} s (\log p)^{3/2 + 1/\gamma}} {\sqrt n} \to 0 $ and $ \Delta_n s (\log p)^{1 + 1 /\gamma} \to 0$, we have that 
		\begin{equation}   \label{le16-1}
			(\log p)^{1/\gamma}  \sup_{t \in (0,\, G^{-1} ( \frac { c_1 q a_n  } { p } ) ] }  \frac { G(t - b_n ) - G(t + b_n) } {  G(t) } \to 0	 
		\end{equation}
		and 
		\begin{equation}  \label{le16-2}
			a_n^{-1} \sum_{j \in \mathcal{H}_1} \mathbb{P} \Big( \widetilde{W}_j < -  G^{-1} ( \frac { c_1 q a_n  } { p } ) + b_n \Big)  \to 0
		\end{equation}
		as $n \to \infty$.
	\end{lemma}
	Lemma \ref{pf-thm5-lemma-3} above shows that Condition \ref{fdr-condition4} is satisfied, whereas Lemma \ref{pf-thm5-lemma-4} implies that Condition \ref{fdr-condition5} is satisfied. Finally, the conclusion of Theorem \ref{thm-debiased-lasso-FDR} can be derived by directly applying the general Theorem \ref{theorem-FDR}. This completes the proof of Theorem \ref{thm-debiased-lasso-FDR}.

	\subsection{Proof of Theorem \ref{thm-FWER}} \label{proof-thm2}
	We first define the corresponding threshold $ \widetilde{T}_v $ for the perfect knockoff statistics $ \{\widetilde{W}_j\}_{j=1}^p $ in the model-X knockoffs inference for the $k$-FWER control as
	\begin{equation*}
		\widetilde{T}_v = \sup \{ t \in \widetilde{\mathcal{W}}: \, \# \{j: - \widetilde{W}_j \geq t\} = v  \},
	\end{equation*}
	where $v$ is defined as in \eqref{eq-4} and $\widetilde{\mathcal{W}} = \{ | \widetilde{W}_1 |, \ldots, | \widetilde{W}_p | \} $. 
	As sketched in Lemmas \ref{FWER-prop1}--\ref{FWER-prop5} below, the main idea of the proof is to show that the threshold $ {T}_v$ based on the approximate knockoff statistics and the threshold $  \widetilde{T}_v $ based on the perfect knockoff statistics are sufficiently close under Condition \ref{fdr-condition1} such that for any $\veps > 0$,  the number of $\widetilde{W}_j$'s that lie between $ {T}_v $ and $\widetilde{T}_v$ is at most $v \veps$ with asymptotic probability one, where $v$ satisfies $v / k \to 1$ as $k \to \infty$.
	Specifically, let $M_v$ be the integer such that
	\begin{equation} \label{eq_Mv}
		\widetilde{T}_{v + M_v} \geq \widetilde{T}_v - 2 b_n > \widetilde{T}_{v + M_v +1}. 
	\end{equation}
	
	Then we can establish a bound for $M_v$ as shown in Lemma \ref{FWER-prop5} below. We first present the three lemmas below that provide an outline of the proof. The proofs of Lemmas \ref{FWER-prop1}--\ref{FWER-prop5} are provided in Sections \ref{new.SecB.4}--\ref{sec:proof-FWER-prop5}, respectively.
 
	\begin{lemma} \label{FWER-prop1}
		Under Condition \ref{fdr-condition1}, we have that 
		\begin{equation}
			\mathbb{P} (   | {T}_v - \widetilde{T}_v | \geq  b_n) \to 0. \label{eq1.new}
		\end{equation}
	\end{lemma}

	\begin{lemma} \label{FWER-prop2}
		Assume that $ k \to \infty $. Then we have that 
		\begin{equation}
			\frac { v } { k }  = 1 + O ( k^{ - 1 /2 }). \label{eq3.new}
		\end{equation}
	\end{lemma}
	
	\ignore{
		\begin{lemma} \label{FWER-prop3}
			Under all the conditions of Theorem \ref{thm-FWER}, we have that 
			\begin{equation}
				\sup_{t \in (0, G^{-1} (\frac {v} {2p}) )} \bigg| \frac { \sum_{j \in \mathcal{H}_0} \mathbbm{1} ( \widetilde{W}_j \geq t ) } {\sum_{j \in \mathcal{H}_0 } \mathbb{P} ( \widetilde{W}_j \geq t ) } - 1 \bigg| = o_p (\epsilon_n)
			\end{equation}
		\end{lemma}
	}
	
	\begin{lemma} \label{FWER-prop5}
		Under all the conditions of Theorem \ref{thm-FWER}, we have that for each $ \veps > 0$, 
		\begin{equation}
			\mathbb{P} ( M_v \leq v \veps  ) \to 1.
		\end{equation}
	\end{lemma}
	
	We are now ready to prove Theorem \ref{thm-FWER}.
	It follows straightforwardly from Lemma \ref{FWER-prop1} that
	\begin{equation*}
		\begin{split}
			\mathbb{P} ( \hat{V} \geq k(1 + 2 \veps)  ) & =  \mathbb{P} \Big(  \sum_{j \in \mathcal{H}_0} \mathbbm{1} (   \hat{W}_j \geq {T}_v  )   \geq k(1 + 2\veps) \Big)  \\
			& \leq \mathbb{P} \Big(  \sum_{j \in \mathcal{H}_0} \mathbbm{1} (   \widetilde{W}_j \geq \widetilde{T}_v  - 2 b_n )   \geq k(1 + 2\veps) \Big)  \\
			& \leq \mathbb{P} \Big(   \sum_{ j \in \mathcal{H}_0} \mathbbm{1} (    \widetilde{W}_j \geq  \widetilde{T}_{v + M_v }   )  \geq k (1 + 2\veps) \Big) \\
			& =  \mathbb{P} \Big(   \sum_{ j \in \mathcal{H}_0} \mathbbm{1} (    
			- \widetilde{W}_j \geq  \widetilde{T}_{v + M_v }   )  \geq k (1 + 2\veps) \Big) \\
			& \leq \mathbb{P} \Big( \sum_{j \in \mathcal{H}_0 } \mathbbm{1} ( - \widetilde{W}_j \geq \widetilde{T}_v) \geq k(1 + 2 \veps) - M_v \Big),
		\end{split}    
	\end{equation*}
	where the second last step above is because of the symmetry of $\widetilde W_j$'s with $j\in \mathcal H_0$ and the last step above is due to 
	$$
	\sum_{ j \in \mathcal{H}_0} \mathbbm{1} (   - \widetilde{W}_j \geq  \widetilde{T}_{v + M_v })-\sum_{ j \in \mathcal{H}_0} \mathbbm{1} (   - \widetilde{W}_j \geq  \widetilde{T}_{v })\leq M_v
	$$
	by the definitions of $\widetilde T_v$ and $M_v$.
	
	Moreover, Lemma \ref{FWER-prop5} above shows that $M_v \leq v \veps $ with asymptotic probability one and Lemma \ref{FWER-prop2} above proves that $v / k = 1 + o(1)$. 
	Then it holds that $2 k \veps > M_v$ with asymptotic probability one. Hence, combining the above results and by the union bound, we can deduce that 
	\begin{equation*}
		\begin{split}
			\mathbb{P} ( \hat{V} \geq k(1 + 2 \veps)  )    & \leq  \mathbb{P} \Big( \sum_{j \in \mathcal{H}_0} \mathbbm{1}( - \widetilde{W}_j \geq  \widetilde{T}_{v } ) \geq k \Big) + o(1) = q + o(1).
		\end{split}
	\end{equation*}
	Consequently, it follows that for each $ \veps > 0$, $$ \limsup_{n \to \infty} \mathbb{P} ( \hat{V} \geq k(1 + 2 \veps)  ) \leq q .$$ This concludes the proof of Theorem \ref{thm-FWER}.


	\subsection{Proof of Theorem \ref{thm-marginal-corr-kFWER}} \label{new.SecA.4}
	The proof of Theorem \ref{thm-marginal-corr-kFWER} is analogous to that of Theorem \ref{thm-marginal-corr-fdr} in Section \ref{new.SecA.3}. We omit the detailed proof here to avoid redundancy.

	\subsection{Proof of Theorem \ref{thm-debiased-lasso-FWER}} \label{new.SecA.6}
	The proof of Theorem \ref{thm-debiased-lasso-FWER} is similar to that of Theorem \ref{thm-debiased-lasso-FDR} in Section \ref{new.SecA.5}. Hence we omit the detailed proof here to avoid redundancy.

	\subsection{Proof of Proposition \ref{prop}} \label{Sec_proof_new_prop1}
	Let $\widehat{\bX}$ and $\widetilde{\bX}$ be matrices generated from the conditional coupling measure $\eta^*$ given $\bX$. 
	By Chebyshev's inequality, we have 
	\begin{align*}
		\mathbb P (\|\widehat{\bX} - \widetilde{\bX} \|_{1, 2}  \geq \Delta_{n}) &=    \mathbb E_{\boldsymbol{X}}[ \mathbb P^*(\|\widehat{\bX} - \widetilde{\bX} \|_{1, 2}  \geq \Delta_{n}|\bX)] \leq \mathbb \e_{\boldsymbol{X}}\Big[ \frac{  \e^* [ \|\hat{\bX} - \widetilde{\bX}\|_{1, 2}\vert \bX]}{\Delta_n}\Big]\\
		&\leq  \e_{\boldsymbol{X}}[C_{\boldsymbol{X}}]c_n\Delta_n^{-1}\rightarrow 0.
	\end{align*}

	\subsection{Proof of Proposition \ref{prop-t}} \label{new.SecA.7}
	From the definitions in \eqref{t-appro-knock} and \eqref{t-perf-knock}, we see that 
	\begin{equation} \label{pf-p3-2}
		\hat{\bX} - \widetilde{\bX} = r \bX \bA +  \bZ \bB + \diag( 1 - \frac{1}{\sqrt{Q_1/\nu}}, \ldots, 1 - \frac{1}{\sqrt{Q_n/\nu}}) \bZ \bC,
	\end{equation}
	where $\bA = \bOmega - \hat\bTheta$, $\bB =(2 r I_p - r^2 \hat\bTheta )^{1/2} -  (2 r I_p - r^2 \bOmega)^{1/2}$, and $\bC = (2 r I_p - r^2 \bOmega)^{1/2}$. 
	In view of assumption \eqref{prop-t-cond1} and the fact that $ \bTheta := [\Cov(X)]^{-1} = \frac{\nu - 2}{\nu  } \bOmega $, it follows from the triangle inequality that with probability $1 - o(1)$, 
	\begin{equation} \label{pf-p3-0}
		\begin{split}
			\| \hat{\bTheta} - \bOmega \|_2 &\leq \| \hat{\bTheta}  - \bTheta \|_2 +  \|  \bTheta - \bOmega \|_2  =    \| \hat{\bTheta}  - \bTheta \|_2  +  {2 \nu^{-1}  \| \bOmega\|_2}  \\
			&\leq C \rho_n \sqrt{\frac{\log p}{n}}  + {2 \nu^{-1} C_l^{-1}}. 
		\end{split}
	\end{equation}
	
	Now we deal with the three terms on the right-hand side of \eqref{pf-p3-2} above separately. 
	First, for the second term above, an application of similar arguments as for \eqref{sec4.1-5} gives that with probability $1 - o(1)$, 
	\begin{equation} \label{pf-p3-1}
		\max_{1 \leq j \leq p} n^{-1}\| (\bZ \bB)_j \|_2^2 \leq 3 \| \bB \|_2^2 /2 \leq C \|\hat{\bTheta} - \bOmega\|_2^2  \leq C \Big(  \frac{ \rho_n^2 \log p} {n} + \nu^{-2} \Big). 
	\end{equation}
	Regarding the first term on the right-hand side of \eqref{pf-p3-2} above, observe that $$(\bX_{i, j} , \bX_{i, l} ) \stackrel{d}{=} (\frac{\eta_{i, j}} {\sqrt{Q_i / \nu} }, \frac{\eta_{i, l}} {\sqrt{Q_i / \nu} }),$$ 
	where $(\eta_{i, 1}, \ldots, \eta_{i, p}) \stackrel{d}{\sim} N({\bf 0}, \bOmega^{-1})$ and $\{Q_{i}\}_{i = 1}^n$ are independent and identically distributed  (i.i.d.) chi-square random variables with $\nu $ degrees of freedom. It holds that for some large constant $C_1 > 0$, 
	\begin{equation} \label{chi-n0}
		\begin{split}
			&\mathbb{P} \bigg(\|  n^{-1} \bX ^{\top} \bX - \bTheta^{-1} \|_{\max} \geq  C_1 \sqrt {\frac{  \log p} {n} } + \nu^{-1/2}  \bigg) \\
			& =  \mathbb{P} \bigg(\max_{1 \leq j, l \leq p}   \bigg| n^{-1}\sum_{i = 1}^n \frac{ \eta_{i, j} \eta_{i, l} } {Q_i / \nu} - \e (\eta_{i, j} \eta_{i, l}) \e ( \frac{\nu}{Q_i  } ) \bigg|  \geq C_1 \sqrt {\frac{  \log p} {n} }+  \nu^{-1/2} \bigg) \\
			& \leq  \mathbb{P} \bigg(\max_{1 \leq j, l \leq p}  \bigg|  n^{-1}\sum_{i = 1}^n \frac{ \nu } {Q_i } (\eta_{i, j} \eta_{i, l} - \e (\eta_{i, j} \eta_{i, l})) \bigg|  \geq C_1 \sqrt {\frac{  \log p} {n} }  \bigg) \\
			& \quad +\mathbb{P} \bigg(\max_{1 \leq j, l \leq p}   \bigg|n^{-1}\sum_{i = 1}^n \e (\eta_{i, j} \eta_{i, l}) \Big( \frac{\nu}{Q_i }-\e ( \frac{\nu}{Q_i } ) \Big)\bigg| \geq  \nu^{-1/2} \bigg).
		\end{split}
	\end{equation}
	
	Before showing the bounds for the two probabilities on the right-hand side of the expression above, we first present some basic results for chi-square random variables. Note that from the property of the chi-square distribution, we have through some immediate calculations that
	\begin{align}
		& \e \Big( \frac{\nu^2}{Q_i^2} \Big)  = \frac{\nu^2} {(\nu - 2) (\nu - 4)}, \\
		& \Var \Big( \frac{\nu} {Q_i } \Big) = \frac{\nu^2 } {(\nu - 2)(\nu - 4)} - (\frac{\nu} {\nu - 2})^2 = O(\nu^{-1}),   \label{chi-n1} \\
		& \Var\Big( \frac{\nu^2} {Q_i^2 }  \Big) = \frac{\nu^4} {(\nu - 2)(\nu - 4)(\nu - 6)(\nu - 8)} -  \Big(\frac{\nu^2 } {(\nu - 2)(\nu - 4)}\Big)^2 = O (\nu^{-1}). \label{chi-n2}
	\end{align}
	Thus, noting that $  \e \Big( \frac{\nu^2}{Q_i^2}\Big)  + \nu^{-1/2}  = \frac{\nu^2}{(\nu - 2) (\nu - 4)}  + \nu^{-1/2} \leq 3$ and  $  \e \Big( \frac{\nu^2}{Q_i^2}\Big)  - \nu^{-1/2} \geq 2/3$ when $\nu \geq 9$, an application of the Markov inequality leads to 
	\begin{equation} \label{chi-n3}
		\begin{split}
			& \mathbb{P} \bigg( n^{-1} \sum_{i = 1}^n \frac{\nu^2}{Q_i^2} \geq  3 \bigg)  +  \mathbb{P} \bigg( n^{-1} \sum_{i = 1}^n \frac{\nu^2}{Q_i^2} \leq 2/3 \bigg) \\
			&  \leq \mathbb{P} \bigg(  n^{-1} \sum_{i = 1}^n \frac{\nu^2}{Q_i^2} \geq   \e \Big( \frac{\nu^2}{Q_i^2}\Big)  + \nu^{-1/2} \bigg) + \mathbb{P} \bigg(  n^{-1} \sum_{i = 1}^n \frac{\nu^2}{Q_i^2} \leq   \e \Big( \frac{\nu^2}{Q_i^2}\Big)  - \nu^{-1/2} \bigg) \\
			& \leq \nu  n^{-1} \Var \Big( \frac{\nu^2} {Q_i^2 }  \Big) = O(n^{-1} ) \to 0.
		\end{split}
	\end{equation}
	
	In addition, noting that $e^{-x/2 } \leq 1  $ and Stirling's formula for the gamma function $\Gamma(x) =  \sqrt{2 \pi / x} (x/ e)^{x} (1 + O(x^{-1}))$  for $x > 0$, we have through applying the density function of the chi-square distribution that for each constant $C > 0$, 
	\begin{equation} \label{chi-n4}
		\begin{split}
			\mathbb{P} \bigg( \max_{1 \leq i \leq n } \frac{\nu}{Q_i} \geq C \sqrt{\frac{n} {\log p}}\bigg) & 
			\leq n \int_{0}^{C^{-1} \nu \sqrt{\frac{\log p}{n}} } \frac{x^{\nu / 2 - 1} e^{- x / 2}} { 2^{\nu / 2} \Gamma(\nu / 2)} \, dx \\
			& \leq \frac{ 2 n  (C^{-1} \nu \sqrt{\frac{\log p} {n}})^{\nu / 2} } { \nu  2^{\nu / 2} \Gamma(\nu / 2)} \\
			& \lesssim n \Big( C^{-2}  \frac{\log p} {n}  \Big)^{\nu / 4}  \frac{\nu ^{\nu /2 }} { \nu 2^{\nu /2} \sqrt{4 \pi / \nu } (\nu / 2 e)^{\nu / 2} } \\
			& = \Big( C^{-2} e^2  \frac{\log p} {n^{1 - 4/ \nu}}  \Big)^{\nu / 4} \frac{ 1} {\sqrt {4 \pi \nu} } \to 0
		\end{split}
	\end{equation}
	when $\log p = o(n^{1 - 4/\nu})$.
	
	Now we are ready to deal with the two probabilities on the right-hand side of \eqref{chi-n0} above. Let us define two events $\mathscr{D}_1 = \{ \max_{1 \leq i \leq n} \frac{\nu} {Q_i}\leq C_2 \sqrt{\frac{n}{\log p}}\}$ for a small constant $C_2 > 0$ and $\mathscr{D}_2 = \{2/3 \leq  n^{-1} \sum_{i = 1}^n \frac{\nu^2} {Q_i^2} \leq  3 \}$. It follows from \eqref{chi-n3} and \eqref{chi-n4} that $ \mathbb{P} (\mathscr{D}_1^c) \to 0 $ and $ \mathbb{P} (\mathscr{D}_2^c) \to 0 $. For the first probability in \eqref{chi-n0} above, since $\eta_{i, j} \eta_{i, l}$ is a sub-exponential random variable and $Q_i \indep \eta_{i, j} \eta_{i, l}$, we can obtain by applying the concentration inequality for the weighted sum of sub-exponential random variables (cf. Corollary 4.2 in \cite{ZhangandChen2021}) that when $C_1$ is large enough and $C_2$ is small enough, 
	\begin{equation} \label{chi-n5}
		\begin{split}
			& \mathbb{P} \bigg(\max_{1 \leq j, l \leq p}  \bigg|  n^{-1}\sum_{i = 1}^n \frac{ \nu } {Q_i } (\eta_{i, j} \eta_{i, l} - \e (\eta_{i, j} \eta_{i, l})) \bigg|  \geq C_1 \sqrt {\frac{  \log p} {n} }  \bigg) \\
			& \leq \mathbb{P}  \bigg(\max_{1 \leq j, l \leq p}  \bigg|  n^{-1}\sum_{i = 1}^n \frac{ \nu } {Q_i } (\eta_{i, j} \eta_{i, l} - \e (\eta_{i, j} \eta_{i, l})) \bigg|  \geq C_1 \sqrt {\frac{  \log p} {n} } , \mathscr{D}_1 \cap \mathscr{D}_2 \bigg) \\
			&\quad+ \mathbb{P} (\mathscr{D}_1^c) +  \mathbb{P} (\mathscr{D}_2^c) \\
			& \leq 2 p^2  \exp \{ - 3 \log p \} + o(1) \to 0.
		\end{split}
	\end{equation}
	Regarding the second probability in \eqref{chi-n0}, since $\max_{1 \leq j, l \leq p} | \e (\eta_{i, j} \eta_{i, l}) | \leq \max_{1 \leq j \leq p} \e(\eta_{i, j}^2) \leq \max_{1 \leq j \leq p} (\bOmega^{-1})_{j, j} \leq C_u$, an application of the Markov inequality and \eqref{chi-n1} yields that 
	\begin{equation} \label{chi-n6}
		\begin{split}
			& \mathbb{P} \bigg(\max_{1 \leq j, l \leq p}   \bigg|n^{-1}\sum_{i = 1}^n \e (\eta_{i, j} \eta_{i, l}) \Big( \frac{\nu}{Q_i }-\e ( \frac{\nu}{Q_i } ) \Big)\bigg| \geq  \nu^{-1/2} \bigg) \\
			& \leq \mathbb{P} \bigg(    \bigg|n^{-1}\sum_i   \Big( \frac{\nu}{Q_i }-\e ( \frac{\nu}{Q_i } ) \Big)\bigg| \geq  C_u^{-1} \nu^{-1/2} \bigg) \\
			& \leq C_u^{-2} \nu n^{-1} \Var (\frac{\nu }{Q_i})  = O(n^{-1}) \to 0.
		\end{split}
	\end{equation}
	
	By plugging \eqref{chi-n5} and \eqref{chi-n6} into \eqref{chi-n0}, we can show that with probability $1 - o(1)$,
	\begin{equation*}
		\max_{\delta: \| \delta\|_0 \leq   \rho_n} \frac{ |\delta^{\top} ( n^{-1} \bX^{\top} \bX - \bTheta^{-1}  ) \delta |} { \| \delta \|_2^2 } \leq C \rho_n \Big( \sqrt{\frac{\log p}{n}} +  \nu^{-1/2} \Big) ,
	\end{equation*}
	which along with the fact $\| \bTheta^{-1} \|_2 = \frac{\nu}{\nu - 2} \|\bOmega^{-1} \|_2 \leq \frac{\nu}{\nu - 2} C_u $ entails that as $ \rho_n = o(\sqrt{n / (\log p)})  $ and $\rho_n = o(\sqrt{\nu})$, 
	\begin{equation}\label{pf-prop1-1}
		\max_{\delta: \| \delta\|_0 \leq  \rho_n} \frac{ \delta^{\top}  \bX^{\top} \bX  \delta } { n \| \delta \|_2^2 } \leq  {C} 
	\end{equation}
	for some constant $ {C} > 0$.
	Using \eqref{pf-p3-0} and the sparsity assumption that $ \max_{1 \leq j \leq p} \|\bOmega_j\|_0 + \|\bOmega_n\|_0 \leq \rho_n $, an application of similar arguments as for \eqref{sec4.1-4} gives that with probability $1 - o(1)$,
	\begin{equation} \label{pf-p3-4}
		\begin{split}
			\max_{1 \leq j \leq p} n^{-1} \|\bX \bA_j  \|_2^2  & = n^{-1}\bA_j^{\top} \bX ^{\top} \bX \bA_j \leq C \max_{1 \leq j \leq p} \|\bA_j \|^2_2   \\
			&= C \| \hat{\bTheta} - \bOmega \|_2^2  \leq  C\Big(  \frac{ \rho_n^2 \log p} {n} + \nu^{-2} \Big).
		\end{split}
	\end{equation}
	
	We now proceed with examining the third term on the right-hand side of \eqref{pf-p3-2} above. Observe that $\bZ \bC_j   \stackrel{d}{\sim} N({\bf 0}, \|\bC_j \|_2^2 I_n)$ and $\max_{1 \leq j \leq p} \|\bC_j \|_2 \leq \| \bC\|_2 \leq 2r $. Hence, it holds for some large constant $C_3 > 0$ that 
	\begin{equation} \label{pf-p3-6}
		\begin{split}
			& \mathbb{P} \bigg( \max_{1 \leq j \leq p } n^{-1} \Big\| \diag( 1 - \frac{1}{\sqrt{Q_1/\nu}}, \ldots, 1 - \frac{1}{\sqrt{Q_n/\nu}}) \bZ \bC_j  \Big\|_2^2 \geq C_3 \nu^{-1}     \bigg)\\
			& =   \mathbb{P} \bigg( \max_{1 \leq j \leq p }  n^{-1} \sum_{i = 1}^n \Big(1 - \frac{ 1} {\sqrt{Q_i / \nu}} \Big)^2 \|\bC_j \|^2 Z_i^2 \geq C_3 \nu^{-1}    \bigg)   \\
			& \leq  \mathbb{P} \bigg( n^{-1} \sum_{i = 1}^n \Big(1 - \frac{ 1} {\sqrt{Q_i / \nu}} \Big)^2   Z_i^2 \geq C_3 \nu^{-1}    / 4r^2  \bigg) ,
		\end{split}
	\end{equation}
	where $\{Z_i \}_{i = 1}^n $ are i.i.d. standard normal random variables that are independent of $\bC$ and $\{Q_i \}_{i = 1}^n$. 
	
	Similar to the calculations in \eqref{chi-n1} and \eqref{chi-n2}, we can deduce that 
	\begin{equation} \label{pf-p3-5}
		\begin{split}
			\e \bigg[ \Big(1 - \frac{1}{\sqrt{Q_i / \nu}} \Big)^2 Z_i^2 \bigg] 
			& = \e (Z_i^2) \e  \bigg[ \Big(1 - \frac{1}{\sqrt{Q_i / \nu}} \Big)^2 \bigg] \\
			& =  1 - \e \Big(  \frac{2} {\sqrt{Q_i / \nu}} \Big) + \e  \Big(  \frac{1} { Q_i / \nu} \Big)  \\
			& = 1 -   \frac{  \sqrt {2 \nu } \Gamma(\frac{\nu - 1}{2})} {\Gamma(\frac{\nu }{2})} + \frac{\nu}{ \nu - 2}
		\end{split}
	\end{equation}
	and 
	\begin{equation}
		\begin{split}
			& \e \bigg[ \Big(1 - \frac{1}{\sqrt{Q_i / \nu}} \Big)^4 Z_i^4 \bigg] \\
			& = 3 \bigg( 1 - \frac{2 \sqrt 2 \nu \Gamma(\frac{\nu - 1}{2})} {\Gamma(\frac{\nu}{2})} + \frac{6 (\nu - 2)}{\nu} - \frac{\sqrt 2 \nu^{3/2} \Gamma(\frac{\nu - 3}{2})}{\Gamma(\frac{\nu}{2})} + \frac{\nu^2}{(\nu - 2) (\nu - 4)}\bigg).
		\end{split}
	\end{equation}
	By applying the asymptotic series of the gamma function 
	$$
	\frac{\Gamma(x + 1/2 )} {\Gamma(x)} = \sqrt {x} \Big(1 - \frac{1}{8 x } + O(x^{-2}) \Big),
	$$
	we can obtain through some direct calculations that 
	\begin{equation} \label{chi-n7}
		\e \bigg[ \Big(1 - \frac{1}{\sqrt{Q_i / \nu}} \Big)^2 Z_i^2 \bigg] = O(\nu^{-1})\ \text{ and } \ \e \bigg[ \Big(1 - \frac{1}{\sqrt{Q_i / \nu}} \Big)^4 Z_i^4 \bigg]  = O(\nu^{-2}).
	\end{equation}
	
	Combining \eqref{pf-p3-6} and \eqref{chi-n7} and applying the Markov inequality, we have that for some large enough constant $C_3 > 0$, 
	\begin{equation} \label{pf-p3-7}
		\begin{split}
			& \mathbb{P} \bigg( \max_{1 \leq j \leq p } n^{-1} \Big\| \diag( 1 - \frac{1}{\sqrt{Q_1/\nu}}, \ldots, 1 - \frac{1}{\sqrt{Q_n/\nu}}) \bZ \bC_j  \Big\|_2^2 \geq C_3  \nu^{-1}      \bigg)\\
			& \leq  \mathbb{P} \bigg( n^{-1} \sum_{i = 1}^n \Big(1 - \frac{ 1} {\sqrt{Q_i / \nu}} \Big)^2   Z_i^2  -\e \bigg[ \Big(1 - \frac{1}{\sqrt{Q_i / \nu}} \Big)^2 Z_i^2 \bigg]  \\
			&\quad \geq C_3 (\nu^{-1}  )  / 4r^2 - O(\nu^{-1})  \bigg) \\
			& \leq C \nu^{-2} n^{-1} \Var \bigg( \Big(1 - \frac{1}{\sqrt{Q_i / \nu}} \Big)^2 Z_i^2  \bigg)\\
			& \leq C \nu^{-2} n^{-1} \e \bigg( \Big( \Big(1 - \frac{1}{\sqrt{Q_i / \nu}} \Big)^4 Z_i^4  \Big) \bigg) \\
			&= O (n^{-1}) \to 0.
		\end{split}
	\end{equation}
	Therefore, a combination of \eqref{pf-p3-2}, \eqref{pf-p3-1}, \eqref{pf-p3-4}, and \eqref{pf-p3-7} yields the desired conclusion in \eqref{re-prop3}. This concludes the proof of Proposition \ref{prop-t}.

	\subsection{Proof of Proposition \ref{prop-gaussian}} \label{new.SecA.8}
	It follows from \eqref{sec4.1-1} and \eqref{sec4.1-2} that
	\begin{equation} \label{sec4.1-0}
		\hat{\bX} - \widetilde{\bX} = r \bX \bA + \bZ \bB,
	\end{equation}
	where $\bA = \bOmega - \hat{\bOmega}$ and $\bB =(2 r I_p - r^2 \hat\bOmega)^{1/2} -  (2 r I_p - r^2 \bOmega)^{1/2}$. By the Gaussianity of $X$, we see that $X_j X_l$ is a sub-exponential random variable and thus for $0< u < C$, 
	\begin{equation*}
		\mathbb{P} ( | n^{-1} \bX_j^{\top} \bX_l - \e (X_j X_l) | \geq u ) \leq 2 \exp \{ - C n u^2 \}.  
	\end{equation*}
	Then we can obtain that 
	\begin{equation*}
		\mathbb{P} \bigg( \max_{1 \leq j \leq p, 1 \leq l \leq p} | n^{-1} \bX_{j} \bX_l - \e (X_j X_l) | \geq C \sqrt{\frac{\log p} {n}} \bigg) = o(1). 
	\end{equation*}
	Consequently, with probability $1 - o(1)$ it holds that  
	\begin{equation*}
		\max_{\delta: \| \delta\|_0 \leq   \rho_n} \frac{ |\delta^{\top} ( n^{-1} \bX^{\top} \bX - \bOmega^{-1}  ) \delta |} { \| \delta \|_2^2 } \leq C \rho_n \sqrt{\frac{\log p}{n}},
	\end{equation*}
	which combined with the assumption that $\| \bOmega^{-1} \|_2 \leq C_u$ leads to
	\begin{equation}\label{pf-prop1-1}
		\max_{\delta: \| \delta\|_0 \leq  \rho_n} \frac{ \delta^{\top}  \bX^{\top} \bX  \delta } { n \| \delta \|_2^2 } \leq C_u + C \rho_n \sqrt{\frac{\log p}{n}} \leq \widetilde{C} 
	\end{equation}
	for some constant $\widetilde{C} > 0$. Since $\| \bA_j \|_0 = \|(\bOmega - \hat{\bOmega})_j\|_0 \leq C \rho_n$ because of the sparsity of $ \bOmega $ and $\hat{\bOmega}$, it follows from \eqref{pf-prop1-1} that with probability $1 - o(1)$, 
	\begin{equation} \label{sec4.1-4}
		\begin{split}
			\max_{1 \leq j \leq p} n^{-1} \| (\bX \bA)_{j} \|_2^2  &  =  \max_{1 \leq j \leq p} n^{-1} \| \bX \bA_{j} \|_2^2   \leq  \max_{1 \leq j \leq p} \widetilde{C} \| \bA_j \|_2^2 \\
			& =  \max_{1 \leq j \leq p} \widetilde{C} \| (\hat\bOmega  - \bOmega)_j \|_2^2 \leq  \max_{1 \leq j \leq p} \widetilde{C}  \| \hat\bOmega  - \bOmega \|_2^2 \\
			& \leq \widetilde{C} \frac{\rho_n^2 \log p} {n},
		\end{split}
	\end{equation}
	where we have used the accuracy assumption in \eqref{eq-5}. 
	
	Next we proceed with analyzing the term $\bZ \bB$. Observe that given $\bB$, $\bZ$ has i.i.d. standard normal components and is independent of $\bB$, and hence
	\begin{equation*}
		\bZ \bB_j|\bB_j  \stackrel{d}{\sim} N( {\bf 0}, \| \bB_j\|_2^2 I_n). 
	\end{equation*} 
	It holds that $\bZ \bB_j|\bB_j \stackrel{d}{=} (Z_1 \| \bB_j\|_2, \ldots, Z_n \| \bB_j \|_2)$ with $\{ Z_i \}_{i = 1}^n$ i.i.d. standard normal random variables. Then we can deduce that 
	\begin{equation}\label{pf-prop1-2}
		\begin{split}
			& \mathbb{P} ( \max_{1 \leq j \leq p } n^{-1} \| (\bZ \bB)_j \|_2^2 \geq 3\| \bB \|_2^2 /2 \big\vert \bB ) \\
			& =   \mathbb{P} ( \max_{1 \leq j \leq } n^{-1} \| \bZ \bB_j \|_2^2 \geq 3\| \bB \|_2^2 /2 \big\vert \bB) \\
			& =  \mathbb{P} \Big( \max_{1 \leq j \leq p} n^{-1} \sum_{i = 1}^n Z_i^2 \| \bB_j \|_2^2 \geq 3 \| \bB \|_2^2 /2 \big\vert \bB\Big) \\
			& \leq  \mathbb{P} \Big(  n^{-1} \sum_{i = 1}^n Z_i^2  \| \bB \|_2^2 \geq 2 \| \bB \|_2^2 \big\vert \bB\Big) \\
			& =  \mathbb{P}\Big(  n^{-1} \sum_{i = 1}^n Z_i^2   \geq 3/2  \Big) \leq e^{- n / 32}\to 0
		\end{split}
	\end{equation}
	as $n\rightarrow \infty$, where we have used the fact that $ \max_{1 \leq j \leq p} \|\bB_j \|_2 \leq \| \bB \|_2$ and the concentration inequality for chi-square random variables that for $0 < t < 1$, 
	\begin{equation*}
		\mathbb{P}\Big( \Big| n^{-1} \sum_{i = 1}^n Z_i^2 - 1 \Big| \geq t \Big) \leq 2 e^{- n t^2 / 8}.
	\end{equation*}
	
	Now we aim to bound $\| \bB \|_2$. For two square matrices $A$ and $B$, it holds that 
	\begin{equation*}
		\begin{split}
			\| A^{1/2} - B^{1/2}\|_2& = \|A^{1/2} (B - A) B^{-1} + (A^{3/2} - B^{3/2}) B^{-1}\|_2 \\
			&\leq \|A^{1/2} (B - A) B^{-1}\|_2 +3 \max \{\| A \|_2^{1/2}, \| B \|_2^{1/2} \} \|A - B \|_2\|B^{-1}\|_2.
		\end{split}
	\end{equation*}
	Applying the above inequality to $\bB$ leads to 
	\begin{equation}
		\begin{split}
			& \|\bB\|_2  \leq \| 2 r I_p - r^2 \hat{\bOmega} \|_2^{1/2} \cdot r^2 \| \hat{\bOmega} - \bOmega \|_2 \cdot  \| 2 r I_p - r^2 {\bOmega} \|^{-1} \\
			& \quad + 3 \max \{\| 2 r I_p - r^2 \hat{\bOmega} \|_2^{1/2}, \| 2 r I_p - r^2 {\bOmega} \|_2^{1/2}   \} \cdot  r^2 \| \hat{\bOmega} - \bOmega \|_2\cdot  \| 2 r I_p - r^2 {\bOmega} \|^{-1}\\
			& \leq C \| \hat\bOmega - \bOmega \|_2. 
		\end{split}
	\end{equation}
	Thus, from \eqref{pf-prop1-2} and assumption \eqref{eq-5}, we can obtain that with probability $1 - o(1)$, 
	\begin{equation}  \label{sec4.1-5}
		\max_{1 \leq j \leq p} n^{-1}\| (\bZ \bB)_j \|_2^2 \leq 3 \| \bB \|_2^2 /2 \leq C \|\hat{\bOmega} - \bOmega\|_2^2  \leq C \frac{\rho_n^2 \log p} {n}. 
	\end{equation}
	Note that 
	$$
	\| \hat{\bX}_j - \widetilde{\bX}_j \|_2 \leq r \| \bX \bA_j \|_2 + \|\bZ \bB_j \|_2.
	$$
	Therefore, in view of \eqref{sec4.1-4} and \eqref{sec4.1-5} we can show that for some constant $C > 0$, 
	\begin{equation}
		\mathbb{P} \bigg(n^{-1/2} \| \hat{\bX}_j - \widetilde{\bX}_j \|_2 \leq C \rho_n \sqrt{\frac{\log p}{n}} \bigg)  \to 1. 
	\end{equation}
	This completes the proof of Proposition \ref{prop-gaussian}.

	\subsection{Proof of Proposition \ref{prop-gaussian-copula}} \label{new.SecA.9}
	In light of the definitions of $\hat{\bX}$ and $\widetilde{\bX}$, we can obtain through the triangle inequality that
	\begin{equation} \label{pf-p2-0}
		\begin{split}
			& n^{-1/2} \max_{1 \leq j \leq p} \| \hat{\bX}_j - \widetilde{\bX}_j \|_2 \\
			& \leq \max_{1 \leq j \leq p} n^{-1/2} \bigg( \sum_{i = 1}^n \big[\hat{F}_j^{-1}(\Phi(\hat{\bU}_{i, j})) - \hat{F}_j^{-1} (\Phi(\widetilde{\bU}_{i, j }))  \big]^2  \bigg)^{1/2}\\
			& \quad + \max_{1 \leq j \leq p} n^{-1/2} \bigg( \sum_{i = 1}^n \big[\hat{F}_j^{-1}(\Phi(\widetilde{\bU}_{i, j})) - {F}_j^{-1} (\Phi(\widetilde{\bU}_{i, j }))  \big]^2  \bigg)^{1/2}. 
		\end{split}
	\end{equation}
	We claim that 
	\begin{align}
		& \mathbb{P} \bigg(\max_{1 \leq j \leq p} n^{-1}   \sum_{i = 1}^n \big[\hat{F}_j^{-1} (\Phi(\hat{\bU}_{i, j})) - \hat{F}_j^{-1}  (\Phi(\widetilde{\bU}_{i, j }))  \big]^2   \geq  \widetilde{C} \Big( \frac{\rho_n^2 \log p}{n}  + \frac{p \rho_n (\log n)^3} {n} \Big) \bigg) \nonumber\\
		&\quad \to 0, \label{pf-p2-1}\\
		& \mathbb{P} \bigg( \max_{1 \leq j \leq p} n^{-1}   \sum_{i = 1}^n \big[\hat{F}_j^{-1} (\Phi(\widetilde{\bU}_{i, j})) - {F}_j^{-1}  (\Phi(\widetilde{\bU}_{i, j }))  \big]^2    \geq \frac{2  M p (\log n)^2 } {n}  \bigg) \to 0, \label{pf-p2-2}
	\end{align}
	which together with \eqref{pf-p2-0} yield the desired conclusion of Proposition \ref{prop-gaussian-copula}. It remains to establish \eqref{pf-p2-1} and \eqref{pf-p2-2}. We will begin with the proof of \eqref{pf-p2-1}. 
	
	\bigskip
	
	\noindent\textbf{Proof of \eqref{pf-p2-1}}. From assumption \eqref{p2-cond2} and the observation that $ \frac{\log n} {n^2} \ll \frac{p \rho_n (\log n)^3} {n} $, it holds that for some large constant $C > 0$, 
	\begin{equation} \label{pf-p2-19}
		\begin{split}
			&  \mathbb{P} \bigg(\max_{1 \leq j \leq p} n^{-1 } \sum_{i = 1}^n \big[\hat{F}_j^{-1} (\Phi(\hat{\bU}_{i, j})) - \hat{F}_j^{-1} (\Phi(\widetilde{\bU}_{i, j }))  \big]^2   \geq C \Big(\frac{\rho_n^2 \log p}{n}   + \frac{p \rho_n (\log n)^3} {n} \Big) \bigg) \\
			& \leq  \mathbb{P} \bigg(\max_{1 \leq j \leq p} n^{-1 } \sum_{i = 1}^n \Big[ | \Phi(\hat{\bU}_{i, j}) - \Phi(\widetilde{\bU}_{i, j})|^2 + (\log n)^2 n^{-2} \\
			& \qquad +  n^{-1 } (\log n )|\Phi(\hat{\bU}_{i, j}) - \Phi(\widetilde{\bU}_{i, j})| \Big]   \geq C \Big(\frac{\rho_n^2 \log p}{n}  + \frac{p \rho_n (\log n)^3} {n} \Big) \bigg) \\ 
			& \quad +  \mathbb{P} \Big( \max_{1 \leq j \leq p} \sup_{ x, y \in (0, 1) } \frac{\big| \hat{F}_j^{-1} (x) - \hat{F}_j^{-1} (y) \big| } {  |x - y| +  ( n^{-1 } (\log n) |x - y| )^{1/2} +   n^{-1} \log n  } \geq M   \Big)\\
			& \leq  \mathbb{P} \bigg(\max_{1 \leq j \leq p} n^{-1 } \sum_{i = 1}^n \Big[ | \Phi(\hat{\bU}_{i, j}) - \Phi(\widetilde{\bU}_{i, j})|^2  \\
			& \qquad +  n^{-1 } (\log n) |\Phi(\hat{\bU}_{i, j}) - \Phi(\widetilde{\bU}_{i, j})| \Big]   \geq {C} \Big(\frac{\rho_n^2 \log p}{n}  + \frac{p \rho_n (\log n)^3} {n} \Big) \bigg) + o(1) \\
			& : = P_1 + o(1).
		\end{split}
	\end{equation}
	
	We next bound term $P_1$ above. Using the fact that $|\Phi(x) - \Phi(y)| \leq \frac{1}{\sqrt{2 \pi}} | x - y |$ and the basic inequality $\sum_{i = 1}^n |a_n| \leq \sqrt n (\sum_{i = 1}^n {a_n}^2)^{1/2}$, we have that 
	\begin{equation}  \label{pf-p2-E1}
		\begin{split} 
			P_1  & \leq  \mathbb{P} \bigg( \max_{1 \leq j \leq p}  \Big( n^{-1}  \| \hat{\bU}_{j}   - \widetilde{\bU}_{ j} \|_2^2 + (\log n) n^{-3/2}   \| \hat{\bU}_{j}   - \widetilde{\bU}_{ j} \|_2    \Big) \\
			&\qquad \geq  {C} \Big(\frac{\rho_n^2 \log p}{n}  + \frac{p \rho_n (\log n)^3} {n} \Big)  \bigg).
		\end{split}
	\end{equation}
	It suffices to consider the bound of $  \max_{1 \leq j \leq p}  n^{-1}  \| \hat{\bU}_{j}   - \widetilde{\bU}_{ j} \|_2^2 $. With the aid of the triangle inequality and the definitions of $\hat{\bU}$ and $\widetilde{\bU}$, it follows that 
	\begin{equation} \label{pf-p2-3}
		\begin{split}
			\max_{1 \leq j \leq p} 
			n^{-1}\| \hat{\bU}_{j}   - \widetilde{\bU}_{ j} \|_2^2 & \leq 3 \max_{1 \leq j \leq p} n^{-1} \| (\hat{\bV} - \widetilde{\bV} ) (I_p - r \hat{\bOmega})_j  \|_2^2 \\
			& \quad+ 3 r^2 \max_{1 \leq j \leq p} n^{-1} \| \widetilde{\bV} ( \hat{\bOmega}_j - \bOmega_j ) \|_2^2 \\
			& \quad + 3 \max_{1 \leq j \leq p} n^{-1} \| \bZ [(2 r I_p - r^2 \hat{\bOmega})^{1/2} - (2 r I_p - r^2 {\bOmega})^{1/2}  ] \|_2^2.
		\end{split}
	\end{equation}
	We will investigate the three terms in the upper bound above separately. Regarding the third term above, under the assumption in \eqref{eq-5} it has been shown in \eqref{sec4.1-5} that with probability $1 - o(1)$, 
	\begin{equation} \label{pf-p2-4}
		\max_{1 \leq j \leq p} n^{-1} \| \bZ [(2 r I_p - r^2 \hat{\bOmega})^{1/2} - (2 r I_p - r^2 {\bOmega})^{1/2}  ] \|_2^2 \leq C \frac{\rho_n^2 \log p}{n} . 
	\end{equation}
	
	As for the second term in the upper bound in \eqref{pf-p2-3}, noting that the rows of $\widetilde\bV$ are i.i.d. and follow the Gaussian distribution $N({\bf 0}, \bOmega^{-1})$, an application of similar arguments as for \eqref{sec4.1-4} gives that with probability $1 - o(1)$, 
	\begin{equation} \label{pf-p2-5}
		\max_{1 \leq j \leq p} n^{-1} \| \widetilde{\bV} ( \hat{\bOmega}_j - \bOmega_j ) \|_2^2  \leq C\frac{\rho_n^2 \log p}{n} .
	\end{equation}
	For the first term in the upper bound in \eqref{pf-p2-3} above, noting that $  \| I_p - r\hat{\bOmega})_j \| \leq \rho_n + 1$ by the sparsity assumption that $ \| \hat\bOmega_j\| \leq \rho_n $, we have that 
	\begin{equation} \label{pf-p2-6}
		\begin{split}
			\max_{1 \leq j \leq p} n^{-1} \| (\hat{\bV} - \widetilde{\bV} ) (I_p - r \hat{\bOmega})_j  \|_2^2 & \leq \max_{J: |J| \leq \rho_n +1 } \|  n^{-1} (\hat{\bV}_J - \widetilde{\bV}_J)^{\top}  (\hat{\bV}_J - \widetilde{\bV}_J) \|_2  \\
			& \quad \times \max_{1 \leq j \leq p}  \| (I_p - r \hat{\bOmega})_j \|_2^2. 
		\end{split}
	\end{equation}
	
	For the second term in the bound above, from the triangle inequality and inequality $ \| \bA_{j} \|_2 \leq \| \bA \|_2 $ for each matrix $\bA$,  it is easy to see that
	\begin{equation*}
		\max_{1 \leq j \leq p} \| (I_p - r \hat{\bOmega})_j \|_2 \leq \| I_p - r \hat{\bOmega} \|_2 \leq \| I_p - r {\bOmega} \|_2 + r \| \hat{\bOmega} - \bOmega \|_2. 
	\end{equation*}
	Thus it follows from assumption \eqref{eq-5} that for a constant $C > 0$, with probability $1 - o(1)$ we have  
	\begin{equation} \label{pf-p2-7}
		\max_{1 \leq j \leq p} \| (I_p - r \hat{\bOmega})_j \|_2 \leq  C.
	\end{equation}
	Regarding the first term on the right-hand side of \eqref{pf-p2-6} above, using the definitions of $\hat{\bV}$ and $\widetilde\bV$, and inequality $\| \bA \|_2 \leq d \|\bA \|_{\max}$ for each square matrix $\bA \in \mathbb{R}^{d \times d}$, we can deduce that 
	\begin{equation} \label{pf-p2-9}
		\begin{split}
			& \max_{J: |J| \leq \rho_n +1 } \|  n^{-1} (\hat{\bV}_J - \widetilde{\bV}_J)^{\top}  (\hat{\bV}_J - \widetilde{\bV}_J) \|_2 \\
			& \leq (\rho_n + 1) \|  n^{-1} (\hat{\bV} - \widetilde{\bV} )^{\top}  (\hat{\bV}  - \widetilde{\bV} ) \|_{\max} \\
			& \leq (\rho_n + 1) \max_{1 \leq j \leq p} n^{-1} \sum_{i = 1}^n |\hat{\bV}_{i, j} - \widetilde{\bV}_{i, j}|^2\\
			& = (\rho_n + 1)  \max_{1 \leq j \leq p} n^{-1} \sum_{i = 1}^n  | \Phi^{-1} (\hat{F}_j (\bX_{i, j} )) -  \Phi^{-1} ( F_j (\bX_{i, j} )) |^2 .
		\end{split}
	\end{equation}
	Denote by $H_{j, n} =[F_j^{-1} (2M n^{-1 } \log n),  F_j^{-1} (1 - 2M n^{-1} \log n )] $ with constant $M$ as given in assumption \eqref{p2-cond4}. We can write that 
	\begin{equation} \label{pf-p2-10}
		\begin{split}
			&\max_{1 \leq j \leq p} n^{-1} \sum_{i = 1}^n  | \Phi^{-1} (\hat{F}_j (\bX_{i, j} )) -  \Phi^{-1} ( F_j (\bX_{i, j} )) |^2 \\
			& =  \max_{1 \leq j \leq p} n^{-1} \sum_{i = 1}^n  | \Phi^{-1} (\hat{F}_j (\bX_{i, j} )) -  \Phi^{-1} ( F_j (\bX_{i, j} )) |^2 \mathbbm{1} (\bX_{i, j}  \in H_{j, n}) \\
			&\quad + \max_{1 \leq j \leq p} n^{-1} \sum_{i = 1}^n  | \Phi^{-1} (\hat{F}_j (\bX_{i, j} )) -  \Phi^{-1} ( F_j (\bX_{i, j} )) |^2 \mathbbm{1} (\bX_{i, j}  \notin H_{j, n}) \\
			& := E_{1} + E_{2}.
		\end{split}
	\end{equation}
	
	Let us first consider term $E_{2}$ above. Observe that 
	\begin{equation} \label{pf-p2-n1}
		\begin{split}
			E_{2} & \leq \max_{1 \leq j \leq p} n^{-1} \sum_{i = 1}^n  | \Phi^{-1} (\hat{F}_j (\bX_{i, j} )) |^2 \mathbbm{1} (\bX_{i, j}  \notin H_{j, n}) \\
			&\quad  + \max_{1 \leq j \leq p} n^{-1} \sum_{i = 1}^n  | \Phi^{-1} ({F}_j (\bX_{i, j} )) |^2 \mathbbm{1} (\bX_{i, j}  \notin H_{j, n}).
		\end{split}
	\end{equation}
	For the first term in the bound above, notice that $$|\Phi^{-1} (\hat{F}_j (\bX_{i, j})) |= O(\sqrt{\log n })$$ due to the assumption that $\frac{1}{2n} \leq F_j(x)  \leq 1 - \frac{1}{2n} $ for each $x\in \supp(X_j)$. Then it follows from the union bound, the Markov inequality, and the definition of $H_{j, n}$ that 
	\begin{equation} \label{pf-p2-n2}
		\begin{split}
			& \mathbb{P} \bigg( \max_{1 \leq j \leq p} n^{-1} \sum_{i = 1}^n  | \Phi^{-1} ({F}_j (\bX_{i, j} )) |^2 \mathbbm{1} (\bX_{i, j}  \notin H_{j, n})  \geq  \frac{p (\log n )^3} {n}  \bigg)  \\
			& \leq \sum_{j = 1}^p  \mathbb{P}  \bigg(   n^{-1} \log n \sum_{i = 1}^n   \mathbbm{1} (\bX_{i, j}  \notin H_{j, n})  \geq \frac{p (\log n )^3} {n} \bigg)  \\
			& \leq \frac{ n } {p (\log n)^2} \sum_{j = 1}^p \mathbb{P} (\bX_{i, j}  \notin H_{j, n}) \\
			& = \frac{p n }{ p (\log n )^2 } \cdot 4 M n^{-1} \log n \\
			&= 4 M (\log n )^{-1} \to 0.
		\end{split}
	\end{equation}
	
	As for the second term in the upper bound in \eqref{pf-p2-n1} above, an application of the Markov inequality and the fact that $F_j(\bX_{i, j} ) $ follows the standard uniform distribution gives that 
	\begin{equation} \label{pf-p2-n3}
		\begin{split}
			& \mathbb{P}  \bigg(\max_{1 \leq j \leq p} n^{-1} \sum_{i = 1}^n  | \Phi^{-1} ({F}_j (\bX_{i, j} )) |^2 \mathbbm{1} (\bX_{i, j}  \notin H_{j, n}) \geq \frac{p (\log n )^3} {n}  \bigg) \\
			& \leq  \frac{n} {p (\log n)^3} \sum_{j = 1}^p \e \Big(| \Phi^{-1} ({F}_j (\bX_{i, j} )) |^2 \mathbbm{1} (\bX_{i, j}  \notin H_{j, n}) \Big) \\
			& =  \frac{2 n} { (\log n)^3}   \int_{-\infty}^{ \Phi^{-1} (\frac{2M\log n}{n} ) } \frac{1}{\sqrt{2\pi}} u^2 e^{-u^2/2} du \\
			& \leq \frac{2n}{(\log n)^3 |\Phi^{-1} (\frac{2M\log n}{n} )| }  \int_{-\infty}^{ \Phi^{-1} (\frac{2M\log n}{n} ) } \frac{1}{\sqrt{2\pi}} |u|^3 e^{-u^2/2} du \\
			& \leq C  \frac{n}{(\log n)^3 |\Phi^{-1} (\frac{2M\log n}{n} ) | }  \cdot \big|  \Phi^{-1} (\frac{2M\log n}{n})\big|^3 \cdot \Phi(\Phi^{-1} (\frac{2M\log n}{n}) ) \\
			& \leq  C (\log n)^{-1} \to 0,
		\end{split}
	\end{equation}
	where in the last step above, we have used the facts that $|\Phi^{-1} (\frac{M \log n} {n}) | \leq C \sqrt{\log n}$, $\int u^3 e^{-u^2 / 2} du = - (u^2 + 2) e^{-u^2/2}$, and $e^{-x^2/2}/\Phi(x) = O(|x|) $ for $x < -2$. Combining \eqref{pf-p2-n1}, \eqref{pf-p2-n2}, and \eqref{pf-p2-n3} yields that with probability $1 - o(1)$, 
	\begin{equation} \label{pf-p2-n4}
		E_{2} \leq \frac{p (\log n )^3}{n}.
	\end{equation}
	
	Next we proceed with studying term $E_{1}$. 
	First, note that when $|\Phi^{-1} (y )| > 2$, it holds that $$  [ \Phi^{-1} (y) ]'  = \frac{1} {\Phi'( \Phi^{-1} (y) )} \leq C \frac{1} {(y \land (1 - y))   |\Phi^{-1} (y)| }$$ due to the fact that $\Phi'(x) /(1 - \Phi(x) ) \geq C   x  $ for $x > 2$ and $\Phi'(x) /  \Phi(x)  \geq C |x| $ for $x < -2$. When $|\Phi^{-1}(y)| \leq 2 $, it is easy to see that $$ [ \Phi^{-1} (y) ]'  = \frac{1} {\Phi'( \Phi^{-1} (y) )} \leq C.$$  Thus, combining the previous two results shows that for $y \in \mathbb{R}$, 
	\begin{equation} \label{pf-p2-8}
		[ \Phi^{-1} (y) ]' \leq  \frac{C} {(y \land (1 - y))  |\Phi^{-1} (y)|  }  \leq  \frac{C} {(y \land (1 - y))   } . 
	\end{equation}
	Let us define an interval $$\delta_j(x) = \bigg[F_j(x) - \sqrt{\frac{M [F_j(x) \land (1 - F_{j}(x)) ] \log n} {n}},   F_j(x) + \sqrt{\frac{M [F_j(x) \land (1 - F_{j}(x)) ]\log n} {n}} \bigg].$$ Observe that under assumption \eqref{p2-cond4}, we have that 
	\begin{equation} \label{pf-p2-13}
		\begin{split}
			&\mathbb{P} ( E_{1}  \geq x)\\
			& \leq \mathbb{P} \bigg( \max_{1 \leq j \leq p}  n^{-1} (\frac{M \log n} {n}) \sum_{i = 1}^n \Big(\sup_{y \in \delta_j(\bX_{i, j} ) } [\Phi^{-1} (y)]' \Big)^2   {F}_j (\bX_{i, j} ) (1  - F_j (\bX_{i, j} )  \\
			&\quad \cdot\mathbbm{1} (\bX_{i, j}  \in H_{j, n})  \geq x\bigg) + o(1).
		\end{split}
	\end{equation}
	
	When $\bX_{i, j}  \in H_{j, n}$, it holds that $F_j(\bX_{i, j} ) \in [2 M n^{-1} \log n, 1 - 2 M n^{-1} \log n]$ and hence 
	$$
	\sup_{y \in \delta(\bX_{i, j} )} \Big| \frac{y} {F ({\bX_{i, j} }) } - 1 \Big| \leq \sqrt{ \frac{M \log n} {n F_{j}(\bX_{i, j} )}} \leq  1/\sqrt 2. 
	$$ 
	Similarly, we have that 
	$$
	\sup_{y \in  \delta(\bX_{i, j} )} \Big| \frac{1 - y} {1 - F ({\bX_{i, j} }) } - 1 \Big| \leq 1/\sqrt 2.
	$$ 
	The above two bounds combined with \eqref{pf-p2-8} yields that for $\bX_{i, j}  \in H_{j, n}$,
	\begin{equation*}
		\sup_{y \in \delta_j(\bX_{i, j} ) } [\Phi^{-1} (y)]' 
		\leq  \sup_{y \in \delta_j(\bX_{i, j} )} \frac{C } { y \land (1 - y)} \leq  \frac{C } {F_j(\bX_{i, j} ) \land (1 - F_j(\bX_{i, j} ))}. 
	\end{equation*}
	In view of the above bound, \eqref{pf-p2-13}, and the fact that $F_j(\bX_{i, j} )$ follows the standard uniform distribution, we can deduce that 
	\begin{equation} \label{pf-p2-n5}
		\begin{split}
			& \mathbb{P} \Big( E_{1} \geq \frac{p (\log n)^3} {n} \Big) \\
			& \leq \mathbb{P}  \bigg(\max_{1 \leq j \leq p} n^{-1} (\frac{M \log n} {n}) 
			\sum_{i = 1}^n  \frac{C   } { F_j(\bX_{i, j} ) \land (1 - F_j(\bX_{i, j} ))    }  \mathbbm{1} (\bX_{i, j}  \in H_{j, n}) \\
			&\qquad\geq \frac{p (\log n)^3} {n} \bigg) + o(1)\\
			& \leq   \frac{C M }{p (\log n)^2 }   \sum_{j = 1}^p \e \Big(  \frac{1   } { F_j(\bX_{i, j} ) \land (1 - F_j(\bX_{i, j} ))    }  \mathbbm{1} (\bX_{i, j}  \in H_{j, n})  \Big)  \\
			& =  \frac{C M }{ (\log n)^2 }   \int_{2 M n^{-1} \log n}^{1 - 2 M n^{-1} \log n} \frac{1} {u\land(1 - u)} du\\
			& \leq  \frac{C M }{ (\log n)^2 } \cdot C \log n \\
			&\leq \frac{C M } {\log n} \to 0.
		\end{split} 
	\end{equation}
	
	A combination of \eqref{pf-p2-9}, \eqref{pf-p2-10}, \eqref{pf-p2-n4}, and \eqref{pf-p2-n5} shows that with probability $1 - o(1)$, 
	\begin{equation} 
		\begin{split}
			& \max_{J: |J| \leq \rho_n +1 } \|  n^{-1} (\hat{\bV}_J - \widetilde{\bV}_J)^{\top}  (\hat{\bV}_J - \widetilde{\bV}_J) \|_2 \leq \frac{C  p \rho_n (\log n )^3 } {n}, 
		\end{split} 
	\end{equation}
	which together with \eqref{pf-p2-3}--\eqref{pf-p2-7} entails that with probability $1 - o(1)$, 
	\begin{equation} \label{pf-p2-15}
		\begin{split}
			n^{-1} \max_{1 \leq j \leq p} \| \hat{\bU}_{j}   - \widetilde{\bU}_{ j} \|_2^2 \leq C \Big( \frac{\rho_n^2 \log p}{n}  + \frac{ p \rho_n (\log n )^3 } {n} \Big)
		\end{split}
	\end{equation}
	and 
	\begin{equation}
		(\log n) n^{-3/2} \max_{1 \leq j \leq p} \| \hat{\bU}_{j}   - \widetilde{\bU}_{ j} \|_2 \leq C  (\log n) n^{-1} \Big(  \rho_n \frac{\log p}{n} + \sqrt{\frac{ p \rho_n (\log n )^3 } {n}} \Big).
	\end{equation}
	Plugging \eqref{pf-p2-15} into \eqref{pf-p2-E1}, it follows that
	\begin{equation} \label{pf-p2-18}
		P_1  \to 0.
	\end{equation} 
	Therefore, substituting \eqref{pf-p2-18} into \eqref{pf-p2-19} derives the desired result \eqref{pf-p2-1}. It remains to establish \eqref{pf-p2-2}. 
	
	\bigskip
	
	\noindent \textbf{Proof of \eqref{pf-p2-2}}. Let us define $I_{n} = [2M n^{-1} \log n, 1 - 2M n^{-1} \log n]$. It holds that 
	\begin{equation} \label{pf-p2-19}
		\begin{split}
			& \mathbb{P} \bigg( \max_{1 \leq j \leq p} n^{-1}   \sum_{i = 1}^n \big[\hat{F}_j^{-1} (\Phi(\widetilde{\bU}_{i, j})) - {F}_j^{-1}  (\Phi(\widetilde{\bU}_{i, j }))  \big]^2    \geq \frac{2  M p (\log n)^2 } {n} \bigg) \\
			& =  \mathbb{P} \bigg( \max_{1 \leq j \leq p} n^{-1}   \sum_{i = 1}^n \big[\hat{F}_j^{-1} (\Phi(\widetilde{\bU}_{i, j})) - {F}_j^{-1}  (\Phi(\widetilde{\bU}_{i, j }))  \big]^2 \mathbbm{1} ( \Phi(\widetilde{\bU}_{i, j}) \in I_{ n})   \\
			&\qquad \geq \frac{ M p (\log n)^2 } {n} \bigg) \\
			& \quad + \mathbb{P} \bigg( \max_{1 \leq j \leq p} n^{-1}   \sum_{i = 1}^n \big[\hat{F}_j^{-1} (\Phi(\widetilde{\bU}_{i, j})) - {F}_j^{-1}  (\Phi(\widetilde{\bU}_{i, j }))  \big]^2 \mathbbm{1} ( \Phi(\widetilde{\bU}_{i, j}) \notin I_{ n})   \\
			&\qquad\quad \geq \frac{ M p  (\log n)^2 } {n}  \bigg). 
		\end{split}
	\end{equation}
	For the first term on the right-hand side of  \eqref{pf-p2-19} above, under assumption \eqref{p2-cond3} we have that 
	\begin{equation} \label{pf-p2-20}
		\begin{split}
			& \mathbb{P} \bigg( \max_{1 \leq j \leq p} n^{-1}   \sum_{i = 1}^n \big[\hat{F}_j^{-1} (\Phi(\widetilde{\bU}_{i, j})) - {F}_j^{-1}  (\Phi(\widetilde{\bU}_{i, j }))  \big]^2 \mathbbm{1} ( \Phi(\widetilde{\bU}_{i, j}) \in I_{ n})   \\
			&\quad\geq  \frac{ M p (\log n)^2 } {n} \bigg) \\
			& \leq \mathbb{P} \bigg( \frac{M \log n } {n} \geq  \frac{ M p (\log n)^2 } {n} \bigg) + o(1) \\
			&= 0 + o(1) \to 0.
		\end{split}
	\end{equation}
	
	Regarding the second term on the right-hand side of \eqref{pf-p2-19} above, observe that $ | F_j^{-1} (\Phi(\widetilde{\bU}_{i, j})) | \leq b $ and $ |\hat{F}_j^{-1} (\Phi(\widetilde{\bU}_{i, j})) | \leq b$ by the assumption $\supp(X_j) \in [-b, b]$.
	In addition, $\Phi(\widetilde{\bU}_{i, j}) $ follows the standard uniform distribution and thus $\mathbb{P} (\Phi(\widetilde{\bU}_{i, j}) \notin I_{n}) = 4 M n^{-1} \log n$. Then we can deduce that 
	\begin{equation} \label{pf-p2-21}
		\begin{split}
			& \mathbb{P} \bigg( \max_{1 \leq j \leq p} n^{-1}   \sum_{i = 1}^n \big[\hat{F}_j^{-1} (\Phi(\widetilde{\bU}_{i, j})) - {F}_j^{-1}  (\Phi(\widetilde{\bU}_{i, j }))  \big]^2 \mathbbm{1} ( \Phi(\widetilde{\bU}_{i, j}) \notin I_{1, n})   \\
			&\quad\geq  \frac{ M p (\log n)^2 } {n}   \bigg) \\
			& \leq  \mathbb{P} \bigg( \max_{1 \leq j \leq p} n^{-1}   \sum_{i = 1}^n  \mathbbm{1} ( \Phi(\widetilde{\bU}_{i, j}) \notin I_{ n})   \geq  \frac{ M p (\log n)^2 } {4 n b^2 }   \bigg) \\
			& \leq \frac{4 n b^2 }{  M p (\log n)^2 } \cdot p \mathbb{P} ( \Phi(\widetilde{\bU}_{i, j} \notin I_{n}) ) \\
			&= \frac{ 16  b^2  } { \log n} \to 0.
		\end{split}
	\end{equation}
	Finally, combining \eqref{pf-p2-19}--\eqref{pf-p2-21} leads to the desired result \eqref{pf-p2-2}. This concludes the proof of Proposition \ref{prop-gaussian-copula}.

	\renewcommand{\thesubsection}{B.\arabic{subsection}}
	
	\section{Proofs of some key lemmas} \label{proof-lemmas}

	\subsection{Proof of Lemma \ref{le-Gaussian-Wasser}} \label{Sec_proof_new_le1}
	
	We claim the following upper bound for $\e [\| \hat{\bX} - \widetilde{\bX} \|_{1, 2}^2 | \bX]$ as presented in \eqref{le-Gaussian-Wasser-1} and lower bound for $\mathbb{W}_{1, 2}(\hat{\mu}^{n}, \widetilde{\mu}^n) $ as shown in \eqref{le-Gaussian-Wasser-2}
	\begin{equation} \label{le-Gaussian-Wasser-1}
		\e [\| \hat{\bX} - \widetilde{\bX} \|_{1, 2}^2 | \bX] \leq 2 \big(1 + \sqrt{2 n^{-1}}\big) (r^2 \lor 1) \max_{1 \leq j \leq p } \big( n^{-1} \bA_j^T \bX^T \bX \bA_j +  \| \bB_j \|_2^2 \big),
	\end{equation}
	\begin{equation}  \label{le-Gaussian-Wasser-2}
		\mathbb{W}_{1, 2}(\hat{\mu}^{n}, \widetilde{\mu}^n) \geq  \max_{1 \leq j \leq p } \Big( n^{-1}r^2 \bA_j^T \bX^T \bX \bA_j +  \big( (2 r - r^2 \hat\bOmega_{j,j})^{1/2} - (2 r - r^2 \bOmega_{j,j})^{1/2}\big)^2 \Big),
	\end{equation} 
	where $\bA = \hat{\bOmega} - \bOmega$ and $\bB = \hat{\bD} - \bD$, and $\bA_j$ and $\bB_j$ stand for the $j$th columns of $\bA$ and $\bB$, respectively. Their proofs are postponed to the end of the proof. In what follows, we will use subscript $j$ to denote the $j$th column of a generic matrix.

	Next we will show that the upper bound in \eqref{le-Gaussian-Wasser-1} can be bounded from above by the lower bound in \eqref{le-Gaussian-Wasser-2} up to a multiplicative constant. 
	Define the eigen-decompositions $ 2 r I_p - r^2 \hat{\bOmega} = \hat{P}^T \hat{\Lambda} \hat{P}$ and $2 r I_p - r^2  {\bOmega} =  {P}^T {\Lambda}  {P} $, where $\hat{\Lambda}$ and $\Lambda$ are diagonal matrices with positive eigenvalues, and $\hat{P} $ and $P$ are the corresponding eigenvector matrices. By definition, we have $\hat{\bD} = \hat{P}^T \hat{\Lambda}^{1/2} \hat{P}$ and $\bD =  {P}^T  {\Lambda}^{1/2} {P}$. 
	For the second term in the upper bound in \eqref{le-Gaussian-Wasser-1},  it holds that
	\begin{equation} \label{eq_upper_bound1}
		\begin{split}
			\| \bB_j \|_2^2 & = \|\hat{\bD}_j - \bD_j\|_2^2\\
			& =   \hat{P}_j^T \hat{\Lambda} \hat{P}_j  + P_j^T \Lambda P_j - 2 \hat{P}_j^T \hat\Lambda^{1/2} \hat{P} P^T \Lambda^{1/2} P_j   \\          
			& = (2 r -  r^2 \hat{\bOmega}_{j, j}) + (2 r - r^2 \bOmega_{j, j}) - 2 \hat{\bD}_j^{T} \bD_j  .
		\end{split}
	\end{equation}
	Moreover, the second term in the lower bound presented in \eqref{le-Gaussian-Wasser-2} can be written as  
	\begin{equation} \label{eq_lower_bound1}
		\begin{split}
			& \big( (2 r - r^2 \hat\bOmega_{j,j})^{1/2} - (2 r - r^2 \bOmega_{j,j})^{1/2}\big)^2 \\
			& = \big( (\hat{P}_j^T \hat{\Lambda} \hat{P}_j )^{1/2} -( {P}_j^T  {\Lambda} {P}_j )^{1/2} \big)^2 \\
			& = (2 r -  r^2 \hat{\bOmega}_{j, j}) + (2 r - r^2 \bOmega_{j, j}) - 2 (\hat{P}_j^T \hat{\Lambda} \hat{P}_j {P}_j^T  {\Lambda} {P}_j )^{1/2} \\
			& = (2 r -  r^2 \hat{\bOmega}_{j, j}) + (2 r - r^2 \bOmega_{j, j}) - 2 \|\hat{\bD}_j\|_2 \|\bD_j \|_2. 
		\end{split}
	\end{equation}
	Therefore, under the assumption that $\|\hat{\bD}_j\|_2 \|\bD_j \|_2 - \hat{\bD}_j ^T \bD_j \leq C \|\hat{\bD}_j - \bD_j \|_2^2$ for a constant $C < 1/2$, we have 
	$$\|\bB_j \|_2^2 \leq \big( (2 r - r^2 \hat\bOmega_{j,j})^{1/2} - (2 r - r^2 \bOmega_{j,j})^{1/2}\big)^2  + 2C  \|\bB_j \|_2^2
	$$
	and hence
	$$
	\|\bB_j \|_2^2 \leq  \frac{1} {1 - 2 C}\big( (2 r - r^2 \hat\bOmega_{j,j})^{1/2} - (2 r - r^2 \bOmega_{j,j})^{1/2}\big)^2. 
	$$
	This combined with \eqref{le-Gaussian-Wasser-1} and \eqref{le-Gaussian-Wasser-2} proves the desired result in the lemma
	\begin{equation}
		\e [\| \hat{\bX} - \widetilde{\bX} \|_{1, 2}^2 | \bX] \leq  \frac{2}{1 - 2 C} \big(1 +  \sqrt{2 n^{-1} }\big) (r^2 \lor 1)  \mathbb{W}_{1, 2}(\hat{\mu}^{n}, \widetilde{\mu}^n). 
	\end{equation}

	It remains to prove \eqref{le-Gaussian-Wasser-1} and \eqref{le-Gaussian-Wasser-2}. We first prove \eqref{le-Gaussian-Wasser-1}. Recall our construction of coupling in \eqref{sec4.1-1} and \eqref{sec4.1-2} that
	\begin{align*}
		\widehat{\bX} & = \bX(I_p - r \hat{\bOmega}) + \bZ \hat\bD, \\
		\widetilde{\bX} & = \bX(I_p - r {\bOmega}) + \bZ \bD,
	\end{align*}
	where $\bZ = (Z_{i, j}) \in \mathbb{R}^{n \times p}$ is independent of $(\bX, \by)$ and consists of i.i.d. standard normal entries $Z_{i, j} \stackrel{d}{\sim}  N(0, 1)$. It immediately follows that  
	\begin{equation*}
		\hat{\bX} - \widetilde{\bX} = - r\bX \bA + \bZ \bB.
	\end{equation*} 
	Therefore, we have 
	\begin{equation} \label{proof-G-W-1}
		\begin{split}
			\e [\| \hat{\bX} - \widetilde{\bX} \|_{1,2}^2 | \bX ]& = \e \Big[   \max_{1 \leq j \leq p} n^{-1} \| \hat{\bX}_j - \widetilde{\bX}_j \|_2^2 \Big| \bX\Big] \\
			& =  \e \Big[ \max_{1 \leq j \leq p} n^{-1} \| r \bX \bA_j + \bZ \bB_j \|_2^2  \Big| \bX \Big] \\
			& \leq 2  \e \Big[  \max_{1 \leq j \leq p}  \big( r^2  n^{-1} \bA_j^T \bX^T \bX \bA_j  + n^{-1} \bB_j^T \bZ^T \bZ \bB_j \big) \Big| \bX \Big] \\
			& \leq 2 \max_{1 \leq j \leq p} ( r^2  n^{-1} \bA_j^T \bX^T \bX \bA_j  +   \|\bB_j \|_2^2 ) \\
			& \quad + 2   \e \Big[  \max_{1 \leq j \leq p}  \big|n^{-1} \bB_j^T \bZ^T \bZ \bB_j -  \|\bB_j \|_2^2 \big| \Big| \bX \Big],
		\end{split}
	\end{equation}
	where the second last inequality follows from the Cauchy--Schwarz inequality. 
	To deal with the second term $  \e \Big[  \max_{1 \leq j \leq p}  \big|n^{-1}\bB_j^T \bZ^T \bZ \bB_j -  \|\bB_j \|_2^2 \big| \Big| \bX \Big]$ in the above upper bound, a key observation is that $\bZ \bB_j \stackrel{d}{=} (\widetilde{Z}_{1} \| \bB_j\|_2, \ldots, \widetilde{Z}_n \| \bB_j\|_2)$, where $\{\widetilde{Z}_i\}$ are i.i.d. standard normal random variables and are independent of all other variables. Hence, it can be obtained that
	\begin{equation}  \label{proof-G-W-2}
		\begin{split}
			\e \Big[  \max_{1 \leq j \leq p}  \big| n^{-1} \bB_j^T \bZ^T \bZ \bB_j -  \|\bB_j \|_2^2 \big| \Big| \bX \Big] & = \e \bigg[ \max_{1 \leq j \leq p}  \| \bB_j \|_2^2 \Big|  n^{-1}   \sum_{i = 1}^n (\widetilde{Z}_i^2 - 1 ) \Big|  \bigg| \bX \bigg] \\
			& = \max_{1 \leq j \leq p}  \|\bB_j \|_2^2 \, \e  \Big[\Big|  n^{-1}   \sum_{i = 1}^n (\widetilde{Z}_i^2 - 1 ) \Big| \Big] \\
			& \leq  \max_{1 \leq j \leq p} \|\bB_j \|_2^2  \, \bigg(\e  \Big[\Big|  n^{-1}   \sum_{i = 1}^n (\widetilde{Z}_i^2 - 1 ) \Big|^2 \Big] \bigg)^{1/2}\\
			& =  \sqrt{\frac{2}{n}}  \max_{1 \leq j \leq p} \|\bB_j \|_2^2, 
		\end{split}
	\end{equation}
	where we have used the fact that $\e[(\widetilde{Z}_i^2 - 1)^2] = 2$. 
	Combining \eqref{proof-G-W-1} and \eqref{proof-G-W-2} yields the desired result \eqref{le-Gaussian-Wasser-1}.
	
	Now we proceed to prove the lower bound in \eqref{le-Gaussian-Wasser-2}. Note that by Jensen's inequality,
	\begin{equation*}
		\begin{split}
			\mathbb{W}_{1, 2}^2(\hat{\mu}^{n}, \widetilde{\mu}^n) & = \inf_{\gamma \in \Gamma(\hat{\mu}^n, \widetilde{\mu}^n)}   \e_{ ( {\scriptsize \vecc(\hat{\bX}), \vecc(\widetilde{\bX}) } ) \stackrel{d}{\sim} \gamma }   \Big(\max_{1 \leq j \leq p} n^{-1} \|\hat{\bX}_j - \widetilde{\bX}_j \|_{2}^2 \Big) \\
			& \geq \inf_{\gamma \in \Gamma(\hat{\mu}^n, \widetilde{\mu}^n)} \max_{1 \leq j \leq p}   \e_{ ( {\scriptsize \vecc(\hat{\bX}), \vecc(\widetilde{\bX}) } ) \stackrel{d}{\sim} \gamma } \big(  n^{-1} \|\hat{\bX}_j - \widetilde{\bX}_j \|_{2}^2 \big). 
		\end{split}   
	\end{equation*}
	Observe that given $\bX$, we have $\hat{\bX}_j \stackrel{d}{\sim} \hat{\nu}_j^n $ and $\widetilde{\bX}_j \stackrel{d}{\sim} \widetilde{\nu}_j^n$, where $\hat{\nu}_j^n$ is the Gaussian distribution $N( \bX(I_p- r\hat{\bOmega})_j, ( 2 r  - r^2 \hat{\bOmega}_{j, j}) I_n )$ and $\widetilde{\nu}_j^n$ is the Gaussian distribution $N( \bX(I_p- r {\bOmega})_j, ( 2 r  - r^2 {\bOmega}_{j, j}) I_n )$. Given $\bX$, let $\Gamma(\hat{\nu}_j^n, \widetilde{\nu}_j^n) $ be the set of all couplings of $\hat{\nu}_j^n$ and $\widetilde{\nu}_j^n$. Note that if $ (  \vecc(\hat{\bX}), \vecc(\widetilde{\bX}) ) \stackrel{d}{\sim} \gamma $ for some $\gamma \in \Gamma(\hat{\mu}^n, \widetilde{\mu}^n)$, then it must hold that $ (  \hat{\bX}_j,  \widetilde{\bX}_j ) \stackrel{d}{\sim} \gamma_j $ for some $\gamma_j \in \Gamma (\hat{\nu}_j^n, \widetilde{\nu}_j^n)$. Therefore, we can obtain that
	\begin{equation} \label{proof-G-W-3}
		\begin{split}
			\mathbb{W}_{1, 2}^2(\hat{\mu}^{n}, \widetilde{\mu}^n) & \geq \inf_{\gamma \in \Gamma(\hat{\mu}^n, \widetilde{\mu}^n)} \max_{1 \leq j \leq p}  \inf_{\gamma_{j} \in \Gamma (\hat{\nu}_j^n, \widetilde{\nu}_j^n) }  \e_{  {\scriptsize  (\hat{\bX}_j, \widetilde{\bX}_j) }  \stackrel{d}{\sim} \gamma_j } \big(n^{-1}  \|\hat{\bX}_j - \widetilde{\bX}_j \|_{2}^2 \big) \\
			& =  \max_{1 \leq j \leq p}  \inf_{\gamma_{j} \in \Gamma (\hat{\nu}_j^n, \widetilde{\nu}_j^n)}  \e_{  {\scriptsize  (\hat{\bX}_j, \widetilde{\bX}_j) }  \stackrel{d}{\sim} \gamma_j } \big( n^{-1}  \|\hat{\bX}_j - \widetilde{\bX}_j \|_{2}^2 \big) \\
			& = \max_{1 \leq j \leq p} n^{-1} \mathbb{W}_2^2 (\hat{\nu}_j^n, \widetilde{\nu}_j^n),
		\end{split}
	\end{equation} 
	where $\mathbb{W}_2^2 (\hat{\nu}_j^n, \widetilde{\nu}_j^n) $ is the squared 2-Wasserstein distance between $\hat{\nu}_j^n$ and $\widetilde{\nu}_j^n$. By the well-known result for the 2-Wasserstein distance for Gaussian measures (\cite{givens1984class}), we have
	\begin{equation} \label{proof-G-W-4}
		\begin{split}
			n^{-1} \mathbb{W}_2^2 (\hat{\nu}_j^n, \widetilde{\nu}_j^n) =  n^{-1} \| r\bX \bA_j \|_2^2 + \big( (2 r - r^2 \hat\bOmega_{j,j})^{1/2} - (2 r - r^2 \bOmega_{j,j})^{1/2}\big)^2.
		\end{split}
	\end{equation}
	Plugging \eqref{proof-G-W-4} into \eqref{proof-G-W-3} derives \eqref{le-Gaussian-Wasser-2}. This completes the proof of Lemma \ref{le-Gaussian-Wasser}.    
	
	\subsection{Proof of Lemma \ref{KL-multi-t}} \label{new.SecB.1}
	Let $g_j (\cdot | \bx_{-j})$ be the conditional density function of $ X_j   | X_{-j} = \bx_{-j} $ for $X = (X_1, \ldots, X_p )^{\top} \stackrel{d}{\sim} t_{\nu} ({\bf 0}, I_p)$ and $h_j(\cdot | \bx_{ -j}) $ the conditional density function of $\hat{X}_j | \hat{X}_{-j} = \bx_{-j}$ for $\hat{X} = (\hat{X}_1, \ldots, \hat{X}_p)^{\top} \stackrel{d}{\sim} N({\bf 0}, \frac{\nu} {\nu - 2}I_p)$. Following the definition in \cite{Barber2020}, we define
	\begin{equation} \label{pf-mt-n1}
		\hat{KL}_j :   = \sum_{i = 1}^n \log \bigg(  \frac{ g_j (\bX_{i, j}  | \bX_{i, -j}) h_j(\hat{\bX}_{i, j} | \bX_{i, j} ) } { h_{j} (\bX_{i, j}  | \bX_{i, - j}) g_{j} (\hat{\bX}_{i, j} | \bX_{i, -j}) } \bigg),
	\end{equation}
	where $\bX = (\bX_{i, j} ) \in \mathbb{R}^{n \times p}$ consists of i.i.d. rows sampled from $t_{\nu} ({\bf 0}, I_p)$ and $\hat{\bX} = (\hat{\bX}_{i, j})\in \mathbb{R}^{n \times p}$ consists of i.i.d. rows sampled from $N({\bf 0}, I_p)$. Note that Theorem 1 in \cite{Barber2020} states that 
	\begin{equation}
		\FDR \leq \min_{\veps \geq 0} \bigg\{q e^{\veps} + \mathbb{P} \bigg(\max_{j \in \mathcal{H}_0} \hat{KL}_j > \veps \bigg) \bigg\}. 
	\end{equation}
	We claim that if $ \frac{n p }{\nu (\nu + p)} \geq C$ for some constant $ C> 0$, there exists some positive constant $\alpha$ such that
	\begin{equation} \label{pf-mt-1}
		\mathbb{P} \bigg( \hat{KL}_{j}  \geq C/4 \bigg) \geq \alpha.  
	\end{equation}
	Then it holds that for $0 < \veps < C/4$, $$ \mathbb{P} \big(\max_{1 \leq j \leq p} \hat{KL}_{j}  \geq \veps \big) \geq \alpha,$$ 
	and thus we cannot obtain the desired asymptotic FDR control $\limsup_{(n, p)} \FDR \leq q$ via applying Theorem 1 in \cite{Barber2020}. By contradiction, to allow $\mathbb{P} \big(\max_{1 \leq j \leq p} \hat{KL}_{j}  \geq \veps \big) \to 0$, we must have that $\frac{np} {\nu (\nu + p)} \to 0$ , which is equivalent to $\nu^2 \gg n \min (n, p)$. 
	Hence, Lemma \ref{KL-multi-t} is proved.  Now it remains to establish \eqref{pf-mt-1}. 
	
	\bigskip
	\noindent {\bf Proof of \eqref{pf-mt-1}}. Note that \cite{ding2016conditional} showed that the conditional density $g_j(\bx_j | \bx_{-j}) $ of the multivariate $t$-distribution satisfies that 
	\begin{equation*}
		g_j(\bX_{i, j}  | \bX_{i, -j}) \propto \bigg( 1 + \frac{\bX_{i, j} ^2} {\nu + \| \bX_{i, -j} \|_2^2 } \bigg)^{- (\nu + p) / 2}.
	\end{equation*}
	It is easy to see that the conditional density $h_j (\bX_{i, j}  | \bX_{i, -j})$ of the standard normal distribution satisfies that 
	\begin{equation*}
		h_j(\bX_{i, j}  | \bX_{i, -j}) \propto \exp \{ - \bX_{i, j} ^2 (\nu - 2) / 2\nu \}.
	\end{equation*}
	Plugging the two expressions above into \eqref{pf-mt-n1} yields that 
	\begin{equation*}
		\begin{split}
			\hat{KL}_{j} & = \sum_{i = 1}^n \bigg[ \frac{\bX_{i, j} ^2(\nu - 2)} {2\nu} - \frac{\nu + p}{2} \log \bigg(1 + \frac{\bX_{i, j} ^2} {\nu +  \| \bX_{i, -j} \|_2^2} \bigg) \\
			& \quad  - \bigg(\frac{\hat{\bX}_{i, j}^2 (\nu - 2)} {2 \nu } - \frac{\nu + p}{2} \log \bigg(1 + \frac{\hat{\bX}_{i, j}^2} {\nu +  \| \bX_{i, -j} \|_2^2} \bigg) \bigg)\bigg].
		\end{split}
	\end{equation*}
	Applying the basic inequality that $|\log (1 + x) - (x - x^2/2)| \leq  x^3 $ for each $x > 0$, we can obtain that 
	\begin{equation} \label{pf-mt-2}
		\begin{split}
			\hat{KL}_{j} & =  R_{1, j} + R_{2, j} + O(R_{3, j}),  
		\end{split}
	\end{equation}
	where 
	\begin{align}
		& R_{1, j} = \sum_{i = 1}^n \bigg[ \frac{\bX_{i, j} ^2 (\nu + p)} {2 (\nu + \|\bX_{i, -j} \|_2^2 )}\bigg( \frac{\nu + \|\bX_{i, -j} \|_2^2}{ \nu + p } \cdot \frac{\nu - 2} {\nu}  - 1 \bigg) - \frac{\hat{\bX}_{i, j}^2 (\nu -2) }{2 \nu } \bigg( 1 - \frac{\nu + p}{\nu + \|\bX_{i, -j} \|_2^2}\bigg) \bigg], \\
		& R_{2, j} =  \sum_{i = 1}^n \frac{\nu + p}{4} \bigg(\frac{\hat{\bX}_{i, j}^4}{(\nu + \|\bX_{i, -j} \|_2^2)^2}  - \frac{\bX_{i, j} ^4}{(\nu + \|\bX_{i, -j} \|_2^2)^2} \bigg), \\
		& R_{3, j} = \sum_{i = 1}^n \frac{\nu + p} {2} \bigg(\frac{\hat{\bX}_{i, j}^6}{(\nu + \|\bX_{i, -j} \|_2^2)^3}  + \frac{\bX_{i, j} ^6}{(\nu + \|\bX_{i, -j} \|_2^2)^3} \bigg). 
	\end{align}
	
	We now calculate the mean and variance of $\hat{KL}_j$ separately. Observe that $\sqrt{\frac{\nu -2} {\nu}}\hat{\bX}_{i, j} \stackrel{d}{\sim} N(0, 1)$,  $(p-1)^{-1}  \| \bX_{i, -j} \|_2^2 \stackrel{d}{\sim} F_{p-1, \nu} $, $\bX_{i, -j} \indep \sqrt{\frac{\nu + p}{\nu +  \| \bX_{i, -j} \|_2^2}} \bX_{i, j} $, and $$\sqrt{\frac{\nu + p-1}{\nu + \| \bX_{i, -j}\|_2^2}} \bX_{i, j}  \stackrel{d}{\sim} t_{\nu + p- 1}$$ as shown in \cite{ding2016conditional}. Using the properties of the multivariate $t$-distribution and $F$-distribution, some straightforward calculations show that 
	\begin{equation} \label{pl-n1}
		\begin{split}
			\e (R_{1, j}) & = \frac{n}{2} \bigg[ \frac{\nu + p}{\nu + p - 3} \bigg( \frac{\nu (\nu + p - 3)}{(\nu - 2)(\nu + p)}\cdot \frac{\nu -2}{ \nu} - 1 \bigg) - \bigg( 1 - \frac{(\nu + 2) (\nu + p)}{\nu (\nu + p - 1)}\bigg) \bigg] \\
			& = n \bigg( \frac{2p}{\nu (\nu + p)} +  O(\nu^{-2}) \bigg),
		\end{split}
	\end{equation}
	\begin{equation} \label{pl-nn1}
		\begin{split}
			\e (R_{2, j}) & = \frac{3 n (\nu + p)}{4 } \bigg[ \frac{1} {(\nu + p - 3) (\nu + p - 5)} - \frac{\nu + 2} {\nu (\nu + p - 1)(\nu + p + 1)}\bigg] \\
			& = O (\frac{n }{\nu (\nu + p)}),
		\end{split}
	\end{equation}
	and 
	\begin{equation} \label{pl-nn2}
		\begin{split}
			\e (R_{3, j}) & \leq C n (\nu + p)^{-2}. 
		\end{split}
	\end{equation}
	Combining \eqref{pl-n1}--\eqref{pl-nn2} yields that when $\nu$ and $p$ are large, 
	\begin{equation} \label{pl-nn3}
		\e (\hat{KL}_j) = \frac{n p} {\nu (\nu + p)} + O(n \nu^{-2})\geq \frac{n p} {2 \nu (\nu + p)}.
	\end{equation}
	
	Next we analyze the variance of $\hat{KL}_j$. Notice that 
	\begin{equation} \label{pl-nn4}
		\begin{split}
			\Var (\hat{KL}_j) & = \e \big( ( \hat{KL}_j - \e \hat{KL}_j )^2 \big) \\
			& \leq C \sum_{i = 1}^n  \e \bigg\{ \bigg[ \frac{\bX_{i, j} ^2 (\nu + p)} {2 (\nu + \|\bX_{i, -j} \|_2^2 )}\bigg( \frac{\nu + \|\bX_{i, -j} \|_2^2}{ \nu + p } \cdot \frac{\nu -2 }{\nu}  - 1 \bigg) \\
			&\qquad- \frac{\hat{\bX}_{i, j}^2 (\nu - 2)  }{2 \nu} \bigg( 1 - \frac{\nu + p}{\nu + \|\bX_{i, -j} \|_2^2}\bigg) \bigg]^2 \bigg\} \\
			& \quad + C \sum_{i = 1}^n \e \bigg[  \frac{(\nu + p)^2}{16} \bigg(\frac{\hat{\bX}_{i, j}^4}{(\nu + \|\bX_{i, -j} \|_2^2)^2}  - \frac{\bX_{i, j} ^4}{(\nu + \|\bX_{i, -j} \|_2^2)^2} \bigg)^2 \bigg] \\
			& \leq  \frac{Cn  p}{\nu (\nu + p)}, 
		\end{split}
	\end{equation}
	where in the last step above, we have used the facts that 
	\begin{align*} 
		& \e \bigg(  \frac{\bX_{i, j} ^4 (\nu + p)^2 } { (\nu + \|\bX_{i, -j} \|_2^2 )^2}   \bigg) \leq C,  \\
		& \e \bigg[ \bigg( \frac{\nu + \|\bX_{i, -j} \|_2^2} {\nu + p} \cdot \frac{\nu - 2}{\nu} - 1  \bigg)^2  \bigg]  = \frac{2 p } {\nu (\nu + p)} + O(\nu^{-2}), \\
		& \e \bigg[ \bigg( 1 - \frac{\nu + p  } {\nu + \|\bX_{i, -j} \|_2^2}   \bigg)^2  \bigg] =  \frac{2 p } {\nu (\nu + p)} + O(\nu^{-2}). 
	\end{align*}
	
	In view of the results on the mean and variance of $\hat{KL}_j$ shown in \eqref{pl-nn2} and \eqref{pl-nn3} above, we see that if $ \frac{np}{\nu (\nu + p)} \geq C $ for some constant $C > 0$, $$\e (\hat{KL}_j ) \geq  \frac{np}{2\nu (\nu + p)} \geq C /2 .$$  
	Therefore, we can obtain through the one-sided Markov inequality that for a small constant $\alpha > 0$ (noting that $\e(\hat{KL}_j) > 2 \alpha \sqrt{\Var (\hat{KL}_j)} $ if $\alpha$ is small),
	\begin{equation} \label{pl-nn4}
		\begin{split}
			\mathbb{P} (\hat{KL}_j \geq C/4) & \geq 
			\mathbb{P} (\hat{KL}_j \geq \e (\hat{KL}_j )/2 ) \\
			& \geq \mathbb{P} \Big( \hat{KL}_j\geq \e (\hat{KL}_j) - \alpha \sqrt{\Var(\hat{KL}_j)} \Big) \\
			& \geq 1 - \frac{\Var(\hat{KL}_j )}{\Var(\hat{KL}_j ) + \alpha^2 \Var(\hat{KL}_j )} \\
			&= \frac{\alpha^2} {1 + \alpha^2},
		\end{split}
	\end{equation}
	which establishes \eqref{pf-mt-1}. This completes the proof of Lemma \ref{KL-multi-t}.

	\subsection{Proof of Lemma \ref{fdr-lemma2}} \label{proof.lem2}
	
	Recall that $ G(t) = p_0^{-1} \sum_{j \in \mathcal{H}_0} \mathbb{P} (\widetilde{W}_j \geq t) $ and $G(t)$ is a decreasing, continuous function. The main idea of the proof is to divide the continuous interval $(0, G^{-1} (\frac{c_1 q a_n} {p})]$ into a diverging number of smaller intervals with end points $\{ t_i \}_{i = 0}^{l_n}$ such that $t_0\geq t_1\geq \cdots \geq t_{l_n}$ and $$|G(t_i)/ G(t_{i+1}) - 1 | \to 0$$ uniformly for $0 \leq i \leq l_n$ as $l_n\rightarrow \infty$. 
	Then the supreme over the continuous interval $(0, G^{-1} (\frac{c_1 q a_n}{p})]$ can be reduced to the supreme over the set of discrete points $\{t_i\}_{i= 0}^{l_n}$ and hence, we can apply the union bound to establish the desired result. Similar arguments have also been used in \cite{Liu2013}, \cite{cai2016large}, and \cite{guo2022threshold}. We detail only the proof of \eqref{prop_b1} here since \eqref{prop_b1_2} can be shown in a similar fashion.
	
	We  start with defining a sequence $  0 \leq z_0 < z_1 < \cdots < z_{l_n} = 1 $ and $$t_i = G^{-1} (z_i),$$ where $ z_0 = \frac {c_1 q a_n} {p}$, $z_i = \frac {c_1 q a_n} {p} + \frac {h_n e^{i ^\gamma} } {p}$, and $l_n = [\log ((p - c_1 q a_n)/h_n)]^{1/\gamma}$ with $0 < \gamma < 1$ and sequence $h_n \to \infty$ satisfying that $h_n /a_n \to 0$. 
	As long as $m_n /a_n = o(1)$, we can choose $$h_n  =  \frac{a_n}  {(a_n / m_n)^{\eta} }$$ for some $\eta  \in (0, 1)$. Then an application of similar technical analysis as in \cite{guo2022threshold} shows that as $a_n \to \infty$,
	\begin{equation}\label{pf-lemma2-G-grid}
		\sup_{0 \leq i \leq l_n}|G(t_i)/G(t_{i+1}) - 1 |  \to 0. 
	\end{equation} 
	For $t \in (0, G( \frac{c_1 q a_n} {p})]$, there exists some $0 \leq i \leq l_n - 1 $ such that $t \in [t_{i+1}, t_i]$.  It follows from the monotonicity of $ \mathbb{P} (\widetilde{W}_j \geq t) $ and $\mathbbm{1} ( \hat{W}_j \geq t) $ that 
	\begin{equation*}  
		\begin{split}
			\bigg| \frac { \sum_{j \in \mathcal{H}_0 } \mathbbm{1} (\hat{W}_j \geq t) } { p_0 G(t) } - 1   \bigg| 
			& \leq  \max \bigg\{ \bigg| \frac{  \sum_{j \in \mathcal{H}_0 } \mathbbm{1} (\hat{W}_j \geq t_{i+1})} {p_0 G(t_i)} - 1  \bigg|, \\
			&\quad\bigg| \frac{ \sum_{j \in \mathcal{H}_0 } \mathbbm{1} (\hat{W}_j \geq t_{i})} {p_0 G(t_{i+1})} - 1  \bigg| \bigg\} .
		\end{split}
	\end{equation*}
	The two terms within the brackets on the right-hand side of the expression above can be bounded similarly and we will provide only the details on how to bound the first term for simplicity. 
	
	With the aid of the fact that $ | x y - 1 | \leq | x -1| |y - 1| + |x - 1| + |y -1| $ for all $x, y \in \mathbb{R}$, we can deduce that 
	\begin{equation*}   
		\begin{split}
			\bigg| \frac{  \sum_{j \in \mathcal{H}_0 } \mathbbm{1} (\hat{W}_j \geq t_{i+1})} {p_0 G(t_i)} - 1  \bigg|  & \leq \bigg|  \frac{\sum_{j \in \mathcal{H}_0 } \mathbbm{1} (\hat{W}_j \geq t_{i+1})} {p_0 G(t_{i+1})} - 1  \bigg| \cdot \sup_{0 \leq i \leq l_n} \bigg| \frac{G(t_{i})} {G(t_{i+1})} -1 \bigg| \\
			& \quad + \bigg|  \frac{\sum_{j \in \mathcal{H}_0 } \mathbbm{1} (\hat{W}_j \geq t_{i+1})} {p_0 G(t_{i+1})} - 1  \bigg| + \sup_{0 \leq i \leq l_n} \bigg| \frac{G(t_{i})} {G(t_{i+1})} -1 \bigg|\\
			& \leq  \bigg|  \frac{\sum_{j \in \mathcal{H}_0 } \mathbbm{1} (\hat{W}_j \geq t_{i+1})} {p_0 G(t_{i+1})} - 1  \bigg|\cdot(1+o(1)) + \sup_{0 \leq i \leq l_n} \bigg| \frac{G(t_{i})} {G(t_{i+1})} -1 \bigg|,
		\end{split}
	\end{equation*}
	where the last step above is because of \eqref{pf-lemma2-G-grid} and the $o(1)$ term is uniformly over all $i$. 
	Combining the above two results and applying \eqref{pf-lemma2-G-grid} again lead to 
	\begin{equation} \label{pf-lemma2-step1} 
		\begin{split}
			& \bigg| \frac { \sum_{j \in \mathcal{H}_0 } \mathbbm{1} (\hat{W}_j \geq t) } { p_0 G(t) } - 1   \bigg| \\
			& \leq \max\bigg\{  \bigg|  \frac{\sum_{j \in \mathcal{H}_0 } \mathbbm{1} (\hat{W}_j \geq t_{i+1})} {p_0 G(t_{i+1})} - 1  \bigg|, \, \bigg|  \frac{\sum_{j \in \mathcal{H}_0 } \mathbbm{1} (\hat{W}_j \geq t_{i})} {p_0 G(t_{i})} - 1  \bigg|     \bigg\} \\
			&\quad\times \big(1 + o(1) \big) + o(1). 
		\end{split}
	\end{equation}
	Thus, to prove the desired result, it is sufficient to show that
	\begin{equation}\label{eq_d_n_to0}
		D_n := \sup_{0 \leq i \leq l_n} \Big| \frac { \sum_{j \in \mathcal{H}_0 } \mathbbm{1} (\hat{W}_j \geq t_i) } { p_0 G(t_i) } - 1  \Big|=o_p(1).
	\end{equation}
	
	We now proceed with establishing \eqref{eq_d_n_to0}. Let us define an event $$\mathcal{B}_3 = \{\max_{1 \leq j \leq p } | \hat{W}_j - \widetilde{W}_j | \leq b_n \}.$$ From Condition \ref{fdr-condition1}, it holds that $\mathbb{P} (\mathcal{B}_3^c) \to 0$. Note that for any two events $A$ and $B$, we have that $\mathbb P(A) \leq \mathbb P(A\cap B) + P(B^c)$. Repeatedly using such inequality, the union bound, and the property that $\mathbb{P} (\mathcal{B}_3^c) \to 0$, we can deduce that for each $\epsilon > 0 $, 
	\begin{equation} \label{eq_d_n}
		\begin{split}
			\mathbb{P} ( D_n \geq \epsilon ) & \leq \sum_{i = 0}^{l_n} \mathbb{P} \Big( \Big| \frac { \sum_{j \in \mathcal{H}_0 } \{ \mathbbm{1} (\hat{W}_j \geq t_i) - \mathbb{P} ( \widetilde{W}_i \geq t_i ) \} } { p_0 G(t_i) }   \Big| \geq \epsilon, \mathcal{B}_3 \Big) + \mathbb{P} (\mathcal{B}_3^c) \\
			& \leq  \sum_{i = 0}^{l_n} \mathbb{P} \Big( \Big| \frac { \sum_{j \in \mathcal{H}_0 } \{ \mathbbm{1} (\widetilde{W}_j \geq t_i) - \mathbb{P} ( \widetilde{W}_i \geq t_i ) \} } { p_0 G(t_i) }   \Big| \geq \epsilon /2 \Big) \\
			& \quad +  \sum_{i = 0}^{l_n} \mathbb{P} \Big(  \Big| \frac { \sum_{j \in  \mathcal{H}_0 } [ \mathbbm{1} (\hat{W}_j \geq t_i) - \mathbbm{1} ( \widetilde{W}_i \geq t_i ) ] } { p_0 G(t_i) }   \Big| \geq \epsilon /2, \mathcal{B}_3\Big) + o(1) \\
			& \leq   \sum_{i = 0}^{l_n} \frac { 4 \e \big[ \big\{\sum_{j \in \mathcal{H}_0 } [ \mathbbm{1} (\widetilde{W}_j \geq t_i) - \mathbb{P} ( \widetilde{W}_i \geq t_i ) ] \big\}^2 \big] } { \epsilon^2 p_0^2 G^2 (t_i) } \\
			& \quad +  \sum_{i = 0}^{l_n} \frac { 2 \sum_{j \in \mathcal{H}_0}   \mathbb{P}\big( t_i - b_n \leq  \widetilde{W}_j \leq t_i + b_n    \big)  } { \epsilon p_0 G(t_i) } + o(1),
		\end{split}
	\end{equation}
	where the last step above is due to the Markov inequality and the fact that $|\mathbbm{1} (\hat{W}_j \geq t_i) - \mathbbm{1} ( \widetilde{W}_i \geq t_i )|\leq \mathbbm{1} (t_i-b_n\leq \widetilde{W}_j \leq t_i+b_n)$ on event $\mathcal B_3$.
	
	We next bound the first two terms on the very right-hand side of \eqref{eq_d_n} above. For the first term, under Condition \ref{fdr-condition4} for the weak dependence between $\{W_j\}$, we have that 
	\begin{equation} \label{pf-le2-1}
		\begin{split}
			& \sum_{i = 0}^{l_n} \frac { 4 \e \big[ \big\{\sum_{j \in \mathcal{H}_0 } [ \mathbbm{1} (\widetilde{W}_j \geq t_i) - \mathbb{P} ( \widetilde{W}_i \geq t_i ) ] \big\}^2 \big] } { \epsilon^2 p_0^2 G^2 (t_i) }  \\ 
			& \leq C \sum_{i = 0}^{l_n}  \frac {   m_n p_0 G(t_i) + o\big( (\log p)^{-1/\gamma} [p_0 G(t_i)]^2 \big)  } { \epsilon^2 p_0^2 G^2 (t_i) } \\
			& =  C  \epsilon^{-2} m_n \sum_{i = 0}^{l_n} \frac{1} {p_0 G(t_i)} + C \epsilon^{-2} o \big(l_n (\log p)^{- 1/\gamma} \big). 
		\end{split}
	\end{equation}
	Moreover, it holds that 
	\begin{equation} \label{pf-le2-3}
		\begin{split}
			\sum_{i = 0}^{l_n} \frac{ 1 } { p_0 G (t_i)} & =   p_0^{-1} \sum_{i = 0}^{l_n} \frac{ 1 } { z_i  }  = \frac{p} {p_0} \sum_{i = 0}^{l_n}   \frac {1} {  c_1 q a_n   + h_n e^{i ^\gamma}  } \\
			&\leq C   h_n^{-1},
		\end{split}    
	\end{equation}
	where the last inequality above is related to the proof of Theorem 3 in \cite{guo2022threshold}. 
	
	In light of the definition of $h_n$ and the assumption of $m_n / a_n \to 0$, we have that $$m_n / h_n = (m_n / a_n)^{1 - \eta} \to 0.$$ Therefore, combining \eqref{pf-le2-1}--\eqref{pf-le2-3}  and the fact that $$l_n =  [\log ((p - c_1 q a_n)/h_n)]^{1/\gamma} \leq (\log p)^{1/\gamma}$$ shows that the first term for the bound in \eqref{eq_d_n} tends to zero as $n \to \infty$. Moreover, since $l_n \leq (\log p)^{1/\gamma}$, the second term on the very right-hand side of \eqref{eq_d_n} above is bounded by
	$$
	\frac {2} {\epsilon}  (\log p)^{1/\gamma} \sup_{t \in (0, G^{-1} (\frac {c_1 q a_n} {p}) ]}  \frac { G(t - b_n ) - G(t + b_n) } { G(t) }, 
	$$
	which converges to zero as $n \to \infty$ under Condition \ref{fdr-condition5}. Finally, we can obtain that for each $\epsilon > 0$,
	\begin{equation}
		\begin{split}
			\mathbb{P} ( D_n > \epsilon ) \to 0,
		\end{split}
	\end{equation}
	which establishes the desired result in \eqref{prop_b1}. This concludes the proof of Lemma \ref{fdr-lemma2}.

	\subsection{Proof of Lemma \ref{fdr-lemma1}} \label{proof-prop1}
	
	We will show that with asymptotic probability one, it holds that for some $0 < c_1 < 1$, 
	\begin{equation}\label{pf-le3-1}
		1 +  \sum_{j = 1}^p  \mathbbm{1} \big( \hat{W}_j < -   G^{-1} (   \frac {c_1 q a_n} { p }  ) \big) \leq q a_n \leq q \sum_{j = 1}^p  \mathbbm{1} \big( \hat{W}_j \geq  G^{-1} (   \frac {c_1 q a_n} { p }   ) \big).
	\end{equation}
	Then from the definition of $T$, we can obtain the desired result of the lemma. We aim to establish \eqref{pf-le3-1}.  The main idea of the proof is to prove that the population counterpart of \eqref{pf-le3-1} holds. Then with an application of Lemma \ref{fdr-lemma2} to both left- and right-hand sides of \eqref{pf-le3-1}, we can connect it to the population counterpart and thus prove that \eqref{pf-le3-1} holds with asymptotic probability one.  
	
	First, it follows from the union bound and the fact that $\mathbb P(A) \leq \mathbb P(A\cap B) + \mathbb P(B^c)$ for any two events $A$ and $B$ that under Conditions \ref{fdr-condition1}--\ref{fdr-condition3}, 
	\begin{equation*}
		\begin{split}
			& \mathbb{P} ( \hat{W}_j < 3 \delta_{n} ~\mbox{for some}~ j \in \mathscr{A}_n) \\
			&\leq  \mathbb{P} ( \hat{W}_j < 3 \delta_{n} ~\mbox{for some}~ j \in \mathscr{A}_n, \max_{1 \leq j \leq p } |\hat{W}_j -  \widetilde{W}_j  | < b_n) + \mathbb{P} (\max_{1 \leq j \leq p } |\hat{W}_j -  \widetilde{W}_j  | \geq b_n ) \\
			&\leq  \mathbb{P} ( \widetilde{W}_j < 3 \delta_{n} + b_n ~\mbox{for some}~ j \in \mathscr{A}_n) + \mathbb{P} (\max_{1 \leq j \leq p } |\hat{W}_j -  \widetilde{W}_j  | \geq b_n ) \\
			&\leq \sum_{j \in \mathscr{A}_n }  \mathbb{P} ( \widetilde{W}_j  - w_j < 3 \delta_n + b_n - w_j ) + o(1)  \\
			& \leq \sum_{j \in \mathscr{A}_n }   \mathbb{P} ( | \widetilde{W}_j - w_j  | > \delta_n )+ o(1)   \\
			& \leq  \sum_{j = 1  }^p    \mathbb{P} ( | \widetilde{W}_j - w_j  | > \delta_n )+ o(1) \to 0 .   
		\end{split}
	\end{equation*}
	Then we have 
	$$\mathbb{P} (\cap_{j \in \mathscr{A}_n} \{\hat{W}_j \geq 3 \delta_n\}) \to 1 $$ and thus with asymptotic probability one,
	\begin{equation}\label{pf-le3-2}
		\sum_{j = 1}^p  \mathbbm{1} ( \hat{W}_j \geq 3 \delta_{ n} )   \geq a_n ,    
	\end{equation}  
	where $a_n = |\mathscr{A}_n|$. 
	
	In addition, since $w_j > - \delta_n $ for $1 \leq j \leq p$ by assumption, we can deduce that 
	\begin{equation} \label{lower-bound-T.new}
		\begin{split}
			\sum_{j = 1}^p \mathbb{P} ( \hat{W}_j < - 3 \delta_n) 
			& \leq \sum_{j = 1}^p   \mathbb{P} ( \hat{W}_j < - 3 \delta_n, \max_{1 \leq j \leq p } |\hat{W}_j -  \widetilde{W}_j  | < b_n  )  \\
			&\quad+ \mathbb{P} (\max_{1 \leq j \leq p } |\hat{W}_j -  \widetilde{W}_j  | \geq b_n ) \\
			& \leq \sum_{j = 1}^p   \mathbb{P} ( \widetilde{W}_j < - 3 \delta_n +  b_n ) + o(1) \\
			& \leq \sum_{j = 1}^p   \mathbb{P} (\widetilde{W}_j  - w_j  \leq - 3 \delta_n + b_n - w_j  ) + o(1) \\
			& \leq \sum_{j = 1}^p   \mathbb{P}  ( | \widetilde{W}_j - w_j | > \delta_n ) + o(1) \to 0 ,
		\end{split}
	\end{equation}
	which yields $ \sum_{j = 1}^p \mathbb{P} ( \hat{W}_j < - 3 \delta_n)\to  0$. 
	Using similar arguments as for  \eqref{lower-bound-T.new}, it holds that $$\sum_{j = 1}^p \mathbb{P} (\widetilde{W}_j \leq - 3 \delta_n ) \to 0.$$ 
	Then we can obtain that 
	\begin{equation*}
		\begin{split}
			G( 3 \delta_n ) & = p_0^{-1} \sum_{j \in \mathcal{H}_0} \mathbb{P} ( \widetilde{W}_j  \leq  -3 \delta_n ) \leq p_0^{-1} \sum_{j = 1}^p  \mathbb{P} ( \widetilde{W}_j  \leq  -3 \delta_n )  \\
			&= o(p_0^{-1}) .
		\end{split}
	\end{equation*}
	Since $  a_n \to \infty $, $p_0 / p \to 1$,  and $G(t)$ is a nonincreasing, continuous function, it follows that $ G(3 \delta_n) \leq \frac { c_1 q a_n } { p } $ and thus $$ G^{-1} ( \frac {c_1 q a_n} { p } ) \leq  3 \delta_n  $$ for some constant $0 < c_1 < 1 $ when  $n $ is sufficiently large.  This together with \eqref{pf-le3-2} entails that with asymptotic probability one,
	$$
	\sum_{j = 1}^p  \mathbbm{1} ( \hat{W}_j \geq  G^{-1} ( \frac {c_1 q a_n} { p } ) )   \geq a_n.
	$$
	This completes the proof of the second inequality in  \eqref{pf-le3-1}.
	
	It remains to establish the first inequality in  \eqref{pf-le3-1}. From the definition of $G(t)$ and Lemma \ref{fdr-lemma2}, it holds that 
	\begin{equation}
		\begin{split}
			\frac { c_1 q a_n } { p } &= p_0^{-1} \sum_{j \in \mathcal{H}_0} \mathbb{P} ( \widetilde{W}_j \leq -  G^{-1} (   \frac {c_1 q a_n} { p } ) ) \\
			& = (1 + o_p(1)) \cdot  p_0^{-1} \sum_{j \in \mathcal{H}_0}  \mathbbm{1} \big( \hat{W}_j < -   G^{-1} (   \frac {c_1 q a_n} { p }  ) \big).
		\end{split}
	\end{equation}
	Then for some constant $c_2$ satisfying $0 < c_1 < c_2 < 1$, we can obtain that with asymptotic probability one, 
	\begin{equation}
		1 +  \sum_{j \in \mathcal{H}_0}  \mathbbm{1} \big( \hat{W}_j < -   G^{-1} (   \frac {c_1 q a_n} { p }  ) \big)  \leq \frac {c_1 q a_n p_0} {p} (1 + o_p(1)) \leq  c_2 q a_n , \label{bound_part1.new}
	\end{equation}
	where we have used the assumption of $p_0/p \to 1$. Further, under \eqref{dist-cond2} in Condition \ref{fdr-condition5}, an application of the union bound yields that 
	\begin{equation}
		\begin{split}
			& \mathbb{P} \Big( \sum_{j \in \mathcal{H}_1} \mathbbm{1} \big( \hat{W}_j < - G^{-1} (\frac {c_1 q a_n} {p}) \big) \geq   (1 - c_2) q a_n   \Big)  \\
			&  \leq \mathbb{P} \Big( \sum_{j \in \mathcal{H}_1} \mathbbm{1} \big( \widetilde{W}_j < - G^{-1} (\frac {c_1 q a_n} {p} ) + b_n \big) \geq   (1 - c_2) q a_n, \, \max_{1 \leq j \leq p } |\hat{W}_j -  \widetilde{W}_j  | < b_n   \Big)  \\
			&\quad+ o(1) \\
			& \leq \frac {1} { (1 - c_2 ) q a_n} \sum_{j \in \mathcal{H}_1} \mathbb{P} \Big( \widetilde{W}_j < - G^{-1} (\frac {c_1 q a_n} {p}) + b_n \Big) + o(1) \to 0,
		\end{split}
	\end{equation}
	which together with \eqref{bound_part1.new} implies that
	\begin{equation}
		1 +  \sum_{j = 1}^p  \mathbbm{1} \big( \hat{W}_j < -   G^{-1} (   \frac {c_1 q a_n} { p }  ) \big) \leq q a_n 
	\end{equation}
	with asymptotic probability one. This proves the first inequality in \eqref{pf-le3-1}, which completes the proof of Lemma \ref{fdr-lemma1}.

	\ignore{\color{red}New proof sketch for lower bound (need to revise to formalize). Now we have proved that with asymptotic probability 1, $T\in (0, G^{-1} (   \frac {c_1 q a_n} { p }  ))$. Denote by this event $\mathcal B_1$.  We will establish the lower bound now. By definition, we have $T\in \mathcal S$ with asymptotic probability one, where
		\begin{align*}
			\mathcal S := \{t\in (0, G^{-1} (   \frac {c_1 q a_n} { p }  )) : \frac{\sum_{j = 1}^p  \mathbbm{1} \big( \hat{W}_j \leq  -t \big)}{1\bigvee \sum_{j = 1}^p  \mathbbm{1} \big( \hat{W}_j \geq    t \big)}\leq q\}.
		\end{align*}
		Note that for any $t\in \mathcal S$, we have 
		\begin{align*}
			\sum_{j \in \mathcal H_0}  \mathbbm 1 \big( \hat{W}_j \leq  -t \big) + \sum_{j\in \mathcal H_1}\mathbbm 1 \big( \hat{W}_j \leq  -t \big)\leq q+ q  \sum_{j \in \mathcal H_0}  \mathbbm 1 \big( \hat{W}_j \geq  t \big) + q\sum_{j\in \mathcal H_1}\mathbbm 1 \big( \hat{W}_j \geq  t \big).
		\end{align*}
		Denote by the event where the inequalities in Lemma 2 hold as $\mathcal B_{2\epsilon}$. Then on $\mathcal B_{2\epsilon}$,
		\begin{align*}
			(1-\epsilon)\sum_{j \in \mathcal H_0}  \mathbb P \big( \widetilde{W}_j \leq  -t \big) \leq q+qs +     (1+\epsilon)q\sum_{j \in \mathcal H_0}  \mathbb P \big( \widetilde{W}_j \geq  t \big).
		\end{align*}
		That is,
		\begin{align*}
			\sum_{j \in \mathcal H_0}  \mathbb P \big( \widetilde{W}_j \leq  -t \big) \leq \frac{q(1+s)}{1-q-\epsilon-q\epsilon},
		\end{align*}
		which yields $t\geq G^{-1}(\frac{q(1+s)}{(1-q-\epsilon-q\epsilon)p})$. That is, on event $\mathcal B_{2\epsilon}\cap \mathcal B$, it holds that
		$ G^{-1}(\frac{q(1+s)}{(1-q-\epsilon-q\epsilon)p})\leq T\leq G^{-1} (   \frac {c_1 q a_n} { p }  )$.
	}

 \subsection{Proof of Lemma \ref{pf-thm3-lemma-1}} \label{new.SecB.7}
	Recall that the perfect and approximate knockoff statistics based on the marginal correlation are defined as $$\widetilde{W}_j = (\sqrt n \|\by \|_2)^{-1} ( | \bX_{j}^{\top} \by | - |\widetilde{\bX}_j^{\top} \by| ) \ \text{ and } \  \hat{W}_j = (\sqrt n \|\by \|_2)^{-1} ( | \bX_{j}^{\top} \by | - |\hat{\bX}_j^{\top} \by| ),$$ respectively. By the triangle inequality, it is easy to see that 
	\begin{equation*}
		\max_{1 \leq j \leq p} | \hat{W}_{ j} - \widetilde{W}_{ j} | \leq \max_{1 \leq j \leq p } (\sqrt n \|\by \|_2)^{-1} | (\hat\bX_j - \widetilde\bX_j)^{\top} \by |.
	\end{equation*}
	Then an application of the Cauchy--Schwarz inequality gives that 
	\begin{equation*}
		\max_{1 \leq j \leq p } | \hat{W}_{ j} - \widetilde{W}_{ j} | \leq (\sqrt n)^{-1} \max_{1 \leq j \leq p } \| \hat{\bX}_j - \widetilde{\bX}_j  \|_2 .
	\end{equation*}
	Thus, the conclusion of Lemma \ref{pf-thm3-lemma-1} can be derived under Condition \ref{accuracy-knockoffs}. This completes the proof of Lemma \ref{pf-thm3-lemma-1}.
	
	\subsection{Proof of Lemma \ref{pf-thm3-lemma-2}} \label{new.SecB.8}
	
	From the definitions of $\widetilde{W}_j$ and $w_j$ and the triangle inequality, it holds that 
	\begin{equation*}
		\begin{split}
			& \mathbb{P} (| \widetilde{W}_j - w_j | \geq \delta_n ) \\
			& \leq \mathbb{P} \bigg( ( n^{-1} \| \by \|^2_2 )^{-1/2}  \Big| n^{-1} ( | \bX_j^{\top} \by | - | \widetilde{\bX}_j^{\top} \by  | ) - ( |\e (X_j Y)| - |\e(\widetilde{X}_j Y)| ) \Big| \geq \delta_n / 2 \bigg) \\
			& \quad + \mathbb{P} \bigg(  \Big| ( n^{-1} \| \by \|^2_2 )^{-1/2} - (\e Y^2)^{-1/2} \Big| \cdot  \Big| |\e (X_j Y)| - |\e(\widetilde{X}_j Y)|  \Big| \geq \delta_n / 2 \bigg)\\
			& := P_1 + P_2.
		\end{split}
	\end{equation*}
	We will aim to show that for $\delta_n \to 0$, 
	\begin{equation}\label{pf-le8-p1}
		P_1 \leq 4 \exp \Big\{ - \frac { n \delta_n^2 \e Y^2 } { 256 \| X_j\|_{\psi_2}^2 \| Y \|_{\psi_2}^2 } \Big\} + \exp\Big\{ - \frac { n (\e Y^2)^2 } {8 \e Y^4 } \Big\} 
	\end{equation}
	and 
	\begin{equation}\label{pf-le8-p2}
		\begin{split}
			P_2 & \leq 2 \exp \Big\{ - \frac { n \delta_n^2 (\e Y^2 )^2 } { 64 | w_j |^2   \|Y \|_{\psi_2}^4   }  \Big\} + \exp \Big\{ - \frac { n ( \e Y^2 )^2 } {8 \e Y^4} \Big\}.
		\end{split}
	\end{equation}
	Then setting $\delta_n = \sqrt{\frac {\log p} {n} } \max\limits_{1 \leq j \leq p} \Big\{   \frac { 16 \sqrt 2 \| X_j \|_{\psi_2} \| Y \|_{\psi_2} } { (\e Y^2)^{1/2} } \lor   \frac { 8\sqrt 2  |w_j| \| Y \|_{\psi_2}^2 } { \e Y^2 } \Big\}$, a combination of the above results leads to the desired conclusion of this lemma.

	We proceed with proving \eqref{pf-le8-p1}.  Since $\| \by\|_2^2 = \sum_{i = 1}^n y_i^2$ is the sum of i.i.d. random variables, an application of  Bernstein's inequality yields that 
	\begin{equation}
		\mathbb{P} ( n^{-1}\| \by \|_2^2 \leq    \e[Y^2] /2 ) \leq  \exp\Big\{ - \frac { n (\e Y^2)^2 } {8 \e Y^4 } \Big\}. \label{eq-lower-bdd}
	\end{equation}
	It follows from the triangle inequality and \eqref{eq-lower-bdd} that
	\begin{equation*}
		\begin{split}
			P_1  &\leq  \mathbb{P} \bigg( \Big| n^{-1} ( | \bX_j^{\top} \by | - | \widetilde{\bX}_j^{\top} \by  | ) - ( |\e (X_j Y)| - |\e(\widetilde{X}_j Y)| ) \Big|  \geq \frac { \delta_n (\e Y^2)^{1/2}  } {2 \sqrt 2} \bigg) \\
			& + \mathbb{P} ( n^{1/2}(\| \by \|_2)^{-1} \geq   \sqrt{2}  (\e[Y^2])^{-1/2} )\\
			&\leq \mathbb{P} \bigg( \frac 1 n \Big| \sum_{i = 1}^n [\bX_{i, j}  y_i - \e(X_j Y)] \Big| \geq \frac { \delta_n (\e Y^2)^{1/2}  } {4 \sqrt 2} \bigg) \\
			& \quad + \mathbb{P} \bigg( \frac 1 n \Big| \sum_{i = 1}^n [\widetilde{\bX}_{i, j} y_i - \e(\widetilde{X}_j Y)] \Big| \geq \frac { \delta_n (\e Y^2)^{1/2}  } {4 \sqrt 2} \bigg) \\
			&\quad+ \exp\Big\{ - \frac { n (\e Y^2)^2 } {8 \e Y^4 } \Big\}.
		\end{split}
	\end{equation*}
	
	We next bound the first two terms on the right-hand side of the expression above.
	Under Condition \ref{marginal_corr_condition1}, we see that $ \bX_{i, j}  y_i $ and $\widetilde{\bX}_{i, j} y_i $ are both sub-exponential random variables, with sub-exponential norms $ \| X_j \|_{\psi_2} \| Y \|_{\psi_2}  $ and $\| X_j \|_{\psi_2} \| Y \|_{\psi_2}$, respectively. Then we can obtain through applying Bernstein's inequality for sub-exponential random variables (see, e.g., Corollary 2.8.3 in \cite{Book_prob}) that when $\delta_n = o(1)$, 
	\begin{equation*}
		\begin{split}
			\mathbb{P} \bigg( \frac 1 n \Big| \sum_{i = 1}^n [\bX_{i, j}  y_i - \e(X_j Y)] \Big| \geq \frac { \delta_n (\e Y^2)^{1/2}  } {4 \sqrt 2} \bigg) 
			\leq 2 \exp \Big\{ - \frac { n \delta_n^2 \e Y^2 } { 256 \| X_j\|_{\psi_2}^2 \| Y \|_{\psi_2}^2 } \Big\}
		\end{split}
	\end{equation*}
	and 
	\begin{equation*} 
		\begin{split}
			\mathbb{P} \bigg( \frac 1 n \Big| \sum_{i = 1}^n [\widetilde{\bX}_{i, j} y_i - \e(\widetilde{X}_j Y)] \Big| \geq \frac { \delta_n (\e Y^2)^{1/2}  } {4 \sqrt 2} \bigg) \leq 2 \exp \Big\{ - \frac { n \delta_n^2 \e Y^2 } { 256 \| X_j\|_{\psi_2}^2 \| Y \|_{\psi_2}^2 } \Big\}.
		\end{split}
	\end{equation*}
	Thus, combining the above three inequalities establishes \eqref{pf-le8-p1}.

	As for term $P_2$, noting that $w_j = (\e Y^2)^{-1/2} ( |\e (X_j Y)| - |\e(\widetilde{X}_j Y)| ) $ and
	$$
	\Big| ( n^{-1} \| \by \|^2_2 )^{-1/2} - (\e Y^2)^{-1/2} \Big| =  \frac{ | n^{-1} \|\by \|_2^2 -\e Y^2 | } { n^{-1/2} \|\by \|_2  (\e Y^2)^{1/2}  ((\e Y^2)^{1/2} + n^{-1/2} \|\by \|_2   )  },
	$$
	we can deduce that 
	\begin{equation}
		\begin{split}
			P_2 & = \mathbb{P} \bigg( |w_j|  \frac{ | n^{-1} \|\by \|_2^2 -\e Y^2 | } { n^{-1/2} \|\by \|_2    ((\e Y^2)^{1/2} + n^{-1/2} \|\by \|_2   )  } \geq \delta_n / 2
			\bigg) \\
			& \leq  \mathbb{P} \bigg( |w_j|  \frac{ | n^{-1} \|\by \|_2^2 -\e Y^2 |} { n^{-1/2} \|\by \|_2    (\e Y^2)^{1/2}    } \geq \delta_n / 2 
			\bigg) \\
			& =  \mathbb{P} \bigg(    | n^{-1} \|\by \|_2^2 -\e Y^2 | \geq \frac{\delta_n        \e Y^2 } {2 \sqrt 2 |w_j| } 
			\bigg) + \mathbb{P} ( n^{-1}\| \by \|_2^2 \leq    \e Y^2 /2 )  .
		\end{split}
	\end{equation}
	The very last term above can be bounded by applying \eqref{eq-lower-bdd}. 
	
	Again we can see that under Condition \ref{marginal_corr_condition1}, $y_i^2$ is a sub-exponential random variable with sub-exponential norm $\| Y \|_{\psi_2}^2$. With the aid of Bernstein's inequality for sub-exponential random variables (Corollary 2.8.3 in \cite{Book_prob}), we can obtain that for $\delta_n = o(1)$, 
	\begin{equation*}
		\begin{split}
			\mathbb{P} \bigg( \frac 1 n \Big| \sum_{i = 1}^n [  y_i^2 - \e( Y^2 )] \Big| \geq \frac{\delta_n        \e Y^2 } {2 \sqrt 2 |w_j| }  \bigg) \leq 2 \exp \Big\{ - \frac { n \delta_n^2 (\e Y^2 )^2 } { 64 | w_j |^2   \|Y \|_{\psi_2}^4   }  \Big\}.
		\end{split}
	\end{equation*}
	Therefore, the bound for term $P_2$ in \eqref{pf-le8-p2} can be shown. This concludes the proof of Lemma \ref{pf-thm3-lemma-2}.

	\subsection{Proof of Lemma \ref{pf-thm3-lemma-3}} \label{new.SecB.9}
	The main idea of the proof is to apply the law of total variance and decompose the total into two terms by conditioning on $(\bX_{\mathcal{H}_1}, \bveps)$, where $\bX_{\mathcal{H}_1}= (\bX_j)_{j \in \mathcal{H}_1}$ and $\bveps = (\varepsilon_1, \ldots, \varepsilon_n)^{\top}$. Specifically, it holds that 
	\begin{equation} \label{pf-lemma9-eq1}
		\begin{split}
			\Var \bigg( \sum_{j \in \mathcal{H}_0} \mathbbm{1} (\widetilde{W}_j \geq t)   \bigg)  
			& = \e \Bigg\{ \e \bigg[ \bigg( \sum_{j \in \mathcal{H}_0} \mathbbm{1} (\widetilde{W}_j \geq t) -  \sum_{j \in \mathcal{H}_0}  \mathbb{P}( \widetilde{W}_j \geq t | \bX_{\mathcal{H}_1}, \bveps ) \bigg)^2 \bigg | \bX_{\mathcal{H}_1}, \bveps \bigg] \Bigg\} \\
			& \quad + \e \Bigg\{ \bigg(\sum_{j \in \mathcal{H}_0}  \mathbb{P}( \widetilde{W}_j \geq t | \bX_{\mathcal{H}_1}, \bveps ) - \sum_{j \in \mathcal{H}_0}  \mathbb{P}( \widetilde{W}_j \geq t) \bigg)^2  \Bigg\} \\
			& := V_1 + V_2. 
		\end{split}
	\end{equation}
	We will bound terms $V_1$ and $V_2$ above separately.
	
	Let us begin with the first term $V_1$. We can expand the square and obtain that 
	\begin{equation} \label{pf-lemma9-eq2}
		\begin{split}
			V_1 & = \sum_{j \in \mathcal{H}_0} \sum_{\ell \in \mathcal{H}_0} \e \Bigg\{ \e \bigg[ \bigg( \mathbbm{1} (\widetilde{W}_j \geq t) -  \mathbb{P}( \widetilde{W}_j \geq t | \bX_{\mathcal{H}_1}, \bveps ) \bigg) \\
			& \hspace{3cm}\times \bigg( \mathbbm{1} (\widetilde{W}_{\ell} \geq t) -  \mathbb{P}( \widetilde{W}_{\ell} \geq t | \bX_{\mathcal{H}_1}, \bveps ) \bigg) \bigg | \bX_{\mathcal{H}_1}, \bveps \bigg] \Bigg\}. 
		\end{split} 
	\end{equation}     
	Observe that conditional on $(\bX_{\mathcal{H}_1}, \bveps)$, it follows from model \eqref{nonpara-model} that $\by$ is deterministic. In addition, $\widetilde{W_j}$ depends only on $ {\bX}_j$ and $\widetilde{\bX}_j$ besides $\by$. Thus, we need only to consider the conditional distribution of $(\bX_j, \widetilde{\bX}_j, \bX_k, \widetilde{\bX}_k) | (\bX_{\mathcal{H}_1}, \bveps)$. We will aim to show that each $\widetilde W_j$ depends on at most $m_n$ 
	random variables in $\{\widetilde W_k: k\in \mathcal H_0\}$. Indeed, it suffices to show that conditional on $(\bX_{\mathcal{H}_1}, \bveps)$, the number of $ (\bX_k, \widetilde{\bX}_k)$'s that are dependent on $(\bX_j, \widetilde{\bX}_j)$ is at most $m_n$. Since the rows of $(\bX, \widetilde{\bX})$ are i.i.d. and are independent of 
	$\bveps$, we need only to consider the distribution of a single row; that is, $(X_j, \widetilde{X}_j, X_k, \widetilde{X}_k) | (X_{\mathcal{H}_1}, \varepsilon) \stackrel{d}{=} (X_j, \widetilde{X}_j, X_k, \widetilde{X}_k) | X_{\mathcal{H}_1}$. 
	
	In view of the multinormal distribution in \eqref{marginal-normal}, it follows that the conditional distribution $(X_j, \widetilde{X}_j, X_k, \widetilde{X}_k) | X_{\mathcal{H}_1} $ is still normal. We can obtain from the conditional distribution that 
	\begin{equation*}
		\begin{split}
			& \Cov \Bigg\{ \Bigg(  \begin{pmatrix}
				X_j    \\
				\widetilde{X}_j  \\
			\end{pmatrix},
			\begin{pmatrix}
				X_k    \\
				\widetilde{X}_k  \\
			\end{pmatrix}                     
			\Bigg) \Bigg|  X_{\mathcal{H}_1} 
			\Bigg\} \\
			&=  \begin{pmatrix}
				\bSigma_{j, k} - \bSigma_{j, \mathcal{H}_1} \bSigma_{\mathcal{H}_1, \mathcal{H}_1}^{-1} \bSigma_{\mathcal{H}_1, k} & \bSigma_{j, k} - \bSigma_{j, \mathcal{H}_1} \bSigma_{\mathcal{H}_1, \mathcal{H}_1}^{-1} \bSigma_{\mathcal{H}_1, k}  \\
				\bSigma_{j, k} - \bSigma_{j, \mathcal{H}_1} \bSigma_{\mathcal{H}_1, \mathcal{H}_1}^{-1} \bSigma_{\mathcal{H}_1, k} & \bSigma_{j, k} - \bSigma_{j, \mathcal{H}_1} \bSigma_{\mathcal{H}_1, \mathcal{H}_1}^{-1} \bSigma_{\mathcal{H}_1, k} \\                 \end{pmatrix}. 
		\end{split}
	\end{equation*}
	In particular, $(X_j, \widetilde{X}_j)$ and $(X_k, \widetilde{X}_k)$ are independent conditional on $ X_{\mathcal{H}_1}$ if and only if 
	$$ \bSigma_{j, k} - \bSigma_{j, \mathcal{H}_1} \bSigma_{\mathcal{H}_1, \mathcal{H}_1}^{-1} \bSigma_{\mathcal{H}_1, k}=0. $$ 
	Thus,  to count the number of dependent pairs of $(X_j, \widetilde{X}_j)$ and $(X_k, \widetilde{X}_k)$ for $j, k \in \mathcal{H}_0$, we need only to count the number of nonzero  $(\bSigma_{j, k} - \bSigma_{j, \mathcal{H}_1} \bSigma_{\mathcal{H}_1, \mathcal{H}_1}^{-1} \bSigma_{\mathcal{H}_1, k})$'s.  Without loss of generality, let us assume that $X = (X_{\mathcal{H}_1}, X_{\mathcal{H}_0})$ and $$\bSigma = \begin{pmatrix}
		\bSigma_{\mathcal{H}_1, \mathcal{H}_1} & \bSigma_{\mathcal{H}_1, \mathcal{H}_0}   \\
		\bSigma_{\mathcal{H}_0, \mathcal{H}_1} & \bSigma_{\mathcal{H}_0, \mathcal{H}_0}  \\
	\end{pmatrix}.$$ 
	Using the formula for the block matrix inverse, it holds that 
	\begin{equation*}
		\bSigma^{-1} =  \begin{pmatrix}
			(\bSigma^{-1})_{11} ~~  & (\bSigma^{-1})_{12}    \\
			(\bSigma^{-1})_{21} ~~ &  \bSigma_{\mathcal{H}_0, \mathcal{H}_0} - \bSigma_{\mathcal{H}_0, \mathcal{H}_1} \bSigma_{\mathcal{H}_1, \mathcal{H}_1}^{-1} \bSigma_{\mathcal{H}_1, \mathcal{H}_0} \\
		\end{pmatrix},
	\end{equation*}
	where $$(\bSigma^{-1})_{11} = \bSigma_{\mathcal{H}_1, \mathcal{H}_1}^{-1} + \bSigma_{\mathcal{H}_1, \mathcal{H}_1}^{-1} \bSigma_{\mathcal{H}_1, \mathcal{H}_0} (\bSigma_{\mathcal{H}_0, \mathcal{H}_0} - \bSigma_{\mathcal{H}_0, \mathcal{H}_1} \bSigma_{\mathcal{H}_1, \mathcal{H}_1}^{-1} \bSigma_{\mathcal{H}_1, \mathcal{H}_0})^{-1} \bSigma_{\mathcal{H}_0, \mathcal{H}_1} \bSigma_{\mathcal{H}_1, \mathcal{H}_1}^{-1} ,$$ $$(\bSigma^{-1})_{12}  =- \bSigma_{\mathcal{H}_1, \mathcal{H}_1}^{-1} \bSigma_{\mathcal{H}_1, \mathcal{H}_0} (\bSigma_{\mathcal{H}_0, \mathcal{H}_0} - \bSigma_{\mathcal{H}_0, \mathcal{H}_1} \bSigma_{\mathcal{H}_1, \mathcal{H}_1}^{-1} \bSigma_{\mathcal{H}_1, \mathcal{H}_0})^{-1}   ,$$ and $(\bSigma^{-1})_{21} = (\bSigma^{-1})_{12}^{\top} $. In addition, Condition \ref{marginal_corr_condition2} assumes that $\max_{1 \leq j \leq p} \| (\bSigma^{-1})_{j} \|_0 \leq m_n$, which indicates that $$ \max_{j \in \mathcal{H}_0}\| ( \bSigma_{\mathcal{H}_0, \mathcal{H}_0} - \bSigma_{\mathcal{H}_0, \mathcal{H}_1} \bSigma_{\mathcal{H}_1, \mathcal{H}_1}^{-1} \bSigma_{\mathcal{H}_1, \mathcal{H}_0} )_j \|_{0} \leq m_n$$ since it is a submatrix of $\bSigma^{-1}$. Hence, we can obtain that for a given $j \in \mathcal{H}_0$,  
	\begin{equation*}
		\sum_{k \in \mathcal{H}_0} \mathbbm{1} \Big( \bSigma_{j, k} - \bSigma_{j, \mathcal{H}_1} \bSigma_{\mathcal{H}_1, \mathcal{H}_1}^{-1} \bSigma_{\mathcal{H}_1, k} = 0 \Big) \leq m_n. 
	\end{equation*}
	
	Consequently, we see that conditional on $(\bX_{\mathcal{H}_1}, \bveps)$, the number of $k \in \mathcal{H}_0$ such that $ (\bX_k, \widetilde{\bX}_k)$  is dependent on $(\bX_j, \widetilde{\bX}_j)$ is at most $m_n$. For $j \in \mathcal{H}_0 $, let us define $$N(j) := \{ k\in \mathcal{H}_0: \widetilde{W}_k \nbigCI \widetilde{W}_j |  (\bX_{\mathcal{H}_1} , \bveps) \}.$$ 
	Then it holds that $| N(j) | \leq m_n$. From \eqref{pf-lemma9-eq2} and the fact that the indicator function takes values between 0 and 1, we can deduce that  
	\begin{equation} \label{pf-lemma9-V1}
		\begin{split}
			V_1 & =  \sum_{j \in \mathcal{H}_0} \sum_{\ell \in N(j)} \e \Bigg\{ \e \Bigg[  \mathbbm{1} (\widetilde{W}_j \geq t) \cdot  \mathbbm{1} (\widetilde{W}_{\ell} \geq t)  \Big| \bX_{\mathcal{H}_1}, \bveps \Bigg] \Bigg\} \\
			& \quad - \sum_{j \in \mathcal{H}_0} \sum_{\ell \in N(j)} \e \Bigg\{ \e \Bigg[  \mathbb{P}( \widetilde{W}_j \geq t | \bX_{\mathcal{H}_1}, \bveps )  \mathbb{P}( \widetilde{W}_{\ell} \geq t \Big| \bX_{\mathcal{H}_1}, \bveps ) \Bigg] \Bigg\} \\
			& \leq  \sum_{j \in \mathcal{H}_0} \sum_{\ell \in N(j)} \e \Big\{ \e \Big[  \mathbbm{1} (\widetilde{W}_j \geq t) \cdot  \mathbbm{1} (\widetilde{W}_{\ell} \geq t)  \Big| \bX_{\mathcal{H}_1}, \bveps \Big] \Big\} \\
			& \leq m_n  \sum_{j \in \mathcal{H}_0} \e \Big\{ \e \Big[  \mathbbm{1} (\widetilde{W}_j \geq t)    \Big| \bX_{\mathcal{H}_1}, \bveps \Big] \Big\} \\
			&  = m_n  \sum_{j \in \mathcal{H}_0}  \mathbb{P} (\widetilde{W}_j \geq t ) = m_n p_0 G(t). 
		\end{split}
	\end{equation}
	
	We next proceed with showing the bound for term $V_2$. We can expand $V_2$ as 
	\begin{equation}
		\begin{split}
			V_2 & = \sum_{j \in \mathcal{H}_0} \sum_{\ell \in \mathcal{H}_0} \e \Big\{ \Big( \mathbb{P} ( \widetilde{W}_j \geq t | \bX_{\mathcal{H}_1}, \bveps ) - \mathbb{P} ( \widetilde{W}_j \geq t  ) \Big) \\
			& \qquad\qquad\qquad \times \Big( \mathbb{P} ( \widetilde{W}_{\ell} \geq t | \bX_{\mathcal{H}_1}, \bveps ) - \mathbb{P} ( \widetilde{W}_{\ell} \geq t  ) \Big) \Big\}.
		\end{split}  
	\end{equation}
	The key idea of the proof is to examine the conditional distribution $\mathbb{P} (\widetilde{W}_j \geq t | \bX_{\mathcal{H}_1}, \bveps )$ and show that given $j \in \mathcal{H}_0$, the number of dependent $\mathbb{P} ( \widetilde{W}_{\ell} \geq t | \bX_{\mathcal{H}_1}, \bveps ) $ is at most $m_n$.  Since $(X, \widetilde{X} )$ is multinormal, it holds that  
	\begin{equation*}
		(X_j, \widetilde{X}_j) \big| ( X_{\mathcal{H}_1}, \veps ) ~ \stackrel{d}{\sim} N \bigg( \begin{pmatrix}
			\bSigma_{j, \mathcal{H}_1} \bSigma_{\mathcal{H}_1, \mathcal{H}_1} ^{-1} X_{\mathcal{H}_1} \\
			\bSigma_{j, \mathcal{H}_1} \bSigma_{\mathcal{H}_1, \mathcal{H}_1} ^{-1} X_{\mathcal{H}_1 } \\
		\end{pmatrix},   
		\Cov_{cond} \bigg),
	\end{equation*}
	where $$ \Cov_{cond}  = \begin{pmatrix}
		\bSigma_{j, j} - \bSigma_{j, \mathcal{H}_1} \bSigma_{\mathcal{H}_1, \mathcal{H}_1} ^{-1} \bSigma_{\mathcal{H}_1, j} 
		&  \bSigma_{j, j} - r - \bSigma_{j, \mathcal{H}_1} \bSigma_{\mathcal{H}_1, \mathcal{H}_1} ^{-1} \bSigma_{\mathcal{H}_1, j}\\
		\bSigma_{j, j} - r - \bSigma_{j, \mathcal{H}_1} \bSigma_{\mathcal{H}_1, \mathcal{H}_1} ^{-1} \bSigma_{\mathcal{H}_1, j}
		& \bSigma_{j, j} - \bSigma_{j, \mathcal{H}_1} \bSigma_{\mathcal{H}_1, \mathcal{H}_1} ^{-1} \bSigma_{\mathcal{H}_1, j} \\
	\end{pmatrix}.  $$
	Since the rows of the augmented data matrix $(\bX, \widetilde{\bX})$ are i.i.d. and $\by $ is deterministic given $( \bX_{\mathcal{H}_1}, \bveps)$, we can obtain that 
	\begin{equation} \label{pf-le9-1}
		\begin{split}
			& \bigg( \frac{   \bX_j^{\top} \by  } {\sqrt n \| \by \|_2},  \frac{  \widetilde{\bX}_j^{\top} \by  } {\sqrt n \| \by \|_2} \bigg) \bigg| ( \bX_{\mathcal{H}_1}, \bveps)  \\
			& \stackrel{d}{\sim} N \bigg((\sqrt n \|\by\|_2)^{-1} 
			\begin{pmatrix}
				\bSigma_{j, \mathcal{H}_1} \bSigma_{\mathcal{H}_1, \mathcal{H}_1} ^{-1} \bX_{\mathcal{H}_1}^{\top} \by \\
				\bSigma_{j, \mathcal{H}_1} \bSigma_{\mathcal{H}_1, \mathcal{H}_1} ^{-1} \bX_{\mathcal{H}_1 }^{\top} \by \\
			\end{pmatrix}, 
			n^{-1} \Cov_{cond} \bigg).
		\end{split}
	\end{equation}
	
	Note that when $\bSigma_{\mathcal{H}_1, j} = {\bf 0}$, the conditional distribution above does not depend on $(\bX_{\mathcal{H}_1}, \bveps)$ and hence any term involving such $j \in \mathcal{H}_0$ in the expansion of $V_2$ will disappear. Denote by $$ N_{dep}  = \{j \in \mathcal{H}_0:  \bSigma_{\mathcal{H}_1, j} \neq {\bf 0}\}.$$ 
	It follows from Condition \ref{marginal_corr_condition2} that $| N_{dep} | \leq m_n$. Then we have that 
	\begin{equation} \label{pf-lemma9-V2}
		\begin{split}
			V_2 & = \sum_{j \in \mathcal{H}_0} \sum_{\ell \in N_{dep}} \e \Big\{ \Big( \mathbb{P} ( \widetilde{W}_j \geq t | \bX_{\mathcal{H}_1}, \bveps ) - \mathbb{P} ( \widetilde{W}_j \geq t  ) \Big) \\
			& \qquad\qquad\qquad \times \Big( \mathbb{P} ( \widetilde{W}_{\ell} \geq t | \bX_{\mathcal{H}_1}, \bveps ) - \mathbb{P} ( \widetilde{W}_{\ell} \geq t  ) \Big) \Big\} \\
			&  \leq \sum_{j \in \mathcal{H}_0} \sum_{\ell \in N_{dep}} \e \Big\{ \mathbb{P} ( \widetilde{W}_j \geq t | \bX_{\mathcal{H}_1}, \bveps ) \mathbb{P} ( \widetilde{W}_{\ell} \geq t | \bX_{\mathcal{H}_1}, \bveps )  \Big\} \\    
			& \leq  \sum_{j \in \mathcal{H}_0} \sum_{\ell \in N_{dep}} \e \Big\{ \mathbb{P} ( \widetilde{W}_j \geq t | \bX_{\mathcal{H}_1}, \bveps ) \Big\} \leq  m_n p_0 G(t). 
		\end{split}
	\end{equation}
	Therefore, substituting \eqref{pf-lemma9-V1} and \eqref{pf-lemma9-V2} into \eqref{pf-lemma9-eq1} yields \eqref{pf-thm3-lemma-3-result}. This completes the proof of Lemma \ref{pf-thm3-lemma-3}.

	\subsection{Proof of Lemma \ref{pf-thm3-lemma-4}} \label{new.SecB.10}
	\noindent\textbf{Proof of \eqref{lemma10-1}}.
	In the proof of Lemma \ref{pf-thm3-lemma-3} in Section \ref{new.SecB.9} (cf. \eqref{pf-le9-1}), we have shown that 
	\begin{equation}
		\bigg( \frac{   \bX_j^{\top} \by  } {  \| \by \|_2},  \frac{  \widetilde{\bX}_j^{\top} \by  } {  \| \by \|_2} \bigg) \bigg| ( \bX_{\mathcal{H}_1}, \bveps)  ~\stackrel{d}{\sim} N \Bigg( 
		\begin{pmatrix}
			\mu_j \\
			\mu_j \\
		\end{pmatrix}, 
		\sigma_j^2 \begin{pmatrix}
			1 & \rho_j  \\
			\rho_j & 1  \\
		\end{pmatrix} \Bigg),
	\end{equation}
	where 
	\begin{equation*}
		\mu_j =  \|\by\|_2^{-1} \bSigma_{j, \mathcal{H}_1} \bSigma_{\mathcal{H}_1, \mathcal{H}_1} ^{-1} \bX_{\mathcal{H}_1}^{\top} \by,
	\end{equation*}
	\begin{equation*}
		\sigma_j^2 = \bSigma_{j, j} - \bSigma_{j, \mathcal{H}_1} \bSigma_{\mathcal{H}_1, \mathcal{H}_1} ^{-1} \bSigma_{\mathcal{H}_1, j}, \ \  \rho_j = 1 - r / \sigma_j^2,
	\end{equation*} 
	and $r$ is as given in \eqref{marginal-normal}. Recall the definition $N_{dep} = \{j \in \mathcal{H}_0:  \bSigma_{\mathcal{H}_1, j} \neq {\bf 0}\}$ in the proof of Lemma \ref{pf-thm3-lemma-3}. It holds that $|N_{dep}| \leq m_n$ in view of Condition \ref{marginal_corr_condition2}. Furthermore, note that $$G(t) \geq c_1 q a_n / p $$ for $t \in (0,\, G^{-1} ( \frac { c_1 q a_n  } { p } ) ] $. Let us define 
	\begin{equation} \label{pf-le10-3}
		R_{n} :=    \sup_{t \in (0,\, G^{-1} ( \frac { c_1 q a_n  } { p } ) ] }  \frac { \sum_{j \in \mathcal{H}_0 \cap N_{dep}^c}  \mathbb{P} (t - \Delta_n \leq \widetilde{W}_j <   t + \Delta_n)  } {  \sum_{j \in \mathcal{H}_0 \cap N_{dep}^c} \mathbb{P} (\widetilde{W}_j \geq t ) }.
	\end{equation}
	
	Then we can write
	\begin{equation} \label{lemma10-eq-2}
		\begin{split}
			& \sup_{t \in (0,\, G^{-1} ( \frac { c_1 q a_n  } { p } ) ] }  \frac { G(t - \Delta_n) - G(t + \Delta_n) } {  G(t) } \\
			& =  \sup_{t \in (0,\, G^{-1} ( \frac { c_1 q a_n  } { p } ) ] }  \frac { \sum_{j \in \mathcal{H}_0 \cap N_{dep}}  \mathbb{P} (t - \Delta_n \leq \widetilde{W}_j <   t + \Delta_n)  } { p_0 G(t) } + R_{n} \\
			& \leq \frac{m_n p  } { c_1 q a_n p_0   } + R_n.
		\end{split}
	\end{equation}
	From the assumptions that $(\log p)^{1/\gamma} m_n / a_n \to 0$ and $p_0 / p \to 1$, we have that $$ (\log p )^{1/\gamma}  \frac{m_n p  } { c_1 q a_n p_0   } \to 0.$$ It remains to establish $(\log p)^{1/\gamma} R_n \to 0$. A key observation is that when $j \in \mathcal{H}_0  \cap N_{dep}^c$, it follows that the conditional distribution 
	\begin{equation}
		\bigg( \frac{   \bX_j^{\top} \by  } { \| \by \|_2},  \frac{  \widetilde{\bX}_j^{\top} \by  } { \| \by \|_2} \bigg) \bigg| ( \bX_{\mathcal{H}_1}, \bveps)  ~\stackrel{d}{\sim} N \Bigg( 
		\begin{pmatrix}
			0\\
			0 \\
		\end{pmatrix}, 
		\begin{pmatrix}
			\bSigma_{j, j}^2 & \bSigma_{j, j}^2 - r  \\
			\bSigma_{j, j}^2 - r & \bSigma_{j, j}^2  \\
		\end{pmatrix} \Bigg),
	\end{equation}
	which does not depend on $( \bX_{\mathcal{H}_1}, \bveps)$. Then we see that the distribution of $\widetilde{W}_j $ does not depend on $( \bX_{\mathcal{H}_1}, \bveps)$ and satisfies that 
	\begin{equation}
		\mathbb{P} ( \sqrt n \widetilde{W}_j \geq t ) =  \mathbb{P}  ( |Z_1| - |Z_2| \geq t ), 
	\end{equation}
	where $(Z_1, Z_2)^{\top}$ is a two-dimensional multinormal random variable with mean $(0, 0)^{\top}$ and covariance matrix $$  \begin{pmatrix}
		\bSigma_{j, j}^2 & \bSigma_{j, j}^2 - r  \\
		\bSigma_{j, j}^2 - r & \bSigma_{j, j}^2  \\
	\end{pmatrix} .$$ 
	
	For $j \in \mathcal{H}_0 \cap N_{dep}^c $ and $t > 0$, the density
	function of $\sqrt n \widetilde{W}_j$ is given by 
	\begin{equation} \label{conditional-density}
		\begin{split}
			f_{\sqrt n \widetilde{W}_j}(t) & = \frac{\sqrt 2} {\sqrt{\pi} c_{2, j}}  \Big[ 1 - \Phi \Big( \frac{t  } {c_{1, j}} \Big) \Big]  \exp\Big\{- \frac{ t^2 } {2 c_{2, j}^2 } \Big\} \\
			& \quad + \frac{ \sqrt 2 } {\sqrt{ \pi} c_{1, j}}  \Big[ 1 - \Phi \Big( \frac{t  } {c_{2, j}} \Big) \Big] \exp\Big\{- \frac{ t  ^2 } {2 c_{1, j}^2 } \Big\} ,
		\end{split}    
	\end{equation}
	where $ c_{1, j} = \sqrt{4 \bSigma_{j,j}^2 - 2 r } $ and $ c_{2, j} =  \sqrt {2r}$. Based on the density function of $\sqrt n  \widetilde{W}_t$ above and the basic inequality that $ 1 - \Phi(x) \leq  e^{-x^2/2} $ for $x \geq 0$, it is easy to see that 
	\begin{equation}\label{pf-le10-1}
		\begin{split}
			\mathbb{P} (  \widetilde{W}_j \geq t  )  & = \mathbb{P} (\sqrt n  \widetilde{W}_j \geq  \sqrt n t  ) \\
			&\leq \int_{\sqrt n t}^{\infty} \frac{\sqrt 2} {\sqrt{ \pi} c_{2, j}} \exp\Big\{- \frac{ x^2 } {2 c_{2, j}^2 } \Big\} dx   +\int_{\sqrt n t}^{\infty} \frac{ \sqrt 2 } {\sqrt{\pi} c_{1, j}} \Phi \Big( \frac{ -x } {c_{2, j}} \Big) dx \\
			&\leq \Big(2 + \frac{ 2 c_{2, j} } {  c_{1, j}} \Big) \Big[1 - \Phi\Big(\frac{\sqrt n t} {c_{2, j}} \Big) \Big].
		\end{split}
	\end{equation}
	Then we can obtain that $$ G(t ) \leq \max_{j \in \mathcal{H}_0} \Big(2 + \frac{ 2 c_{2, j} } {  c_{1, j}} \Big) \Big[1 - \Phi\Big(\frac{\sqrt n t} { c_{2, j}} \Big) \Big]  .$$ Setting $t = G^{-1} ( \frac{c_1 q a_n} { p } )$ in the  inequality above yields that $$ G^{-1} ( \frac{c_1 q a_n} { p } ) = O( \sqrt {\frac{\log p} {n}} )$$ when $ C_1 < r < \bSigma_{j, j}^2 < C_2$ with some absolute constants $C_1> 0$ and $C_2> 0$ for each $j \in \mathcal{H}_0 $. 
	
	We will bound the ratio in $R_n$ by considering two ranges of $t\in (0, 4n^{-1/2} \max_{j \in \mathcal{H}_0} c_{1, j} \lor c_{2, j})$ and $t\in [4n^{-1/2} \max_{j \in \mathcal{H}_0} c_{1, j} \lor c_{2, j}, G^{-1}(c_1qa_n/p)]$ separately. When $t$ falls into  the first range, in view of \eqref{conditional-density} the denominator $G(t)$ in the ratio in $R_n$ is of a constant order, while the numerator is uniformly bounded from above by $O(\sqrt n \Delta_n)$ over all $t$ in this range because the density $f_{\sqrt n \widetilde W_j}(t)$ is bounded from above by a constant.  
	
	We now consider the ratio in $R_n$ in the second range of $t\in [4n^{-1/2} \max_{j \in \mathcal{H}_0} c_{1, j} \lor c_{2, j}, G^{-1}(c_1qa_n/p)]$. We will bound the numerator and denominator in \eqref{dist-cond1} separately in this range. It follows from \eqref{conditional-density} and the mean value theorem that there exists some $ \xi \in (\sqrt n t - \sqrt n \Delta_n, \sqrt n t + \sqrt n \Delta_n) $ such that 
	\begin{equation*}
		\begin{split}
			&\mathbb{P} (\sqrt n t - \sqrt n \Delta_n \leq  \sqrt n \widetilde{W}_j \leq  \sqrt n t + \sqrt n \Delta_n) \\
			& = 2 \sqrt n \Delta_n \bigg\{ \frac{\sqrt 2} {\sqrt { \pi} c_{2, j}}   \Big[  1 - \Phi\Big( \frac{ \xi  } {c_{1, j}} \Big)    \Big] \exp \Big\{ - \frac{ \xi^2 } { 2 c_{2, j}^2 } \Big\} \\
			& \quad + \frac{\sqrt 2} {\sqrt { \pi} c_{1, j}}   \exp \Big\{- \frac{ \xi^2 }{2 c_{1, j}^2} \Big\}    \Big[1 - \Phi \Big( \frac{\xi} {c_{2, j}} \Big) \Big] \bigg\}.
		\end{split}
	\end{equation*}
	Moreover, since $ \sqrt n t \leq \sqrt n G^{-1} (\frac{c_1 q a_n} {p}) = O( \sqrt {\log p}  ) $ and $ \Delta_n \sqrt {n \log p}  \to 0 $, we can obtain through some direct calculations that
	$$
	\Bigg| \frac{ 1 - \Phi\Big( \frac{ \xi  } {c_{1, j}} \Big) } {  1 - \Phi\Big( \frac{ \sqrt n t } {c_{1, j}} \Big)    } - 1 \Bigg| \leq C \sqrt n t \cdot \sqrt n \Delta_n = O (  \Delta_n \sqrt{n \log p}).
	$$
	Similarly, it holds that 
	$$
	\Bigg| \frac{ \exp \Big\{- \frac{ \xi^2 }{2 c_{1, j}^2} \Big\}  } {  \exp \Big\{- \frac{ (\sqrt  n t)^2 }{2 c_{1, j}^2} \Big\}  } - 1 \Bigg| \leq C \sqrt n t \cdot \sqrt n \Delta_n = O ( \Delta_n \sqrt{n \log p}).
	$$
	Combining the above three inequalities yields that when $\Delta_n \sqrt {n  \log p} \to 0 $, 
	\begin{equation}\label{pf-le10-2}
		\begin{split}
			&\mathbb{P} (t - \Delta_n \leq \widetilde{W}_j <   t + \Delta_n)\\
			& = \mathbb{P} ( \sqrt n t - \sqrt n \Delta_n \leq  \sqrt n \widetilde{W}_j \leq \sqrt n t + \sqrt n \Delta_n) \\
			& \leq C \sqrt n \Delta_n [1 + O (\sqrt n \Delta_n \log p)] \Bigg\{ \frac{\sqrt 2} {\sqrt{\pi} c_{2, j}}  \Big[ 1 - \Phi \Big( \frac{ \sqrt n t  } {c_{1, j}} \Big) \Big]  \exp\Big\{- \frac{ (\sqrt n t )^2 } {2 c_{2, j}^2 } \Big\} \\
			& \quad + \frac{ \sqrt 2 } {\sqrt{ \pi} c_{1, j}}  \Big[ 1 - \Phi \Big( \frac{ \sqrt n t  } {c_{2, j}} \Big) \Big] \exp\Big\{- \frac{ (\sqrt n t )  ^2 } {2 c_{1, j}^2 } \Big\}  \Bigg\}.
		\end{split}
	\end{equation}
	
	Next we need to deal with the denominator $ \mathbb{P} (\sqrt n \widetilde{W}_j \geq t) $. Via integration by parts, we can deduce that for $t \in [4n^{-1/2} \max_{j \in \mathcal{H}_0} c_{1, j} \lor c_{2, j}, G^{-1}(c_1qa_n/p)]$, 
	\begin{equation}  \label{distribution-eq}
		\begin{split}
			\mathbb{P} (\sqrt n \widetilde{W}_j \geq \sqrt n t) & =  2 \Big[ 1 - \Phi \Big(  \frac{ \sqrt n t   } { c_{1, j}} \Big)    \Big] \Big[ 1 - \Phi \Big( \frac{\sqrt n t} {c_{2, j}} \Big) \Big] \\
			& \geq {C} \bigg\{ (\sqrt n t)^{-1}  \Big[ 1 - \Phi \Big(  \frac{ \sqrt n t  } { c_{1, j}} \Big)   \Big] \exp \Big\{ - \frac{ (\sqrt n t )^2 } {2 c_{2, j}^2 } \Big\} \\
			& \quad + (\sqrt n t)^{-1}  \Big[ 1 - \Phi \Big(  \frac{ \sqrt n t  } { c_{2, j}} \Big)   \Big] \exp \Big\{ - \frac{ (\sqrt n t )^2 } {2 c_{1, j}^2 } \Big\} \bigg\} \\
			& \geq \widetilde{C} (\sqrt n t)^{-1} f_{\sqrt n \widetilde{W}_j} (\sqrt n t) ,
		\end{split}
	\end{equation}
	where we have used the definition of the density in \eqref{conditional-density} and the fact that $$1 - \Phi(x) \geq 0.75 x^{-1} e^{- x^2/ 2}$$ for $x \geq 4$, and $\widetilde{C}$ is some constant depending on $c_{1, j}$ and $c_{2, j}$. 
	
	Combining \eqref{pf-le10-2} and \eqref{distribution-eq} and using some direct calculations, we can obtain the bound for the ratio in $R_n$ in the second range
	\begin{equation} \label{pf-le10-n1}
		\begin{split}
			& \sup_{t\in [4n^{-1/2} \max_{j \in \mathcal{H}_0} c_{1, j} \lor c_{2, j}, G^{-1}(c_1qa_n/p))}  \frac { \sum_{j \in \mathcal{H}_0 \cap N_{dep}^c}  \mathbb{P} (t - \Delta_n \leq \widetilde{W}_j <   t + \Delta_n)  } {  \sum_{j \in \mathcal{H}_0 \cap N_{dep}^c} \mathbb{P} (\widetilde{W}_j \geq t ) }  \\
			& \qquad\qquad\qquad \leq \widetilde{C} \sqrt n \Delta_n\cdot \sqrt n G^{-1} ( \frac { c_1 q a_n  } { p } ) = O (\sqrt n \Delta_n \sqrt {\log p}). 
		\end{split}
	\end{equation}
	This together with the result for the first range proven previously leads to 
	\begin{equation} \label{lemma-10-eq1}
		\begin{split}
			R_n =  O (\sqrt n \Delta_n \sqrt {\log p}). 
		\end{split}
	\end{equation}
	Finally, plugging \eqref{lemma-10-eq1} into \eqref{lemma10-eq-2} yields \eqref{lemma10-1} because $(\log p)^{1/\gamma} m_n / a_n \to 0 $ and $$\sqrt n \Delta_n (\log p)^{1/2 + 1/\gamma} \to 0.$$
	
	\bigskip
	\noindent\textbf{Proof of \eqref{lemma10-2}}. 
	Recall from Condition \ref{marginal_corr_condition4} that $$ p_1^{-1} \sum_{j \in \mathcal{H}_1}  \mathbb{P} ( \widetilde{W}_j <  - t ) \leq G(t)   $$ for $t \in (0, C \sqrt{n^{-1} \log p})$ with $C$ some large constant. Also, note that $$\Delta_n = o(G^{-1} (\frac{c_1 q a_n } {p}))$$ since $\sqrt n \Delta_n \to 0$ by assumption and $G^{-1} (\frac{c_1 q a_n} {p}) = O(\sqrt {n^{-1} \log p})$ as shown in the proof of \eqref{lemma10-1}. It follows from some direct calculations that 
	\begin{equation} \label{pf-lemma-10-eq3}
		\begin{split}
			& a_n ^{-1} \sum_{j \in \mathcal{H}_1}  \mathbb{P}  \Big( \widetilde{W}_j <  -  G^{-1} ( \frac { c_1 q a_n  } { p } ) + \Delta_n \Big) \\
			& \leq  a_n^{-1 } (p - p_0) G\Big(  G^{-1} ( \frac { c_1 q a_n  } { p }) - \Delta_n  \Big) \\
			& = \frac{ c_1 q  (p - p_0) } { p }  + a_n^{-1}(p - p_0)  | G' ( \xi ) |\Delta_n, 
		\end{split}
	\end{equation} 
	where $ \xi $ is some number lying between $  G^{-1} ( \frac { c_1 q a_n  } { p })$ and $  G^{-1} ( \frac { c_1 q a_n  } { p }) - \Delta_n$.  From \eqref{conditional-density} and $f_{\sqrt n \widetilde{W}_j} (\sqrt n \xi )\leq C$ with $C>0$ some constant,  we can deduce that 
	\begin{equation*}
		\begin{split}
			| G' (\xi)| & = \sum_{j \in \mathcal{H}_0} p_0^{-1} \sqrt n  f_{\sqrt n \widetilde{W}_j} (\sqrt n \xi )\\
			&\leq \frac{C \sqrt n  m_n } {p_0} + p_0^{-1}\sum_{j \in \mathcal{H}_0 \cap N_{dep}^c }   \sqrt n f_{\sqrt n \widetilde{W}_j} (\sqrt n \xi ) \\
			& \leq  \frac{C \sqrt n m_n } {p_0} + C p_0^{-1} \sqrt n \cdot \sqrt n G( \frac{c_1 q a_n } {p}) \sum_{H_0 \cap N_{dep}^c} \mathbb{P} \Big( \widetilde{W}_j \geq G( \frac{c_1 q a_n } {p}) \Big) \\
			& \leq \frac{C \sqrt n m_n } {p_0} + C p_0^{-1} \sqrt {n \log p }\, p_0 \frac{c_1 q a_n } {p},
		\end{split}
	\end{equation*}
	where the second last step above is due to  \eqref{distribution-eq}.
	
	Therefore, substituting the bound above into \eqref{pf-lemma-10-eq3} gives that 
	\begin{equation*} 
		\begin{split}
			&a_n ^{-1} \sum_{j \in \mathcal{H}_1}  \mathbb{P}  \Big( \widetilde{W}_j <  -  G^{-1} ( \frac { c_1 q a_n  } { p } + \Delta_n \Big) \\
			& \leq  \frac{c_1 q (p - p_0)} {p} + \frac{C \Delta_n  \sqrt n m_n (p - p_0 )} {a_n p_0 }\\
			& \quad + \frac{C \Delta_n \sqrt {n \log p } \, q (p - p_0)} {p} \\
			& \to 0,
		\end{split}
	\end{equation*} 
	where we have used the assumption that $p_0 / p \to 1 $, $\Delta_n \sqrt{n \log p } \to 0 $, and $m_n / a_n  \to 0 $. This derives \eqref{lemma10-2}, which concludes the proof of Lemma \ref{pf-thm3-lemma-4}.

	\subsection{Proof of Lemma \ref{pf-thm5-lemma-1}} \label{new.SecB.11}
	The main intuition of the proof is that when the approximate augmented data matrix $\hat{\bX}^{\augg}$ is close to its perfect counterpart $\widetilde{\bX}^{\augg}$, the corresponding Lasso estimators would be close as well. From the definitions of $\widetilde{\bbeta}_j $ in \eqref{eq-debiased-lasso} and $\hat{\bbeta}_j$ in \eqref{eq-debiased-lasso-approx}, it holds that 
	\begin{equation} \label{eq-lemma13-0}
		\begin{split}
			\max_{1 \leq j \leq 2p } | \widetilde{\beta}_j  - \hat{\beta}_j | & \leq \max_{1 \leq j \leq 2p } |  \widetilde{\beta}_j^{\init}  - \hat{\beta}_j^{\init} | \\
			& \quad +  \max_{1 \leq j \leq 2p } \bigg| \frac{ \widetilde\bz_j^{\top} \big(\by - \widetilde{\bX}^{\augg} \widetilde{\bbeta}^{\init} \big) } {\widetilde\bz_j^{\top} \widetilde\bX^{\augg}_j}  -  \frac{ \hat\bz_j^{\top} \big(\by - \hat{\bX}^{\augg} \hat{\bbeta}^{\init} \big) } {\hat\bz_j^{\top} \hat\bX^{\augg}_j} \bigg|.
		\end{split}
	\end{equation}
	We will aim to prove that for some large enough constant $C$,
	\begin{equation} \label{eq-lemma13-4}
		\mathbb{P} \bigg( \| \widetilde{\bbeta}^{\init} - \hat{\bbeta}^{\init}  \|_2 \leq C \Delta_n s  \sqrt{\frac{\log p} {n}} \bigg) \to  1,
	\end{equation}
	\begin{equation}\label{eq-lemma13-4-part2}
		\mathbb{P} \bigg( \max_{1 \leq j \leq 2p } \bigg| \frac{ \widetilde\bz_j^{\top} \big(\by - \widetilde{\bX}^{\augg} \widetilde{\bbeta}^{\init} \big) } {\widetilde\bz_j^{\top} \widetilde\bX^{\augg}_j}  -  \frac{ \hat\bz_j^{\top} \big(\by - \hat{\bX}^{\augg} \hat{\bbeta}^{\init} \big) } {\hat\bz_j^{\top} \hat\bX^{\augg}_j} \bigg|\leq C \Delta_n s\sqrt{\frac{ \log p} {n}} \bigg)\to  1.
	\end{equation}
	Then combining the two results above can establish the desired conclusion of Lemma \ref{pf-thm5-lemma-1}. We next proceed with proving \eqref{eq-lemma13-4} and \eqref{eq-lemma13-4-part2}.
	
	\bigskip
	\noindent\textbf{Proof of \eqref{eq-lemma13-4}}.
	It follows from the Karush--Kuhn--Tucker (KKT) condition that 
	\begin{align}
		n^{-1} [\widetilde\bX^{\augg}]^{\top}\widetilde\bX^{\augg} (\widetilde\bbeta^{\init} - \bbeta^{\augg} ) & = n^{-1} [\widetilde\bX^{\augg}]^{\top} \bveps - \lambda \widetilde{\boldsymbol{\zeta}}, \label{KKT-1} \\
		n^{-1} [\hat\bX^{\augg}]^{\top}\hat\bX^{\augg} (\hat\bbeta^{\init} - \bbeta^{\augg} ) & = n^{-1} [\hat\bX^{\augg}]^{\top} \bveps - \lambda \hat{\boldsymbol{\zeta}}, \label{KKT-2}
	\end{align}
	where $\widetilde{\boldsymbol{\zeta}} = (\widetilde\zeta_1, \ldots, \widetilde\zeta_{2p})$ and $\hat{\boldsymbol{\zeta}} = (\hat\zeta_1, \ldots, \hat\zeta_{2p})$ with
	\begin{equation*}
		\widetilde\zeta_j = \left\{ \begin{array}{cc}
			\sgn(\widetilde{\beta}^{\init}_j) & \mbox{ if} ~ \widetilde{\beta}_j^{\init} \neq 0,  \\
			\in [-1, 1] &  \mbox{ if} ~ \widetilde{\beta}_j^{\init} = 0,
		\end{array}
		\right.
		\quad \mbox{and} \quad 
		\hat\zeta_j = \left\{ \begin{array}{cc}
			\sgn(\hat{\beta}_j^{\init}) & \mbox{ if} ~ \hat{\beta}_j^{\init} \neq 0,  \\
			\in [-1, 1] &  \mbox{ if} ~ \hat{\beta}_j^{\init} = 0.
		\end{array}
		\right.
	\end{equation*}
	Taking the difference between \eqref{KKT-1} and \eqref{KKT-2} above leads to 
	\begin{equation*}
		\begin{split}
			& n^{-1} [\widetilde\bX^{\augg}]^{\top}\widetilde\bX^{\augg} (\widetilde{\bbeta}^{\init} - \hat{\bbeta}^{\init} )  + n^{-1} \Big([\widetilde\bX^{\augg}]^{\top}\widetilde\bX^{\augg}  - [\hat\bX^{\augg}]^{\top}\hat\bX^{\augg} \Big) (\hat{\bbeta}^{\init} - \widetilde{\bbeta}^{\init}) \\ 
			& = - n^{-1} \Big([\widetilde\bX^{\augg}]^{\top}\widetilde\bX^{\augg}  - [\hat\bX^{\augg}]^{\top}\hat\bX^{\augg} \Big) ( \widetilde{\bbeta}^{\init} - \bbeta^{\augg} ) \\
			&\quad+ n^{-1} \Big( \widetilde\bX^{\augg}  - \hat\bX^{\augg}  \Big)^{\top} \bveps - \lambda (\widetilde{\boldsymbol{\zeta}} - \hat{\boldsymbol{\zeta}}). 
		\end{split}
	\end{equation*}
	Furthermore, multiplying both sides of the equation above by $ (\widetilde{\bbeta}^{\init} - \hat{\bbeta}^{\init} )^{\top} $ yields that 
	\begin{equation}
		\begin{split}\label{eq:upper-bound}
			& n^{-1} \|\widetilde\bX^{\augg} (\widetilde{\bbeta}^{\init} - \hat{\bbeta}^{\init} )\|_2^2  \\
			& = n^{-1}( \widetilde{\bbeta}^{\init} - \hat{\bbeta}^{\init})^{\top} \Big([\widetilde\bX^{\augg}]^{\top}\widetilde\bX^{\augg}  - [\hat\bX^{\augg}]^{\top}\hat\bX^{\augg} \Big) ( \widetilde{\bbeta}^{\init} - \hat{\bbeta}^{\init})  \\
			& \quad  - n^{-1} ( \widetilde{\bbeta}^{\init} - \hat{\bbeta}^{\init})^{\top}  \Big([\widetilde\bX^{\augg}]^{\top}\widetilde\bX^{\augg}  - [\hat\bX^{\augg}]^{\top}\hat\bX^{\augg} \Big) ( \widetilde{\bbeta}^{\init} - \bbeta^{\augg} )  \\
			& \quad + n^{-1} ( \widetilde{\bbeta}^{\init} - \hat{\bbeta}^{\init})^{\top}\Big( \widetilde\bX^{\augg}  - \hat\bX^{\augg}  \Big)^{\top} \bveps - \lambda  ( \widetilde{\bbeta}^{\init} - \hat{\bbeta}^{\init})^{\top} (\widetilde{\boldsymbol{\zeta}} - \hat{\boldsymbol{\zeta}}). 
		\end{split}
	\end{equation}
	
	We claim that the last term on the right-hand side of the expression above satisfies that $$ ( \widetilde{\bbeta}^{\init} - \hat{\bbeta}^{\init})^{\top} (\widetilde{\boldsymbol{\zeta}} - \hat{\boldsymbol{\zeta}}) \geq 0 .$$ To understand this, observe that when both $\widetilde\beta_j^{\init} $ and $\hat{\beta}_j^{\init} $ are nonzero or zero, it is easy to see that $$(\widetilde{\beta}_j^{\init} - \hat{\beta}_j^{\init}) (\widetilde{\zeta}_j  - \hat{\zeta}_j ) \geq 0.$$ When either of $\widetilde\beta_j^{\init} $ and $\hat{\beta}_j^{\init} $ is zero, without loss of generality let us assume that $\widetilde\beta_j^{\init} = 0 $ and $\hat{\beta}_j^{\init} \neq 0$. When $\widetilde{\beta}_j^{\init} = 0 $  and $\hat{\beta}_j^{\init} > 0$, it follows that $\widetilde{\zeta}_j \leq 1 = \hat{\zeta}_j$ and hence $$(\widetilde{\beta}_j^{\init} - \hat{\beta}_j^{\init}) (\widetilde{\zeta}_j - \hat{\zeta}_j) = - \hat{\beta}_j^{\init} ((\widetilde{\zeta}_j - \hat{\zeta}_j)) \geq 0.$$ Similarly, we can show that  $$(\widetilde{\beta}_j^{\init} - \hat{\beta}_j^{\init}) (\widetilde{\zeta}_j - \hat{\zeta}_j) \geq 0$$ when $\widetilde{\beta}_j^{\init}=0$ and $\hat{\beta}_j^{\init} < 0$. 
	Thus,  the last term on the right-hand side of \eqref{eq:upper-bound} above satisfies that $$ -( \widetilde{\bbeta}^{\init} - \hat{\bbeta}^{\init})^{\top} (\widetilde{\boldsymbol{\zeta}} - \hat{\boldsymbol{\zeta}}) \leq 0 .$$
	We next examine the three terms on the right-hand side of the earlier expression above separately.
	
	First, we observe that 
	\begin{align*}
		&\Big\| n^{-1}  [\widetilde\bX^{\augg}]^{\top}\widetilde\bX^{\augg}  - [\hat\bX^{\augg}]^{\top}\hat\bX^{\augg} \Big\|_{\max}\\
		&\leq    \Big\| n^{-1}  [\widetilde\bX^{\augg}]^{\top}(\widetilde\bX^{\augg}  - \hat\bX^{\augg}) \Big\|_{\max}+ \|n^{-1} ( \widetilde\bX^{\augg} - \hat\bX^{\augg})]^{\top}\hat\bX^{\augg} \|_{\max}\\
		& \leq \max_j\|n^{-1/2}\widetilde\bX_j^{\augg}\|_2\max_j\|n^{-1/2}(\widetilde\bX_j^{\augg}  - \hat\bX_j^{\augg})\|_2
		\\
		&\quad+ \max_j\|n^{-1/2}\hat\bX_j^{\augg}\|_2\max_j\|n^{-1/2}(\widetilde\bX_j^{\augg}  - \hat\bX_j^{\augg})\|_2.
	\end{align*}
	Under Condition \ref{accuracy-knockoffs} and the sub-Gaussian assumption for $\bX$, it can be shown that
	\begin{equation} \label{eq-lemma13-2}
		\mathbb{P} \bigg( \Big\| n^{-1}  [\widetilde\bX^{\augg}]^{\top}\widetilde\bX^{\augg}  - [\hat\bX^{\augg}]^{\top}\hat\bX^{\augg} \Big\|_{\max} \geq C \Delta_n   \bigg) \to 0
	\end{equation}
	for some constant $C > 0$. From the sparsity of $\widetilde{\bbeta}$ and $\hat{\bbeta}$ in Condition \ref{debiased-lasso-condition1}, we have that with probability $1 - o(1)$, the first term on the right-hand side of \eqref{eq:upper-bound} can be bounded as 
	\begin{equation} \label{eq-lemma13-1}
		\begin{split}
			& n^{-1}\bigg|( \widetilde{\bbeta}^{\init} - \hat{\bbeta}^{\init})^{\top} \Big([\widetilde\bX^{\augg}]^{\top}\widetilde\bX^{\augg}  - [\hat\bX^{\augg}]^{\top}\hat\bX^{\augg} \Big) ( \widetilde{\bbeta}^{\init} - \hat{\bbeta}^{\init})\bigg| \\
			&\leq C \Delta_n s \| \widetilde{\bbeta}^{\init} - \hat{\bbeta}^{\init}  \|_2^2.
		\end{split}
	\end{equation}
	By the Cauchy--Schwarz inequality, we can bound the second term on the right-hand side of \eqref{eq:upper-bound} as
	\begin{align*}
		&\bigg|n^{-1} ( \widetilde{\bbeta}^{\init} - \hat{\bbeta}^{\init})^{\top}  \Big([\widetilde\bX^{\augg}]^{\top}\widetilde\bX^{\augg}  - [\hat\bX^{\augg}]^{\top}\hat\bX^{\augg} \Big) ( \widetilde{\bbeta}^{\init} - \bbeta^{\augg} )\bigg|\\
		&\leq \|\widetilde{\bbeta}^{\init} - \hat{\bbeta}^{\init}\|_2   \Big\|n^{-1}\Big([\widetilde\bX^{\augg}]^{\top}\widetilde\bX^{\augg}  - [\hat\bX^{\augg}]^{\top}\hat\bX^{\augg} \Big) ( \widetilde{\bbeta}^{\init} - \bbeta^{\augg} ) \Big\|_2.
	\end{align*}
	
	Finally, with the aid of Condition \ref{debiased-lasso-condition1} on sparsity and  Condition \ref{debiased-lasso-condition2} on the restrictive eigenvalues, the left-hand side of \eqref{eq:upper-bound} can be lower bounded by $c_1\| \widetilde{\bbeta}^{\init} - \hat{\bbeta}^{\init}  \|_2^2$. Combining all the results above and applying the Cauchy--Schwarz inequality to the second and third terms on the right-hand side of \eqref{eq:upper-bound}, we can deduce that as $\Delta_n s \to 0 $, the representation in \eqref{eq:upper-bound} entails that with probability $1 - o(1)$,
	\begin{equation} \label{eq-lemma13-3}
		\begin{split}
			\| \widetilde{\bbeta}^{\init} - \hat{\bbeta}^{\init}  \|_2  & \lesssim \Big\| n^{-1} \Big([\widetilde\bX^{\augg}]^{\top}\widetilde\bX^{\augg}  - [\hat\bX^{\augg}]^{\top}\hat\bX^{\augg} \Big) ( \widetilde{\bbeta}^{\init} - \bbeta^{\augg}) \Big  \|_2\\
			& \quad
			+  \max_{J: |J| \leq C s} \Big\| n^{-1}\Big( \widetilde\bX^{\augg}_J  - \hat\bX^{\augg}_J  \Big)^{\top} \bveps \Big\|_2 := I_1 + I_2.
		\end{split}   
	\end{equation}
	
	We will bound the two terms $I_1$ and $I_2$ above separately. It follows from \eqref{eq-lemma13-2}, the sparsity of $\widetilde{\bbeta}$ and $ {\bbeta}^{\augg}$, and Lemma \ref{pf-thm5-lemma_00} that with probability $1 - o(1)$, 
	\begin{equation} \label{lemma13-I1}
		I_1 \leq C \Delta_n s^{1/2} \| \widetilde\bbeta^{\init} - \bbeta^{\augg}\|_2 \leq C \Delta_n s  \sqrt{\frac{\log p} {n}}.
	\end{equation} 
	As for  term $I_2$, conditional on $ (\widetilde\bX^{\augg},  \hat\bX^{\augg} )$ we have that for each $1 \leq j \leq 2p$, 
	\begin{equation*}
		n^{-1/2} \Big( \widetilde\bX^{\augg}_j  - \hat\bX^{\augg}_j  \Big)^{\top} \bveps ~ \stackrel{d}{\sim} N \Big(0, n^{-1}  \Big\|\widetilde\bX^{\augg}_j  - \hat\bX^{\augg}_j \Big\|_2^2   \Big).
	\end{equation*}
	Thus, it holds that 
	\begin{equation*}
		\begin{split}
			& \mathbb{P} \bigg( I_2\geq C \sigma \Delta_n \sqrt{ \frac{s \log n } {n } }  \bigg) \\
			& \leq \mathbb{P} \bigg( s  \max_{1 \leq j \leq 2p } \Big( n^{-1/2}\Big( \widetilde\bX^{\augg}_j  - \hat\bX^{\augg}_j  \Big)^{\top} \bveps \Big)^2 \geq C^2 \sigma^2 \Delta_n^2  s \log n \bigg) \\
			& = \mathbb{P} \bigg(   \max_{1 \leq j \leq 2p}   n^{-1/2}\Big\| \widetilde\bX^{\augg}_j  - \hat\bX^{\augg}_j    \Big\|_2 |Z| \geq C \sigma \Delta_n  \sqrt {\log n} \bigg),
		\end{split}  
	\end{equation*}
	where $Z \stackrel{d}{\sim} N(0, \sigma^2) $ is independent of $\widetilde{\bX}^{\augg}$ and $
	\hat{\bX}^{\augg}$.
	
	Moreover, Condition \ref{accuracy-knockoffs} implies that $$\max_{1 \leq j \leq 2p} n^{-1/2} \| \widetilde\bX^{\augg}_j  - \hat\bX^{\augg}_j  \|_2 \leq \Delta_n$$ with probability $1 - o(1)$. Then using the union bound, we can obtain that for some constant $C > \sqrt 2$, 
	\begin{equation} \label{lemma13-I2}
		\begin{split}
			& \mathbb{P} \bigg( I_2 \geq C \sigma \Delta_n \sqrt{ \frac{s \log n } {n } }  \bigg) \leq \mathbb{P} \Big( |Z| > C \sigma \sqrt{\log n} \Big) \\
			&\qquad + \mathbb P(\max_{1 \leq j \leq 2p} \| \widetilde\bX^{\augg}_j  - \hat\bX^{\augg}_j  \|_2 \geq \Delta_n) \to 0.
		\end{split}
	\end{equation}
	Consequently, substituting \eqref{lemma13-I1} and \eqref{lemma13-I2} into \eqref{eq-lemma13-3} leads to \eqref{eq-lemma13-4}. Further, applying \eqref{eq:upper-bound} again with the bounds in \eqref{eq-lemma13-3}, \eqref{lemma13-I1}, \eqref{lemma13-I2}, and \eqref{eq-lemma13-4} yields that 
	\begin{equation} \label{eq-lemma13-4-0}
		\mathbb{P} \bigg(n^{-1/2} \|\widetilde{\bX}^{\augg} (\widetilde{\bbeta}^{\init} - \hat{\bbeta}^{\init} ) \|_2 \leq C \Delta_n s  \sqrt{\frac{\log p} {n}} \bigg) \to  1.
	\end{equation}
	
	\bigskip
	\noindent\textbf{Proof of  \eqref{eq-lemma13-4-part2}.} 
	Let us first state three results \eqref{eq-lemma13-6}, \eqref{eq-lemma13-4-1}, and \eqref{eq-lemma13-7} below that will be used repeatedly in our proof. 
	With similar arguments as for \eqref{eq-lemma13-4} and \eqref{eq-lemma13-4-0} and the union bound, we can deduce that under Conditions \ref{debiased-lasso-condition1}--\ref{debiased-lasso-condition3}, 
	\begin{equation} \label{eq-lemma13-6}
		\mathbb{P} \bigg( \max_{1 \leq j \leq 2p }  \| \widetilde{\bgamma}_j - \hat{\bgamma}_j \|_2 \leq C \Big( m_n^{1/2} \Delta_n + \Delta_n m_n \sqrt{\frac{\log p} {n}} \Big) \leq C m_n^{1/2} \Delta_n \bigg)\to 1,
	\end{equation}
	\begin{equation} \label{eq-lemma13-4-1}
		\mathbb{P} \bigg(n^{-1/2}\max_j \|\widetilde{\bX}_{-j}^{\augg} ( \widetilde{\bgamma}_j - \hat{\bgamma}_j )\|_2 \leq C  m_n^{1/2} \Delta_n \bigg) \to  1,
	\end{equation}
	where we have used $ \sqrt{\frac{m_n \log p} {n}} \to 0$ for showing \eqref{eq-lemma13-6}. Observe that for  $1 \leq j \leq 2p$, 
	\begin{equation*}
		\begin{split}
			\| n^{-1/2} ( \widetilde{\bz}_j  - \hat{\bz}_j ) \|_2 & \leq \| n^{-1/2} ( \widetilde{\bX}_j^{\augg} - \hat{\bX}_j^{\augg}) \|_2 +  \|n^{-1/2} \widetilde{\bX}_{-j}^{\augg} ( \widetilde{\bgamma}_j - \hat{\bgamma}_j ) \|_2 \\
			& \quad + \| n^{-1/2} (\widetilde{\bX}_{-j}^{\augg} - \hat{\bX}_{-j}^{\augg}) \bgamma_j  \|_2 \\
			& \quad +  \| n^{-1/2} (\widetilde{\bX}_{-j}^{\augg} - \hat{\bX}_{-j}^{\augg}) (\hat{\bgamma}_j - \bgamma_j)  \|_2.
		\end{split}
	\end{equation*}
	Then it follows from the sparsity of $ \mathcal{S}_j  = \supp(\bgamma_j) \cup \supp(\widetilde{\bgamma}_j) \cup \supp(\hat{\bgamma}_j)$, the sub-Gaussianity of $X_j$, and the bound in \eqref{eq-lemma13-6} that with probability $1 - o(p^{-1})$,  
	\begin{equation} \label{eq-lemma13-7}
		\begin{split}
			& \max_{1 \leq j \leq 2p} \| n^{-1/2} ( \widetilde{\bz}_j  - \hat{\bz}_j ) \|_2 \\
			& \leq C \bigg(\Delta_n  + \Delta_n  m_n^{1/2}   +   \Delta_n m_n^{1/2} \max_{1 \leq j \leq 2p} \| \bgamma_j\|_2 + m_n \Delta_n \sqrt{\frac{\log p} {n}} \bigg) \\
			& \leq C  \Delta_n m_n^{1/2}.
		\end{split}
	\end{equation}
	
	We are now ready to establish \eqref{eq-lemma13-4-part2}. In particular, we have the decomposition for the main term in \eqref{eq-lemma13-4-part2} 
	\begin{equation} \label{eq-lemma13-5}
		\begin{split}
			&  \max_{1 \leq j \leq 2p } \bigg| \frac{ \widetilde\bz_j^{\top} \big(\by - \widetilde{\bX}^{\augg} \widetilde{\bbeta}^{\init} \big) } {\widetilde\bz_j^{\top} \widetilde\bX^{\augg}_j}  -  \frac{ \hat\bz_j^{\top} \big(\by - \hat{\bX}^{\augg} \hat{\bbeta}^{\init} \big) } {\hat\bz_j^{\top} \hat\bX^{\augg}_j} \bigg| \\
			& \leq   \max_{1 \leq j \leq 2p } \bigg| \frac{ (\widetilde\bz_j - \hat{\bz}_j )^{\top} \big(\by - \widetilde{\bX}^{\augg} \widetilde{\bbeta}^{\init} \big) } {\widetilde\bz_j^{\top} \widetilde\bX^{\augg}_j} \bigg| +  \max_{1 \leq j \leq 2p } \bigg| \frac{ \hat\bz_j^{\top} \big(\hat{\bX}^{\augg} \hat{\bbeta}^{\init} - \widetilde{\bX}^{\augg} \widetilde{\bbeta}^{\init} \big) } {\widetilde\bz_j^{\top} \widetilde\bX^{\augg}_j} \bigg|   \\
			& \quad +  \max_{1 \leq j \leq 2p } \bigg|  \hat\bz_j^{\top} \big(\by - \hat{\bX}^{\augg} \hat{\bbeta}^{\init} \big)  \Big( \frac {1} { \hat\bz_j^{\top} \hat\bX^{\augg}_j } -  \frac {1} { \widetilde\bz_j^{\top} \widetilde\bX^{\augg}_j } \Big)\bigg| := P_1 + P_2 + P_3.
		\end{split}
	\end{equation} 
	We will investigate the three terms $P_1$, $P_2$, and $P_3$ above separately. Let us first deal with term $P_1$. Note that
	\begin{equation} \label{eq-lemma13-10}
		P_1 \leq  \max_{1 \leq j \leq 2p } \bigg| \frac{ (\widetilde\bz_j - \hat{\bz}_j )^{\top}   \widetilde{\bX}^{\augg} \big( \widetilde{\bbeta}^{\init} - \bbeta^{\augg} \big) } {\widetilde\bz_j^{\top} \widetilde\bX^{\augg}_j} \bigg| +   \max_{1 \leq j \leq 2p } \bigg| \frac{ (\widetilde\bz_j - \hat{\bz}_j )^{\top}    \bveps  } {\widetilde\bz_j^{\top} \widetilde\bX^{\augg}_j} \bigg|.  
	\end{equation}
	
	Since $\bveps \stackrel{d}{\sim} N({\bf 0}, I_n)$ and is independent of design matrix $\bX$, it holds that conditional on design matrix $\bX$, 
	\begin{equation*}
		\frac{ (\widetilde\bz_j - \hat{\bz}_j )^{\top}    \bveps  } {\widetilde\bz_j^{\top} \widetilde\bX^{\augg}_j} \stackrel{d}{\sim} N \bigg( 0, \,  \frac{ \|  \widetilde\bz_j - \hat{\bz}_j \|_2^2 } { [\widetilde\bz_j^{\top} \widetilde\bX^{\augg}_j ]^2  }  \bigg).
	\end{equation*}
	This together with the bounds in \eqref{eq-lemma13-7} and \eqref{eq-normal-error-2} leads to 
	\begin{equation} \label{eq-lemma13-8}
		\begin{split}
			& \mathbb{P} \bigg( \max_{1 \leq j \leq 2p } \bigg| \frac{ (\widetilde\bz_j - \hat{\bz}_j )^{\top}    \bveps  } {\widetilde\bz_j^{\top} \widetilde\bX^{\augg}_j} \bigg|   >  C m_n^{1/2} \Delta_n \sqrt{\frac{\log p} {n}}    \bigg) \\
			& = \sum_{j = 1}^{2p} \mathbb{P} \bigg( \frac{ \|  \widetilde\bz_j - \hat{\bz}_j \|_2 } { |\widetilde\bz_j^{\top} \widetilde\bX^{\augg}_j |  }  \cdot | Z |   >  C m_n^{1/2} \Delta_n \sqrt{\frac{\log p} {n}}  \bigg)   \\
			& \leq  \sum_{j = 1}^{2p} \mathbb{P} \bigg( \frac{ \|  \widetilde\bz_j - \hat{\bz}_j \|_2 } { n }  \cdot | Z |   >  C m_n^{1/2} \Delta_n \sqrt{\frac{\log p} {n}}  \bigg)  + o(1) \\
			& \leq  \sum_{j = 1}^{2p}   \mathbb{P} (|Z| > C \sqrt{\log p}) + o(1) = o(1),
		\end{split}
	\end{equation}
	where $Z \stackrel{d}{\sim} N(0, \sigma^2) $ is independent of $\widetilde{\bX}^{\augg}$ and $
	\hat{\bX}^{\augg}$, and $C$ is some large constant that may take different value at each appearance. 
	
	In addition, from \eqref{eq-normal-error-2}, the Cauchy--Schwarz inequality,  Lemma \ref{pf-thm5-lemma_00}, and \eqref{eq-lemma13-7}, we can deduce that with probability $1 - o(1)$,  
	\begin{equation} \label{eq-lemma13-9}
		\begin{split}
			\max_{1 \leq j \leq 2p } \bigg| \frac{ (\widetilde\bz_j - \hat{\bz}_j )^{\top}   \widetilde{\bX}^{\augg} \big( \widetilde{\bbeta}^{\init} - \bbeta^{\augg} \big) } {\widetilde\bz_j^{\top} \widetilde\bX^{\augg}_j} \bigg| &  \leq  \max_{1 \leq j \leq 2p }  \frac{ \|\widetilde\bz_j - \hat{\bz}_j \|_2  \| \widetilde{\bX}^{\augg} (\bbeta^{\augg} - \widetilde{\bbeta}^{\init})  \|_2  } { | \widetilde\bz_j^{\top} \widetilde\bX^{\augg}_j | } \\
			& \leq C \Delta_n  m_n^{1/2}  \sqrt{\frac{s \log p} {n}}.
		\end{split}
	\end{equation}
	Substituting \eqref{eq-lemma13-8} and \eqref{eq-lemma13-9} into \eqref{eq-lemma13-10} yields that with probability $1 - o(1)$, 
	\begin{equation} \label{eq-le13-P1}
		P_1 \leq C \Delta_n m_n^{1/2} \sqrt{\frac{s \log p} {n}}.
	\end{equation}
	
	We next turn to the bound for term $P_2$. It is easy to see that 
	\begin{equation} \label{eq-le13-P2}
		\begin{split}
			P_2 
			& \leq  \max_{1 \leq j \leq 2p }     \bigg| \frac{  \widetilde\bz_j^{\top}   \widetilde{\bX}^{\augg} \big(\widetilde{\bbeta}^{\init} - \hat{\bbeta}^{\init} \big)  } {\widetilde\bz_j^{\top} \widetilde\bX^{\augg}_j} \bigg| +  \max_{1 \leq j \leq 2p }     \bigg| \frac{  \widetilde\bz_j^{\top}  ( \widetilde{\bX}^{\augg}  - \hat{\bX}^{\augg} ) \hat{\bbeta}^{\init}  } {\widetilde\bz_j^{\top} \widetilde\bX^{\augg}_j} \bigg| \\
			& \quad +  \max_{1 \leq j \leq 2p }     \bigg| \frac{  (\hat{\bz}_j - \widetilde\bz_j )^{\top}   \widetilde{\bX}^{\augg} \big(\widetilde{\bbeta}^{\init} - \hat{\bbeta}^{\init} \big)  } {\widetilde\bz_j^{\top} \widetilde\bX^{\augg}_j} \bigg| \\
			& \quad +  \max_{1 \leq j \leq 2p }     \bigg| \frac{  (\hat{\bz}_j - \widetilde\bz_j )^{\top}  ( \widetilde{\bX}^{\augg}  - \hat{\bX}^{\augg} ) \hat{\bbeta}^{\init}  } {\widetilde\bz_j^{\top} \widetilde\bX^{\augg}_j} \bigg| \\
			&:= P_{21} + P_{22} + P_{23} + P_{24}. 
		\end{split}
	\end{equation}
	Regarding term $P_{21}$, in view of  \eqref{eq-lemma13-4} and the definition of $\widetilde\bz_j$, we have that with probability $1 - o(1)$, 
	\begin{equation} \label{eq-lemma13-11}
		\begin{split}
			P_{21} & \leq   \max_{1 \leq j \leq 2p }   | \widetilde{\beta}^{\init}_j - \hat{\beta}_j^{\init} | +   \max_{1 \leq j \leq 2p }   \bigg| \frac{  \widetilde\bz_j^{\top}   \widetilde{\bX}^{\augg}_{-j} \big(\widetilde{\bbeta}^{\init}_{-j} - \hat{\bbeta}^{\init}_{-j} \big)  } {\widetilde\bz_j^{\top} \widetilde\bX^{\augg}_j} \bigg| \\
			& \leq  C \Delta_n s  \sqrt{\frac{\log p} {n}} + \max_{1 \leq j \leq 2p }   \bigg| \frac{  ( \be_j + \widetilde{\bX}^{\augg}_{-j} ( \bgamma_j - \widetilde{\bgamma}_j ) )   ^{\top}   \widetilde{\bX}^{\augg}_{-j} \big(\widetilde{\bbeta}^{\init}_{-j} - \hat{\bbeta}^{\init}_{-j} \big)  } {\widetilde\bz_j^{\top} \widetilde\bX^{\augg}_j} \bigg| \\
			& \leq  C \Delta_n s  \sqrt{\frac{\log p} {n}} + \max_{1 \leq j \leq 2p }  \bigg| \frac{   \be_j^{\top}   \widetilde{\bX}^{\augg}_{-j} \big(\widetilde{\bbeta}^{\init}_{-j} - \hat{\bbeta}^{\init}_{-j} \big)  } {\widetilde\bz_j^{\top} \widetilde\bX^{\augg}_j} \bigg| \\
			& \quad + \max_{1 \leq j \leq 2p }  \bigg| \frac{ [\widetilde{\bX}^{\augg}_{-j} ( \bgamma_j - \widetilde{\bgamma}_j ) ]^{\top}   \widetilde{\bX}^{\augg}_{-j} \big(\widetilde{\bbeta}^{\init}_{-j} - \hat{\bbeta}^{\init}_{-j} \big)  } {\widetilde\bz_j^{\top} \widetilde\bX^{\augg}_j} \bigg|.
		\end{split}
	\end{equation}
	We will bound the last two terms on the very right-hand side of the expression above separately.  
	
	Since for $\ell \neq j$, $ n^{-1} \e [ \be_j^{\top}   \widetilde{\bX}^{\augg}_{\ell}] = 0$ due to zero correlation between $\be_j$ and $\bX_{-j}^{\augg}$, and $\be_j$ and $\widetilde{\bX}_{\ell}^{\augg}$ both have i.i.d. sub-Gaussian entries, we can show that for $\ell \neq j$,
	\begin{equation} 
		\begin{split}
			\mathbb{P} \bigg(  \max_{1 \leq j \leq 2p } \max_{\ell \neq j}  n^{-1}   | \be_j^{\top} \bX_{\ell}^{\augg} | \geq C \sqrt{\frac{\log p} {n}} \bigg) \leq C p^{-1} \to 0.
		\end{split}
	\end{equation} 
	This combined with \eqref{eq-normal-error-2}, the sparsity assumption that $|J| = |\supp(\bbeta) \cup \supp(\widetilde{\bbeta}) \cup \supp(\hat{\bbeta})| \lesssim s$, and the result in \eqref{eq-lemma13-4} yields that with probability $1 - o(1)$, the second term on the very right-hand side of \eqref{eq-lemma13-11} above can be bounded as
	\begin{equation} \label{eq-lemma13-12}
		\begin{split}
			& \max_{1 \leq j \leq 2p }  \bigg| \frac{   \be_j^{\top}   \widetilde{\bX}^{\augg}_{-j} \big(\widetilde{\bbeta}^{\init}_{-j} - \hat{\bbeta}^{\init}_{-j} \big)  } {\widetilde\bz_j^{\top} \widetilde\bX^{\augg}_j} \bigg| \\
			& \leq C n^{-1} \max_{1 \leq j \leq 2p }  \max_{J': | J' | \lesssim s } \|  \be_j^{\top} \widetilde{\bX}_{ J'\setminus\{j\} }^{\augg} \|_2 \cdot \| \widetilde{\bbeta}^{\init}_{J'\setminus\{j\}} - \hat{\bbeta}^{\init}_{J'\setminus\{j\}} \|_2   \\     
			& \leq C \sqrt{\frac{ s \log p} {n}} \cdot \Delta_n s  \sqrt{\frac{\log p} {n}} \leq C \Delta_n s  \sqrt{\frac{\log p} {n}},
		\end{split}
	\end{equation}
	where the last inequality above holds due to the assumption that $\sqrt{  \frac{ s \log p} {n}} \to 0$.  By the Cauchy--Schwarz inequality, we can deduce that with probability $1 - o(1)$, the third term on the very right-hand side of \eqref{eq-lemma13-11} above can be bounded as 
	\begin{equation} \label{pf-lem13-1}
		\begin{split}
			& \max_{1 \leq j \leq 2p }  \bigg| \frac{ [\widetilde{\bX}^{\augg} ( \bgamma_j - \widetilde{\bgamma}_j ) ]^{\top}   \widetilde{\bX}^{\augg}_{-j} \big(\widetilde{\bbeta}^{\init}_{-j} - \hat{\bbeta}^{\init}_{-j} \big)  } {\widetilde\bz_j^{\top} \widetilde\bX^{\augg}_j} \bigg| \\
			& \leq C n^{-1} \max_{1 \leq j \leq 2p }  \|  \widetilde{\bX}^{\augg} ( \bgamma_j - \widetilde{\bgamma}_j )  \|_2 \cdot  \| \widetilde{\bX}^{\augg}_{-j} \big(\widetilde{\bbeta}^{\init}_{-j} - \hat{\bbeta}^{\init}_{-j} \big)\|_2. 
		\end{split}
	\end{equation}
	
	An application of Lemma \ref{pf-thm5-lemma_01}, \eqref{eq-lemma13-4}, and the sub-Gaussian assumption of $X_j$ gives that with probability $1 - o(1)$, the second term on the right-hand side above can be bounded as 
	\begin{equation}\label{pf-lem13-2}
		\begin{split}
			& \max_{1 \leq j \leq 2p} n^{ -1/2} \| \widetilde{\bX}^{\augg}_{-j} \big(\widetilde{\bbeta}^{\init}_{-j} - \hat{\bbeta}^{\init}_{-j} \big)\|_2  \\
			&\leq n^{ -1/2} \| \widetilde{\bX}^{\augg}  \big(\widetilde{\bbeta}^{\init}  - \hat{\bbeta}^{\init} \big)\|_2 \\
			& \quad +    \max_{1 \leq j \leq 2p} n^{ -1/2} \| \widetilde{\bX}^{\augg}_j \|_2 | \widetilde{\beta}_j  - \hat{\beta}_j| \\
			& \leq C \Delta_n s \sqrt{\frac{\log p}{n}}.
		\end{split}     
	\end{equation}
	Then plugging \eqref{pf-lem13-2} into \eqref{pf-lem13-1} yields that 
	\begin{equation} \label{eq-lemma13-13}
		\begin{split}
			&\max_{1 \leq j \leq 2p }  \bigg| \frac{ [\widetilde{\bX}^{\augg} ( \bgamma_j - \widetilde{\bgamma}_j ) ]^{\top}   \widetilde{\bX}^{\augg}_{-j} \big(\widetilde{\bbeta}^{\init}_{-j} - \hat{\bbeta}^{\init}_{-j} \big)  } {\widetilde\bz_j^{\top} \widetilde\bX^{\augg}_j} \bigg|  \\
			&\leq C \sqrt{\frac{m_n \log p} {n}} \cdot C  \Delta_n s  \sqrt{\frac{\log p} {n}} \leq C \Delta_n s \sqrt{\frac{\log p} {n}}, 
		\end{split}
	\end{equation}
	where the last inequality above is due to the assumption that $  \sqrt{\frac{s\log p} {n}} \to 0$ and $m_n \lesssim s $. Hence, it follows from substituting \eqref{eq-lemma13-12} and \eqref{eq-lemma13-13} into \eqref{eq-lemma13-11} that with probability $1 - o(1)$, 
	\begin{equation} \label{eq-lemma13-18}
		P_{21} \leq C \Delta_n s  \sqrt{\frac{\log p} {n}}. 
	\end{equation}
	
	We next proceed with considering term $P_{22}$ introduced in \eqref{eq-le13-P2}. Observe that $$\widetilde{\bX}^{\augg} - \hat{\bX}^{\augg} = [\bf{0},  \, \widetilde{\bX} - \hat{\bX}]$$ and $ \bbeta^{\augg} = (\bbeta^{\top} ,  {\bf 0}^{\top})^{\top}$. Then it holds that $$ ( \widetilde{\bX}^{\augg}  - \hat{\bX}^{\augg} )  \bbeta^{\augg} = \bf{0}.$$ 
	From \eqref{eq-normal-error-2} and the Cauchy--Schwarz inequality, we can deduce that 
	\begin{equation*}
		\begin{split}
			P_{22} & \leq  \max_{1 \leq j \leq 2p }     \bigg| \frac{  \widetilde\bz_j^{\top}  ( \widetilde{\bX}^{\augg}  - \hat{\bX}^{\augg} )  \bbeta^{\augg}  } {\widetilde\bz_j^{\top} \widetilde\bX^{\augg}_j} \bigg| +  \max_{1 \leq j \leq 2p }     \bigg| \frac{  \widetilde\bz_j^{\top}  ( \widetilde{\bX}^{\augg}  - \hat{\bX}^{\augg} ) (\hat{\bbeta}^{\init} - \bbeta^{\augg}  ) } {\widetilde\bz_j^{\top} \widetilde\bX^{\augg}_j} \bigg|\\
			& \leq C n^{-1} \max_{1 \leq j \leq 2p} \| \widetilde{\bz}_j\|_2 \cdot \|  ( \widetilde{\bX}^{\augg}  - \hat{\bX}^{\augg} ) (\hat{\bbeta}^{\init} - \bbeta^{\augg}  )  \|_2.
		\end{split}
	\end{equation*}
	Moreover, we have $\widetilde{\bz}_j = \be_j + \widetilde{\bX}_{-j}^{\augg} (\bgamma_j - \widetilde{\bgamma}_j) $. Since the components of $\be_j$ are i.i.d. sub-Gaussian random variables, it is easy to see that $$\mathbb{P} (\max_{1 \leq j \leq 2p} \|n^{-1/2} \be_j \|_2  \geq C) \to 0 $$ for some large enough constant $C > 0$. Further, it follows from the sub-Gaussianity of $X_j$  and the sparsity of  $ \bgamma_j $ and $\widetilde{\bgamma}_j$ that 
	\begin{equation*}
		\begin{split}
			\max_{1 \leq j \leq 2p}  n^{-1/2}  \| \widetilde{\bX}_{-j}^{\augg} (\bgamma_j - \widetilde{\bgamma}_j) \|_2 & \leq C m_n^{1/2} \sqrt{\frac{m_n \log p} {n}} \\ & \leq C m_n \sqrt{\frac{\log p} {n}} \to 0.
		\end{split}
	\end{equation*}
	Thus, when $m_n \sqrt{\frac{\log p} {n}} \to 0$ we have
	\begin{equation} \label{eq-lemma13-19}
		\mathbb{P} ( n^{-1/2} \max_{1 \leq j \leq 2p } \| \widetilde{\bz}_j \|_2 \geq C) \to 0 . 
	\end{equation}
	
	Similarly, based on Lemma \ref{pf-thm5-lemma_00} and the sparsity of $\hat{\bbeta}^{\init}$ and $\bbeta^{\augg}$, it holds that with probability $1-o(1)$,
	\begin{equation} \label{eq-lemma13-16}
		\begin{split}
			&   n^{-1/2} \|  ( \widetilde{\bX}^{\augg}  - \hat{\bX}^{\augg} ) (\hat{\bbeta}^{\init} - \bbeta^{\augg}  )  \|_2     \\
			& \leq  \max_{J': |J'| \lesssim s} \Big( \sum_{j \in J'} n^{-1}   \|\widetilde{\bX}^{\augg}_j  - \hat{\bX}^{\augg}_j \|_2^2 \Big)^{1/2} \cdot  \| \hat{\bbeta}^{\init}_{J'} - \bbeta_{J'}^{\augg}  \|_2   \\
			& \leq C  s^{1/2} \Delta_n  \cdot ( \sqrt{\frac{s \log p} {n}} + \Delta_n s  \sqrt{\frac{ \log p } {n}} ) \\
			&\leq C  \Delta_n s \sqrt{\frac{\log p} {n}},
		\end{split}   
	\end{equation}
	where the last inequality above holds due to $\Delta_n s^{1/2} \to 0$. 
	Consequently, combining the above three inequalities shows that with probability $1 - o(1)$, 
	\begin{equation} \label{eq-lemma13-14}
		P_{22} \leq C  \Delta_n s\sqrt{\frac{\log p}{n}}.
	\end{equation}
	
	We now deal with term $P_{23}$ in \eqref{eq-le13-P2}. In view of the Cauchy--Schwarz inequality and $\Delta_n m_n^{1/2} \to 0$,  \eqref{eq-normal-error-2}, \eqref{eq-lemma13-7}, and \eqref{eq-lemma13-4-0}, we can obtain that with probability $1 - o(1)$, 
	\begin{equation} \label{eq-lemma13-15}
		\begin{split}
			P_{23} & \leq  \max_{1 \leq j \leq 2p }    \frac{  \| \hat{\bz}_j - \widetilde\bz_j \|_2   } {\widetilde\bz_j^{\top} \widetilde\bX^{\augg}_j}   \cdot  \|  \widetilde{\bX}^{\augg} \big(\widetilde{\bbeta}^{\init} - \hat{\bbeta}^{\init} ) \|_2 \\
			& \leq C \Delta_n m_n^{1/2} \cdot  \Delta_n s  \sqrt{\frac{\log p}{n}} \\
			&\leq C \Delta_n s \sqrt{\frac{\log p}{n}}.
		\end{split}   
	\end{equation}
	As for term $P_{24}$, since  $( \widetilde{\bX}^{\augg}  - \hat{\bX}^{\augg} ) \bbeta = {\bf 0}  $ it follows that with probability $1 - o(1)$,
	\begin{equation} \label{eq-lemma13-17}
		\begin{split}
			P_{24} & =   \max_{1 \leq j \leq 2p }     \bigg| \frac{  (\hat{\bz}_j - \widetilde\bz_j )^{\top}  ( \widetilde{\bX}^{\augg}  - \hat{\bX}^{\augg} ) (\hat{\bbeta}^{\init}  - \bbeta^{\augg}) } {\widetilde\bz_j^{\top} \widetilde\bX^{\augg}_j } \bigg| \\
			& \leq  \max_{1 \leq j \leq 2p }    \frac{  \| \hat{\bz}_j - \widetilde\bz_j \|_2   } {\widetilde\bz_j^{\top} \widetilde\bX^{\augg}_j}   \cdot \|  ( \widetilde{\bX}^{\augg}  - \hat{\bX}^{\augg} ) (\hat{\bbeta}^{\init} - \bbeta ^{\augg} )  \|_2 \\
			& \leq C \Delta_n m_n^{1/2} \cdot \Delta_n s \sqrt{\frac{\log p}{n}} \\
			&\leq C \Delta_n s  \sqrt{\frac{\log p}{n}},
		\end{split}    
	\end{equation}
	where we have applied the bounds in \eqref{eq-lemma13-7}, \eqref{eq-lemma13-16}, and \eqref{eq-normal-error-2}. Consequently, plugging \eqref{eq-lemma13-18}, \eqref{eq-lemma13-14}, \eqref{eq-lemma13-15}, and \eqref{eq-lemma13-17} into \eqref{eq-le13-P2} yields that with probability $1 - o(1)$,
	\begin{equation} \label{eq-le13-P2-bdd}
		P_{2} \leq C \Delta_n s  \sqrt{\frac{\log p}{n}}.
	\end{equation}
	
	Now we proceed with dealing with term $P_3$. Note that 
	\begin{equation} \label{eq-le13-P3}
		P_3 \leq \max_{1 \leq j \leq 2p } | \hat{\bz}_j ^{\top}  ( \by - \hat{\bX}^{\augg} \hat{\bbeta}^{\init} ) |  \cdot \frac{  \big| \hat\bz_j^{\top} \hat\bX^{\augg}_j  -   { \widetilde\bz_j^{\top} \widetilde\bX^{\augg}_j } \big| } { \big| \hat\bz_j^{\top} \hat\bX^{\augg}_j  \big| \cdot \big|  { \widetilde\bz_j^{\top} \widetilde\bX^{\augg}_j } \big| }.
	\end{equation}
	From \eqref{eq-lemma13-7} and \eqref{eq-lemma13-19}, we can see that with probability $1 - o(1)$,
	\begin{equation} \label{eq-lemma13-22}
		\begin{split}
			\max_{1 \leq j \leq 2p } n^{-1/2} \| \hat{\bz}_j \|_2 & \leq  \max_{1 \leq j \leq 2p } n^{-1/2} \| \widetilde{\bz}_j \|_2 +  \max_{1 \leq j \leq 2p } n^{-1/2} \| \widetilde{\bz}_j -  \hat{\bz}_j \|_2 \\
			& \leq C + C m_n^{1/2} \Delta_n \leq C. 
		\end{split}
	\end{equation}
	It follows from \eqref{eq-lemma13-7}, Condition \ref{accuracy-knockoffs}, and the sub-Gaussian distribution of $\widetilde{\bX}_j^{\augg}$ that with probability $1 - o(1)$, 
	\begin{align}
		& n^{-1} | (\hat{\bz}_j - \widetilde\bz_j)^{\top} \widetilde\bX^{\augg}_j | \leq C \Delta_n m_n^{1/2},  \label{le13-num-1} \\
		& n^{-1}  | \hat\bz_j^{\top} (\widetilde\bX^{\augg}_j - \hat{\bX}_j^{\augg})| \leq C \Delta_n. \label{le13-num-2}
	\end{align}    
	Then with the aid of \eqref{eq-normal-error-2}, we can show that with probability $1 - o(1)$, 
	\begin{equation} \label{le13-den}
		\begin{split}
			& \min_{1 \leq j \leq 2p} n^{-1} | \hat\bz_j^{\top} \hat\bX^{\augg}_j | \\
			& \geq  \min_{1 \leq j \leq 2p}  n^{-1}| \widetilde\bz_j^{\top} \widetilde\bX^{\augg}_j | -  \max_{1 \leq j \leq 2p} \Big( n^{-1} | (\hat{\bz}_j - \widetilde\bz_j)^{\top} \widetilde\bX^{\augg}_j | - n^{-1}  | \hat\bz_j^{\top} (\widetilde\bX^{\augg}_j - \hat{\bX}_j^{\augg})| \Big)\\
			& \geq C - Cm_n \Delta_n - C \Delta_n \\
			&\geq C 
		\end{split}
	\end{equation}
	as $m_n \Delta_n \to 0$.
	
	As for the second component on the right-hand side of \eqref{eq-le13-P3} above, combining the results in \eqref{le13-num-1}, \eqref{le13-num-2}, and \eqref{le13-den} gives that with probability $1 - o(1)$, 
	\begin{equation} \label{eq-lemma13-21}
		\begin{split}
			\max_{1 \leq j \leq 2p} \frac{  \big| \hat\bz_j^{\top} \hat\bX^{\augg}_j  -   { \widetilde\bz_j^{\top} \widetilde\bX^{\augg}_j } \big| } { \big| \hat\bz_j^{\top} \hat\bX^{\augg}_j  \big| \cdot \big|  { \widetilde\bz_j^{\top} \widetilde\bX^{\augg}_j } \big| } 
			& \leq \max_{1 \leq j \leq 2p}  \frac{  \big| ( \widetilde{\bz}_j - \hat\bz_j  )^{\top} \widetilde\bX^{\augg}_j \big| } { \big| \hat\bz_j^{\top} \hat\bX^{\augg}_j  \big| \cdot \big|  { \widetilde\bz_j^{\top} \widetilde\bX^{\augg}_j } \big| } \\
			& \quad +  \max_{1 \leq j \leq 2p}  \frac{  \big| \hat\bz_j^{\top} (\widetilde{\bX}_j^{\augg} - \hat{\bX}_j^{\augg})  \big| } { \big| \hat\bz_j^{\top} \hat\bX^{\augg}_j  \big| \cdot \big|  { \widetilde\bz_j^{\top} \widetilde\bX^{\augg}_j } \big| } \\
			& \leq C n^{-1} ( m_n^{1/2} \Delta_n + \Delta_n ).
		\end{split}
	\end{equation}
	Regarding the first component on the right-hand side in \eqref{eq-le13-P3}, from $(\widetilde{\bX}^{\augg} - \hat{\bX}^{\augg}) 
	\bbeta = {\bf 0}$ we can deduce that 
	\begin{equation*}
		\begin{split}
			\max_{1 \leq j \leq 2p }   n^{-1} \big| \hat{\bz}_j ^{\top}  ( \by - \hat{\bX}^{\augg} \hat{\bbeta}^{\init} ) \big| & \leq \max_{1 \leq j \leq 2p }   n^{-1} \big|  \hat{\bz}_j ^{\top}\bveps | + \max_{1 \leq j \leq 2p }   n^{-1} \big|   \hat{\bz}_j ^{\top}\widetilde{\bX}^{\augg} ( {\bbeta}^{\augg}  -  \hat{\bbeta}^{\init} ) \big|    \\
			&\quad   +  \max_{1 \leq j \leq 2p }  n^{-1} \big|  \hat{\bz}_j ^{\top}  (\widetilde{\bX}^{\augg} - \hat{\bX}^{\augg})  \hat{\bbeta}^{\init}  \big| .
		\end{split}
	\end{equation*}
	Since $\bveps \stackrel{d}{\sim} N({\bf 0}, \sigma^2I_n )$, it is easy to see that for the standard normal random variable $Z$, 
	\begin{equation*}
		\begin{split}
			\mathbb{P} \bigg(\max_{1 \leq j \leq 2p } n^{-1} \big|  \hat{\bz}_j ^{\top}\bveps | > C \sqrt{\frac{\log p}{n}} \bigg) 
			& =   \mathbb{P} \bigg(\max_{1 \leq j \leq 2p } n^{-1}  \| \hat{\bz}_j\|_2 \cdot  |Z| >  C \sqrt{\frac{\log p}{n}}  \bigg) \\
			& \leq \mathbb{P} (|Z| > C \sqrt{\log p}) \to 0.
		\end{split}
	\end{equation*}
	
	Further, by Lemma \ref{pf-thm5-lemma_00}, the sub-Gaussianity of $ X_j $, and the sparsity of ${\bbeta}^{\augg}$ and $ \hat{\bbeta}^{\init}$, we can obtain that with probability $1 - o(1)$,
	\begin{align*}
		n^{-1} | \hat{\bz}_j^{\top}\widetilde{\bX}^{\augg} 
		( {\bbeta}^{\augg}  -  \hat{\bbeta}^{\init} ) |  
		& \leq n^{-1} | \hat{\bz}_j^{\top} \widetilde{\bX}^{\augg} 
		( {\bbeta}^{\augg}  -  \widetilde{\bbeta}^{\init} ) | + n^{-1} |  \hat{\bz}_j^{\top}\widetilde{\bX}^{\augg} 
		( \widetilde{\bbeta}^{\init}  -  \hat{\bbeta}^{\init} ) | \\
		& \leq C\bigg(   \sqrt{ \frac{ s\log p } {n}  } + \Delta_n s  \sqrt{\frac{\log p } {n}} \bigg) \\
		&\leq C   \sqrt{ \frac{ s\log p } {n}  }.
	\end{align*}
	Similarly, since $(\widetilde{\bX}^{\augg} - \hat{\bX}^{\augg}) 
	\bbeta = {\bf 0}$, it holds that with probability $1 - o(1)$,
	\begin{align*}
		n^{-1} |  \hat{\bz}_j^{\top} (\widetilde{\bX}^{\augg} - \hat{\bX}^{\augg})  \hat{\bbeta}^{\init}  | & = n^{-1} |  \hat{\bz}_j^{\top} (\widetilde{\bX}^{\augg} - \hat{\bX}^{\augg}) (  \hat{\bbeta}^{\init} -\bbeta^{\augg} ) | \\
		& \leq C \Delta_n s^{1/2} \cdot  \sqrt{\frac{ s \log p}{n}}  \\
		&\leq C   \sqrt{\frac{ s \log p}{n}}.
	\end{align*}
	Consequently, by $m_n\lesssim s$ in Condition \ref{debiased-lasso-condition1} we have that with probability $1 - o(1)$, 
	\begin{equation} \label{eq-le13-P3-bdd}
		P_3 \leq C m_n^{1/2} \Delta_n  \cdot \sqrt{\frac{s\log p}{n}} \leq C \Delta_n s  \sqrt{\frac{\log p}{n}}. 
	\end{equation}
	Finally, a combination of \eqref{eq-lemma13-5}, \eqref{eq-le13-P1}, \eqref{eq-le13-P2-bdd}, and \eqref{eq-le13-P3-bdd} establishes \eqref{eq-lemma13-4-part2}. 
	This completes the proof of Lemma \ref{pf-thm5-lemma-1}.

	\subsection{Proof of Lemma \ref{pf-thm5-lemma-2}} \label{new.SecB.12}
	Using the definitions of $\widetilde{W}_j$ and $w_j$ and the triangle inequality, we see that 
	\begin{equation} \label{pf-le14-1} 
		\begin{split}
			& \sum_{j = 1}^p \mathbb{P} (| \widetilde{W}_j - w_j | \geq C \sqrt{n^{-1} \log p }) \\
			& \leq  \sum_{j = 1}^p  \mathbb{P} \Big( \sqrt n  \big| |\widetilde{\beta}_j - \beta_j | - |\widetilde{\beta}_{j + p} - \beta_{j + p}|  \big|  \geq C \sqrt{\log p }  \Big) \\
			& \leq  \sum_{j = 1}^p \bigg[ \mathbb{P} \Big( \sqrt n  |\widetilde{\beta}_j - \beta_j | \geq C \sqrt{\log p } / 2 \Big) +  \mathbb{P} \Big( \sqrt n |\widetilde{\beta}_{j + p} - \beta_{j + p}|    \geq C \sqrt{\log p } /2 \Big) \bigg].
		\end{split}
	\end{equation}
	The main idea of the proof is to exploit the decomposition in \eqref{normal-decomp} and the observation that the main term therein follows the normal distribution. Let us start with bounding the error term in \eqref{normal-decomp}. We claim that with probability $1 - o (p^{-1})$,  
	\begin{equation} \label{bound-normal-error}
		\max_{1 \leq j \leq 2p} \Bigg|  \sum_{k \neq j} \frac{ \sqrt n  \widetilde\bz_j^{\top} \widetilde\bX^{\augg}_k (\beta_k^{\augg} - \widetilde{\beta}_k^{\init})} { \widetilde\bz_j^{\top} \widetilde\bX^{\augg}_j } \Bigg|  \leq \frac{C  m_n^{1/2} s  \log p } { \sqrt  n}. 
	\end{equation}
	From the fact that $\beta^{\augg}_{j + p} = 0$ for $1 \leq j \leq p$ and the bound in \eqref{bound-normal-error}, since $\frac{  m_n^{1/2} s  \log p } { \sqrt  n} \ll \sqrt{\log p} $ we can deduce through the union bound that 
	\begin{equation}
		\begin{split}
			& \sum_{j = 1}^p \mathbb{P} \Big( \sqrt n  |\widetilde{\beta}_j - \beta_j | \geq C \sqrt{\log p } / 2 \Big) \\
			& \leq  \sum_{j = 1}^p \mathbb{P} \Big(  \frac{ |\widetilde\bz_j^{\top} \bveps | } { \|\widetilde\bz_j \|_2 } \cdot \sqrt n \tau_j  \geq C \sqrt{\log p } / 3 \Big) \\
			& \quad +  \sum_{j = 1}^p \mathbb{P} \bigg(  \max_{1 \leq j \leq 2p} \Bigg|  \sum_{k \neq j} \frac{ \sqrt n  \widetilde\bz_j^{\top} \widetilde\bX^{\augg}_k (\beta_k - \widetilde{\beta}_k^{\init})} { \widetilde\bz_j^{\top} \widetilde\bX^{\augg}_j } \Bigg| > \frac{C  m_n^{1/2} s  \log p } { \sqrt  n} \bigg) \\
			& \leq  \sum_{j = 1}^p \mathbb{P} \Big(  \frac{ |\widetilde\bz_j^{\top} \bveps | } { \|\widetilde\bz_j \|_2 } \cdot \sqrt n \tau_j  \geq C \sqrt{\log p } / 3 \Big) + o(1).
		\end{split}
	\end{equation}
	
	Recall the result \eqref{bdd-tau} in Lemma \ref{pf-thm5-lemma_01} and that $\frac{\bz_j^{\top} \bveps} { \|\bz_j \|_2 } \sim N(0, \sigma^2)$.  
	As $\frac{m_n \log p} {n} = o(1)$,  it holds that for some large constant $C> 0$, 
	\begin{equation*}
		\begin{split}
			&\sum_{j = 1}^p \mathbb{P} \Big(  \frac{\widetilde\bz_j^{\top} \bveps} { \|\widetilde\bz_j \|_2 } \cdot \sqrt n \tau_j  \geq C \sqrt{\log p } / 3 \Big)\\ & \leq \sum_{j = 1}^p \mathbb{P} \Big(  \frac{\widetilde\bz_j^{\top} \bveps} { \|\widetilde\bz_j \|_2 }    \geq \widetilde{C} \sqrt{\log p } \Big) \\
			&=  p \exp \{ - \widetilde{C}^2 \log p / 2 \} \to 0.
		\end{split}
	\end{equation*}
	Similarly, we can show that 
	\begin{equation}
		\sum_{j = 1}^p \mathbb{P} \Big(  \frac{\widetilde\bz_{j + p}^{\top} \bveps} { \|\widetilde\bz_{j+p} \|_2 } \cdot \sqrt n \tau_j  \geq C \sqrt{\log p } \Big) \to 0.
	\end{equation}
	Plugging the two inequalities above into \eqref{pf-le14-1} leads to the desired result in Lemma \ref{pf-thm5-lemma-2}. It remains to establish  \eqref{bound-normal-error}.
	
	\bigskip
	\noindent \textbf{Proof of \eqref{bound-normal-error}}. Observe that for $k \neq j$, 
	\begin{equation}   \label{pf-le14-4}
		\begin{split}
			n^{-1}\widetilde\bz_j^{\top} \widetilde\bX^{\augg}_k & = n^{-1}(\widetilde\bX^{\augg}_j - \widetilde\bX^{\augg}_{-j} \widetilde{\bgamma}_j) ^{\top}\widetilde\bX^{\augg}_k \\
			& = n^{-1} \be_j ^{\top}\widetilde\bX^{\augg}_k + n^{-1} (\bgamma_j - \widetilde{\bgamma}_j )^{\top} (\widetilde\bX^{\augg}_{-j} )^{\top} \widetilde\bX^{\augg}_k.
		\end{split}      
	\end{equation}
	Since $\be_j$ and $\widetilde\bX_k^{\augg} $ are uncorrelated, it follows from the sub-Gaussian assumption in Condition \ref{debiased-lasso-condition3} that for some constant $C > 0$,
	$$
	\mathbb{P} \Big(  n^{-1} | \be_j ^{\top}\widetilde\bX^{\augg}_k |  \geq C \sqrt{\frac{\log p } {n}} \Big) \leq 2 p^{-3}.
	$$
	In light of lemma \ref{pf-thm5-lemma_01} and the sub-Gaussian assumption on $\widetilde{\bX}_j$, we can deduce that with probability $1 - o(p^{-3})$,
	\begin{equation}
		\begin{split}
			|   n^{-1} (\bgamma_j - \widetilde{\bgamma}_j )^{\top} (\widetilde\bX^{\augg}_{-j} )^{\top} \widetilde\bX^{\augg}_k | & \leq \| n^{-1/2} \widetilde\bX^{\augg}_{-j} (\bgamma_j - \widetilde{\bgamma}_j) \|_2 \|n^{-1/2} \widetilde{\bX}_k^{\augg} \|_2 \\
			& \leq C \sqrt{\frac{m_n \log p} {n}}.
		\end{split}
	\end{equation}
	Plugging the above two results into \eqref{pf-le14-4}, when $m_n\log p = o(n)$ an application of the union bound shows that with probability $ 1 - o( p^{-1})$, 
	\begin{equation} \label{eq-normal-error-1}
		\begin{split}
			\max_ {1 \leq j \leq p } \max_{k \neq j} n^{-1} | \widetilde\bz_j^{\top} \widetilde\bX^{\augg}_k | & \leq C \sqrt{\frac{\log p } {n}}  +  C  \sqrt{\frac{m_n\log p } {n}} \\
			&\leq C  \sqrt{\frac{ m_n \log p } {n}}. 
		\end{split}
	\end{equation}
	
	Similarly,  when $  \sqrt{ \frac{\log p } {n} } = o(1)$,  we can show that there exists some constant $C>0$ such that with probability $1 - o(p^{-1})$,
	\begin{equation}  \label{eq-normal-error-2}
		\min_{1 \leq j \leq p} n^{-1} \widetilde\bz_j^{\top} \widetilde\bX^{\augg}_j \geq C. 
	\end{equation}
	Consequently, plugging \eqref{eq-normal-error-1}, \eqref{eq-normal-error-2}, and \eqref{lasso-1} into Lemma \ref{pf-thm5-lemma_00} yields that with probability $1 - o (p^{-1})$,  
	\begin{equation} 
		\begin{split}
			& \max_{1 \leq j \leq p} \Bigg|  \sum_{k \neq j} \frac{ \sqrt n  \widetilde\bz_j^{\top} \widetilde\bX^{\augg}_k (\beta_k - \widetilde{\beta}_k^{\init})} { \widetilde\bz_j^{\top} \widetilde\bX^{\augg}_j } \Bigg| \\
			& \leq \sqrt n \frac{ \max_{1 \leq j \leq p} \max_{k \neq j } | \widetilde\bz_j^{\top} \widetilde\bX^{\augg}_k | } {\min_{1 \leq j \leq p} | \widetilde\bz_j^{\top} \widetilde\bX^{\augg}_j | } \cdot \|\bbeta^{\augg} - \widetilde{\bbeta}^{\init}  \|_1\\
			&  \leq C  \sqrt{ m_n \log p  } \cdot  s \sqrt{\frac{\log p}{n}} = \frac{C  m_n^{1/2} s  \log p } { \sqrt  n}, 
		\end{split}
	\end{equation}
	which establishes \eqref{bound-normal-error}. This concludes the proof of Lemma \ref{pf-thm5-lemma-2}. 
	
	\subsection{Proof of Lemma \ref{pf-thm5-lemma-3}} \label{new.SecB.13}
	
	The intuition of the proof is that the sparsity of $\bOmega^{A}$ implies the weak dependence among the components of the knockoff statistic vector $\widetilde{\bW} = (\widetilde{W}_1, \ldots, \widetilde{W}_p)$, which entails the weak dependence among the indicator functions $\mathbbm{1} (\widetilde{W}_j > t)$'s. For $1 \leq j \leq p$, let us define $$N_j = \{ l \in \mathcal{H}_0: \bOmega_{j, l}^A \neq 0 \}.$$ From the sparsity assumption  on $\bOmega^A$ in Condition \ref{debiased-lasso-condition1}, we see that $|N_j| \leq m_n$ for any $1 \leq j\leq p$. Then we can obtain through expanding the variance that
	\begin{equation} \label{pf-le15-6}
		\begin{split}
			\Var{ \big( \sum_{j \in \mathcal{H}_0} \mathbbm{1} (\widetilde{W}_j > t)  } \big) & =  \sum_{j \in \mathcal{H}_0} \sum_{\substack{l \in N_j^c \cap \mathcal{H}_0 \\ l \neq j}} \Big( \mathbb{P} (\widetilde{W}_j \geq t, \widetilde{W}_l \geq t) - \mathbb{P} (\widetilde{W}_j \geq t) \mathbb{P} (\widetilde{W}_l \geq t)    \Big) \\
			& \quad + \sum_{j \in \mathcal{H}_0} \sum_{l \in N_j \cup \{ j \}} \Big( \mathbb{P} (\widetilde{W}_j \geq t, \widetilde{W}_l \geq t) - \mathbb{P} (\widetilde{W}_j \geq t) \mathbb{P} (\widetilde{W}_l \geq t) \Big) \\
			& := V_1(t) + V_2(t).
		\end{split}    
	\end{equation}
	We will deal with terms $V_1(t)$ and $V_2(t)$ above  separately. 
	
	Regarding the second term $V_2(t)$, it follows from $ |N_j \cup \{ j\}| \leq m_n + 1  $ that 
	\begin{equation} \label{pf-le15-5}
		\begin{split}
			\sup_{t \in (0,\, G^{-1} ( \frac { c_1 q a_n  } { p } ) ] } \frac{ V_2 (t) } {p_0 G(t) }  & \leq  \sup_{t \in (0,\, G^{-1} ( \frac { c_1 q a_n  } { p } ) ] } \frac{ \sum_{j \in \mathcal{H}_0} \sum_{l \in N_j \cup \{ j \}} \mathbb{P} (\widetilde{W}_j \geq t) }  {\sum_{j \in \mathcal{H}_0}  \mathbb{P} (\widetilde{W}_j \geq t) } \\
			& \leq   \sup_{t \in (0,\, G^{-1} ( \frac { c_1 q a_n  } { p } ) ] } \frac{  \sum_{j \in \mathcal{H}_0}  (m_n + 1) \mathbb{P} (\widetilde{W}_j \geq t)  } { \sum_{j \in \mathcal{H}_0}  \mathbb{P} (\widetilde{W}_j \geq t)  } \\
			&\leq m_n + 1.
		\end{split}
	\end{equation}
	We claim that as $  \frac{ m_n^{1/2}  s (\log p)^{3/2 + 1/\gamma } } { \sqrt n }  \to 0 $, 
	\begin{equation} \label{pf-le15-n1}
		(\log p)^{1/\gamma}   \sup_{t \in (0,\, G^{-1} ( \frac { c_1 q a_n  } { p } ) ] }  \frac{V_1 (t) } { [p_0  G (t)]^2  } \to 0. 
	\end{equation} 
	Therefore, combining \eqref{pf-le15-6}, \eqref{pf-le15-5}, and \eqref{pf-le15-n1} leads to the desired result of Lemma \ref{pf-thm5-lemma-3}. It remains to establish \eqref{pf-le15-n1}.
	
	\bigskip
	\noindent\textbf{Proof of \eqref{pf-le15-n1}}. 
	Let $\{\eta_j\}_{j = 1}^p$ be a sequence of independent random variables with $\eta_j$ having density function given by
	\begin{equation}  \label{pf-le15-n3}
		\begin{split}
			h_j (t) &=  \frac{ \sqrt{2} } {\sqrt {\pi } a_{j}} \big[1 - \Phi( b_{j}^{-1} t ) \big] \exp \big\{ - t^2 / (2 v_{ j}^2) \big\} \\
			& \quad + \frac{ \sqrt{2} } {\sqrt {\pi } b_{j}} \big[1 - \Phi( v_{ j}^{-1} t ) \big] \exp \big\{ - t^2 / (2 b_{ j}^2) \big\},  
		\end{split}
	\end{equation}
	where $ v_j = \sqrt{ 2 (\e e_j^2)^{-1} (1  - \corr(e_j, e_{j+p})) }$ and $ b_j =\sqrt{ 2 (\e e_j^2)^{-1} (1  + \corr(e_j, e_{j+p})) } $.
	For $1 \leq j \leq 2p$, let us define $ \xi_j = \sqrt n \tau_j \cdot \frac{ \widetilde{\bz}_j^{\top} \bveps } {\| \widetilde{\bz}_j \|_2 } $. The essential step in the proof is to show that for $l \in N_j^{c} \cap \mathcal{H}_0$,  $$(  |\xi_j| - |\xi_{j+p}|,  |\xi_l| - |\xi_{l+p}|) \stackrel{d}{\to} (\eta_j, \eta_{l}).$$ We proceed with proving such result. 
	Define $\delta_n = C \frac{m_n^{1/2} s \log p}{\sqrt n }$. We claim that for $l \neq j$ and $l \in N_j^{c} \cap \mathcal{H}_0$,
	\begin{equation} \label{pf-le15-10}
		\begin{split}
			& \mathbb{P} ( \widetilde{W}_j \geq t, \widetilde{W}_l \geq t )  \\
			& \leq \mathbb{P} (   \eta_j \geq \sqrt n t - \delta_n ) \mathbb{P} (   \eta_l \geq \sqrt n t - \delta_n )  \bigg(1 + O \Big( \sqrt{\frac{m_n (\log p)^3} {n}}  \Big)\bigg) + O(p^{-3}),
		\end{split}
	\end{equation}
	\begin{equation} \label{pf-le15-11}
		\begin{split}
			& \mathbb{P} ( \widetilde{W}_j \geq t, \widetilde{W}_l \geq t )  \\
			& \geq \mathbb{P} (   \eta_j \geq \sqrt n t + \delta_n ) \mathbb{P} (   \eta_l \geq  \sqrt n t + \delta_n )  \bigg(1 + O \Big( \sqrt{\frac{m_n (\log p)^3} {n}}  \Big)\bigg)   + O(p^{-3}),
		\end{split}
	\end{equation}
	\begin{align} \label{pf-le15-12} 
		& \mathbb{P} ( \widetilde{W}_j \geq t)  \geq \mathbb{P} (   \eta_j \geq \sqrt n t + \delta_n ) \bigg(1 + O \Big( \sqrt{\frac{m_n (\log p)^3} {n}}  \Big)\bigg) + O(p^{-3}),  
	\end{align}
	\begin{equation}  \label{pf-le15-13}
		\mathbb{P} ( \widetilde{W}_j \geq t)  \leq \mathbb{P} (   \eta_j \geq \sqrt n t - \delta_n ) \bigg(1 + O \Big( \sqrt{\frac{m_n (\log p)^3} {n}}  \Big)\bigg) + O(p^{-3}). 
	\end{equation}
	
	The proofs for \eqref{pf-le15-10}--\eqref{pf-le15-13} above are analogous. Without loss of generality, we will present only the proof of \eqref{pf-le15-10} and postpone it to the end of the proof for Lemma \ref{pf-thm5-lemma-3}. In view of \eqref{pf-le15-10}--\eqref{pf-le15-13} above and the definition of $V_1 (t)$ in \eqref{pf-le15-6}, we can deduce that 
	\begin{equation} \label{pf-le15-n12}
		\begin{split}
			& V_1 (t)  = \sum_{j \in \mathcal{H}_0} \sum_{\substack{l \in N_j^c \cap \mathcal{H}_0 \\ l \neq j}}  \Big( \mathbb{P} (\widetilde{W}_j \geq t, \widetilde{W}_l \geq t) - \mathbb{P} (\widetilde{W}_j \geq t) \mathbb{P} (\widetilde{W}_l \geq t)    \Big) \\
			& \leq  \sum_{j \in \mathcal{H}_0} \sum_{\substack{  l \neq j}} \Bigg\{  \mathbb{P} (   \eta_j \geq \sqrt n t - \delta_n ) \mathbb{P} (   \eta_l \geq \sqrt n t - \delta_n )   \bigg(1 + O \Big( \sqrt{\frac{m_n (\log p)^3} {n}}  \Big)  \\
			& \quad -  \mathbb{P} (   \eta_j \geq \sqrt n t + \delta_n ) \mathbb{P} (   \eta_j \geq \sqrt n t + \delta_n )  \bigg(1 + O \Big( \sqrt{\frac{m_n (\log p)^3} {n}}  \Big)\bigg)
			\Bigg\} + O(p^{-1}) \\
			& = \sum_{j \in \mathcal{H}_0} \sum_{\substack{ l \neq j}} 
			\mathbb{P} (\sqrt n t - \delta_n \leq \eta_j \leq \sqrt n t + \delta_n) \mathbb{P} (   \eta_l \geq \sqrt n t - \delta_n ) \\
			&  \quad+ \sum_{j \in \mathcal{H}_0} \sum_{\substack{ l \neq j}} \mathbb{P} (   \eta_j \geq \sqrt n t - \delta_n )  \mathbb{P} (\sqrt n t - \delta_n \leq \eta_l \leq \sqrt n t + \delta_n)  \\
			&  \quad+ \sum_{j \in \mathcal{H}_0} \sum_{\substack{ l \neq j}} \mathbb{P} (   \eta_j \geq \sqrt n t - \delta_n )  \mathbb{P} (   \eta_l \geq \sqrt n t - \delta_n ) \cdot O\Big( \sqrt {\frac{m_n (\log p)^3 } {n}  } \Big)
			+ O(p^{-1}) \\
			& :=  V_{11}(t) + V_{12}(t) + V_{13}(t) + O(p^{-1}). 
		\end{split} 
	\end{equation}
	
	Recall that $p_0 G(t) = \sum_{j \in \mathcal{H}_0} \mathbb{P} (\widetilde{W}_j \geq t)$. Then it follows from the definition of $V_{11}(t)$ and \eqref{pf-le15-12} that 
	\begin{equation} \label{pf-le15-n2}
		\begin{split}
			&  \frac{V_{11}(t)} { [p_0 G(t)]^2 }  \leq  \frac{\sum_{j \in \mathcal{H}_0} \sum_{  l \neq j} \mathbb{P} (\sqrt n t - \delta_n \leq \eta_j \leq \sqrt n t + \delta_n) \mathbb{P} (   \eta_l \geq \sqrt n t - \delta_n )  } {\big[ \sum_{j \in \mathcal{H}_0} \mathbb{P} (   \eta_j \geq \sqrt n t + \delta_n ) \big(1 + O ( \sqrt{\frac{m_n (\log p)^3} {n}}  )\big) + O(p^{-2}) \big]^2}.
		\end{split}
	\end{equation}
	We will consider two ranges $t \in (0, 4 n^{-1/2} \max_{1 \leq j \leq p} (v_j \lor b_j))$ and $t \in [4 n^{-1/2} \max_{1 \leq j \leq p} 
	(v_j \lor b_j), G^{-1 } (\frac{c_1 q a_n} {p}) ]$ separately. 
	For the first range $t \in (0, 4 n^{-1/2} \max_{1 \leq j \leq p} (v_j \lor b_j))$, we can see that $\sqrt n t$ is upper bounded by a constant. Since $\delta_n = o(1)$ by the assumption that $  \frac{ m_n^{1/2}  s (\log p)^{3/2 + 1/\gamma } } { \sqrt n }  \to 0 $, it follows that $\sqrt n t + \delta_n$ and $\sqrt n t - \delta_n$ are both of a constant order.  
	Hence, by the definition of the density function $h_j(\cdot)$ of $\eta_j$ shown in \eqref{pf-le15-n3}, $ \max_{1 \leq j \leq p} h_j(u)$ is bounded by a constant for $u \in [\sqrt n t - \delta_n, \sqrt n t + \delta_n]$,  and $$C_1 \leq  \min_{1\leq j \leq p} \mathbb{P} (\eta_j \geq \sqrt n t + \delta_n) \leq \max_{1 \leq j \leq p} \mathbb{P} (\eta_j \geq \sqrt n t - \delta_n) \leq C_2$$ for some positive constants $C_1 < C_2$. Thus, it is easy to see that 
	\begin{equation} \label{pf-le15-n8}
		\begin{split}
			&  \sup_{t \in (0, 4 n^{-1/2} \max_{1 \leq j \leq p}  (v_j \lor b_j))} \frac{V_{11}(t)} { [p_0 G(t)]^2 } \\
			& \leq C \frac{ p_0^2 \delta_n \max_{1\leq j \leq p} \sup_{u \in [\sqrt n t - \delta_n, \sqrt n t + \delta_n]} h_{j} (u) \max_{1 \leq j \leq p} \mathbb{P} (\eta_j \geq \sqrt n t - \delta_n) } { p_0^2 [\min_{1 \leq j \leq p} \mathbb{P} (\eta_j \geq \sqrt n t + \delta_n) ]^2 }\\
			& \leq C \delta_n = C \frac{m_n^{1/2} s \log p}{\sqrt n }.
		\end{split}
	\end{equation}
	
	We proceed with considering the second range $t \in [4 n^{-1/2} \max_{1 \leq j \leq p}  (v_j \lor b_j), G^{-1 } (\frac{c_1 q a_n} {p}) )$. An application of similar arguments as for \eqref{pf-le10-n1} shows that 
	\begin{equation} \label{pf-le15-n5}
		\begin{split}
			& \max_{1 \leq j \leq p} \sup_{t \in [4 n^{-1/2} \max_{1 \leq j \leq p}  (v_j \lor b_j)), G^{-1 } (\frac{c_1 q a_n} {p})]}  \frac{  \sum_{j \in \mathcal{H}_0}   \mathbb{P} ( \sqrt n   t - \delta_n \leq \eta_j \leq \sqrt n t + \delta_n)    } {  \sum_{j \in \mathcal{H}_0} \mathbb{P} (   \eta_j \geq \sqrt n t + \delta_n )   } \\
			& \leq C \sqrt n G^{-1} (\frac{c_1 q a_n} {p}) \cdot \delta_n. 
		\end{split}
	\end{equation}
	Moreover, it follows from plugging $t = G^{-1}(\frac{c_1 q a_n} {p}) $ into \eqref{pf-le15-13} and taking summation over $j \in \mathcal{H}_0$ that 
	\begin{equation} \label{pf-le15-n4}
		\begin{split}
			\frac{c_1 q a_n p_0} {p} & \leq \sum_{j \in \mathcal{H}_0} \mathbb{P} (\eta_j \geq \sqrt n  G^{-1}(\frac{c_1 q a_n} {p}) - \delta_n) \bigg(1 + O \Big( \sqrt{\frac{m_n (\log p)^3} {n}}  \Big)\bigg) \\
			&\quad+ O(p^{-3}).
		\end{split}
	\end{equation}
	Then from the density function $h_j(t)$ for $\eta_j$, we can obtain through some direct calculations that 
	\begin{equation} \label{pf-le15-15}
		\mathbb{P} (   \eta_j \geq t)  = 2 [1 - \Phi(v_j^{-1} t)] [1 - \Phi(b_j^{-1} t)].
	\end{equation}
	
	Further, combining \eqref{pf-le15-n4} and \eqref{pf-le15-15} yields that $$  G^{-1}(\frac{c_1 q a_n} {p}) = O(\sqrt{\frac{\log p}{n}} ).$$ Substituting this bound into \eqref{pf-le15-n5} implies that
	\begin{equation} \label{pf-le15-n7}
		\begin{split}
			& \max_{1 \leq j \leq p} \sup_{t \in [4 n^{-1/2} \max_{1 \leq j \leq p}  (v_j \lor b_j)), G^{-1 } (\frac{c_1 q a_n} {p})]}  \frac{  \sum_{j \in \mathcal{H}_0}   \mathbb{P} ( \sqrt n   t - \delta_n \leq \eta_j \leq \sqrt n t + \delta_n)    } {  \sum_{j \in \mathcal{H}_0} \mathbb{P} (   \eta_j \geq \sqrt n t + \delta_n )   } \\
			& \leq C \frac{ m_n^{1/2} s (\log p)^{3/2}} {\sqrt n },
		\end{split}
	\end{equation}
	where in the last inequality above we have utilized the definition of $\delta_n$. Thus as $\frac{ m_n^{1/2} s (\log p)^{3/2}} {\sqrt n } \to 0$, it holds that 
	\begin{equation} \label{pf-le15-n6}
		\begin{split}
			& \max_{1 \leq j \leq p} \sup_{t \in [4 n^{-1/2} \max_{1 \leq j \leq p}  (v_j \lor b_j)), G^{-1 } (\frac{c_1 q a_n} {p})]} \bigg| \frac{  \sum_{j \in \mathcal{H}_0}   \mathbb{P} ( \eta_j \geq \sqrt n   t - \delta_n   )  } {  \sum_{j \in \mathcal{H}_0} \mathbb{P} (   \eta_j \geq \sqrt n t + \delta_n )   }  - 1 \bigg| \\
			& \leq C  \frac{ m_n^{1/2} s (\log p)^{3/2}} {\sqrt n } \to 0.
		\end{split}
	\end{equation}
	
	Since $p_0 G(t) \geq c_1 q a_n p_0 / p \to \infty $ for $0\leq t \leq G^{-1 } (\frac{c_1 q a_n} {p})$, it follows from taking summation over $j \in \mathcal{H}_0$ on both sides of \eqref{pf-le15-13} that as $ m_n^{1/2} (\log p)^{3/2} / \sqrt n  \to 0$, 
	\begin{equation*}  
		\sum_{j \in \mathcal{H}_0} \mathbb{P} (\eta_j \geq \sqrt n t - \delta_n) \geq  C \Big( \frac{c_1 q a_n p_0} { p} + O(p^{-2} ) \Big)  \to \infty,
	\end{equation*}
	which along with \eqref{pf-le15-n6} implies that 
	\begin{equation*}  
		\sum_{j \in \mathcal{H}_0} \mathbb{P} (\eta_j \geq \sqrt n t + \delta_n) \geq  C \Big( \frac{c_1 q a_n p_0} { p} + O(p^{-2} ) \Big)  \to \infty.
	\end{equation*}
	Combining this with \eqref{pf-le15-n7}, we can further bound the ratio in \eqref{pf-le15-n2} in the second range of  $t \in [4 n^{-1/2} \max_{1 \leq j \leq p}  (v_j \lor b_j)), G^{-1 } (\frac{c_1 q a_n} {p}))$ as 
	\begin{equation*}
		\begin{split}
			& \sup_{t \in [4 n^{-1/2} \max_{1 \leq j \leq p}  (v_j \lor b_j)), G^{-1 } (\frac{c_1 q a_n} {p})]} \frac{V_{11}(t)} { [p_0 G(t)]^2 } \\
			& \leq \Bigg\{ \frac{ \big[\sum_{j \in \mathcal{H}_0}   \mathbb{P} ( \sqrt n t - \delta_n \leq \eta_j \leq \sqrt n t + \delta_n) \big]^2  } { \big[  \sum_{j \in \mathcal{H}_0} \mathbb{P} (   \eta_j \geq \sqrt n t + \delta_n ) \big]^2  }  + \frac{  \sum_{j \in \mathcal{H}_0}   \mathbb{P} ( \sqrt n t - \delta_n \leq \eta_j \leq \sqrt n t + \delta_n)    } {  \sum_{j \in \mathcal{H}_0} \mathbb{P} (   \eta_j \geq \sqrt n t + \delta_n )   }\Bigg\} \\
			& \quad \times \bigg(1 + O \Big( \sqrt{\frac{m_n (\log p)^3} {n}} + p^{-2}  \Big)\bigg) \\
			& \leq C \frac{ m_n^{1/2} s (\log p)^{3/2}} {\sqrt n }.
		\end{split}
	\end{equation*}
	Hence, we see from the above result and \eqref{pf-le15-n8} that  
	\begin{equation}  \label{pf-le15-n9}
		\sup_{t \in (0, G^{-1}(\frac{c_1 q a_n}{p}))} \frac{V_{11}(t)} { [p_0 G(t)]^2 } \leq C  \frac{ m_n^{1/2} s (\log p)^{3/2}} {\sqrt n }. 
	\end{equation}

	In a similar manner, we can deduce that 
	\begin{equation} \label{pf-le15-n10}
		\sup_{t \in (0, G^{-1}(\frac{c_1 q a_n}{p}))} \frac{V_{12}(t)} { [p_0 G(t)]^2 } \leq C  \frac{ m_n^{1/2} s (\log p)^{3/2}} {\sqrt n }
	\end{equation}
	and 
	\begin{equation} \label{pf-le15-n11}
		\sup_{t \in (0, G^{-1}(\frac{c_1 q a_n}{p}))} \frac{V_{13}(t)} { [p_0 G(t)]^2 } \leq C \sqrt{ \frac{m_n (\log p)^3} {n} }.
	\end{equation}
	Combining \eqref{pf-le15-n12} and \eqref{pf-le15-n9}--\eqref{pf-le15-n11} yields  \eqref{pf-le15-n1} as $  \frac{ m_n^{1/2}  s (\log p)^{3/2 + 1/\gamma } } { \sqrt n }  \to 0$. This completes the proof of \eqref{pf-le15-n1}. It remains to establish \eqref{pf-le15-10}.
	
	\bigskip
	\noindent\textbf{Proof of \eqref{pf-le15-10}}. Note that for $j \in \mathcal{H}_0$, it holds that $\beta_j^{\augg} = \beta_{j + p}^{\augg} = 0$ under the setting of the linear model. Then it follows that  $$\widetilde{W}_j  = | \widetilde{\beta}_j | - |\widetilde{\beta}_{j+p}| = | \widetilde{\beta}_j - \beta_j^{\augg} | - | \widetilde{\beta}_{j + p} - \beta_{j+p}^{\augg} |.$$
	For $1 \leq j \leq 2p$, let us define $ \xi_j = \sqrt n \tau_j \cdot \frac{ \widetilde{\bz}_j^{\top} \bveps } {\| \widetilde{\bz}_j \|_2 } $.  
	In view of the expression in \eqref{normal-decomp} and the bound of the remainder term established in \eqref{bound-normal-error}, an application of the total probability inequality gives that 
	\begin{equation} \label{pf-le15-4}
		\begin{split}
			& \mathbb{P} \big( \widetilde{W}_j \geq t, \widetilde{W}_l \geq t \big)  \leq  \mathbb{P} \big(  | \xi_j  | -   |  \xi_{j + p}| \geq \sqrt n t - \delta_n, \, | \xi_l  | -   |  \xi_{l + p}|  \geq \sqrt n t - \delta_n \big)  \\
			& \qquad + \mathbb{P} \bigg(  \max_{1 \leq j \leq 2p} \Bigg|  \sum_{k \neq j} \frac{ \sqrt n  \widetilde\bz_j^{\top} \widetilde\bX^{\augg}_k (\beta_k - \widetilde{\beta}_k^{\init})} { \widetilde\bz_j^{\top} \widetilde\bX^{\augg}_j } \Bigg| >  \delta_n \bigg) \\
			& =  \mathbb{P} \big(  | \xi_j  | -   |  \xi_{j + p}| \geq \sqrt n t - \delta_n, \, | \xi_l  | -   |  \xi_{l + p}|  \geq \sqrt n t - \delta_n \big)  + O (p^{-3}).
		\end{split} 
	\end{equation}
	It suffices to consider probability $ \mathbb{P} \big(  | \xi_j  | -   |  \xi_{j + p}| \geq   t - \delta_n, \, | \xi_l  | -   |  \xi_{l + p}|  \geq  t - \delta_n \big) $ for  $t \in (0, \sqrt n G^{-1} (\frac{c_1 q a_n} {p}) ]$. A useful observation is that 
	\begin{equation} \label{pf-le15-7}
		\begin{split}
			& \mathbb{P} \big(  | \xi_j  | -   |  \xi_{j + p}| \geq t - \delta_n, \, | \xi_l  | -   |  \xi_{l + p}|  \geq t - \delta_n \big) \\
			& \leq \mathbb{P} \Big(  | \xi_j  | -   |  \xi_{j + p}| \geq t - \delta_n, \, | \xi_l  | -   |  \xi_{l + p}|  \geq t - \delta_n, \\
			& \hspace{1cm} \max \{ | \xi_j  |,  |  \xi_{j + p}|, | \xi_l  |,  |  \xi_{l + p}| \}  \leq C \sqrt{\log p }\Big)  \\
			& \quad + \mathbb{P} \big( \max \{ | \xi_j  |,  |  \xi_{j + p}|, | \xi_l  |,  |  \xi_{l + p}|   \} > C \sqrt{\log p} \big) \\
			&:= P_{1} + P_{2}.
		\end{split}
	\end{equation}
	We will consider terms $P_1$ and $P_2$ above separately. 
	
	Let us first deal with term $P_2$. From the definition of $\xi_j$,  \eqref{bdd-tau} in Lemma \ref{pf-thm5-lemma_01}, and the fact that $\frac{ \widetilde{\bz}_j^{\top} \bveps } {\| \widetilde{\bz}_j \|_2 } \stackrel{d}{\sim} N(0, 1)$, we can obtain through the union bound that as $ \frac{ m_n \log p }{n}  \to 0 $ and for some large constant $C > 4 (\e e_j^2)^{-1/2} $, 
	\begin{equation*}
		\begin{split}
			\mathbb{P} ( |\xi_j| \geq C \sqrt {\log p} ) & \leq \mathbb{P} \bigg(  \bigg|  \frac{ \widetilde{\bz}_j^{\top} \bveps } {\| \widetilde{\bz}_j \|_2 } \bigg| \geq 2 C (\e e_j^2)^{1/2} \sqrt {\log p} / 3 \bigg) \\
			&\quad+ \mathbb{P} (\sqrt n \tau_j \geq  3 (\e e_j^2)^{-1/2} /2 )  \\
			& = O (p^{-3}). 
		\end{split}
	\end{equation*}
	Hence, the inequality above implies that
	\begin{equation} \label{pf-le15-n24}
		P_{2} = O( p^{-3}). 
	\end{equation}
	
	We next proceed with analyzing term $P_1$. Given $\widetilde{\bX}^{\augg}$, denote by $f_{\xi, \xi_{j+p}} (x, y)$ the density of $(\xi_i, \xi_{j+p})$ and $f_{\xi_{l}, \xi_{l+p} |(\xi_{j}, \xi_{j+p}) } (u, w | x, y)$ the conditional density of $ (\xi_l, \xi_{l+p} )  | (\xi_{j}, \xi_{j+p})  $. Then probability $P_2$ can be written as 
	\begin{equation} \label{pf-le15-n16}
		\begin{split}
			&  \mathbb{P} \bigg(  | \xi_j  | -   |  \xi_{j + p}| \geq t - \delta_n, \, | \xi_l  | -   |  \xi_{l + p}|  \geq t - \delta_n, \\
			&\quad \max \{ | \xi_j  |,  |  \xi_{j + p}|, | \xi_l  |,  |  \xi_{l + p}| \leq C \sqrt{\log p } \bigg) \\
			& = \e_{\widetilde\bX^{\augg}} \Bigg[\int_{\substack{|x| - |y| \geq t - \delta_n \\ |x| \leq C \sqrt{\log p} \\ |y| \leq C \sqrt{\log p}} } f_{\xi, \xi_{j+p}} (x, y)  \\
			&\quad \cdot \int_{\substack{|u| - |w| \geq t - \delta_n \\ |u| \leq C \sqrt{\log p} \\ |w| \leq C \sqrt{\log p}}} f_{\xi_{l}, \xi_{l+p} |(\xi_{j}, \xi_{j+p}) } (u, w | x, y)  \,du\, dv\, dx\, dy \Bigg].
		\end{split}
	\end{equation}
	Since $\bveps \stackrel{d}{\sim} N({\bf 0}, I_n)$ and is independent of $\widetilde{\bX}^{\augg}$, it is easy to see that for $j \neq l$, conditional on $\widetilde{\bX}^{\augg}$ we have  
	\begin{equation*}
		(\xi_j, \xi_{j + p}, \xi_l, \xi_{l + p})^{\top} \big| \widetilde{\bX}^{\augg} \stackrel{d} {\sim} N ({\bf 0}, \bV), 
	\end{equation*}
	where the covariance matrix is given by  $\bV = \begin{pmatrix} \bV_{11} \bV_{12} \\ \bV_{21} \bV_{22}\end{pmatrix}$ with  
	\begin{equation*}  
		\begin{split}
			& \bV_{11} = 
			\begin{pmatrix}
				n \tau_j^2 
				& \frac{n \widetilde{\bz}_j^{\top} \widetilde{\bz}_{j + p}} { |\widetilde{\bz}_j^{\top} \widetilde{\bX}_j^{\augg}|  |\widetilde{\bz}_{j + p}^{\top} \widetilde{\bX}_{j + p}^{\augg}| }  \\
				\frac{n \widetilde{\bz}_j^{\top} \widetilde{\bz}_{j + p}} { |\widetilde{\bz}_j^{\top} \widetilde{\bX}_j^{\augg}|  |\widetilde{\bz}_{j + p}^{\top} \widetilde{\bX}_{j + p}^{\augg}| }   
				& n \tau_{j + p}^2
			\end{pmatrix}, \\
			& \bV_{12} = \bV_{21}^{\top} = 
			\begin{pmatrix}
				\frac{n \widetilde{\bz}_j^{\top} \widetilde{\bz}_{l}} { |\widetilde{\bz}_j^{\top} \widetilde{\bX}_j^{\augg}|  |\widetilde{\bz}_{l}^{\top} \widetilde{\bX}_{l}^{\augg}|   } 
				& \frac{n \widetilde{\bz}_j^{\top} \widetilde{\bz}_{l + p}} { |\widetilde{\bz}_j^{\top} \widetilde{\bX}_j^{\augg}|  |\widetilde{\bz}_{l + p}^{\top} \widetilde{\bX}_{l + p}^{\augg}|   } \\ 
				\frac{n \widetilde{\bz}_l^{\top} \widetilde{\bz}_{j + p}} { |\widetilde{\bz}_l^{\top} \widetilde{\bX}_l^{\augg}|  |\widetilde{\bz}_{j + p}^{\top} \widetilde{\bX}_{j + p}^{\augg}|   }
				&  \frac{n \widetilde{\bz}_{l +p}^{\top} \widetilde{\bz}_{j + p}} { |\widetilde{\bz}_{l + p}^{\top} \widetilde{\bX}_{l +p}^{\augg}|  |\widetilde{\bz}_{j + p}^{\top} \widetilde{\bX}_{j + p}^{\augg}|   }
			\end{pmatrix},  \\
			& \bV_{22} = 
			\begin{pmatrix} 
				n \tau_l^2
				&  \frac{n \widetilde{\bz}_l^{\top} \widetilde{\bz}_{l + p}} { |\widetilde{\bz}_l^{\top} \widetilde{\bX}_l^{\augg}|  |\widetilde{\bz}_{l + p}^{\top} \widetilde{\bX}_{l + p}^{\augg}|   }\\
				\frac{n \widetilde{\bz}_l^{\top} \widetilde{\bz}_{l + p}} { |\widetilde{\bz}_l^{\top} \widetilde{\bX}_l^{\augg}|  |\widetilde{\bz}_{l + p}^{\top} \widetilde{\bX}_{l + p}^{\augg}|   }
				& n \tau_{l + p}^2
			\end{pmatrix}.
		\end{split}
	\end{equation*} 
	
	It follows from the conditional distribution of the multivariate normal distribution that given $\widetilde{\bX}^{\augg}$, 
	\begin{equation} \label{pf-le15-n21}
		\begin{split}
			& f_{\xi_{l}, \xi_{l+p} |(\xi_{j}, \xi_{j+p}) } (u, v | x, y)\\
			& = \frac{1 }  {2\pi | \bV_{22} - \bV_{21} \bV_{11}^{-1} \bV_{12} |^{1/2}} \times \\
			&\quad \exp \Bigg\{ - \frac{1} {2} \left[ \left( \begin{matrix} 
				u\\ v
			\end{matrix} \right) - \bV_{21} \bV_{11}^{-1} \left( \begin{matrix} 
				x\\ y
			\end{matrix} \right) \right]^{\top} (\bV_{22} - \bV_{21} \bV_{11}^{-1} \bV_{12})^{-1} \\
			&\quad \cdot \left[ \left( \begin{matrix} 
				u\\ v
			\end{matrix} \right) - \bV_{21} \bV_{11}^{-1} \left( \begin{matrix} 
				x\\ y
			\end{matrix} \right) \right]   \Bigg\}.
		\end{split}
	\end{equation}
	For $l \neq j$ and $l \in N_{j}^c$, it holds that $$ \e (e_j, e_l) = \frac{\bOmega_{j, l}^A} {\bOmega_{j, j}^A \bOmega_{l, l}^A} = 0.$$ Since $\bOmega_{j, l}^A = \bOmega_{j, l+p}^A = \bOmega_{j + p, l}^A = \bOmega_{j+p, l+p}^A$ due to the symmetric structure of $\bOmega$, we also have $$\e (e_j, e_{l+p}) = \e (e_{j+p}, e_{l}) = \e (e_{j+p}, e_{l+p} ) = 0$$ for $l \neq j$ and $l \in N_{j}^c$. 
	Then it follows from \eqref{bdd-zz} in Lemma \ref{pf-thm5-lemma_01} that for $l \neq j$ and $l \in N_{j}^c$, with probability $ 1- O(p^{-3})$  
	\begin{equation} \label{pf-le15-n14}
		\begin{split}
			& n^{-1} \widetilde{\bz}_j^{\top} \widetilde{\bz}_l  \leq C \sqrt{\frac{m_n \log p}{n}}, \ \   n^{-1} \widetilde{\bz}_j^{\top} \widetilde{\bz}_{l+p}  \leq C \sqrt{\frac{m_n \log p}{n}},  \\
			& n^{-1} \widetilde{\bz}_{j+p}^{\top} \widetilde{\bz}_{l}  \leq C \sqrt{\frac{m_n \log p}{n}}, \ \  n^{-1} \widetilde{\bz}_{j+p}^{\top} \widetilde{\bz}_{l+p}  \leq C \sqrt{\frac{m_n \log p}{n}}.
		\end{split}
	\end{equation}
	Similarly, for $1 \leq j \leq 2p$ we can show that with probability $1 - O(p^{-3})$, 
	\begin{equation}    \label{pf-le15-n15}
		n^{-1} \widetilde\bz_j^{\top} \widetilde\bX^{\augg}_j \geq C. 
	\end{equation}
	\ignore{ 
		\begin{equation*}
			\begin{split}
				\widetilde{\bz}_j^{\top} \widetilde{\bz}_l & = (\widetilde{\bX}_j^{\augg} - \widetilde{\bX}_{-j}^{\augg} \widetilde\bgamma_j)^{\top}  (\widetilde{\bX}_l^{\augg} - \widetilde{\bX}_{-l}^{\augg} \widetilde\bgamma_l) \\
				& = ( \be_j +  \widetilde{\bX}^{\augg}_{-j} (\bgamma_j - \widetilde{\bgamma}_j) )^{\top} ( \be_l +  \widetilde{\bX}_{-l}^{\augg} (\bgamma_l - \widetilde{\bgamma}_l) ) \\
				& =  \be_j^{\top} \be_l + \be_j^{\top}  \widetilde{\bX}_{-l}^{\augg} (\bgamma_l - \widetilde{\bgamma}_l) +   \be_l^{\top}  \widetilde{\bX}_{-j}^{\augg} (\bgamma_j - \widetilde{\bgamma}_j) \\
				& \hspace{2cm} + [  \widetilde{\bX}_{-j}^{\augg} (\bgamma_j - \widetilde{\bgamma}_j) ]^{\top}  \widetilde{\bX}_{-l}^{\augg} (\bgamma_l - \widetilde{\bgamma}_l).
			\end{split}
		\end{equation*}
		For $l \neq j$ and $l \in N_{j}^c$, we know that $\e [e_j e_l] = \Cov (e_j, e_l) = \frac{\bOmega_{j, l}^A} {\bOmega_{j, j}^A \bOmega_{l, l}^A} = 0$. In addition, since $e_j$ and $e_l$ are sub-Gaussian random variables, we can obtain by applying Bernstein?s inequality for sub-exponential random variables that  
		\begin{equation*}
			\mathbb{P} (n^{-1} | \be_j^{\top} \be_l  | \geq C \sqrt{\frac{\log p} {n}}) = O(p^{-3}).
		\end{equation*}
		Moreover,  in view of Lemma \ref{pf-thm5-lemma_01}, sparsity of $\bgamma_j$ and $\widetilde{\bgamma}_j$ and the fact that $\Cov (e_j, X_l) = 0$ for $j \neq l$, we have with probability $1 - O(p^{-3})$
		\begin{equation*}
			\begin{split}
				n^{-1} \big| \be_j^{\top}  \widetilde{\bX}_{-l}^{\augg} (\bgamma_l - \widetilde{\bgamma}_l) \big| & \leq n^{-1} \big| \be_j^{\top}  \widetilde{\bX}_{j}^{\augg} (\bgamma_{l, j} - \widetilde{\bgamma}_l) \big| +  n^{-1} \Big| \sum_{k \neq j, l} \be_j^{\top}  \widetilde{\bX}_{k}^{\augg} (\bgamma_{l,k} - \widetilde{\bgamma}_{l, k}) \Big| \\
				& \leq n^{-1} \big| \be_j^{\top}  \widetilde{\bX}_{j}^{\augg} (\bgamma_{l, j} - \widetilde{\bgamma}_l) \big| \\
				& \qquad +  n^{-1} \Big[\max_{J': |J'| \lesssim m_n}  \sum_{\substack{k \neq j, l\\k \in J'}} (\be_j^{\top}  \widetilde{\bX}_{k}^{\augg} )^2  \sum_{\substack{k \neq j, l\\k \in J'} } (\bgamma_{l,k} - \widetilde{\bgamma}_{l, k})^2 \Big] ^2  \\
				& \leq C\sqrt{\frac{m_n \log p}{n}} +C \sqrt{\frac{m_n \log p}{n}} \cdot \sqrt{\frac{m_n \log p}{n}} \leq C \sqrt {\frac{m_n \log p}{n}}. 
			\end{split}    
		\end{equation*}
		Similarly, it holds that with probability $1 - O(p^{-3})$,
		\begin{equation*}
			n^{-1} \big| \be_l^{\top}  \widetilde{\bX}_{-j}^{\augg} (\bgamma_j - \widetilde{\bgamma}_j) \big| \leq C \sqrt {\frac{m_n \log p}{n}}. 
		\end{equation*}
		Additionally, we have that with probability $1 - O(p^{-3})$,
		\begin{equation*}
			\big|  [  \widetilde{\bX}_{-j}^{\augg} (\bgamma_j - \widetilde{\bgamma}_j) ]^{\top}  \widetilde{\bX}_{-l}^{\augg} (\bgamma_l - \widetilde{\bgamma}_l) \big| \leq C \sqrt {\frac{m_n \log p}{n}} \cdot \sqrt {\frac{m_n \log p}{n}} \leq C \sqrt{\frac{m_n \log p}{n}}. 
		\end{equation*}
		Therefore, for $  l \neq j $ and $l \in N_j^c$,  it follows that with probability $1 - O(p^{-3})$,
		\begin{equation*}
			n^{-1} \widetilde{\bz}_j^{\top} \widetilde{\bz}_l  \leq C \sqrt{\frac{m_n \log p}{n}}. 
		\end{equation*}
	}
	Then from \eqref{pf-le15-n14}, \eqref{pf-le15-n15}, and the definition of $\bV_{12}$,  we can obtain that with probability $1 - O(p^{-3})$, 
	\begin{equation} \label{eq-le15-1}
		\| \bV_{12} \|_{\max} \leq C \sqrt{\frac{m_n \log p}{n}}. 
	\end{equation}

	We have shown in \eqref{eq-le15-1} that $$ \| \bV_{12} \|_{\max} \leq C \sqrt{\frac{m_n \log p}{n}}$$ with probability $1 - O(p^{-3})$. Similarly, when $\e{e_j^2} \e e_{j+p}^2 - (\e[e_j e_{j+p}] )^2  > C $  for some constant $C > 0$, it can be shown that $ | V_{11} | \geq C $ and $|V_{22}| \geq C$ with probability $ 1 - O(p^{-3})$. Let us define an event 
	\begin{equation*}
		\begin{split}
			\mathcal{C} &= \bigg\{ \widetilde{\bX}^{\augg}: \| \bV_{12} \|_{\max} \leq C_1 \sqrt{\frac{m_n \log p}{n}}, \, |\bV_{22}| \geq C_2, \, |\bV_{11}| \geq C_2, \\
			&\qquad \|\bV_{11} \|_{\max} \leq C_3, \, \|\bV_{22} \|_{\max} \leq C_3\bigg\}.
		\end{split}
	\end{equation*}
	We have shown that $\mathbb{P} (\mathcal{C}) \geq 1- O(p^{-3})$. 
	Then it is straightforward to see that conditional on event $\mathcal{C}$, we have 
	\begin{equation} \label{pf-le15-n17}
		\frac{1 }  {2\pi | \bV_{22} - \bV_{21} \bV_{11}^{-1} \bV_{12} |^{1/2}} = \frac{1} {2 \pi | V_{22} |^{-1/2}} \Big(1 + O \big( \frac{m_n \log p}{n} \big) \Big)
	\end{equation}
	and
	\begin{equation} \label{pf-le15-n18}
		\| \bV_{22}^{-1} - (\bV_{22} - \bV_{21} \bV_{11}^{-1} \bV_{12})^{-1}  \|_{\max} \leq C \frac{m_n \log p}{n}.
	\end{equation}
	In addition, given event $\mathcal{C}$ and the range that $|x| \leq C \sqrt{\log p}$ and $|y| \leq C \sqrt {\log p}$, it holds that  
	\begin{equation} \label{pf-le15-n19}
		\left\|  \bV_{21} \bV_{11}^{-1} \left( \begin{matrix} 
			x\\ y
		\end{matrix} \right) \right\|_2 \leq C \sqrt{\frac{m_n}{n}} \log p.
	\end{equation}
	
	Further, given event $\mathcal{C}$ and that $ \max \{|u|, |w|, |x|, |y|\} \leq C \sqrt{\log p} $, it follows from \eqref{pf-le15-n17}--\eqref{pf-le15-n19} that as $ \frac{m_n (\log p)^3}{n} = o(1) $, 
	\begin{equation} \label{pf-le15-n20}
		\begin{split}
			& \Bigg|  \left[ \left( \begin{matrix} 
				u\\ w
			\end{matrix} \right) - \bV_{21} \bV_{11}^{-1} \left( \begin{matrix} 
				x\\ y
			\end{matrix} \right) \right]^{\top} (\bV_{22} - \bV_{21} \bV_{11}^{-1} \bV_{12})^{-1} \left[ \left( \begin{matrix} 
				u\\ w
			\end{matrix} \right) - \bV_{21} \bV_{11}^{-1} \left( \begin{matrix} 
				x\\ y
			\end{matrix} \right) \right] \\
			& \quad - \left( \begin{matrix} 
				u\\ w
			\end{matrix} \right)^{\top}  \bV_{22}^{-1} \left( \begin{matrix} 
				u\\ w
			\end{matrix} \right) \Bigg|  \leq C \sqrt{\frac{m_n (\log p)^3}{n}}. 
		\end{split}    
	\end{equation}
	Hence, substituting the bounds in \eqref{pf-le15-n17} and \eqref{pf-le15-n20} into \eqref{pf-le15-n21} yields that as $ \frac{m_n (\log p)^3}{n} = o(1) $, 
	\begin{equation} \label{pf-le15-n22}
		\begin{split}
			& f_{\xi_{l}, \xi_{l+p} |(\xi_{j}, \xi_{j+p}) } (u, w | x, y)\\
			& = \frac{1 }  {2\pi | \bV_{22} |^{1/2}} \exp  \Bigg\{ - \frac 1 2 \left( \begin{matrix} 
				u\\ w
			\end{matrix} \right)^{\top}  \bV_{22}^{-1} \left( \begin{matrix} 
				u\\ w
			\end{matrix} \right)  \Bigg\} \cdot  \bigg(1 + O \Big( \sqrt{\frac{m_n (\log p)^3} {n} } \Big)\bigg) \\
			& = f_{\xi_l, \xi_{l+p}} (u, w)  \bigg(1 + O \Big( \sqrt{\frac{m_n (\log p)^3} {n} } \Big)\bigg),
		\end{split}
	\end{equation}
	which entails that $(\xi_{l}, \xi_{l+p})$ is asymptotically independent of $(\xi_j, \xi_{j+p})$ for $l \neq j$ and $l \in N_j^c$. By plugging \eqref{pf-le15-n22} into \eqref{pf-le15-n16}, we can deduce that 
	\begin{equation} \label{pf-le15-8}
		\begin{split}
			P_1
			& \leq \mathbb{E} \Big\{\mathbbm{1} (\mathcal{C})    \mathbb{P} \big( |\xi_j| - |\xi_{j+p}| \geq t - \delta_n, \max\{ |\xi_j|, |\xi_{j+p}|\} \leq C \sqrt {\log p} \,|\, \widetilde{\bX}^{\augg}  \big) \\
			& \hspace{1cm} \times \mathbb{P} \big( |\xi_l| - |\xi_{l+p}| \geq t - \delta_n, \max\{ |\xi_l|, |\xi_{l+p}|\} \leq C \sqrt {\log p} \,|\, \widetilde{\bX}^{\augg}  \big)\Big\} \\
			& \quad \times \bigg(1 + O \Big( \sqrt{\frac{m_n (\log p)^3} {n} } \Big)\bigg)  +   \mathbb{P} (\mathcal{C}^c ),
		\end{split}
	\end{equation}
	where $  \mathbb{P} (\mathcal{C}^c ) = O(p^{-3})$. 
	
	We next show that given $\widetilde{\bX}^{\augg}$, $ |\xi_j| - |\xi_{j+p}| $ converges in distribution to $\eta_j$. 
	Given $\widetilde{\bX}^{\augg}$, we see that $$(\xi_j, \xi_{j+p}) \stackrel{d}{\sim} N({\bf 0}, \bV_{11}).$$ Without ambiguity, let us denote by $$\bV_{11} = \begin{pmatrix}
		\sigma_{1, n}^2 & \rho_n \sigma_{1, n} \sigma_{2, n}\\
		\rho_n \sigma_{1, n} \sigma_{2, n} & \sigma_{2, n}^2
	\end{pmatrix}$$ for simpler notation, where $\sigma_{1, n}^2 = n \tau_j^2$, $\sigma_{2, n}^2 = n \tau_{j + p}^2$, and $\rho_n = {\widetilde\bz_j^{\top} \widetilde\bz_{j+p}} / (\| \widetilde\bz_j\|_2 \| \widetilde{\bz}_{j+p}\|_2 ) $.
	We define an event 
	\begin{equation*}
		\begin{split}
			\mathcal{E} &= \bigg\{|\sigma_{1, n}^2 - (\e e_j^2)^{-1} | \leq C \sqrt{\frac{m_n \log p}{n}}, \, |\sigma_{2, n}^2 - (\e e_{j+p}^2)^{-1} | \leq C \sqrt{\frac{m_n \log p}{n}}, \\
			&\quad \mbox{ and } \big| \rho_n -  \corr(e_j, e_{j+p})  \big| \leq C \sqrt{\frac{m_n \log p}{n}} \bigg\}. 
		\end{split}
	\end{equation*}
	It follows from Lemma \ref{pf-thm5-lemma_01} that $\mathbb{P} (\mathcal{E}) \geq 1 - O(p^{-3}) $. Some straightforward calculations show that for $t > 0$, given $\widetilde{\bX}^{\augg}$  the density of $|\xi_j| - |\xi_{j+p}| $ can be written as 
	\begin{equation} \label{pf-le15-n23}
		\begin{split}
			f_{|\xi_j| - |\xi_{j+p}|} (t) & = \frac{ \sqrt{2} } {\sqrt {\pi }a_{1, n}} \big[1 - \Phi( a_{2, n}^{-1} t ) \big] \exp \big\{ - t^2 / (2 a_{1, n}^2) \big\} \\
			& \qquad + \frac{ \sqrt{2} } {\sqrt {\pi }a_{3, n}} \big[1 - \Phi( a_{4,n}^{-1} t ) \big] \exp \big\{ - t^2 / (2 a_{3, n}^2) \big\}, 
		\end{split}   
	\end{equation}
	where 
	\begin{align*}
		& a_{1, n} = \sqrt{\sigma_{1, n}^2 + \sigma_{2, n}^2 - 2 \rho_n \sigma_{1, n} \sigma_{2, n} }, \ \  a_{2, n} = \frac {  \sigma_{1, n} \sigma_{2, n} a_{1, n} \sqrt{ (1 - \rho_{n}^2)} } { \sigma_{2, n}^2 - \rho_n \sigma_{1, n} \sigma_{2, n} },\\
		& a_{3, n} = \sqrt{\sigma_{1, n}^2 + \sigma_{2, n}^2 + 2 \rho_n \sigma_{1, n} \sigma_{2, n}}, \ \  a_{4, n} = \frac{   \sigma_{1, n} \sigma_{2, n} a_{3, n} \sqrt{ (1 - \rho_n^2)} }{ \sigma_{2, n}^2 + \rho_n \sigma_{1, n} \sigma_{2, n} } .
	\end{align*} 
	
	Recall the notation  
	$$ v_j = \sqrt{ 2 (\e e_j^2)^{-1} (1  - \corr(e_j, e_{j+p})) }$$ 
	and 
	$$ b_j =\sqrt{ 2 (\e e_j^2)^{-1} (1  + \corr(e_j, e_{j+p})) }. $$ 
	It holds that $\e(e_j^2 ) = (\bOmega_{j, j}^A)^{-1} = (\bOmega^A_{j+p, j+p})^{-1} = \e (e_{j+p}^2)$ due to the symmetry of $\bOmega^A$. On event $\mathcal{E}$, we have that 
	\begin{align*}
		&  | a_{1, n} / v_j  - 1   | \leq C \sqrt{\frac{m_n \log p}{n}}, \ \   |  a_{2, n} / b_j - 1   | \leq C \sqrt{\frac{m_n \log p}{n}},  \\
		&  |a_{3, n} / b_j - 1 | \leq C \sqrt{\frac{m_n \log p}{n}}, \ \  |   a_{4, n} / v_j - 1  \Big | \leq C \sqrt{\frac{m_n \log p}{n}}.
	\end{align*}
	Thus, in view of the definition of $h_j(t) $ in \eqref{pf-le15-n3} and \eqref{pf-le15-n23}, it follows that as $|t| \leq C\sqrt{\log p}$,  
	\begin{equation*}
		\begin{split}
			f_{|\xi_j| - |\xi_{j+p}|} (t) & = h_j (t)  \bigg(1 + O \Big( \sqrt{\frac{m_n (\log p)^3} {n}}  \Big)\bigg).
		\end{split}   
	\end{equation*}
	With the aid of the above result, we can deduce that on event $\mathcal{E}$, 
	\begin{equation} \label{pf-le15-2}
		\begin{split}
			& \mathbb{P} \big( |\xi_j| - |\xi_{j+p}| \geq t - \delta_n, \max\{ |\xi_j|, |\xi_{j+p}|\} \leq C \sqrt {\log p} \,|\, \widetilde{\bX}^{\augg}  \big) \\
			& \leq \mathbb{P} \big( |\xi_j| - |\xi_{j+p}| \geq t - \delta_n,   
			|\xi_j| - |\xi_{j+p}|   \leq C \sqrt {\log p} \,|\, \widetilde{\bX}^{\augg}  \big) \\
			& \leq \bigg(\int_{ t - \delta_n }^{ C \sqrt{\log p}} h_j (u)\, du \bigg) \bigg(1 + O \Big( \sqrt{\frac{m_n (\log p)^3} {n}}  \Big)\bigg) \\
			& = \mathbb{P} (  t - \delta_n \leq   \eta_j \leq C \sqrt {\log p} ) \bigg(1 + O \Big( \sqrt{\frac{m_n (\log p)^3} {n}}  \Big)\bigg) \\
			& = \big[\mathbb{P} (  \eta_j \geq  t - \delta_n ) - \mathbb{P} (\eta_j > C \sqrt{\log p} ) \big] \bigg(1 + O \Big( \sqrt{\frac{m_n (\log p)^3} {n}}  \Big)\bigg).
		\end{split}  
	\end{equation}
	
	Moreover, in light of \eqref{pf-le15-15} it is easy to see that $$\mathbb{P} (\eta_j > C \sqrt{\log p} ) = O(p^{-3}) $$ for some large constant $C$, which together with \eqref{pf-le15-2} leads to 
	\begin{equation} \label{pf-le15-2}
		\begin{split}
			& \mathbb{P} \big( |\xi_j| - |\xi_{j+p}| \geq t - \delta_n, \max\{ |\xi_j|, |\xi_{j+p}|\} \leq C \sqrt {\log p} \,|\, \widetilde{\bX}^{\augg}  \big) \\
			& \leq  \mathbb{P} (  \eta_j \geq  t - \delta_n )   \bigg(1 + O \Big( \sqrt{\frac{m_n (\log p)^3} {n}}  \Big)\bigg) +  O (p^{-3}).
		\end{split}  
	\end{equation}
	Plugging \eqref{pf-le15-2} into \eqref{pf-le15-8} shows that 
	\begin{equation} \label{pf-le15-n25}
		P_1 \leq  \mathbb{P} (  \eta_j \geq  t - \delta_n )  \mathbb{P} (  \eta_l \geq  t - \delta_n )   \bigg(1 + O \Big( \sqrt{\frac{m_n (\log p)^3} {n}}  \Big)\bigg) + O(p^{-3}). 
	\end{equation}
	Finally, combining \eqref{pf-le15-4}, \eqref{pf-le15-7}, \eqref{pf-le15-n24}, and \eqref{pf-le15-n25} yields \eqref{pf-le15-10}. Similarly, we can also establish \eqref{pf-le15-11}--\eqref{pf-le15-13}. This completes the proof of Lemma \ref{pf-thm5-lemma-3}.

	\subsection{Proof of Lemma \ref{pf-thm5-lemma-4}} \label{new.SecB.14}
	
	Let us first prove \eqref{le16-1}. In the proof of Lemma \ref{pf-thm5-lemma-3} in Section \ref{new.SecB.13}, we have established the lower bound and upper bound for $\mathbb{P} (\widetilde{W}_j \geq t)$ in \eqref{pf-le15-12} and \eqref{pf-le15-13}, respectively. Recall the definitions that $ \delta_n = C \frac{ m_n^{1/2} s \log p} {\sqrt n} $ and $b_n =  C \Delta_n s \sqrt{\frac{\log p} {n}}$. For the numerator and denominator in \eqref{le16-1},  we can write that 
	\begin{equation} \label{pf-le16-n1}
		\begin{split}
			& p_0( G(t - b_n) - G(t + b_n) )   =   \sum_{j \in \mathcal{H}_0} \big[ \mathbb{P} (\widetilde{W}_j \geq t - b_n ) - \mathbb{P} (\widetilde{W}_j \geq t + b_n ) \big]  \\
			& \leq \sum_{j \in \mathcal{H}_0}  \mathbb{P} (   \eta_j \geq \sqrt n t - \sqrt n b_n -\delta_n ) \big(1 + O \big( \sqrt{\frac{m_n (\log p)^3} {n}}  \big)\big) \\
			& \quad - \sum_{j \in \mathcal{H}_0}  \mathbb{P} (   \eta_j \geq \sqrt n t + \sqrt n b_n + \delta_n ) \big(1 + O \big( \sqrt{\frac{m_n (\log p)^3} {n}}  \big)\big) + O(p^{-2})  \\
			& \leq  \sum_{j \in \mathcal{H}_0} \mathbb{P} ( \sqrt n t - \sqrt n b_n - \delta_n \leq  \eta_j \leq \sqrt n t + \sqrt n b_n + \delta_n  ) \\
			& \quad + \sum_{j \in \mathcal{H}_0} \mathbb{P} (   \eta_j \geq \sqrt n t - \sqrt n b_n -\delta_n )  \cdot O \big( \sqrt{\frac{m_n (\log p)^3} {n}}  \big)  + O(p^{-2})
		\end{split}    
	\end{equation}
	and 
	\begin{equation} \label{pf-le16-n2}
		p_0 G(t) \geq \sum_{j \in \mathcal{H}_0} \mathbb{P} (\eta_j \geq \sqrt n t  + \delta_n )  \big(1 + O \big( \sqrt{\frac{m_n (\log p)^3} {n}}  \big)\big)  + O(p^{-2}), 
	\end{equation}
	respectively.
	
	It follows from \eqref{pf-le16-n1}--\eqref{pf-le16-n2}, similar arguments as for \eqref{pf-le15-n9}, and $G^{-1} (\frac{c_1 q a_n} {p}) = O(\sqrt {\frac{\log p}{n}})$ in the proof of Lemma \ref{pf-thm5-lemma-3} that as $ \sqrt n G^{-1} (\frac{c_1 q a_n} {p}) (\sqrt n b_n+ \delta_n ) \to 0$, 
	\begin{equation*}
		\begin{split}
			\sup_{t \in (0,  G^{-1} (\frac{c_1 q a_n} {p})]}  \frac { G(t - b_n ) - G(t + b_n) } {  G(t) } &  \leq C \sqrt{\log p} (\sqrt n b_n  + \delta_n) + C \sqrt{\frac{m_n (\log p)^3} {n}} \\
			& \leq C \Big( \frac{m_n^{1/2} s (\log p)^{3/2}} {\sqrt n } + \Delta_n s \log p \Big).
		\end{split}
	\end{equation*}
	Thus, we see that when $    \frac{m_n^{1/2} s (\log p)^{3/2 + 1/\gamma}} {\sqrt n} + \Delta_n s (\log p)^{1 + 1 /\gamma} \to 0$, the desired result \eqref{le16-1} holds. 
	
	We next proceed with establishing \eqref{le16-2}. In view of Condition \ref{marginal_corr_condition4}, it holds that $$ p_1^{-1} \sum_{j \in \mathcal{H}_1}  \mathbb{P} ( \widetilde{W}_j <  - t ) \leq G(t)   $$ for $t = O(\sqrt{n^{-1}\log p})$. Moreover, we have $$b_n = C \Delta_n s \sqrt{\frac{\log p} {n}} = o(G^{-1}(\frac{c_1 q a_n }{p})) $$ due to the assumption $\Delta_n s \to 0$ and $G^{-1} (\frac{c_1 q a_n }{p}) = O (\sqrt{\frac{\log p} {n}} )$. 
	Then it follows that 
	\begin{equation} \label{pf-le16-1}
		\begin{split}
			& a_n ^{-1} \sum_{j \in \mathcal{H}_1}  \mathbb{P}  \Big( \widetilde{W}_j <  -  G^{-1} ( \frac { c_1 q a_n  } { p } ) + b_n \Big)\\
			& \leq  a_n^{-1 } (p - p_0) G\Big(   G^{-1} ( \frac { c_1 q a_n  } { p }) - b_n  \Big) \\
			& = \frac{ c_1 q  (p - p_0) } { p }  + a_n^{-1}(p - p_0)  \bigg[ G \Big( G^{-1} ( \frac { c_1 q a_n  } { p }) - b_n  \Big) - G \Big( G^{-1} ( \frac { c_1 q a_n  } { p }) \Big)  \bigg]. 
		\end{split}
	\end{equation} 
	For notational simplicity, let us define $$t_n = G^{-1} ( \frac { c_1 q a_n  } { p }).$$  With the aid of the upper and lower bounds for $\mathbb{P} (\widetilde{W}_j \geq t)$ given in \eqref{pf-le15-12} and \eqref{pf-le15-13}, we can deduce that
	\begin{equation} \label{pf-le16-n3}
		\begin{split}
			& G  ( t_n  - b_n   ) - G  ( t_n )  
			\\
			& \leq p_0^{-1} \sum_{j \in \mathcal{H}_0} \mathbb{P} ( \eta_j \geq \sqrt n t_n - \sqrt n b_n - \delta_n    ) \bigg(1 + O\Big( \sqrt{ \frac{m_n (\log p)^3 }{n} }\Big) \bigg) \\
			& \quad -  p_0^{-1}\sum_{j \in \mathcal{H}_0} 
			\mathbb{P} ( \eta_j \geq \sqrt n t_n  + \delta_n    ) \bigg(1 + O\Big(  \sqrt{ \frac{m_n (\log p)^3 }{n} } \Big) \bigg) + O(p^{-2}) \\
			& = p_0^{-1} \sum_{j \in \mathcal{H}_0} \mathbb{P} (\sqrt n t_n - \sqrt n b_n - \delta_n \leq  \eta_j \leq \sqrt n t_n + \delta_n  ) \\
			& \quad + p_0^{-1} \sum_{j \in \mathcal{H}_0}   \mathbb{P} ( \eta_j \geq \sqrt n t_n - \sqrt n b_n - \delta_n    ) \cdot O\Big(  \sqrt{ \frac{m_n (\log p)^3 }{n} }\Big) + O(p^{-2}).
		\end{split}
	\end{equation}
	
	An application of similar arguments as for  \eqref{pf-le10-n1} leads to 
	\begin{equation}  \label{pf-le16-n4}
		\begin{split}
			& \frac{\mathbb{P} (\sqrt n t_n - \sqrt n b_n - \delta_n \leq  \eta_j 
				\leq \sqrt n t_n + \delta_n  ) } { \mathbb{P} (\eta_j \geq \sqrt n t_n + \delta_n ) } \\
			& \leq C \sqrt {n} t_n ( \sqrt n b_n + \delta_n ) \\
			& \leq C  \Big( \frac{m_n^{1/2} s (\log p)^{3/2}} {\sqrt n } + \Delta_n s \log p \Big)
		\end{split}
	\end{equation}
	and 
	\begin{equation}  \label{pf-le16-n5}
		\bigg| \frac{ \mathbb{P} ( \eta_j \geq \sqrt n t_n - \sqrt n b_n - \delta_n    ) } {\mathbb{P} (\eta_j \geq \sqrt n t_n + \delta_n )  } - 1 \bigg| \leq C  \Big( \frac{m_n^{1/2} s (\log p)^{3/2}} {\sqrt n } + \Delta_n s \log p \Big).
	\end{equation}
	It follows from the lower bound in \eqref{pf-le15-12} and $G(t_n) = G(G^{-1}(\frac{c_1 q a_n} {p})) = \frac{c_1 q a_n} {p}$ that as $\frac{m_n (\log p)^3} {n} \to 0$, 
	\begin{equation}  \label{pf-le16-n6}
		p_0^{-1} \sum_{j \in \mathcal{H}_0} \mathbb{P} (\eta_j \geq \sqrt n t_n + \delta_n ) \leq C (\frac{c_1 q a_n} {p} + O(p^{-3}) ) \leq C \frac{c_1 q a_n} {p}.
	\end{equation}
	
	Therefore, combining \eqref{pf-le16-n3}--\eqref{pf-le16-n6} shows that 
	\begin{equation*}
		\begin{split}
			G  ( t_n  - b_n   ) - G  ( t_n ) & \leq C  \Big( \frac{m_n^{1/2} s (\log p)^{3/2}} {\sqrt n } + \Delta_n s \log p \Big) \cdot \frac{c_1 q a_n } {p} + C  \sqrt{ \frac{m_n (\log p)^3 }{n} } \cdot \frac{c_1 q a_n } {p} \\
			& \quad  + O(p^{-2})\\
			& \leq  C  \Big( \frac{m_n^{1/2} s (\log p)^{3/2}} {\sqrt n } + \Delta_n s \log p \Big) \cdot \frac{c_1 q a_n } {p}  + O(p^{-2}). 
		\end{split}
	\end{equation*}
	Finally, substituting the above bound into \eqref{pf-le16-1} yields that as  $    \frac{m_n^{1/2} s (\log p)^{3/2 }} {\sqrt n} + \Delta_n s (\log p) \to 0$, 
	\begin{equation*} 
		\begin{split}
			& a_n ^{-1} \sum_{j \in \mathcal{H}_1}  \mathbb{P}  \Big( \widetilde{W}_j <  -  G^{-1} ( \frac { c_1 q a_n  } { p } + \Delta_n \Big) \\
			& \leq  \frac{c_1 q (p - p_0)} {p} + C  \Big( \frac{m_n^{1/2} s (\log p)^{3/2}} {\sqrt n } + \Delta_n s \log p \Big) \cdot \frac{c_1 q (p - p_0)} {p} \\
			&\quad+ O(\frac{ p - p_0 } { a_n  p^{2}}) \\
			& \to 0,
		\end{split}
	\end{equation*} 
	where we have used the assumption that $p_0/p \to 1$. This establishes \eqref{le16-2}, which concludes the proof of Lemma \ref{pf-thm5-lemma-4}.

	\subsection{Proof of Lemma \ref{FWER-prop1}} \label{new.SecB.4}
	The proof of this lemma relies on the definitions of $  {T}_v $ and $\widetilde{T}_v$, with the intuition that $ \widetilde{T}_v  $ resembles the $v$th order statistic of $ - \widetilde{W}_j $, while $ {T}_v $ resembles the $v$th order statistic of $ -\hat{W}_j $. Intuitively, this means that if the distance between $ \widetilde{W}_j $ and $ \hat{W}_j $ is bounded by $ b_n $, the distance between the corresponding order statistics should also be bounded by $ b_n $. We will formalize such argument next.
	
	Let us define an event $$ \mathscr{C} := \{ \max_{1 \leq j \leq p} | \hat{W}_j 
	- \widetilde{W}_j|  \leq b_n\} .$$ Condition \ref{fdr-condition1} assumes that $\mathbb{P} (\mathscr{C}) \to 1$. Denote by 
	\begin{equation}
		\hat{S}_v  = \big\{ 1 \leq j \leq p : - \hat{W}_j \geq {T}_v \big\} 
	\end{equation}
	and 
	\begin{equation}
		\widetilde{S}_v  = \big\{ 1 \leq j \leq p: - \widetilde{W}_j \geq \widetilde{T}_v \big\} .
	\end{equation}
	Observe that $ | \hat{S}_v | = v $ and $ | \widetilde{S}_v | = v  $ by the definitions of $  {T}_v $ and $ \widetilde{T}_v $. If $ j_0 \in \hat{S}_v $, on event $\mathscr{C}$ we have that 
	\begin{equation}
		- \widetilde{W}_{j_0}  = - \hat{W}_{j_0} + ( \hat{W}_{j_0} -  \widetilde{W}_{j_0} ) \geq  {T}_v  - b_n,
	\end{equation}
	which entails that $  \sum_{j = 1}^p \mathbbm{1} ( - \widetilde{W}_j \geq  {T}_n - b_n ) \geq v $. Moreover, since $ \widetilde{T}_v  $ satisfies $ \sum_{j = 1}^p \mathbbm{1}( - \widetilde{W}_j \geq \widetilde{T}_v ) = v  $, it follows that $$ \widetilde{T}_v \geq {T}_v - b_n  $$ by the monotonicity of the indicator function. 
	Similarly, we can also show that $$  {T}_v  \geq \widetilde{T}_v  - b_n $$ on event $\mathscr{C}$. Thus, \eqref{eq1.new} is derived. This concludes the proof of Lemma \ref{FWER-prop1}.

	\subsection{Proof of Lemma \ref{FWER-prop2}} \label{new.SecB.5}
	Note that $ k $ is the number of failures before $v$ successes in a binomial process with success probability $\frac{1}{2}$. The major intuition of the desired result \eqref{eq3.new} is that by the law of large numbers,  the number of failures and successes should become asymptotically comparable as the number of trials tends to infinity. Let $ D_{k + v - 1 } $ be a binomial random variable with distribution $ B ( k + v - 1,  \frac 12 ) $ and $ L_v  $ the negative binomial random variable with distribution $ NB(v, \frac 12 ) $. Observe that \eqref{eq-4} is equivalent to $ \mathbb{P} ( L_v \geq k  ) \leq  q$. According to the relationship between the negative binomial distribution and binomial distribution, we have that 
	\begin{equation}
		\begin{split}
			\mathbb{P} ( L_v \geq k  ) & =  1 - \mathbb{P} ( L_v \leq k - 1  ) \\
			& = 1 -   \mathbb{P}  ( D_{k + v - 1 }  \geq v ) \\
			& =  \mathbb{P}  ( D_{k + v - 1 }  \leq v - 1 ).
		\end{split}
	\end{equation}
	
	By the central limit theorem, it holds that when $ k + v \to \infty $,
	\begin{equation*}
		\mathbb{P}  ( D_{k + v - 1 }  \leq v - 1 ) =  \Phi  \Big( \frac  {   v - 1 -   k   }    { \sqrt { k + v - 1 }  }  \Big) + o(1).
	\end{equation*}
	Therefore, \eqref{eq-4} implies that 
	\begin{equation}
		\frac  {    v - 1 -   k  }    { \sqrt { k + v - 1 }  }  \leq  \Phi ^{-1 } (q -  o(1) ). \label{eq5.new}
	\end{equation}
	In addition, since $ v $ is the largest integer such that \eqref{eq-4} holds, we have that $$ \mathbb{P} ( L_{v +1} \geq k) > q .$$ 
	Using similar arguments as for \eqref{eq5.new}, it follows that as $ k + v \to \infty $,
	\begin{equation*}
		\mathbb{P} ( L_{v + 1 } \geq k  ) = \mathbb{P} ( D_{k + v} \leq v  ) = \Phi \Big(  \frac { v - k  } { \sqrt{ k + v  } }  \Big) + o(1)
	\end{equation*}
	and hence
	\begin{equation}
		\frac { v - k  } { \sqrt {k + v } } \geq \Phi^{-1} ( q - o(1) ),
	\end{equation}
	which along with \eqref{eq5.new} leads to  \eqref{eq3.new}. This completes the proof of Lemma \ref{FWER-prop2}.

	\subsection{Proof of Lemma \ref{FWER-prop5}} \label{sec:proof-FWER-prop5}
	The proof of this lemma consists of two steps. We will first establish the tight bounds below for $\widetilde{T}_v$. In the second step, noting that $\widetilde{T}_{v + M_v + 1} <  \widetilde{T}_v - 2 b_n \leq \widetilde{T}_{v + M_v}$ by the definition of $M_v $ in \eqref{eq_Mv}, we will show that $M_v$ is bounded as long as $b_n$ is sufficiently small.
	\begin{lemma} \label{FWER-prop4}
		For $0< \veps < 1/8$,  under Conditions \ref{fdr-condition1}, \ref{fwer-condition1}, and \ref{kFWER-cond1} we have that 
		\begin{equation}
			\mathbb{P} \Big( G^{-1} \big( \frac {v(1+\veps)} {p_0} \big) <  \widetilde{T}_v < G^{-1} \big(\frac {v(1 - \veps)} {p_0} \big) \Big) \to 1.
		\end{equation}
	\end{lemma}
	
	\begin{lemma} \label{lemma-b_n}
		Under Condition \ref{kFWER-cond1}, we have that 
		\begin{equation*}
			2 b_n < G^{-1} \big( \frac {v(1 + \veps)} {p_0} \big) - G^{-1} \big( \frac{v (1 + 3 \veps) (1 - \veps) } {p_0} \big). 
		\end{equation*}
	\end{lemma}
	
	Using similar arguments as in the proof of Lemma \ref{FWER-prop4} below, we can show that under Conditions  \ref{fdr-condition1}, \ref{fwer-condition1}, and \ref{kFWER-cond1}, 
	\begin{equation}
		\mathbb{P} \Big( G^{-1} \big( \frac {v (1 + 3 \veps)(1 + \veps)} {p_0} \big) <  \widetilde{T}_{v (1 + 3 \veps)} < G^{-1} \big(\frac {(v (1 + 3 \veps))(1 - \veps)} {p_0} \big) \Big) \to 1.
	\end{equation}
	Then it follows that $$ \widetilde{T}_{v (1 + 3 \veps)} < G^{-1} (\frac{ v (1 + 3 \veps) (1 - \veps) } {p_0}) < G^{-1} ( \frac{v (1 + \veps)} {p_0} ) <  \widetilde{T}_v  .$$ Additionally, applying Lemmas \ref{FWER-prop4} and \ref{lemma-b_n} together with the definition of $\widetilde T_v$ gives that with asymptotic probability one, 
	\begin{equation*}
		\begin{split}
			\widetilde{T}_{v + M_v }  & \geq \widetilde{T}_v - 2 b_n \\ &\geq G^{-1} ( \frac{v(1 + \veps)} {p_0} ) - \Big[ G^{-1} ( \frac {v(1 + \veps)} {p_0} ) - G^{-1} ( \frac{v (1 + 3 \veps) (1 - \veps)} {p_0} ) \Big] \\
			& =  G^{-1} ( \frac{ v (1 + 3 \veps) (1 - \veps)} {p_0} ) > \widetilde{T}_{v (1 + 3 \veps)}.
		\end{split}
	\end{equation*} 
	Therefore, we can obtain that $$\mathbb{P} (M_v <  3 v \veps) \to 1$$ since $\widetilde{T}_v$ is decreasing with respect to $v$. This will conclude the proof of Lemma \ref{FWER-prop5}. 
	
	We will present the formal proofs of Lemmas \ref{FWER-prop4} and \ref{lemma-b_n} below.
	
	\bigskip
	\noindent\textit{Proof of Lemma \ref{FWER-prop4}}. 
	The main idea of the proof is to establish the convergence of the empirical distribution of $\{\widetilde{W}_j\}$ that $\sum_{j \in \mathcal{H}_0} \mathbbm{1}(\widetilde{W}_j \geq t) $ is close to $\sum_{j \in \mathcal{H}_0} \mathbb{P}(\widetilde{W}_j \geq t) $. Using similar arguments as in the proof of Lemma \ref{fdr-lemma2} in Section \ref{proof.lem2}, we can obtain that when $m_n/k \to 0$ (which combined with Lemma \ref{FWER-prop2} implies that $m_n / v \to 0$),
	\begin{equation}\label{pf-le17-1}
		\sup_{t \in (G^{-1}(\frac{3k} {2p}), G^{-1}(\frac{k} {2p}) ) }  \bigg| \frac{\sum_{j \in \mathcal{H}_0} \mathbbm{1}(\widetilde{W}_j \leq - t) } {\sum_{j \in \mathcal{H}_0} \mathbb{P}(\widetilde{W}_j \leq - t)} - 1 \bigg| = o_p(1).
	\end{equation}
	Since $\sum_{j \in \mathcal{H}_0} \mathbb{P}(\widetilde{W}_j \leq - G^{-1}(\frac{v (1 + \veps) } {p_0}) = v (1 + \veps) $, we see from \eqref{pf-le17-1} that 
	\begin{equation}\label{pf-le17-2}
		\begin{split}
			\sum_{j=1}^p \mathbbm{1}(-\widetilde{W}_j \geq  G^{-1}(\frac{v (1 + \veps) } {p_0}))  & \geq \sum_{j \in \mathcal{H}_0} \mathbbm{1}(\widetilde{W}_j \leq - G^{-1}(\frac{v (1 + \veps) } {p_0})) \\  
			&= v (1 + \veps) (1 + o_p(1)) > v
		\end{split}
	\end{equation} 
	holds with asymptotic probability one.  Hence, from the definition of $\widetilde T_v$, we have that 
	\begin{equation} \label{T_v_lowerbound}
		\mathbb{P} \Big(\widetilde{T}_v > G^{-1} ( \frac{v (1 + \veps)} {p_0} ) \Big) \to 1.
	\end{equation}
	
	We next prove the upper bound for $\widetilde T_v$. Note that $\sum_{j =1}^p \mathbbm{1} (\widetilde{W}_j \leq - \widetilde{T}_v) = v$. We will aim to show that with asymptotic probability one, 
	\begin{equation}\label{pf-le17-3}
		\sum_{j \in \mathcal{H}_1} \mathbbm{1} ( \widetilde{W}_j \leq - \widetilde{T}_v ) < v\veps / 2.
	\end{equation}  
	Then with asymptotic probability one, it holds that
	\begin{equation}
		\sum_{j \in \mathcal{H}_0} \mathbbm{1} (\widetilde{W}_j \leq - \widetilde{T}_v)  \geq v (1 - \veps/2). 
	\end{equation}
	On the other hand, applying \eqref{pf-le17-1} and similar argument as for \eqref{pf-le17-2}, we can obtain that with asymptotic probability one, 
	\begin{equation} \label{eq_upper_1}
		\sum_{j \in \mathcal{H}_0} \mathbbm{1}(\widetilde{W}_j \leq - G^{-1}(\frac{v (1 - \epsilon_n) } {p_0})  < v (1 - \veps/2) .
	\end{equation}
	Combining the above two results shows that with asymptotic probability one, $$\widetilde T_v\leq G^{-1}(\frac{v (1 - \veps) } {p_0}), $$ which completes the proof for the upper bound. 
	
	It remains to establish \eqref{pf-le17-3}. Since $ p_0/ p \to  1 $ and $v/k \to 1$ (cf. Lemma \ref{FWER-prop2}), we have that $$ G^{-1} (\frac {3 k } {2 p}) <  G^{-1} ( \frac{v (1 + \veps)} {p_0} ) $$ when $n$ and $p$ are sufficiently large and $0 < \veps < 1/8$. Then from \eqref{T_v_lowerbound}, it holds that $G^{-1} (\frac {3 k } {2 p}) \leq \widetilde T_v$ and hence with asymptotic probability one, 
	\begin{equation} \label{pf-le17-4}
		\sum_{j \in \mathcal{H}_1} \mathbbm{1} (\widetilde{W}_j \leq - \widetilde{T}_v) \leq \sum_{j \in \mathcal{H}_1} \mathbbm{1} (\widetilde{W}_j < - G^{-1} (\frac {3 k } {2 p})) .
	\end{equation}
	Moreover, an application of the Markov inequality, Lemma \ref{FWER-prop2}, and \eqref{KWER-cond1-2} in Condition \ref{kFWER-cond1} yields that as $n \to \infty$,
	\begin{equation}
		\begin{split}
			& \mathbb{P} \Big(\sum_{j \in \mathcal{H}_1} \mathbbm{1} (\widetilde{W}_j < - G^{-1} (\frac {3 k } {2 p})) > v \veps/2 \Big) \\
			& \leq \frac{2} { v \veps} \sum_{j \in \mathcal{H}_1} \mathbb{P} \Big(\widetilde{W}_j < - G^{-1} (\frac {3 k } {2 p}) \Big) \to 0.
		\end{split}
	\end{equation}
	Therefore, \eqref{pf-le17-3} is derived in view of \eqref{pf-le17-4}.
	This completes the proof of Lemma \ref{FWER-prop4}.
	
	\bigskip
	\noindent\textit{Proof of Lemma \ref{lemma-b_n}}. 
	Let us observe that 
	\begin{equation}
		\frac{v (1 + 3 \veps) (1 - \veps)} {p_0}  -  \frac {v(1 + \veps)} {p_0} = \frac{v} {p_0} ( \veps - 3 \veps^2).
	\end{equation}
	By the assumptions that $p_0/p\rightarrow 1$ and $m_n/k\rightarrow 0$, and applying Lemma \ref{FWER-prop2} and the observation above, it follows that when $k$ and $p$ are sufficiently large, 
	\begin{equation}
		\frac{v (1 + 3 \veps) (1 - \veps)} {p_0}  -  \frac {v(1 + \veps)} {p_0} \geq  \frac{ k  \veps  } { 2 p } .
	\end{equation}
	Note that assumption \eqref{kWER-cond1-1} in Condition \ref{kFWER-cond1} entails that
	\begin{equation}
		\sup_{t \in ( G^{-1} (\frac{3k} {2p}), G^{-1}(\frac{k} {2p})) } [ G(t - b_n ) - G(t + b_n) ] = o( \frac{ k } {p} ).
	\end{equation}
	Combining the above two results and Lemma \ref{FWER-prop2}, we can obtain that 
	\begin{equation}
		\frac{v (1 + 3 \veps) (1 - \veps)} {p_0}  -  \frac {v(1 + \veps)} {p_0}   \gg \sup_{t \in ( G^{-1} (\frac{3k} {2p}), G^{-1}(\frac{k} {2p})) } [ G(t - b_n ) - G(t + b_n) ].
	\end{equation}
	Notice that 
	$$G^{-1}(\frac{v (1 + 3 \veps) (1 - \veps)} {p_0} ) \in ( G^{-1} (\frac{3k} {2p}), G^{-1}(\frac{k} {2p})) $$ 
	and 
	$$ G^{-1} (\frac {v(1 + \veps)} {p_0} ) \in  ( G^{-1} (\frac{3k} {2p}), G^{-1}(\frac{k} {2p})) $$ when $k$ and $p$ are sufficiently large. Therefore, using proof by contradiction and the monotonicity of function $G(\cdot)$, we can establish the desired result of Lemma \ref{lemma-b_n}. This concludes the proof of Lemma \ref{lemma-b_n}.
	
\subsection{Lemma \ref{Lemma-corr-indep} and its proof} \label{secB.16-Lemma-indep-corr}
Following the definitions in Section \ref{new.Sec3.2},  let us consider the marginal correlation approximate knockoff statistics defined as  $\hat{W}_j = (\sqrt n \| \by \|_2)^{-1} (|\bX_j^T \by| - |\hat{\bX}_j^T \by|) $ and the coupled perfect knockoff statistics given by $\widetilde{W}_j =(\sqrt n \| \by \|_2)^{-1} (|\bX_j^T \by| - |\widetilde{\bX}_j^T \by|)  $ with $1 \leq j \leq p$. When features $X_1, \ldots, X_p$ are independent, we can obtain the following sharper bound of order $\Delta_n \sqrt{\frac{\log p} {n}}$ for the coupling accuracy of the knockoff statistics, compared to the general bound $\Delta_n$ in \eqref{marginal-w-closeness}.

\begin{lemma} \label{Lemma-corr-indep}
Assume that features $\{X_j\}_{j =1}^p$ are independent and follow a Gaussian distribution $X_j \stackrel{d}{\sim} N(0, \sigma_j^2)$. Let the approximate and coupled knockoff variable matrices be defined as 
\begin{equation*}
              \widehat{\bX}  =\bZ \, \diag( \hat{\sigma}_1, \ldots,  \hat{\sigma}_p) ~~ \mbox{and} ~~ \widetilde{\bX}  =  \bZ \, \diag(  {\sigma}_1, \ldots,   {\sigma}_p),
          \end{equation*}
where $\bZ = (\bZ_{ij}) \in \mathbb{R}^{n \times p}$ has i.i.d. standard normal entries and is independent of $(\bX, \by)$, and $\hat{\sigma}_j$ is the estimator of $\sigma_j$, which can be learned in sample. Then under Condition \ref{accuracy-knockoffs}, we have that when $\log p = o(n)$, 
    \begin{equation} \label{mar-corr-indep}
        \max_{1 \leq j \leq p} |\widehat{W}_j - \widetilde{W}_j| \leq 4 \Delta_n \sqrt{\frac{\log p} {n}}.
    \end{equation}
\end{lemma}

\medskip
	\noindent\textit{Proof of Lemma \ref{Lemma-corr-indep}}. 
         We first show that Condition \ref{accuracy-knockoffs} leads to $\mathbb{P} (\max_{1 \leq j \leq p }|\hat{\sigma}_j - \sigma_j | \leq 2 \Delta_n ) \to 1$. First note that $\mathbb{P} (\min_{1 \leq j \leq p} n^{-1/2} \| \bZ_j\|_2 > 1/2) \to 1$ since $\log p = o(n)$, due to the concentration inequality for the sum of i.i.d $\chi_1^2$ random variables. 
         If $\max_{1 \leq j \leq p} n^{-1/2}\| \hat{\bX}_j - \widetilde{\bX}_j  \|_2 \leq \Delta_n $ with probability approaching one, then we have $\Delta_n \geq  \max_{1 \leq j \leq p} n^{-1/2} |\hat\sigma_j - \sigma_j| \| \bZ_j\|_2 \geq \max_{1 \leq j \leq p} |\hat{\sigma}_j - \sigma_j| \min_{1 \leq j \leq p} n^{-1/2} \|\bZ_j \|_2 > \frac {1}{2} \max_{1 \leq j \leq p} |\hat{\sigma}_j - \sigma_j| $ with asymptotic probability one. This proves that Condition \ref{accuracy-knockoffs} leads to $$\mathbb{P} (\max_{1 \leq j \leq p }|\hat{\sigma}_j - \sigma_j | < 2 \Delta_n ) \to 1.$$
         
          Then we can deduce that 
          \begin{equation*}
              \begin{split}
                  &\mathbb{P} \Big( \max_{1 \leq j \leq p} |\widehat{W}_j - \widetilde{W}_j| \geq 4 \Delta_n \sqrt{\frac{\log p} {n}} \Big) \\
                  & \leq  \mathbb{P} \Big(  \max_{1 \leq j \leq p} (\sqrt n \| \by\|_2)^{-1} \big| (\widehat{\bX}_j^T - \widetilde{\bX}_j)^T \by \big| \geq 4 \Delta_n \sqrt{\frac{\log p} {n}} \Big) \\
                  & \leq \mathbb{P} \Big(  \max_{1 \leq j \leq p} |\hat{\sigma}_j - \sigma_{j}|  \max_{1 \leq j \leq p} (\sqrt n \| \by\|_2)^{-1} \big| \bZ_j^T  \by \big| \geq 4 \Delta_n \sqrt{\frac{\log p} {n}} \Big) \\
                  & \leq \mathbb{P} \Big( \max_{1 \leq j \leq p}   |\hat{\sigma}_j - \sigma_{j}| \geq 2 \Delta_n \Big)  +  \sum_{1 \leq j \leq p}  \mathbb{P} \Big( (\sqrt n \| \by\|_2)^{-1} \big| \bZ_j^T  \by \big| \geq 2 \sqrt{\frac{\log p} {n}} \Big).
              \end{split}
          \end{equation*}
          Observing that $(\sqrt n \| \by\|_2)^{-1} \bZ_j^T  \by  \stackrel{d}{\sim} N(0, n^{-1})$ and $  \max_{1 \leq j \leq p} |\hat{\sigma}_j - \sigma_{j}| \leq 2 \Delta_n $ with asymptotic probability one, we have
          \begin{equation*}
              \begin{split}
                  \mathbb{P} ( \max_{1 \leq j \leq p} |\widehat{W}_j - \widetilde{W}_j| \geq 4 \Delta_n \sqrt{\frac{\log p} {n}} )   \leq o(1) + p^{-1} \to 0.
              \end{split}
          \end{equation*}
          This completes the proof of Lemma \ref{Lemma-corr-indep}.

	\renewcommand{\thesubsection}{C.\arabic{subsection}}
	
	\section{RCD with debiased Lasso in GLM}  \label{Supp_Sec_C}
	
	In this section, we extend the results in Section \ref{new.Sec3.3} to the setting of the generalized linear model (GLM) 
	$$
	\e [Y |  X ]  =  g^{-1} ( X^T  \balpha^0) ,
	$$
	where $\balpha^0 = (\alpha_j^0)_{1 \leq j \leq p} \in \mathbb{R}^p$ is the true regression coefficient vector and $g$ is the link function. {Assume that feature vector $X = (X_1, \ldots, X_p)^{\top}$ has zero mean.} 
	Define $\widetilde{X}^{\augg} = (X^T, \widetilde{X}^T)^T \in \mathbb{R}^{2p}$ and $\widehat{X}^{\augg} = (X^T, \widehat{X}^T)^T \in \mathbb{R}^{2p}$, where $\widetilde{X}$ and $\widehat{X}$ are the perfect knockoffs and the approximate knockoffs for $X$, respectively. Denote by $ \bbeta^0  = ((\balpha^{0})^{\top},  {\bf 0}_p^{\top})^{\top} \in \mathbb{R}^{2p}$ the augmented true parameter vector. 
	
	Consider the negative log-likelihood function $\rho(y; a): a \mapsto \mathbb{R}$ defined as $\rho(y;a) = - y a + b(a)$, up to a constant independent of the unknown parameters, where $b(\cdot)$ is a known strictly convex and twice continuously differentiable function.
	Define the loss function $\rho_{\bbeta} (Y; \widetilde{X}^{\augg}) = \rho(Y; (\widetilde{X}^{\augg})^T \bbeta )   $. Denote by $\dot{\rho}_{\bbeta} := \frac{\partial} {\partial \bbeta} \rho_{\bbeta}$ and $\ddot{\rho}_{\bbeta} :=  \frac{\partial^2} {\partial \bbeta \partial \bbeta^T} \rho_{\bbeta}$ the partial derivatives. Note that $\dot{\rho}_{\bbeta} = \dot{\rho}(Y; (\widetilde{X}^{\augg})^T \bbeta ) \widetilde{X}^{\augg} $ and $\dot{\rho}_{\bbeta} = \ddot{\rho}(Y; (\widetilde{X}^{\augg})^T \bbeta ) \widetilde{X}^{\augg} (\widetilde{X}^{\augg})^{T}$.

	Let $\widehat{\bb} = (\hat{b}_j)_{1 \leq j \leq 2p} $ be the debiased estimator for the GLM given in \cite{VanGLM2014} based on the augmented design matrix $\hat{\bX}^{\augg}:= [\bX, \hat{\bX}] \in \mathbb{R}^{n \times 2p}$, where $\hat{\bX}$ is the approximate knockoff variable matrix. Assume that Condition \ref{accuracy-knockoffs} is satisfied and $\widetilde \bX$ is the coupled perfect knockoffs variable matrix.  
	Similarly, define $\widetilde{\bX}^{\augg} := [\bX, \widetilde{\bX}] \in \mathbb{R}^{n \times 2p}$. Then $\hat{\bb}$ can be coupled with the debiased Lasso estimator denoted as $\widetilde\bb =(\widetilde{b}_j)_{1 \leq j \leq 2p}\in \mathbb{R}^{2p}$ based on $\widetilde{\bX}^{\augg}$. The regression coefficient difference knockoff statistics can be defined as 
	\begin{equation}
		\hat{W}_j = |\hat{b}_j| - |\hat{b}_{j+p}| \ \text{ and } \ \widetilde{W}_j = |\widetilde{b}_j| - |\widetilde{b}_{j+p}|, \quad 1 \leq j \leq p 
	\end{equation}
	for the approximate and the coupled perfect knockoffs procedures, respectively..  
	
	We provide the explicit definition of the debiased Lasso estimator to assist future presentation. For each $1\leq j \leq 2p$, the debiased Lasso estimator $\widehat{\bb} = (\hat{b}_j)_{1 \leq j \leq 2p} $ is a one-step bias correction from the Lasso estimator $\hat\bbeta =(\hat{\beta}_j)_{1 \leq j \leq 2p}\in \mathbb{R}^{2p}$. First, the Lasso estimator is given by 
	\begin{equation} \label{GLM-lasso-hat}
		\widehat{\bbeta} = \argminA_{\bbeta \in \mathbb{R}^{2p}} \Big\{  n^{-1} \sum_{i = 1}^n \rho_{\bbeta} (y_i; \widehat{\bX}^{\augg}_{i, \cdot})   + \lambda \|\bbeta \|_1 \Big\},
	\end{equation}
	where  $\widehat{\bX}^{\augg}_{i, \cdot }$ is the $i$th row (observation) of the augmented design matrix $ \widehat{\bX}^{\augg} $. To obtain the debiased Lasso estimator, define
	\begin{equation*}
		\widehat{\bSigma} = n^{-1} \sum_{i = 1}^n \ddot{\rho}_{\hat{\bbeta}}(y_i; \widehat{\bX}^{\augg}_{i, \cdot} )  
		\ignore{= n^{-1} \sum_{i = 1}^n \ddot{\rho} (y_i; \widehat{\bX}^{\augg}_{i, \cdot}  \widehat{\bbeta} ) (\widehat{\bX}_{i, \cdot}^{\augg} )^T \widehat{\bX}_{i, \cdot}^{\augg} }
		= n^{-1} (\hat{\bX}^{\augg})^T \hat\bD \hat{\bX}^{\augg},
	\end{equation*}
	where $\hat\bD =\diag(\ddot{\rho} (y_1; \widehat{\bX}^{\augg}_{1, \cdot}  \widehat{\bbeta} ), \ldots, \ddot{\rho} (y_n; \widehat{\bX}^{\augg}_{n, \cdot}  \widehat{\bbeta} )) \in \mathbb{R}^{n \times n} $ is a diagonal matrix. Further, for $ 1 \leq j \leq 2p$, define
	\begin{equation*}
		\hat{\bgamma}_j =   \argminA_{\bgamma \in \mathbb{R}^{2p-1} } (\widehat{\bSigma}_{j, j} - 2 \widehat{\bSigma}_{j, - j} \bgamma + \bgamma^T \widehat{\bSigma}_{-j, -j } \bgamma + 2 \lambda_j \| \bgamma\|_1 ), 
	\end{equation*}
	where $\lambda$ and $\{ \lambda_j\}_{j = 1}^{2p}$ are the nonnegative regularization parameters. In addition, let 
	\begin{equation*}
		\widehat{\tau}_j^2 = \widehat{\bSigma}_{j, j} - \widehat{\bSigma}_{j, - j} \widehat{\bgamma}_{j}. 
	\end{equation*}
	Then the debiased Lasso estimator for GLM (\cite{VanGLM2014}) based on the approximate augmented design matrix $\widehat{\bX}^{\augg}$ is defined as 
	\begin{equation} \label{eq-debiased-lasso-approx-GLM}
		\hat{b}_j = \hat{\beta}_j - \frac{ n^{-1}  \dot{\brho}_{\hat{\bbeta}}^T (\widehat{\bX}^{\augg}_{j} -   \widehat{\bX}^{\augg}_{-j } \hat{\bgamma}_j )   } {\hat{\tau}_j^2 } ,   \quad 1 \leq j \leq p,
	\end{equation}
	where $ \dot{\brho}_{\hat{\bbeta}} := (\dot{\rho} (y_1; \widehat{\bX}^{\augg}_{1, \cdot }  \hat{\bbeta} ), \ldots, \dot{\rho} (y_n; \widehat{\bX}^{\augg}_{n, \cdot } \hat{\bbeta} )) \in \mathbb{R}^n$. 
	
	Analogously, the coupled debiased Lasso estimator $\widetilde\bbeta = (\widetilde{\beta}_j)_{1 \leq j \leq 2p}$ based on the perfect augmented design matrix $\widetilde{\bX}^{\augg}$ can be defined componentwisely as
	\begin{equation} \label{eq-debiased-lasso-GLM}
		\widetilde{b}_j = \widetilde{\beta}_j  - \frac{ n^{-1}  \dot{\brho}_{\widetilde{\bbeta}}^T ( \widetilde{\bX}^{\augg}_{j} -  \widetilde{\bX}^{\augg}_{-j } \widetilde{\bgamma}_j )   } {\widetilde{\tau}_j^2 } ,  
	\end{equation}
	where  $ \dot{\brho}_{\widetilde{\bbeta}} := (\dot{\rho} (y_1; \widetilde{\bX}^{\augg}_{1, \cdot }  \widetilde{\bbeta} ), \ldots, \dot{\rho} (y_n; \widetilde{\bX}^{\augg}_{n, \cdot } \widetilde{\bbeta} )) \in \mathbb{R}^n$,
	\begin{equation} \label{GLM-lasso-tilde}
		\widetilde{\bbeta} = \argminA_{\bbeta \in \mathbb{R}^{2p}} \Big\{  n^{-1} \sum_{i = 1}^n \rho_{\bbeta} (y_i; \widetilde{\bX}^{\augg}_{i, \cdot})   + \lambda \|\bbeta \|_1 \Big\},
	\end{equation}
	\begin{equation*}
		\widetilde{\bgamma}_j =   \argminA_{\bgamma \in \mathbb{R}^{2p-1} } (\widetilde{\bSigma}_{j, j} - 2 \widetilde{\bSigma}_{j, - j} \bgamma + \bgamma^T \widetilde{\bSigma}_{-j, -j } \bgamma + 2 \lambda_j \| \bgamma\|_1 ), \quad   \widetilde{\tau}_j^2 = \widetilde{\bSigma}_{j, j} - \widetilde{\bSigma}_{j, - i} \widetilde{\bgamma}_{j}, 
	\end{equation*}
	and 
	\begin{equation*}
		\widetilde{\bSigma} = n^{-1} \sum_{i = 1}^n \ddot{\rho}_{\widetilde{\bbeta}}(y_i; \widetilde{\bX}^{\augg}_{i, \cdot} ) =  n^{-1} (\widetilde{\bX}^{\augg})^T \widetilde\bD \widetilde{\bX}^{\augg}.
	\end{equation*}
	In the above, $\widetilde\bD =\diag(\ddot{\rho} (y_1; \widetilde{\bX}^{\augg}_{1, \cdot}  \widetilde{\bbeta} ), \ldots, \ddot{\rho} (y_n; \widetilde{\bX}^{\augg}_{n, \cdot}  \widetilde{\bbeta} )) \in \mathbb{R}^{n \times n}$ is a diagonal matrix.

	It is important to emphasize that the \textit{same} regularization parameters $\lambda$ and $\lambda_j$'s in defining $\widehat \bb$ should be used as in defining $\widetilde\bb $ in \eqref{eq-debiased-lasso-GLM} so that their constructions differ only by the used design matrix; this plays a key role in applying our coupling technique.

	Indeed, we prove in Lemma \ref{pf-GLM-lemma-1} that the coupling technique together with Condition \ref{accuracy-knockoffs} and some other regularity conditions ensures that with asymptotic probability one,
	\begin{equation}\label{eq: LCD-paring-GLM}
		\max_{1\leq j\leq 2p}|\widetilde{b}_j-\widehat{b}_j|\lesssim  \Delta_n s \sqrt{ \frac{\log p}{n} } +  \frac{ s^{3/2} \log p} {n} .   
	\end{equation}
	The above result guarantees that $\hat W_j$'s and $\widetilde W_j$'s are also uniformly close over $1\leq j\leq p$ with  $\max_{1\leq j\leq p}|\hat W_j - \widetilde W_j|\lesssim \Delta_ns\sqrt{(\log p)/n} + s^{3/2} (\log p) / n$. As long as $s\Delta_n\rightarrow 0$ and $s^{3/2} \sqrt{(\log p)/ n} \to 0$, this upper bound has a smaller order than the concentration rate $\delta_n$ of $\widetilde W_j$ (cf.  Condition \ref{fdr-condition2}), because here $\delta_n \sim \sqrt{n^{-1}\log p}$ as shown in our Lemma \ref{pf-GLM-lemma-2}. As commented after Theorem \ref{thm-marginal-corr-fdr}, the assumption that the coupling rate of $\max_{1\leq j\leq p}|\widetilde W_j-\hat W_j|$ is of a smaller order than the concentration rate $\delta_n$ plays a key role in establishing our theory on the asymptotic FDR control.
	
	We next introduce some additional notation and formally present the regularity conditions specific to this section. Let $\bD = \diag(  \ddot{\rho}(y_1; \widetilde\bX_{1, \cdot}^{\augg} \bbeta^0), \ldots,  \ddot{\rho}(y_n; \widetilde\bX_{n, \cdot}^{\augg} \bbeta^0)) \in \mathbb{R}^{n \times n}$ be a diagonal matrix  and $  {\bU}  = \bD^{1/2 } \widetilde{\bX}^{\augg} $ the weighted perfect design matrix. We define $\bSigma  = n^{-1} \e[  {\bU} ^T  {\bU}  ] = n^{-1} \e [(\widetilde{\bX}^{\augg})^T \bD \widetilde{\bX}^{\augg} ]$. 
	Let $\bOmega  = \bSigma^{-1}$ and $ \bgamma_{j}   = (\gamma_{j, l} )_{l \neq j}$ with $\gamma_{j, l}  = -  {\bOmega_{ j, l}} /{\bOmega_{j, j}}$. 
	\ignore{It has been shown in  \cite{peng2009partial} that  the residuals
		\begin{equation*}
			e_j  = \widetilde \bU^{\bbeta^0}_j  - \widetilde{\bU}^{\bbeta^0}_{-j} \bgamma_j^{\bbeta^0}
		\end{equation*}
		satisfy that $\Cov ( e_{j}, \widetilde{\bU}_{-j}^{\bbeta^0} )  = {\textbf 0}$, $\Var ( e_{j} ) = 1/ \bOmega^{\bbeta^0}_{j, j} $, and  $  \Cov(e_{j},  e_{l} ) = \frac {\bOmega^{\bbeta^0}_{j, l}} {\bOmega^{\bbeta^0}_{j, j} \bOmega^{\bbeta^0}_{l, l} } $.
	}
	For $1 \leq j \leq 2p$, denote by $ \mathcal{S}_j  = \supp(\bgamma_j ) \cup \supp(\widetilde{\bgamma}_j) \cup \supp(\hat{\bgamma}_j)$. 
	Let $J = \supp(\bbeta^{0}) \cup \supp(\widetilde{\bbeta}) \cup \supp(\hat{\bbeta})$ and $s := \| \bbeta^{0} \|_0 = \| \balpha^0 \|_0 = o(n)$.
	We make the technical assumptions below. 
	
	\begin{condition} \label{cond-GLM-00}
		For a large constant $r >0$, it holds with probability $1 - O(p^{-r})$ that  
		\begin{align}
			\| \widetilde{\bbeta}  - \bbeta^{0} \|_1 \leq C s \sqrt{ \frac{ \log p} {n} }, \label{lasso-1} \\
			\| \widetilde{\bbeta }  - \bbeta^{0} \|_2 \leq C \sqrt{ \frac{ s\log p} {n} },  \\
			\| \widetilde{\bX}^{\augg} (\widetilde{\bbeta}  - \bbeta^{0} ) \|_2 \leq C \sqrt { s\log p} \label{lasso-x-error}.
		\end{align}
	\end{condition}
	
	\begin{condition} \label{cond-GLM-01}
		For a large constant $r >0$, it holds with probability $1 - O(p^{-r})$ that 
		\begin{align}
			\max_{1 \leq  j \leq 2p } \| \widetilde{\bgamma}_j - \bgamma_j  \|_1 \leq C (s + m_n) \sqrt{ \frac{ \log p} {n} } ,   \\
			\max_{1 \leq  j \leq 2p } \|  \widetilde{\bgamma}_j - \bgamma_j   \|_2 \leq  C \sqrt{ \frac{ (s + m_n) \log p} {n} },  \\
			\max_{1 \leq j \leq 2p }  \| \widetilde{\bX}_{-j}^{\augg} (\widetilde{\bgamma}_j - \bgamma_j ) \|_2 \leq C \sqrt { (s + m_n) \log p}, \label{coefficients-x-error} \\
			\max_{1 \leq j \leq 2p} |  \widetilde\tau_j^2 -  \bOmega_{j, j}^{-1} | \leq C \sqrt{\frac{(s + m_n) \log p}{n}},  \label{bdd-tau}
		\end{align}
		\begin{equation}
			\begin{split}
				&\max_{1 \leq j,l \leq 2p}\Big|n^{-1}(\widetilde{\bX}_j^{\augg} - \widetilde{\bX}_{-j}^{\augg} \widetilde{\bgamma}_j)^T  \bD (\widetilde{\bX}_l^{\augg} - \widetilde{\bX}_{-l}^{\augg} \widetilde{\bgamma}_l)  - \frac{\bOmega_{j, l}} {\bOmega_{j, j} \bOmega_{l, l} } \Big| \\
				& \leq C \sqrt{\frac{ (s + m_n) \log p} {n}}, \label{bdd-zz}
			\end{split}
		\end{equation}
		where $m_n $ is the sparsity level of $\bOmega$ defined in Condition \ref{GLM-cond-spar}.
	\end{condition}
	
	Conditions \ref{cond-GLM-00} and \ref{cond-GLM-01} are well-known results about the the consistency of the GLM Lasso estimator and hold under some regularity conditions (\cite{VanGLM2014}).

	\begin{condition}[Loss function] \label{GLM-cond-Loss}
		The derivatives $\dot\rho(y; a) := \frac{\partial}{\partial a} \rho(y; a) $ and $\ddot\rho(y; a) := \frac{\partial^2}{\partial a^2} \rho(y; a)$ exist for all $(y, a)$, and for some $\delta$-neighborhood with $\delta > 0 $, $ \ddot{\rho} (y; a)$ is Lipschitz such that 
		\begin{equation*}
			\max_{a_0 \in \{X^T \boldsymbol{\alpha}^0\} } \sup_{|a - a_0 | \lor |a' - a_0| \leq \delta} \sup_{y \in \mathcal{Y}} \frac{|\ddot{\rho} (y; a) - \ddot{\rho} (y; a')|}{ |a - a'| } \leq C_4,
		\end{equation*}
		where $\mathcal{Y}$ is the space in which the response variable $Y$ lives. 
		In addition, the derivatives are bounded such that for constants $K_1, K_2 > 0 $, 
		\begin{equation*}
			\max_{a_0 \in \{X^T \boldsymbol\alpha^0\} } \sup_{y \in \mathcal{Y}}  |\dot\rho(y; a_{0}) | \leq K_1, \quad \max_{a_0 \in \{X^T \boldsymbol\alpha^0\} } \sup_{y \in \mathcal{Y}}  |\ddot\rho(y; a) | \leq K_2, \quad \min_{a_0 \in \{X^T \boldsymbol\alpha^0\} } \min_{y \in \mathcal{Y}}  |\ddot\rho(y; a) | \geq K_3.
		\end{equation*} 
		
	\end{condition}
	
	\begin{condition}[Sparsity] \label{GLM-cond-spar}
		(i) For some constant $C_5 > 0$, $\mathbb{P} (|J| \leq C_5 s) \to 1 $. \\
		(ii) For some sequence $ m_n \lesssim s$, it holds that $ \max_{1 \leq j \leq 2 p} \| \bOmega_j \|_0 \leq m_n $ and $\mathbb{P} ( \max_{1 \leq j \leq 2p } |\mathcal{S}_j| \leq C_6 m_n ) \to 1 $ with some constant $C_6>0$. \\
		(iii) $\max_{1 \leq j \leq 2p} \|\bgamma_j  \|_2 \leq C_7$ and $C_8 < \lambda_{\min} (\bOmega  ) \leq \lambda_{\max} (\bOmega ) < C_9$ with some positive constants $C_7$, $C_8$, and $C_9$.
	\end{condition}
	
	\begin{condition}[Compatibility] \label{GLM-cond-comp}
		Assume that with probability $1 - o(1)$,
		\begin{equation}
			\min_{\| \bbeta \|_0 \leq C_9 s} \frac { \bbeta^T  {\bU} ^T  {\bU} \bbeta } {n \|\bbeta\|_2^2} \geq \kappa_1
		\end{equation}
		for some large enough constant $C_9>0$ and a small constant $\kappa_1 > 0$. 
	\end{condition}
	
	\begin{condition}[boundedness] \label{GLM-cond-subG}
		Assume that $\|\widetilde{\bX} \|_{\infty} = \max_{i, j} |\widetilde\bX_{i, j}| \leq M$ for a constant $M > 0$. In addition, $\| \widetilde{\bX}^{\augg}_{-j} \bgamma_j \|_{\infty} \leq M $. 
	\end{condition}
	Note that the boundedness assumption in Condition \ref{GLM-cond-subG} is for technical simplicity; it can be replaced with a less stringent sub-Gaussian condition and the results in Theorem \ref{thm-debiased-lasso-FDR-GLM} remain to hold. 
	
	\begin{condition}[Signal strength] \label{GLM-cond-Signal}
		Let $\mathscr{A}_n = \{j \in \mathcal{H}_1: |\beta_j^0| \gg  \sqrt{n^{-1} \log p}  \}$ and it holds that $a_n :=  |\mathscr{A}_n| \to \infty$. 
	\end{condition}
	
	We are now ready to state our results on the FDR control for the approximate knockoffs inference based on the debiased Lasso coefficients for GLM.
	
	\begin{theorem} \label{thm-debiased-lasso-FDR-GLM}
		Assume that Conditions \ref{accuracy-knockoffs}, \ref{marginal_corr_condition4}, and \ref{cond-GLM-00}--\ref{GLM-cond-Signal} hold, $m_n / a_n \to 0 $, and $    \frac{s^{3/2} (\log p)^{3/2 + 1/\gamma}} {\sqrt n} + \Delta_n s (\log p)^{1 + 1 /\gamma} \to 0$ for some constant $0 < \gamma < 1$. Then we have
		\begin{equation*}
			\limsup_{n \to \infty} \FDR \leq  q. 
		\end{equation*}
	\end{theorem}
	
	\subsection{Proof of Theorem \ref{thm-debiased-lasso-FDR-GLM}}
	The main idea of the proof is to directly apply Theorem \ref{theorem-FDR} by verifying Conditions \ref{fdr-condition1}--\ref{fdr-condition5} for the knockoff statistics constructed from the debiased Lasso coefficients under the GLM. There are two key observations. The first one is that the Lasso estimators based on the approximate knockoffs and the perfect coupling counterpart should be close if the design matrices $\hat{\bX}^{\augg} $ and $\widetilde{\bX}^{\augg}$ are close to each other. 
	The second key observation is that the debiased Lasso coefficients are asymptotically normal (\cite{VanGLM2014}). Let $\dot\brho_{\bbeta^0} := (\dot{\rho} (y_1; \widehat{\bX}^{\augg}_{1, \cdot }   {\bbeta^0} ), \ldots, \dot{\rho} (y_n; \widehat{\bX}^{\augg}_{n, \cdot }  {\bbeta}^0 )) = (\dot{\rho} (y_1; \widetilde{\bX}^{\augg}_{1, \cdot }   {\bbeta^0} ), \ldots, \dot{\rho} (y_n; \widetilde{\bX}^{\augg}_{n, \cdot }  {\bbeta}^0 ))\in \mathbb{R}^n$.
	It follows from the Taylor expansion that $\dot\rho(y_i; \widetilde{\bX}^{\augg}_{i, \cdot} \bbeta^0) - \dot\rho(y_i; \widetilde{\bX}^{\augg}_{i, \cdot} \widetilde\bbeta ) =  \ddot{\rho} (y_i; \xi) \widetilde{\bX}_{i, \cdot}^{\augg}(\bbeta^0 - \widetilde{\bbeta}  ) $ for some $\xi$ locating between $\widetilde{\bX}^{\augg}_{i, \cdot} \bbeta^0$ and $\widetilde{\bX}^{\augg}_{i, \cdot} \widetilde\bbeta $. By Condition \ref{GLM-cond-Loss}, we can obtain that
	\begin{equation}
		\big| \dot\rho(y_i; \widetilde{\bX}^{\augg}_{i, \cdot} \bbeta^0) - \dot\rho(y_i; \widetilde{\bX}^{\augg}_{i, \cdot} \widetilde\bbeta ) -  \ddot\rho(y_i; \widetilde{\bX}^{\augg}_{i, \cdot} \widetilde\bbeta ) \widetilde{\bX}_{i, \cdot}^{\augg}(\bbeta^0 - \widetilde{\bbeta}  ) \big| \leq C_4 [\widetilde{\bX}_{i, \cdot}^{\augg}(\bbeta^0 - \widetilde{\bbeta}  ) ]^2.
	\end{equation}
	In view of \eqref{eq-debiased-lasso-approx-GLM}, the debiased Lasso coefficient can be written as 
	\begin{equation} \label{normal-decomp_GLM}
		\begin{split}
			\sqrt{n} (\widetilde{b}_j - \beta_j^{0} ) & =   \sqrt n (\widetilde{\beta}_j - \beta^0_j) - \frac{n^{-1/2} \dot\brho_{\widetilde{\bbeta}}^T (\widetilde{\bX}_j^{\augg} - \widetilde{\bX}_{-j}^{\augg} \widetilde{\bgamma}_j) } {\widetilde{\tau}_j^2} \\
			& = - \frac{n^{-1/2} \dot\brho_{ {\bbeta^0}}^T (\widetilde{\bX}_j^{\augg} - \widetilde{\bX}_{-j}^{\augg} \widetilde{\bgamma}_j) } {\widetilde{\tau}_j^2} + \sqrt n ({\widetilde{\beta}_j - \beta^0_j}) \\
			& \quad  + \frac{n^{-1/2} (\widetilde{\bD} \widetilde{\bX}^{\augg} (\bbeta^0 - \widetilde\bbeta))^T (\widetilde{\bX}_j^{\augg} - \widetilde{\bX}_{-j}^{\augg} \widetilde{\bgamma}_j) } {\widetilde{\tau}_j^2} + \frac{ C_4 n^{-1/2} \widetilde\bR | \widetilde{\bX}_j^{\augg} - \widetilde{\bX}_{-j}^{\augg} \widetilde{\bgamma}_j | } {\widetilde{\tau}_j^2}\\
			& = - \frac{n^{-1/2} \dot\brho_{ {\bbeta^0}}^T (\widetilde{\bX}_j^{\augg} - \widetilde{\bX}_{-j}^{\augg} \widetilde{\bgamma}_j) } {\widetilde{\tau}_j^2}  + \frac{ C_4 n^{-1/2} \widetilde\bR | \widetilde{\bX}_j^{\augg} - \widetilde{\bX}_{-j}^{\augg} \widetilde{\bgamma}_j | } {\widetilde{\tau}_j^2} \\
			& \quad + \frac{n^{-1/2} (\bbeta^0_{-j} - \widetilde\bbeta_{-j})^T (\widetilde{\bX}^{\augg}_{-j})^{T} \widetilde{\bD} (\widetilde{\bX}_j^{\augg} - \widetilde{\bX}_{-j}^{\augg} \widetilde{\bgamma}_j) } {\widetilde{\tau}_j^2}, 
		\end{split}
	\end{equation}
	where $\widetilde\bR = ([\widetilde{\bX}_{1, \cdot}^{\augg}(\bbeta^0 - \widetilde{\bbeta}  ) ]^2, \ldots, [\widetilde{\bX}_{n, \cdot}^{\augg}(\bbeta^0 - \widetilde{\bbeta}  ) ]^2)$, and we have used the equality $ \widetilde\tau_j^2 =  \widetilde{\bSigma}_{j, j} - \widetilde{\bSigma}_{j, - j} \widetilde{\bgamma}_{j} = (\widetilde{\bX}_j^{\augg} )^T \widetilde{\bD} (\widetilde{\bX}_j^{\augg} - \widetilde{\bX}_{-j}^{\augg} \widetilde{\bgamma}_j ) $.
	
	By the property of GLM, we have  $\e [\dot\rho(y_i; \widetilde{\bX}^{\augg}_{i, \cdot}\bbeta^0)  | \widetilde\bX^{\augg}] = 0$, and hence $\e [\dot\brho_{ {\bbeta^0}}^T (\widetilde{\bX}_j^{\augg} - \widetilde{\bX}_{-j}^{\augg} \widetilde{\bgamma}_j) ] = 0 $. In addition, it holds that 
 \begin{align*}
 \Var [n^{-1/2}\dot\brho_{ {\bbeta^0}}^T (\widetilde{\bX}_j^{\augg} - \widetilde{\bX}_{-j}^{\augg} \widetilde{\bgamma}_j) | \widetilde{\bX}^{\augg}] & = n^{-1}(\widetilde{\bX}_j^{\augg} - \widetilde{\bX}_{-j}^{\augg} \widetilde{\bgamma}_j)^T \bD (\widetilde{\bX}_j^{\augg} - \widetilde{\bX}_{-j}^{\augg} \widetilde{\bgamma}_j) \\
 & \approx \widetilde\tau_j^2.
 \end{align*}
	Thus, as the remainders in \eqref{normal-decomp_GLM} are asymptotically negligible, the debiased Lasso estimator is asymptotically normal in the sense that  
	\begin{equation}
		\sqrt{n} \widetilde\tau_j (\widetilde{b}_j - \beta_j^0) \stackrel{d}{\to} N(0, 1). 
	\end{equation}
	Our proof will build mainly on such intuition. Throughout the proof below, constant $C$ may take different values from line to line.

	The four lemmas below outline the proof for verifying the general Conditions \ref{fdr-condition1}--\ref{fdr-condition5}. Proofs of Lemma \ref{pf-GLM-lemma-1}--\ref{pf-GLM-lemma-4} are provided in Sections \ref{pf-of-lemma-glm-1}--\ref{pf-of-lemma-glm-4}, respectively.
	
	\begin{lemma} \label{pf-GLM-lemma-1} 
		Assume that Conditions \ref{accuracy-knockoffs} and \ref{cond-GLM-00}--\ref{GLM-cond-subG} are satisfied. Then as $\Delta_n s^{1/2}   \to 0$ and $  s \sqrt{\frac{ \log p} {n}}  \to 0$, we have that 
		\begin{equation}
			\mathbb{P} \bigg( \max_{1 \leq j \leq 2p} | \widetilde{b}_j - \hat{b}_j | \geq C  \Big( \Delta_n s \sqrt{ \frac{\log p}{n} } +  \frac{ s^{3/2} \log p} {n}\Big)   \bigg) \to 0.
		\end{equation}
	\end{lemma}
	Lemma \ref{pf-GLM-lemma-1} above indicates that Condition \ref{fdr-condition1} is satisfied with convergence rate $b_n := C ( \Delta_n s \sqrt{\frac{\log p} {n}} + \frac{ s^{3/2} \log p} {n} )$. 
	Let us define $w_j = |\beta_j^0| $.
	\begin{lemma} \label{pf-GLM-lemma-2}
		Assume that Conditions \ref{cond-GLM-00}--\ref{GLM-cond-subG} are satisfied. Then as $   s^{3/2} \sqrt{\frac{  \log p} {n}} \to 0 $, we have that for some $C > 0$, $ \sum_{j= 1}^p \mathbb{P} (|\widetilde{W}_j - w_j | \geq C \sqrt{n^{-1}\log p} ) \to 0$. 
	\end{lemma}

	Lemma \ref{pf-GLM-lemma-2} above shows that Condition \ref{fdr-condition2} related to the concentration rate of $\widetilde{W}_j$ is satisfied with $\delta_n = C \sqrt{n^{-1} \log p} $.  In addition, it holds that $ b_n \ll C \sqrt{n^{-1} \log p}  $ due to the assumptions $\Delta_n s \to 0 $ and $  s \sqrt{\frac{ \log p} {n}}  \to 0$ in Theorem \ref{thm-debiased-lasso-FDR-GLM}. In addition, in light of the definition of $w_j$, under Condition \ref{GLM-cond-Signal} we have that the general Condition \ref{fdr-condition3} on the signal strength is also satisfied. We next turn to the verification of Conditions \ref{fdr-condition4}--\ref{fdr-condition5}.

	\begin{lemma} \label{pf-GLM-lemma-3}
		Assume that Conditions \ref{cond-GLM-00}--\ref{GLM-cond-subG} are satisfied. Then as $  \frac{  s^{3/2} (\log p)^{3/2 + 1/\gamma } } { \sqrt n }  \to 0 $, we have that $ \Var{ \big( \sum_{j \in \mathcal{H}_0} \mathbbm{1} (\widetilde{W}_j > t)  } \big) \leq  V_1 (t) + V_2 (t) $, where for some $ 0 < \gamma < 1$ and $ 0< c_1 < 1 $,  
		\begin{equation} 
			(\log p )^{1/\gamma}  \sup_{t \in (0,\, G^{-1} ( \frac { c_1 q a_n  } { p } ) ] }  \frac{V_1 (t) } { [p_0  G (t)]^2  } \to 0 
		\end{equation}
		and 
		\begin{equation} 
			\sup_{t \in (0,\, G^{-1} ( \frac { c_1 q a_n  } { p } ) ] }   \frac{ V_2(t)   } { p_0  G (t) }  \lesssim  m_n.
		\end{equation} 
		
	\end{lemma}

	\begin{lemma} \label{pf-GLM-lemma-4}
		Assume that Conditions \ref{accuracy-knockoffs}, \ref{marginal_corr_condition4}, and \ref{cond-GLM-00}--\ref{GLM-cond-subG} are satisfied. Then when $    \frac{  s^{3/2} (\log p)^{3/2 + 1/\gamma}} {\sqrt n} \to 0 $ and $ \Delta_n s (\log p)^{1 + 1 /\gamma} \to 0$, we have that 
		\begin{equation}   \label{le16-1}
			(\log p)^{1/\gamma}  \sup_{t \in (0,\, G^{-1} ( \frac { c_1 q a_n  } { p } ) ] }  \frac { G(t - b_n ) - G(t + b_n) } {  G(t) } \to 0	 
		\end{equation}
		and 
		\begin{equation}  \label{le16-2}
			a_n^{-1} \sum_{j \in \mathcal{H}_1} \mathbb{P} \Big( \widetilde{W}_j < -  G^{-1} ( \frac { c_1 q a_n  } { p } ) + b_n \Big)  \to 0
		\end{equation}
		as $n \to \infty$.
	\end{lemma}
	Lemma \ref{pf-GLM-lemma-3} above shows that Condition \ref{fdr-condition4} is satisfied, whereas Lemma \ref{pf-GLM-lemma-4} implies that Condition \ref{fdr-condition5} is satisfied. Finally, the conclusion of Theorem \ref{thm-debiased-lasso-FDR-GLM} can be derived by directly applying the general Theorem \ref{theorem-FDR}. This completes the proof of Theorem \ref{thm-debiased-lasso-FDR-GLM}.

	\subsection{Proof of Lemma \ref{pf-GLM-lemma-1}} \label{pf-of-lemma-glm-1}
	The proof is analogous to that of Lemma \ref{pf-thm5-lemma-1}. The main idea is to apply the KKT condition to the GLM Lasso and then use Condition \ref{accuracy-knockoffs}. From the definitions of $\hat{b}_j$ in \eqref{eq-debiased-lasso-approx-GLM} and the coupled counterpart $\widetilde{b}_j$ in \eqref{eq-debiased-lasso-GLM}, we have that 
	\begin{equation}
		\begin{split}
			\max_{1 \leq j \leq 2p} |\hat{b}_j - \widetilde{b}_j | & \leq  \max_{1 \leq j \leq 2p} |\hat{\beta}_j - \widetilde{\beta}_j | \\
			& +  \max_{1 \leq j \leq 2p} \bigg| \frac{ n^{-1}  \dot{\brho}_{\hat{\bbeta}}^T (\widehat{\bX}^{\augg}_{j} -   \widehat{\bX}^{\augg}_{-j } \hat{\bgamma}_j )   } {\hat{\tau}_j^2 } -  \frac{ n^{-1} | \dot{\brho}_{\widetilde{\bbeta}}^T (\widetilde{\bX}^{\augg}_{j} -   \widetilde{\bX}^{\augg}_{-j } \widetilde{\bgamma}_j )   } {\widetilde{\tau}_j^2 } \bigg|. 
		\end{split}
	\end{equation}
	We will show that for some constant $C > 0$, it holds that 
	\begin{equation} \label{pf-accuracy-glm-goal1} 
		\mathbb{P} \bigg( \|\hat{\bbeta} - \widetilde{\bbeta} \|_2 \leq C \Big( \Delta_n s \sqrt{ \frac{\log p}{n} } +  \frac{ s^{3/2} \log p} {n} \Big) \bigg) \to 1, 
	\end{equation}
	\begin{equation} \label{pf-accuracy-glm-goal2} 
		\begin{split}
			\mathbb{P} \bigg( \max_{1 \leq j \leq 2p} \bigg|  \frac{ n^{-1}  \dot{\brho}_{\hat{\bbeta}}^T (\widehat{\bX}^{\augg}_{j} -   \widehat{\bX}^{\augg}_{-j } \hat{\bgamma}_j )   } {\hat{\tau}_j^2 } & -  \frac{ n^{-1} | \dot{\brho}_{\widetilde{\bbeta}}^T (\widetilde{\bX}^{\augg}_{j} -   \widetilde{\bX}^{\augg}_{-j } \widetilde{\bgamma}_j )   } {\widetilde{\tau}_j^2 } \bigg| \\
			& \qquad \leq C  \Big( \Delta_n s \sqrt{ \frac{\log p}{n} } +  \frac{ s^{3/2} \log p} {n}\Big)  \bigg) \to 1.
		\end{split}
	\end{equation}
	Then combining the two results above can establish the desired conclusion of Lemma \ref{pf-GLM-lemma-1}. We next proceed with proving \eqref{pf-accuracy-glm-goal1} and \eqref{pf-accuracy-glm-goal2}.
	
	\bigskip
	
	\noindent \textbf{Proof of \eqref{pf-accuracy-glm-goal1}.} Recall the definitions of Lasso estimators $\hat{\bbeta}$ in \eqref{GLM-lasso-hat} and $\widetilde{\bbeta}$ in \eqref{GLM-lasso-tilde}. It follows from the KKT condition that
	\begin{align}
		n^{-1} \sum_{i = 1}^n \dot{\rho}(y_i; \hat{\bX}^{\augg}_{i, \cdot} \hat\bbeta) (\hat{\bX}_{i, \cdot}^{\augg})^T + \lambda \hat{\bzeta} = 0, \label{KKT_GLM_1}\\
		n^{-1} \sum_{i = 1}^n \dot{\rho}(y_i; \widetilde{\bX}^{\augg}_{i, \cdot} \widetilde\bbeta) (\widetilde{\bX}_{i, \cdot}^{\augg})^T + \lambda \widetilde{\bzeta} = 0 \label{KKT_GLM_2}, 
	\end{align}
	where $\widetilde{\boldsymbol{\zeta}} = (\widetilde\zeta_1, \ldots, \widetilde\zeta_{2p})$ and $\hat{\boldsymbol{\zeta}} = (\hat\zeta_1, \ldots, \hat\zeta_{2p})$ with
	\begin{equation*}
		\widetilde\zeta_j = \left\{ \begin{array}{cc}
			\sgn(\widetilde{\beta}_j) & \mbox{ if} ~ \widetilde{\beta}_j  \neq 0,  \\
			\in [-1, 1] &  \mbox{ if} ~ \widetilde{\beta}_j  = 0,
		\end{array}
		\right.
		\quad \mbox{and} \quad 
		\hat\zeta_j = \left\{ \begin{array}{cc}
			\sgn(\hat{\beta}_j ) & \mbox{ if} ~ \hat{\beta}_j  \neq 0,  \\
			\in [-1, 1] &  \mbox{ if} ~ \hat{\beta}_j  = 0.
		\end{array}
		\right.
	\end{equation*}
	Taking the difference between \eqref{KKT_GLM_1} and \eqref{KKT_GLM_1} above  and multiplying both sides by $\hat{\bbeta} - \widetilde{\bbeta}$ lead to 
	\begin{equation*}
		\begin{split}
			& n^{-1} \sum_{i = 1}^n \dot{\rho}(y_i; \hat{\bX}^{\augg}_{i, \cdot} \hat\bbeta) (\hat{\bX}_{i, \cdot}^{\augg})(\hat{\bbeta} - \widetilde{\bbeta}) -n^{-1} \sum_{i = 1}^n \dot{\rho}(y_i; \widetilde{\bX}^{\augg}_{i, \cdot} \widetilde\bbeta) (\widetilde{\bX}_{i, \cdot}^{\augg}) (\hat{\bbeta} - \widetilde{\bbeta}) \\
			& = - \lambda (\widehat{\boldsymbol{\zeta}} - \widetilde{\boldsymbol{\zeta}})^{T} (\hat{\bbeta} - \widetilde{\bbeta}) \leq 0. 
		\end{split}
	\end{equation*}
	Further applying the Taylor expansion for function $\dot{\rho}$ and Condition \ref{GLM-cond-Loss} yields
	\begin{equation}
		\begin{split}
			&  n^{-1} \sum_{i=1}^n \big[ \dot{\rho}(y_i; \widehat{\bX}_{i, \cdot}^{\augg} \bbeta^0) + \ddot{\rho}(y_i; \widehat{\bX}_{i, \cdot}^{\augg} \bbeta^0) \hat{\bX}_{i, \cdot}^{\augg}(\hat{\bbeta} - \bbeta^0 ) \big] \hat{\bX}_{i, \cdot}^{\augg} (\hat{\bbeta} - \bbeta^0 ) \\
			& - n^{-1} \sum_{i=1}^n \big[ \dot{\rho}(y_i; \widetilde{\bX}_{i, \cdot}^{\augg} \bbeta^0) + \ddot{\rho}(y_i; \widetilde{\bX}_{i, \cdot}^{\augg} \bbeta^0) \widetilde{\bX}_{i, \cdot}^{\augg}(\widetilde{\bbeta} - \bbeta^0 ) \big] \widetilde{\bX}_{i, \cdot}^{\augg} (\widetilde{\bbeta} - \bbeta^0 )  \\
			& \leq C_4 n^{-1} \sum_{i = 1}^n \big| \widetilde{\bX}_{i, \cdot} ^{\augg} (\widetilde{\bbeta} - \bbeta^0) \big|^3, 
		\end{split}
	\end{equation}
	which can be equivalently written in the matrix form as 
	\begin{equation} \label{GLM-accuracy-decom}
		\begin{split}
			&n^{-1} (\hat{\bbeta} - \widetilde{\bbeta})^{T} (\hat{\bX}^{\augg} - \widetilde{\bX}^{\augg})^T \dot{\brho}_{\bbeta^0} +  n^{-1} (\hat{\bbeta} - \widetilde{\bbeta})^{T} (\widetilde{\bX}^{\augg})^T \bD \widetilde{\bX}^{\augg} (\hat{\bbeta} - \widetilde{\bbeta}) \\
			& + n^{-1} (\hat{\bbeta} - \widetilde{\bbeta})^{T} [(\widehat{\bX}^{\augg})^T \bD \widehat{\bX}^{\augg} 
			- (\widetilde{\bX}^{\augg})^T \bD \widetilde{\bX}^{\augg}] (\hat{\bbeta} - \widetilde{\bbeta}) \\
			& + n^{-1} (\hat{\bbeta} - \widetilde{\bbeta})^{T} [(\widehat{\bX}^{\augg})^T \bD \widehat{\bX}^{\augg} 
			- (\widetilde{\bX}^{\augg})^T \bD \widetilde{\bX}^{\augg}] (\widetilde{\bbeta} - {\bbeta^0})\\
			& \leq C_4 n^{-1} \sum_{i = 1}^n \big| \widetilde{\bX}_{i, \cdot} ^{\augg} (\widetilde{\bbeta} - \bbeta^0) \big|^3.
		\end{split}
	\end{equation}
	
 Note that by Condition \ref{GLM-cond-spar}, $ | \supp(\hat{\bbeta}) \cup \supp(\widetilde{\bbeta}) \cup\supp({\bbeta^0}) |  \leq C s$  with probability approaching one. Thus, by a similar technique of proving \eqref{eq-lemma13-3}, we can obtain from Conditions \ref{GLM-cond-spar} and \ref{GLM-cond-comp} that
	\begin{equation}
		\begin{split}
			\|\hat{\bbeta} - \widetilde{\bbeta}  \|_{2} & \lesssim    \max_{J: |J| \leq Cs}  \big\| n^{-1} (\hat{\bX}^{\augg}_{J} - \widetilde{\bX}^{\augg}_J)^T \dot{\brho}_{\bbeta^0} \big\|_2 \\
			& \quad  +  \max_{J: |J| \leq Cs} \big\|n^{-1}[(\widehat{\bX}^{\augg}_J)^T \bD \widehat{\bX}^{\augg} 
			- (\widetilde{\bX}^{\augg}_J)^T \bD \widetilde{\bX}^{\augg}] (\widetilde{\bbeta} - {\bbeta^0}) \big\|_2 \\
			& \quad + \max_{J: |J| \leq Cs} n^{-1} \sum_{i = 1}^n   \big(\widetilde{\bX}_{i, \cdot} ^{\augg} (\widetilde{\bbeta} - \bbeta^0) \big)^2 \big\| \widetilde{\bX}_{i, J} \big\|_2
			:= R_1 + R_2 + R_3. 
		\end{split}
	\end{equation}
	Observe that given $(\bX, \widetilde{\bX})$, $\dot{\rho}_{\bbeta^0}$ is a vector consisting of i.i.d bounded random variables with zero mean and bounded variance. Following the same technique of proving \eqref{lemma13-I1} and \eqref{lemma13-I2}, we can obtain that  
	\begin{equation} \label{GLM_accuracy_R_1}
		\mathbb{P} \bigg( R_1 \leq C \Delta_n \sqrt{\frac{s \log n} {n} } \bigg)  \to 1 
	\end{equation}
	and 
	\begin{equation} \label{GLM_accuracy_R_2}
		\mathbb{P} \bigg( R_2 \leq C \Delta_n s \sqrt{\frac{ \log p} {n} } \bigg)  \to 1. 
	\end{equation}
	
 Regarding $R_3$, it follows from Conditions \ref{cond-GLM-00} and \ref{GLM-cond-subG} that with probability $1 - o(1)$, 
	\begin{equation} \label{GLM_accuracy_R_3}
		\begin{split}
			R_3 \leq C \sqrt s M n^{-1} \| \widetilde{\bX}^{\augg} (\widetilde{\bbeta} - \bbeta^0) \|_2^2 \leq C \sqrt s M n^{-1} s \log p \leq  C \frac{s^{3/2} \log p }{n}.
		\end{split}
	\end{equation}
	Combining \eqref{GLM_accuracy_R_1}--\eqref{GLM_accuracy_R_3} derives \eqref{pf-accuracy-glm-goal1}. Further, applying \eqref{GLM-accuracy-decom} again with the bounds in \eqref{GLM_accuracy_R_1}--\eqref{GLM_accuracy_R_3}  and \eqref{pf-accuracy-glm-goal1} yields that 
	\begin{equation} \label{GLM-X-accuracy}
		\mathbb{P} \bigg(n^{-1/2} \| \bD^{1/2} \widetilde{\bX}^{\augg} (\widehat{\bbeta} - \widetilde\bbeta) \|_2 \leq C\Big( \Delta_n s \sqrt{ \frac{\log p}{n} } +  \frac{ s^{3/2} \log p} {n} \Big)  \bigg) \to 1.
	\end{equation}
	Therefore, it follows by Condition \ref{GLM-cond-Loss} that
	\begin{equation} \label{GLM-X-accuracy-1}
		\mathbb{P} \bigg(n^{-1/2} \| \widetilde{\bX}^{\augg} (\widehat{\bbeta} - \widetilde\bbeta) \|_2 \leq C \Big( \Delta_n s \sqrt{ \frac{\log p}{n} } +  \frac{ s^{3/2} \log p} {n} \Big)  \bigg) \to 1.
	\end{equation}
	
	\bigskip
	
	\noindent \textbf{Proof of \eqref{pf-accuracy-glm-goal2}.} Observe that $\hat{\bgamma}_j$ and $\widetilde{\bgamma}_j$ can be equivalently written as 
	\begin{equation}
		\hat{\bgamma}_j = \argminA_{\bgamma \in \mathbb{R}^{2p-1}} \| \hat{\bD}^{1/2}\hat{\bX}^{\augg}_j - \hat{\bD}^{1/2} \hat{\bX}^{\augg}_{-j} \bgamma \|_2^2  + \lambda_j \|\bgamma \|_1.
	\end{equation}
	In addition, it can be obtained from Conditions \ref{cond-GLM-00}, \ref{GLM-cond-Loss}, \ref{GLM-cond-spar}, and \ref{GLM-cond-subG} that with probability $ 1 - o(1)$, 
	$$
	|\ddot{\rho}(y_i; \widetilde{\bX}_{i, \cdot} \widetilde\bbeta) - \ddot{\rho}(y_i; \widetilde{\bX}_{i, \cdot} \bbeta^0)| \leq C |\widetilde{\bX}_{i, \cdot} (\widetilde\bbeta - \bbeta^0)| \leq C M \sqrt {s} \sqrt{s \frac{\log p}{n}} = C M s \sqrt{\frac{\log p}{n}} \to 0.
	$$
	Hence, under Conditions \ref{accuracy-knockoffs}, \ref{cond-GLM-00}, \ref{GLM-cond-Loss}, \ref{GLM-cond-spar}, and \ref{GLM-cond-subG}, we have that with probability $1 - o(1)$, 
	\begin{equation} \label{GLM-accuracy-newdesign}
		\begin{split}
			& n^{-1/2} \| \hat{\bD}^{1/2} \widehat{\bX}^{\augg}_j - \widetilde{\bD}^{1/2} \widetilde{\bX}^{\augg}_j \|_2  \\
			& \leq n^{-1/2} \| (\hat{\bD}^{1/2} - \widetilde{\bD}^{1/2})\widehat{\bX}^{\augg}_j \|_2 + n^{-1/2} \| \widetilde{\bD}^{1/2} (\widehat{\bX}^{\augg}_j - \widetilde{\bX}^{\augg}_j) \|_2 \\
			& \leq C M n^{-1/2} \| \widehat{\bX}^{\augg} \hat{\bbeta} - \widetilde{\bX}^{\augg} \widetilde{\bbeta} \|_2 +  n^{-1/2} \|  {\bD}^{1/2} (\widehat{\bX}^{\augg}_j - \widetilde{\bX}^{\augg}_j) \|_2 \\
			& \quad + n^{-1/2} \| (\widetilde{\bD}^{1/2} - \bD^{1/2})(\widehat{\bX}^{\augg}_j - \widetilde{\bX}^{\augg}_j) \|_2 \\
			& \leq C M n^{-1/2} \| (\hat{\bX}^{\augg} - \widetilde{\bX}^{\augg} ) (\hat{\bbeta} - {\bbeta}^0) \|_2 +  C M n^{-1/2} \|\widetilde{\bX}^{\augg} (\widehat{\bbeta} - \widetilde{\bbeta} ) \|_2 \\
			& \quad + n^{-1/2} \|  {\bD}^{1/2} (\widehat{\bX}^{\augg}_j - \widetilde{\bX}^{\augg}_j) \|_2 + n^{-1/2} \| (\widetilde{\bD}^{1/2} - \bD^{1/2})(\widehat{\bX}^{\augg}_j - \widetilde{\bX}^{\augg}_j) \|_2 \\
			& \lesssim  \Delta_n s \sqrt{\frac{\log p}{n}} + \Delta_n \Big(1 + s \sqrt{\frac{\log p}{n}} \Big) \lesssim \Delta_n.
		\end{split}    
	\end{equation}
	Consequently, using similar argument as for \eqref{pf-accuracy-glm-goal1} and \eqref{GLM-X-accuracy}, we can obtain that 
	\begin{equation}
		\mathbb{P} \bigg( \max_{1 \leq j \leq 2p } \| \widetilde{\bgamma}_j - \hat{\bgamma}_j \|_2 \leq   C m_n^{1/2} \Delta_n \bigg) \to 1,
	\end{equation}
	\begin{equation}
		\mathbb{P} \bigg(n^{-1/2} \max_{1 \leq j \leq 2p} n^{-1/2} \|  \widetilde{\bX}_{-j}^{\augg} (\widetilde{\bgamma}_j - \hat{\bgamma}_j ) \|_2 \leq C m_n^{1/2} \Delta_n   \bigg) \to 1.
	\end{equation}
	Moreover, by similar arguments as for \eqref{eq-lemma13-7}, \eqref{eq-normal-error-1}, and \eqref{eq-normal-error-2}, we can deduce that with probability $1 - o(p^{-1})$, 
	\begin{equation}\label{GLM-Lemma20-bound-z}
		\| \hat{\bX}_{j}^{\augg} - \hat{\bX}_{-j}^{\augg} \widehat\bgamma_j - (\widetilde{\bX}_{j}^{\augg} - \widetilde{\bX}_{-j}^{\augg} \widetilde\bgamma_j ) \|_2  \lesssim \Delta_n m_n^{1/2},
	\end{equation}
	\begin{equation} \label{GLM-tau-j-lower}
		\min_{1 \leq j \leq p} \widetilde{\tau}_j^2 = \min_{1 \leq j \leq p} 
		n^{-1} (\widetilde{\bX}_j^{\augg})^T \widetilde{\bD} (\widetilde{\bX}_j^{\augg} - \widetilde{\bX}_{-j}^{\augg} \widetilde{\bgamma}_j)  \geq C ,
	\end{equation} 
	and 
	\begin{equation} \label{GLM-tau-jk-upper}
		\max_{1 \leq j \leq p}  \max_{k\neq j}  
		n^{-1} \big|(\widetilde{\bX}_k^{\augg})^T \widetilde{\bD} (\widetilde{\bX}_j^{\augg} - \widetilde{\bX}_{-j}^{\augg} \widetilde{\bgamma}_j) \big| \leq C \sqrt{\frac{(m_n + s) \log p}{n}}.
	\end{equation}
	
	Now we are ready to establish \eqref{pf-accuracy-glm-goal2}. Specifically, the main term in \eqref{pf-accuracy-glm-goal2} can be decomposed into the three terms below 
	\begin{equation} \label{GLM-Lemma20-Decom-1}
		\begin{split}
			& \max_{1 \leq j \leq 2p} \bigg| \frac{n^{-1} (\dot{\brho}_{\widehat{\bbeta}} - \dot{\brho}_{\widetilde{\bbeta}})^T (\widetilde{\bX}_{j}^{\augg} - \widetilde{\bX}_{-j}^{\augg} \widetilde\bgamma_j )} { \widetilde{\tau}_j^2 }  \bigg| \\
			&\quad + \max_{1 \leq j \leq 2p} \bigg| \frac{n^{-1}  \dot{\brho}_{\widehat{\bbeta}}^T \big(\hat{\bX}_{j}^{\augg} - \hat{\bX}_{-j}^{\augg} \widehat\bgamma_j - (\widetilde{\bX}_{j}^{\augg} - \widetilde{\bX}_{-j}^{\augg} \widetilde\bgamma_j ) \big)} { \widetilde{\tau}_j^2 }  \bigg| \\
			& \quad +\max_{1 \leq j \leq 2p} \bigg| n^{-1} \dot{\brho}_{\widehat{\bbeta}}^T (\widehat{\bX}_{j}^{\augg} - \widehat{\bX}_{-j}^{\augg} \widehat\bgamma_j ) \bigg(\frac{1}{\widetilde{\tau}_j^2} - \frac{1}{\widehat{\tau}_j^2} \bigg) \bigg| := I_1 + I_2 + I_3.
		\end{split}
	\end{equation}
	We will deal with the three terms $I_1$, $I_2$, and $I_3$ separately. First for $I_1$, it follows from Condition \ref{GLM-cond-Loss} that
	\begin{equation} \label{GLM-I_1-decom}
		\begin{split}
			I_1 & \leq \max_{1 \leq j \leq 2p} \bigg|  \frac{n^{-1} [ (\widehat{\bX}^{\augg} \widehat{\bbeta} - \widetilde{\bX}^{\augg} \widetilde{\bbeta}) ]^T \widetilde{\bD} (\widetilde{\bX}_{j}^{\augg} - \widetilde{\bX}_{-j}^{\augg} \widetilde\bgamma_j ) } { \widetilde{\tau}_j^2 }   \bigg| \\
			& \quad + C \max_{1 \leq j \leq 2p} \bigg|  \frac{n^{-1} \sum_{i = 1}^n (\widehat{\bX}_{i, \cdot}^{\augg} \widehat{\bbeta} - \widetilde{\bX}_{i, \cdot}^{\augg} \widetilde{\bbeta})^2 |\widetilde{\bX}_{i, j}^{\augg} - \widetilde{\bX}_{i, -j}^{\augg} \widetilde\bgamma_j | } { \widetilde{\tau}_j^2 }   \bigg| := I_{11} + I_{12}.
		\end{split}
	\end{equation}
	Regarding $I_{12}$, in view of Conditions \ref{cond-GLM-01} and \ref{GLM-cond-subG},  and \eqref{GLM-tau-j-lower},  we have 
	\begin{equation} \label{GLM-I_12}
		\begin{split}
			I_{12} & \leq C \max_{1 \leq j \leq 2p}  n^{-1} \sum_{i = 1}^n (\widehat{\bX}_{i, \cdot}^{\augg} \widehat{\bbeta} - \widetilde{\bX}_{i, \cdot}^{\augg} \widetilde{\bbeta})^2 |  \widetilde{\bX}_{i, j}^{\augg} - \widetilde{\bX}_{i, -j}^{\augg}   \bgamma_j |  \\
			& \quad  + C \max_{1 \leq j \leq 2p}  n^{-1} \sum_{i = 1}^n (\widehat{\bX}_{i, \cdot}^{\augg} \widehat{\bbeta} - \widetilde{\bX}_{i, \cdot}^{\augg} \widetilde{\bbeta})^2 | \widetilde{\bX}_{i, -j}^{\augg} (\widetilde{\bgamma}_j - \bgamma_j ) |   \\
			& \leq C M \bigg( 1 + m_n^{1/2} \sqrt{\frac{(s + m_n) \log p} {n}} \bigg)   n^{-1} \| \widehat{\bX}^{\augg} \widehat{\bbeta} - \widetilde{\bX}^{\augg} \widetilde{\bbeta} \|_2^2 \\
			& \leq C M n^{-1 }\| \widehat{\bX}^{\augg} \widehat{\bbeta} - \widetilde{\bX}^{\augg} \widetilde{\bbeta} \|_2^2,
		\end{split}
	\end{equation}
	where we have applied the assumption that $s \sqrt{\frac{\log p}{n}} \to 0$ and $m_n \lesssim s$. In addition, noting that $(\widehat{\bX}^{\augg} - \widetilde{\bX}^{\augg} ) \bbeta^0 = {\bf 0}$ by definition, we obtain from \eqref{GLM-X-accuracy-1} and Condition \ref{cond-GLM-00} that with probaility $1 - o(1)$, 
	\begin{equation} \label{GLM-Lemma20-eq1}
		\begin{split}
			n^{-1} \| \widehat{\bX}^{\augg} \widehat{\bbeta} - \widetilde{\bX}^{\augg} \widetilde{\bbeta} \|_2^2 & \leq n^{-1} \| \widetilde{\bX} (\widetilde{\bbeta} - \widehat{\bbeta}) \|_2^2 + n^{-1} \| (\widehat{\bX}^{\augg} - \widetilde{\bX}^{\augg}) (\hat{\bbeta} - \bbeta^0) \|_2^2 \\
			& \lesssim \frac{\Delta_n^2 s^2 \log p } {n} + \frac{s^{3} (\log p) ^2 }{n^2} + \Delta_n^2 s \frac{s \log p} {n} \\
			& \lesssim \frac{\Delta_n^2 s^2 \log p } {n} + \frac{s^{3} (\log p) ^2 }{n^2} ,
		\end{split}
	\end{equation}
	which together with \eqref{GLM-I_12} yields 
	\begin{equation} \label{GLM-Lemma20-I12}
		I_{12} \leq C \bigg(\frac{\Delta_n^2 s^2 \log p } {n} + \frac{s^{3} (\log p) ^2 }{n^2} \bigg).
	\end{equation}
 
	Now we proceed with examining $I_{11}$. Observe that it admits the decomposition 
	\begin{equation} \label{GLM-Lemma20-I_11-decom}
		\begin{split}
			I_{11} & \leq \max_{1 \leq j \leq 2p} \bigg|  \frac{n^{-1} (\widetilde{\bbeta} - \widehat{\bbeta}) ^T (\widetilde{\bX}^{\augg} ) ^T\widetilde{\bD} (\widetilde{\bX}_{j}^{\augg} - \widetilde{\bX}_{-j}^{\augg} \widetilde\bgamma_j ) } { \widetilde{\tau}_j^2 }  \bigg| \\
			& \quad +  \max_{1 \leq j \leq 2p} \bigg|  \frac{n^{-1}  \widehat{\bbeta}^T (\widetilde{\bX}^{\augg} - \widehat{\bX}^{\augg}) ^T\widetilde{\bD} (\widetilde{\bX}_{j}^{\augg} - \widetilde{\bX}_{-j}^{\augg} \widetilde\bgamma_j ) } { \widetilde{\tau}_j^2 }  \bigg| \\
			& \leq \max_{1 \leq j \leq 2p}  | \widetilde{\beta}_j - \widehat{\beta}_j | + \max_{1 \leq j \leq 2p}  \Big|   n^{-1} (\widetilde{\bbeta}_{-j} - \widehat{\bbeta}_{-j}) ^T (\widetilde{\bX}^{\augg}_{-j} ) ^T\widetilde{\bD} (\widetilde{\bX}_{j}^{\augg} - \widetilde{\bX}_{-j}^{\augg} \widetilde\bgamma_j )    \Big| \\
			& \quad + \max_{1 \leq j \leq 2p}   n^{-1}  \Big| \widehat{\bbeta} ^T (\widetilde{\bX}^{\augg} - \widehat{\bX}^{\augg}) ^T\widetilde{\bD} (\widetilde{\bX}_{j}^{\augg} - \widetilde{\bX}_{-j}^{\augg} \widetilde\bgamma_j ) \Big| \\
			& := I_{111} + I_{112} + I_{113}.
		\end{split}
	\end{equation}
	As for $I_{112}$,  it follows from \eqref{pf-accuracy-glm-goal1} and \eqref{GLM-tau-jk-upper} that with probability $1 - o(1)$,
	\begin{equation} \label{GLM-Lemma20-I_112}
		\begin{split}
			I_{112} & \leq \max_{1 \leq j \leq 2p} \max_{J: |J| \leq C s }    n^{-1} \big\|\widetilde{\bbeta}_{-j} - \widehat{\bbeta}_{-j} \big\|_2 
			\big\| (\widetilde{\bX}^{\augg}_{J \setminus \{j\}} ) ^T\widetilde{\bD} (\widetilde{\bX}_{j}^{\augg} - \widetilde{\bX}_{-j}^{\augg} \widetilde\bgamma_j )   \big\|_2 \\
			& \leq C s^{1/2} \big\|\widetilde{\bbeta}  - \widehat{\bbeta}  \big\|_2  \max_{1 \leq j \leq 2p} \max_{k \neq j }   n^{-1}   \big|(\widetilde{\bX}_k^{\augg})^T \widetilde{\bD} (\widetilde{\bX}_j^{\augg} - \widetilde{\bX}_{-j}^{\augg} \widetilde{\bgamma}_j) \big| \\
			& \leq s \sqrt{\frac{\log p}{n}} \Big( \Delta_n s \sqrt{ \frac{\log p}{n} } +  \frac{ s^{3/2} \log p} {n} \Big) \lesssim   \Delta_n s \sqrt{ \frac{\log p}{n} } +  \frac{ s^{3/2} \log p} {n} ,
		\end{split}
	\end{equation}
	where we have used the assumption that $ s \sqrt{\frac{\log p}{n}} \to 0$. 
 
 Regarding $I_{113}$, noting that $(\widehat{\bX}^{\augg} - \widetilde{\bX}^{\augg} ) \bbeta^0 = {\bf 0}$, by a similar argument as for \eqref{eq-lemma13-14}, we have that with probability $1 - o(1)$,
	\begin{equation} \label{GLM-Lemma20-I_113}
		\begin{split}
			I_{113} & =  \max_{1 \leq j \leq 2p}   n^{-1}  \Big| (\widehat{\bbeta} - \bbeta^0)^T (\widetilde{\bX}^{\augg} - \widehat{\bX}^{\augg})^T \widetilde{\bD} (\widetilde{\bX}_{j}^{\augg} - \widetilde{\bX}_{-j}^{\augg} \widetilde\bgamma_j ) \Big| \\
			& \leq n^{-1/2} \|(\widetilde{\bX}^{\augg} - \widehat{\bX}^{\augg}) (\widehat{\bbeta} - \bbeta^0) \|_2  \max_{1 \leq j \leq 2p} n^{-1/2} \| \widetilde{\bD} (\widetilde{\bX}_{j}^{\augg} - \widetilde{\bX}_{-j}^{\augg} \widetilde\bgamma_j ) \|_2 \\
			& \lesssim \Delta_n s \sqrt{\frac{\log p}{n}}.
		\end{split}
	\end{equation}
	Combining \eqref{pf-accuracy-glm-goal1}, \eqref{GLM-Lemma20-I_11-decom}, \eqref{GLM-Lemma20-I_112}, and \eqref{GLM-Lemma20-I_113}, we can derive that with probability $1 - o(1)$,
	\begin{equation} \label{GLM-Lemma20-I_11-bound}
		I_{11} \lesssim  \Delta_n s \sqrt{ \frac{\log p}{n} } +  \frac{ s^{3/2} \log p} {n},  
	\end{equation}
	which together with \eqref{GLM-I_1-decom} and \eqref{GLM-Lemma20-I12} gives that 
	\begin{equation} \label{GLM-Lemma20-I_1-bound}
		\mathbb{P} \bigg( I_{1} \leq C \Big( \Delta_n s \sqrt{ \frac{\log p}{n} } +  \frac{ s^{3/2} \log p} {n}\Big) \bigg) \to 1.
	\end{equation}
	
	Next we turn to $I_{2}$ in \eqref{GLM-Lemma20-Decom-1}. Applying the Taylor expansion and Condition \ref{GLM-cond-Loss}, we can obtain that 
	\begin{equation} \label{GLM-Lemma20-I_2-decom}
		\begin{split}
			I_{2} & \lesssim \max_{1 \leq j \leq 2p} \Big| n^{-1} \dot{\brho}_{\bbeta^0} ^T \big(\hat{\bX}_{j}^{\augg} - \hat{\bX}_{-j}^{\augg} \widehat\bgamma_j - (\widetilde{\bX}_{j}^{\augg} - \widetilde{\bX}_{-j}^{\augg} \widetilde\bgamma_j ) \big) \Big| \\
			& \quad + \max_{1 \leq j \leq 2p}  \Big| n^{-1}  (\hat{\bbeta} - \bbeta^0)^T  (\hat{\bX}^{\augg})^T \bD \big(\hat{\bX}_{j}^{\augg} - \hat{\bX}_{-j}^{\augg} \widehat\bgamma_j - (\widetilde{\bX}_{j}^{\augg} - \widetilde{\bX}_{-j}^{\augg} \widetilde\bgamma_j ) \big) \Big| \\
			& \quad +  \max_{1 \leq j \leq 2p} n^{-1} \sum_{i = 1}^n [\widehat{\bX}_{i, \cdot}^{\augg} ( \widehat{\bbeta} - {\bbeta}^0 ) ]^2 \big|\hat{\bX}_{i, j}^{\augg} - \hat{\bX}_{i, -j}^{\augg} \widehat\bgamma_{i, j} - (\widetilde{\bX}_{i, j}^{\augg} - \widetilde{\bX}_{i, -j}^{\augg} \widetilde\bgamma_j ) \big| \\
			& := I_{21} + I_{22} + I_{23}.
		\end{split}
	\end{equation}
	Note that $\e [ \dot{\brho}_{\bbeta^0} | (\bX , \widetilde{\bX}, \widehat{\bX} )]  = {\bf 0}$ and with probability $1 - o(p^{-1})$, 
	\begin{equation*} 
		\begin{split}
			& \Var ( n^{-1} \dot{\brho}_{\bbeta^0} ^T \big(\hat{\bX}_{j}^{\augg} - \hat{\bX}_{-j}^{\augg} \widehat\bgamma_j - (\widetilde{\bX}_{j}^{\augg} - \widetilde{\bX}_{-j}^{\augg} \widetilde\bgamma_j ) \big) | (\bX , \widetilde{\bX}, \widehat{\bX} )) \\
			& = n^{-2} \big(\hat{\bX}_{j}^{\augg} - \hat{\bX}_{-j}^{\augg} \widehat\bgamma_j - (\widetilde{\bX}_{j}^{\augg} - \widetilde{\bX}_{-j}^{\augg} \widetilde\bgamma_j ) \big) ^T \bD \big(\hat{\bX}_{j}^{\augg} - \hat{\bX}_{-j}^{\augg} \widehat\bgamma_j - (\widetilde{\bX}_{j}^{\augg} - \widetilde{\bX}_{-j}^{\augg} \widetilde\bgamma_j ) \big) \\
			& \lesssim n^{-2} \| \hat{\bX}^{\augg}_j - \widetilde{\bX}^{\augg}_j +   \widetilde{\bX}^{\augg} _{-j} (\widetilde{\bgamma}_j - \widehat{\bgamma}_j ) +  (\widetilde{\bX}_{-j}^{\augg} - \widehat{\bX}_{-j}^{\augg} ) \widehat{\bgamma}_j \|_2^2 \\
			& \lesssim n^{-1} ( \Delta_n^2 +  \Delta_n^2 m_n + \Delta_n^2 m_n )  \lesssim n^{-1} \Delta_n^2 m_n.
		\end{split}
	\end{equation*}
	Since the components of $\dot{\brho}_{\bbeta^0}$ are all bounded by $K_1$ under Condition \ref{GLM-cond-Loss}, we see that $ n^{-1} \dot{\brho}_{\bbeta^0} ^T \big(\hat{\bX}_{j}^{\augg} - \hat{\bX}_{-j}^{\augg} \widehat\bgamma_j - (\widetilde{\bX}_{j}^{\augg} - \widetilde{\bX}_{-j}^{\augg} \widetilde\bgamma_j ) \big)  $ is sub-Gaussian, which entails that with probability $1 - o(1)$,
	\begin{equation} \label{GLM-Lemma20-I_21-bound}
		I_{21} \leq C \Delta_n \sqrt{\frac{m_n \log p}{n}}.
	\end{equation} 
	
 In the same manner of proving \eqref{eq-lemma13-9}, we can show that with probability $1 - o(1)$, 
	\begin{equation} \label{GLM-Lemma20-I_22-bound}
		I_{22} \lesssim \sqrt{\frac{s \log p}{n}}  \Delta_n m_n^{1/2} \lesssim \Delta_n s \sqrt{\frac{\log p}{n}}.
	\end{equation}
	Regarding $I_{23}$, it holds that with probability $1 - o(1)$, 
	\begin{equation} \label{GLM-Lemma20-I_23-bound}
		\begin{split}
			I_{23}  & \leq n^{-1}  \| \hat{\bX}^{\augg} (\hat{\bbeta} - \bbeta^0)  \|_2^2 \| \hat{\bX}_{j}^{\augg} - \hat{\bX}_{-j}^{\augg} \widehat\bgamma_j - (\widetilde{\bX}_{j}^{\augg} - \widetilde{\bX}_{-j}^{\augg} \widetilde\bgamma_j ) \|_2 \\
			& \leq \frac{s \log p}{n}  \Delta_n m_n^{1/2} \lesssim \Delta_n s \sqrt{\frac{\log p}{n}}.
		\end{split}
	\end{equation}
	A combination of \eqref{GLM-Lemma20-I_2-decom}--\eqref{GLM-Lemma20-I_23-bound} leads to  
	\begin{equation} \label{GLM-Lemma20-I_2-bound}
		\mathbb{P} \bigg( I_2 \leq C  \Delta_n s \sqrt{\frac{\log p}{n}} \bigg) \to 1.
	\end{equation}
	
	Now we proceed to deal with $I_3$ in \eqref{GLM-Lemma20-Decom-1}. Note that 
	\begin{equation} \label{GLM-Lemma20-I_3-decom}
		I_3 \leq \max_{1 \leq j \leq 2p}  \big| n^{-1} \dot{\brho}_{\widehat{\bbeta}}^T (\widehat{\bX}_{j}^{\augg} - \widehat{\bX}_{-j}^{\augg} \widehat\bgamma_j ) \big| \cdot \max_{1 \leq j \leq 2p}   \frac{ | \widehat{\tau}_j^2 - \widetilde{\tau}_j^2 | }{| \widetilde{\tau}_j^2 \widehat{\tau}_j^2 | } := I_{31} \cdot I_{32} .
	\end{equation}
	It can be seen  that 
	\begin{equation*}  
		\begin{split}
			I_{31} & \lesssim \max_{1 \leq j \leq 2p}  \big| n^{-1} \dot{\brho}_{\bbeta^0}^T (\widehat{\bX}_{j}^{\augg} - \widehat{\bX}_{-j}^{\augg} \widehat\bgamma_j ) \big| + \max_{1 \leq j \leq 2p}  \big| n^{-1} (\dot{\brho}_{\hat\bbeta} - \dot{\brho}_{\bbeta^0} )^T (\widehat{\bX}_{j}^{\augg} - \widehat{\bX}_{-j}^{\augg} \widehat\bgamma_j ) \big|.
		\end{split}
	\end{equation*}
	By a similar argument as for \eqref{GLM-Lemma20-I_21-bound}, we can show that with probability $1 - o(1)$,
	\begin{equation*} 
		\max_{1 \leq j \leq 2p}  \big| n^{-1} \dot{\brho}_{\bbeta^0}^T (\widehat{\bX}_{j}^{\augg} - \widehat{\bX}_{-j}^{\augg} \widehat\bgamma_j ) \big|  \lesssim \sqrt{\frac{\log p}{n}}.
	\end{equation*}
	In addition, we have under Condition \ref{GLM-cond-Loss} that with probability $1 - o(1)$,
	\begin{equation*}  
		\begin{split}
			& \max_{1 \leq j \leq 2p}  \big| n^{-1} (\dot{\brho}_{\hat\bbeta} - \dot{\brho}_{\bbeta^0} )^T (\widehat{\bX}_{j}^{\augg} - \widehat{\bX}_{-j}^{\augg} \widehat\bgamma_j ) \big| \\
			& \lesssim \max_{1 \leq j \leq 2p} |n^{-1} (\widehat{\bX}^{\augg}(\hat{\bbeta} - \bbeta^0))^T \bD (\widehat{\bX}_{j}^{\augg} - \widehat{\bX}_{-j}^{\augg} \widehat\bgamma_j) | \\
			& \lesssim \max_{1 \leq j \leq 2p} n^{-1} \| \widehat{\bX}^{\augg}(\hat{\bbeta} - \bbeta^0)\|_2 \|\widehat{\bX}_{j}^{\augg} - \widehat{\bX}_{-j}^{\augg} \widehat\bgamma_j \|_2 \lesssim \sqrt{\frac{s \log p}{n}}.
		\end{split}    
	\end{equation*}
	Thus, we can obtain that with probability $1 - o(1)$, 
	\begin{equation} \label{GLM-Lemma20-I_31-bound}
		I_{31} \lesssim \sqrt{\frac{s \log p}{n}}.
	\end{equation}
	
 As for $I_{32}$, by definition it holds that 
	\begin{equation} \label{GLM-Lemma20-I_32-decom}
		\begin{split}
			| \hat{\tau}_j^2 - \widetilde{\tau}_j^2 | & = n^{-1}\big| (\hat{\bX}^{\augg}_j)^T \hat{\bD} (\hat{\bX}_j^{\augg} - \hat{\bX}^{\augg}_{-j} \hat{\bgamma}_j) - (\widetilde{\bX}^{\augg}_j)^T\widetilde{\bD} (\widetilde{\bX}_j^{\augg} - \widetilde{\bX}^{\augg}_{-j} \widetilde{\bgamma}_j) \big|\\
			& \leq n^{-1}\big| (\hat{\bX}_j^{\augg} - \widetilde{\bX}_j^{\augg})^T \hat{\bD}(\hat{\bX}_j^{\augg} - \hat{\bX}^{\augg}_{-j} \hat{\bgamma}_j) \big|  \\
			& \quad + n^{-1} \big| (\widetilde{\bX}_j^{\augg})^T (\hat{\bD} - \widetilde{\bD}) (\hat{\bX}_j^{\augg} - \hat{\bX}^{\augg}_{-j} \hat{\bgamma}_j) \big| \\
			& \quad + n^{-1} \big| (\widetilde{\bX}_j^{\augg})^T \widetilde{\bD} ( \hat{\bX}_j^{\augg} - \hat{\bX}^{\augg}_{-j} \hat{\bgamma}_j - (\widetilde{\bX}_j^{\augg} - \widetilde{\bX}^{\augg}_{-j} \widetilde{\bgamma}_j) ) \big|.
		\end{split}
	\end{equation}
	Furthermore, it can be shown that with probability $1 - o(1)$,
	\begin{equation} \label{GLM-Lemma20-I_32-1}
		\begin{split}
			& n^{-1}\big| (\hat{\bX}_j^{\augg} - \widetilde{\bX}_j^{\augg})^T \hat{\bD}(\hat{\bX}_j^{\augg} - \hat{\bX}^{\augg}_{-j} \hat{\bgamma}_j) \big| \\
			& \lesssim n^{-1}\| \hat{\bX}_j^{\augg} - \widetilde{\bX}_j^{\augg} \|_2 \|\hat{\bX}_j^{\augg} - \hat{\bX}^{\augg}_{-j} \hat{\bgamma}_j \|_2 \\
   & \lesssim \Delta_n
		\end{split}
	\end{equation}
	and
	\begin{equation}  \label{GLM-Lemma20-I_32-2}
		\begin{split}
			& n^{-1} \big| (\widetilde{\bX}_j^{\augg})^T (\hat{\bD} - \widetilde{\bD}) (\hat{\bX}_j^{\augg} - \hat{\bX}^{\augg}_{-j} \hat{\bgamma}_j) \big| \\
			& \leq n^{-1} \sum_{i = 1}^n | \hat{\bX}_{i, \cdot}^{\augg} \hat{\bbeta} - \widetilde{\bX}_{i, \cdot}^{\augg} \widetilde{\bbeta} | |\widetilde{\bX}_{i, j} | |\hat{\bX}_{i, j}^{\augg} - \hat{\bX}_{i, -j}^{\augg}\hat{\bgamma}_j | \\
			& \leq n^{-1} M \sum_{i = 1}^n | \hat{\bX}_{i, \cdot}^{\augg} \hat{\bbeta} - \widetilde{\bX}_{i, \cdot}^{\augg} \widetilde{\bbeta} |   |\hat{\bX}_{i, j}^{\augg} - \hat{\bX}_{i, -j}^{\augg}\hat{\bgamma}_j | \\
			& \leq n^{-1} M \| \hat{\bX}^{\augg} \hat{\bbeta} - \widetilde{\bX}^{\augg} \widetilde{\bbeta}  \|_2 \|\hat{\bX}_{ j}^{\augg} - \hat{\bX}_{ -j}^{\augg}\hat{\bgamma}_j \|_2 \\
   & \lesssim  \Delta_n s \sqrt{ \frac{\log p}{n} } +  \frac{ s^{3/2} \log p} {n},
		\end{split}
	\end{equation}
	where we have applied the bound obtained in \eqref{GLM-Lemma20-eq1}.
	
	Moreover, we have that 
	\begin{equation} \label{GLM-Lemma20-I_32-2}
		\begin{split}
			& n^{-1} \big| (\widetilde{\bX}_j^{\augg})^T \widetilde{\bD} ( \hat{\bX}_j^{\augg} - \hat{\bX}^{\augg}_{-j} \hat{\bgamma}_j - (\widetilde{\bX}_j^{\augg} - \widetilde{\bX}^{\augg}_{-j} \widetilde{\bgamma}_j) ) \big| \\
			& \lesssim n^{-1} \| \widetilde{\bX}_j^{\augg}\|_2 \| \hat{\bX}_j^{\augg} - \hat{\bX}^{\augg}_{-j} \hat{\bgamma}_j - (\widetilde{\bX}_j^{\augg} - \widetilde{\bX}^{\augg}_{-j} \widetilde{\bgamma}_j)  \|_2 \\
   & \lesssim \Delta_n m_n^{1/2},
		\end{split}
	\end{equation}
	which together with \eqref{GLM-tau-j-lower}, \eqref{GLM-Lemma20-I_32-decom}, \eqref{GLM-Lemma20-I_32-1}, and \eqref{GLM-Lemma20-I_32-2} yields that with probability $1 - o(1)$, 
	\begin{equation} \label{GLM-Lemma20-I_32-bound}
		I_{32} = \max_{1 \leq j \leq 2p}   \frac{ | \widehat{\tau}_j^2 - \widetilde{\tau}_j^2 | }{| \widetilde{\tau}_j^2 \widehat{\tau}_j^2 | }  \lesssim \Delta_n m_n^{1/2} + \frac{s^{3/2} \log p}{n}.
	\end{equation}
	Combining \eqref{GLM-Lemma20-I_3-decom}, \eqref{GLM-Lemma20-I_31-bound}, and \eqref{GLM-Lemma20-I_32-bound} leads to 
	\begin{equation} \label{GLM-Lemma20-I_3-bound}
		\mathbb{P} \bigg( I_3 \leq C \Big( \Delta_n s \sqrt{ \frac{\log p}{n} } +  \frac{ s^{3/2} \log p} {n}\Big)  \bigg) \to 1.
	\end{equation}
	Consequently, substituting \eqref{GLM-Lemma20-I_1-bound}, \eqref{GLM-Lemma20-I_2-bound}, and \eqref{GLM-Lemma20-I_3-bound} into \eqref{GLM-Lemma20-Decom-1} gives the desired result \eqref{pf-accuracy-glm-goal2}. This completes the proof of Lemma \ref{pf-GLM-lemma-1}.

	\subsection{Proof of Lemma \ref{pf-GLM-lemma-2}} \label{pf-of-lemma-glm-2}
	The main idea of the proof is to bound the remainders in the decomposition of $ \sqrt{n} (\widetilde{b}_j - \beta_j^0) $ as presented in \eqref{normal-decomp_GLM} and use the fact that the main term is sub-Gaussian. Note that by the triangle inequality and the fact that $w_j = |\beta_j^0| = |\beta_j^0| - |\beta_{j+p}^0|$, it holds that 
	\begin{equation} \label{GLM-Lemma21-decom1}
		\begin{split}
			& \sum_{j= 1}^p \mathbb{P} \big(|\widetilde{W}_j - w_j | \geq C \sqrt{n^{-1} \log p} \big) \\
			& \leq \sum_{j = 1}^p \Big[\mathbb{P} (\sqrt{n} |\widetilde{b}_j - \beta_j^0| \geq C\sqrt{  \log p} /2 ) +  \mathbb{P} (\sqrt{n} |\widetilde{b}_{j + p} - \beta_{j+p}^0| \geq C\sqrt{  \log p} /2  ) \Big] \\
			& = \sum_{j = 1}^{2p} \mathbb{P} (\sqrt{n} |\widetilde{b}_j - \beta_j^0| \geq C\sqrt{  \log p} /2 ). 
		\end{split}
	\end{equation}
	For the second remainder in \eqref{normal-decomp_GLM}, applying the bounds in \eqref{GLM-tau-jk-upper}, \eqref{GLM-tau-j-lower}, and  \eqref{lasso-1}, we have that with probability $1 - o(p^{-3})$,
	\begin{equation} \label{GLM-lemma21-remainder1}
		\begin{split}  
			& \max_{1 \leq j \leq p} \frac{n^{-1/2} (\bbeta^0_{-j} - \widetilde\bbeta_{-j})^T (\widetilde{\bX}^{\augg}_{-j})^{T} \widetilde{\bD} (\widetilde{\bX}_j^{\augg} - \widetilde{\bX}_{-j}^{\augg} \widetilde{\bgamma}_j) } {\widetilde{\tau}_j^2}  \\
			& \leq  \max_{1 \leq j \leq p} n^{-1/2} \sum_{k \neq j} \frac{n^{-1/2} |(\widetilde{\bX}_k^{\augg})^T \widetilde{\bD} (\widetilde{\bX}^{\augg}_j - \widetilde{\bX}^{\augg}_{-j} \widetilde{\bgamma}_j ) |  \widetilde{\beta}_k  - \beta_k^0 | }{\widetilde{\tau}_j^2} \\
			& \leq  \max_{1 \leq j \leq p} \max_{k \neq j}  n^{-1/2} |(\widetilde{\bX}_k^{\augg})^T \widetilde{\bD} (\widetilde{\bX}^{\augg}_j - \widetilde{\bX}^{\augg}_{-j} \widetilde{\bgamma}_j ) | \|\widetilde\bbeta - \bbeta^0 \|_1 \\
   & \lesssim  \frac{s^{3/2} \log p}{\sqrt n}. 
		\end{split}
	\end{equation}
	Regarding the first remainder in \eqref{normal-decomp_GLM}, applying \eqref{lasso-x-error}, \eqref{coefficients-x-error}, \eqref{GLM-tau-j-lower}, and the fact that $\|\widetilde{\bX}_j - \widetilde{\bX}_{-j} {\bgamma}_j\| \leq M $ for some $M >0$, we can obtain that with probability $1 - o(p^{-3})$, 
	\begin{equation} \label{GLM-lemma21-remainder2}
		\frac{ n^{-1/2} \widetilde\bR | \widetilde{\bX}_j^{\augg} - \widetilde{\bX}_{-j}^{\augg} \widetilde{\bgamma}_j | } {\widetilde{\tau}_j^2} \lesssim n^{-1/2} \| \widetilde{\bX}^{\augg} (\widetilde{\bbeta} - \bbeta^0) \|_2^2 \lesssim \frac{s \log p}{\sqrt{n}}.
	\end{equation}
	Further, observe that the main term $ n^{-1/2} \dot{\brho}_{\bbeta^0} (\widetilde{\bX}^{\augg}_j - \widetilde{\bX}^{\augg} \widetilde{\bgamma}_j) $ in \eqref{normal-decomp_GLM} is sub-Gaussian since $\|  \widetilde{\bX}^{\augg}_j - \widetilde{\bX}^{\augg} \widetilde{\bgamma}_j \|_{\infty} \leq M$ and $\| \dot{\brho}_{\bbeta^0} \|_{\infty} \leq M$ for some constant $M > 0$. Moreover, it holds that 
 $$\Var( n^{-1/2} \dot{\brho}_{\bbeta^0} (\widetilde{\bX}^{\augg}_j - \widetilde{\bX}^{\augg} \widetilde{\bgamma}_j)\vert  \widetilde{\bX}^{\augg} ) =n^{-1}   (\widetilde{\bX}^{\augg}_j - \widetilde{\bX}^{\augg} \widetilde{\bgamma}_j) ^T \bD  (\widetilde{\bX}^{\augg}_j - \widetilde{\bX}^{\augg} \widetilde{\bgamma}_j) \leq M$$ for some constant $M > 0$. Therefore, using similar arguments as in the proof of Lemma \ref{pf-thm5-lemma-2}, we can establish the desired result in Lemma \ref{pf-GLM-lemma-2}.

	\subsection{Proof of Lemma \ref{pf-GLM-lemma-3}} \label{pf-of-lemma-glm-3}
	We will apply the moderate deviation result (i.e., the rate of convergence) for multivariate normal approximation (\cite{saulis1992probabilities}), and the remaining proof can proceed by the same technique as used for proving Lemma \ref{pf-thm5-lemma-3}. 
	From the decomposition for $\sqrt{n} (\widetilde{b}_j - \beta_j^0)$ outlined in \eqref{normal-decomp_GLM} and the bounds in \eqref{GLM-lemma21-remainder1} and \eqref{GLM-lemma21-remainder2}, it is seen that the main term is $ \xi_j:=- \frac{  \dot\brho_{ {\bbeta^0}}^T (\widetilde{\bX}_j^{\augg} - \widetilde{\bX}_{-j}^{\augg} \widetilde{\bgamma}_j) } { \sqrt n \widetilde{\tau}_j^2}$ and the two remainders in \eqref{normal-decomp_GLM} are bounded by $ C \frac{s^{3/2} \log p}{\sqrt n} $ with probability $1 - o(p^{-1})$. 
	Denote by $\widetilde{\bz}_j = \widetilde{\bX}_j^{\augg} - \widetilde{\bX}_{-j}^{\augg} \widetilde{\bgamma}_j$ for $1 \leq j \leq 2p$. 
	Observe that given $\widetilde\bX^{\augg}$,  $(\xi_j, \xi_{j+ p}, \xi_{l}, \xi_{l+p})^T \stackrel{d}{\sim} N({\bf 0}, \bV)$, where the covariance matrix $\bV$ is given by  $\bV = \begin{pmatrix} \bV_{11} \bV_{12} \\ \bV_{21} \bV_{22}\end{pmatrix}$ with  
	\begin{equation*}  
		\begin{split}
			& \bV_{11} = 
			\begin{pmatrix}
				\frac{ \widetilde{\bz}_j^T \bD \widetilde{\bz}_j }{n \widetilde{\tau}_j^4} 
				& \frac{ \widetilde{\bz}_j^T \bD \widetilde{\bz}_{j+p} }{n \widetilde{\tau}_{j}^2 \widetilde{\tau}_{j+p}^2}  \\
				\frac{ \widetilde{\bz}_j^T \bD \widetilde{\bz}_{j+p} }{n \widetilde{\tau}_{j}^2 \widetilde{\tau}_{j+p}^2} 
				& \frac{ \widetilde{\bz}_{j+p}^T \bD \widetilde{\bz}_{j+p} }{n \widetilde{\tau}_{j+p}^4}
			\end{pmatrix}, 
			\quad \bV_{12} = \bV_{21}^{\top} = 
			\begin{pmatrix}
				\frac{ \widetilde{\bz}_j^T \bD \widetilde{\bz}_l }{n \widetilde{\tau}_j^2 \widetilde{\tau}_l^2} 
				& \frac{ \widetilde{\bz}_j^T \bD \widetilde{\bz}_{l+p} }{n \widetilde{\tau}_{j}^2 \widetilde{\tau}_{l+p}^2}  \\
				\frac{ \widetilde{\bz}_{j+p}^T \bD \widetilde{\bz}_{l} }{n \widetilde{\tau}_{j+p}^2 \widetilde{\tau}_{l}^2} 
				& \frac{ \widetilde{\bz}_{j+p}^T \bD \widetilde{\bz}_{l+p} }{n \widetilde{\tau}_{j+p}^2 \widetilde{\tau}_{l+p}^2}
			\end{pmatrix},  \\
			& \bV_{22} = 
			\begin{pmatrix} 
				\frac{ \widetilde{\bz}_l^T \bD \widetilde{\bz}_l }{n \widetilde{\tau}_l^4 } 
				& \frac{ \widetilde{\bz}_l^T \bD \widetilde{\bz}_{l+p} }{n \widetilde{\tau}_{l}^2 \widetilde{\tau}_{l+p}^2}  \\
				\frac{ \widetilde{\bz}_l^T \bD \widetilde{\bz}_{l+p} }{n \widetilde{\tau}_{l}^2 \widetilde{\tau}_{l+p}^2}
				& \frac{ \widetilde{\bz}_{j+p}^T \bD \widetilde{\bz}_{l+p} }{n \widetilde{\tau}_{l+p}^4}
			\end{pmatrix}.
		\end{split}
	\end{equation*} 
	Let us define the event 
	\begin{equation*}
		\mathcal{E} := \bigg\{ \max_{1 \leq j, l \leq 2p} \Big| n^{-1} \widetilde{\bz}_j \bD \widetilde{\bz}_l - \frac{\bOmega_{j, l}}{\bOmega_{j, j} \bOmega_{l, l}} \Big| \leq C \sqrt{\frac{s \log p}{ n}} \bigg\}.
	\end{equation*}
 By Condition \ref{cond-GLM-01}, we see that $\mathbb{P} (\mathcal{E}) \geq 1 - o(p^{-3})$. 
	
 Let $(Z_1, Z_2, Z_3, Z_4)^T \stackrel{d}{\sim} N({\bf 0}, \bV)$.
	Given $\widetilde\bX^{\augg}$ and event $\mathcal{E}$, it follows from the rate of convergence (i.e., the moderate deviation theorem) for multivariate normal approximate (e.g. Theorem 1 in \cite{saulis1992probabilities}) that for any $1 \leq j \neq l \leq 2 p$, 
	\begin{equation} \label{multi-norm-MD}
		\bigg| \frac{ \mathbb{P} (\xi_{j+p} \geq 0, \xi_j - \xi_{j+p} \geq t, \xi_{l+p} \geq 0, \xi_l - \xi_{l+p} \geq  t) } { \mathbb{P} (Z_2 \geq 0, Z_{1} - Z_2 \geq t, Z_4 \geq 0, Z_3 - Z_4 \geq t) } - 1 \bigg| \leq C \frac{1 + t^3}{ \sqrt n } 
	\end{equation}
	uniformly for $t \in [0, C \sqrt{\log p}]$ when $\log p = o(n^{1/3})$. Noting that $\mathbb{P} (|\xi_j| - |\xi_{j+p}| \geq t, |\xi_{l} - \xi_{l+p}| \geq t)$ can be decomposed into 16 probabilities that are similar to the numerator in \eqref{multi-norm-MD}, we can deduce that for any $1 \leq j \neq l \leq 2p $,
	\begin{equation} \label{multi-norm-MD-overall}
		\bigg|\frac{ \mathbb{P} (|\xi_j| - |\xi_{j+p}| \geq t, |\xi_{l} - \xi_{l+p}| \geq t) } {\mathbb{P} (|Z_1| - |Z_2| \geq t, |Z_3| - |Z_4| \geq t) } - 1 \bigg| \leq C\frac{1 + t^3}{\sqrt n}
	\end{equation}
	uniformly for $t \in [0, C \sqrt{\log p}]$ when $\log p = o(n^{1/3})$. Analogously, we can show that for any $1 \leq j \neq 2p$,
	\begin{equation} \label{multi-norm-MD-partial}
		\bigg|\frac{ \mathbb{P} (|\xi_j| - |\xi_{j+p}| \geq t) } {\mathbb{P} (|Z_1| - |Z_2| \geq t) } - 1 \bigg| \leq C\frac{1 + t^3}{\sqrt n}
	\end{equation}
	uniformly for $t \in [0, C \sqrt{\log p}]$ when $\log p = o(n^{1/3})$. Therefore, following exactly the same procedure for proving Lemma \ref{pf-thm5-lemma-3}, we can establish Lemma \ref{pf-GLM-lemma-3}. To avoid redundancy, we omit the proof details here.

\subsection{Proof of Lemma \ref{pf-GLM-lemma-4}} \label{pf-of-lemma-glm-4}
By the moderate deviation result in \eqref{multi-norm-MD-partial}, the probability $\mathbb{P} (\widetilde{W}_j \geq t) \approx \mathbb{P} (|\xi_j| - |\xi_{j+p}| \geq t)$ can be approximated by the probability $\mathbb{P } (|Z_1| - |Z_2| \geq t) $ of normal distribution with controlled relative rate of convergence as $t \leq C \sqrt{\log p}$. By the same technique for proving Lemma \ref{pf-thm5-lemma-4}, but just with slightly different definitions that $\delta_n = \frac{s^{3/2} \log p}{ \sqrt n}$ and $b_n = C( \Delta_n s \sqrt{\frac{\log p}{n}} + \frac{s^{3/2} \log p}{n} )$, we can establish the desired results in Lemma \ref{pf-GLM-lemma-4}. To avoid redundancy, we omit the proof details here.
 
\end{document}